\documentclass[reqno,oneside]{amsart}
\usepackage{etex} 
\usepackage[margin=3cm]{geometry}
\usepackage{amsmath}
\usepackage{amssymb}
\usepackage{tikz}
\usetikzlibrary{calc}
\usepackage{amsbsy}
\usepackage{amsfonts}
\usepackage{dsfont}
\usepackage[utf8]{inputenc}
\usepackage{enumerate} 

\usepackage{listings}

\usepackage[ruled,vlined]{algorithm2e}
\usepackage[dvips]{epsfig}
\usepackage{colortbl}
\usepackage{algcompatible}
\usepackage{tabularx}
\usepackage{afterpage}
\usepackage{lscape}
\usepackage{multirow}
\usepackage{multicol}
\usepackage{arydshln}  
\usepackage{cancel}
\usepackage{makecell}
\usepackage{booktabs}
\usepackage{hhline}

\usepackage{amsfonts,amscd}
\usepackage[T1]{fontenc}
\usepackage{lmodern}
\usepackage{nicefrac}
\usepackage{graphics,caption}
\usepackage[labelformat=simple]{subcaption}

\usepackage{arydshln}
\usepackage{lipsum}
\usepackage{amsfonts}
\usepackage{graphicx}
\usepackage{epstopdf}
\usepackage{hyperref}
\usepackage{mathtools}
\usepackage{autonum}
\usepackage[numbers,sort&compress]{natbib}
\usepackage{stmaryrd,xparse}
\usepackage{doi}
\usepackage{listings}
\usepackage{pgf}
\usepackage{tikz}
\usepackage{tikz-uml}
\usetikzlibrary[arrows,positioning,patterns,calc]
\usepackage{parcolumns}
\usepackage{lscape}
\usepackage{rotating}

\include{definitions}

\newtheorem{theorem}{Theorem}[section]

\newtheorem{definition}[theorem]{Definition}

\newtheorem{example}[theorem]{Example}

\def\FEMPAR{{\texttt{FEMPAR}}}
\def\R{\mathbb{R}}
\def\N{\mathbb{N}}
\def\Bit{\mathbb{B}}

\def\trialsph{\trialsp_h}
\def\testsph{\testsp_h}
\def\georef{{\hat K}}
\def\geophy{K}
\def\geomap{\boldsymbol{\Phi}}
\def\shapetest#1{\phi^{#1}}
\def\shapetrial#1{\psi^{#1}}
\def\shapetestref#1{\hat{\phi}^{#1}}

\def\fsp{\mathcal{X}}
\def\fesp{\fsp_h}
\def\lfesp{\mathcal{V}}
\def\reffesp{\hat{\mathcal{V}}}
\def\funmap{\Psi}

\def\momentsref{{\hat \moments}}

\def\moment#1{\sigma_{#1}}
\def\moments{\Sigma}

\def\triang{\mathcal{T}_h}

\def\Hdiv{H({\rm div},\Omega)}
\def\Hcurl{H(\boldsymbol{\rm curl},\Omega)}

\def\fematrix{\mathbf{A}}
\def\ferhs{\mathbf{f}}
\def\fesol{\mathbf{u}}
\def\mid{\boldsymbol{\alpha}}

\DeclareTextFontCommand{\mytexttt}{\ttfamily\hyphenchar\font=45\relax}

\def\compijk{{\boldsymbol{\alpha}}}

\usepackage{acronym}
\acrodef{FE}[FE]{finite element}
\acrodef{DOF}[DOF]{degree of freedom}
\acrodefplural{DOF}[DOFs]{degrees of freedom}
\acrodef{TBP}[TBP]{type-bound procedure}
\acrodef{PDE}[PDE]{partial differential equation}
\acrodef{DD}[DD]{domain decomposition}
\acrodef{OO}[OO]{object-oriented}
\acrodef{DG}[DG]{discontinuous galerkin}
\acrodef{MPI}[MPI]{message passing interface}
\acrodef{HPC}[HPC]{high performance computing}

\begin{document}

\title[\FEMPAR{}: an object-oriented finite element framework]{{\mytexttt{FEMPAR}}: An object-oriented parallel finite element framework}

\author[S. Badia]{Santiago Badia$^{1,2}$}

\author[A. F. Mart\'in]{Alberto F. Mart\'in$^{1,2}$}

\author[J. Principe]{Javier Principe$^{2,3}$}

\address{$^{1}$ Department of Civil and Environmental Engineering. Universitat Polit\`ecnica de Catalunya, Jordi Girona 1-3, Edifici C1, 08034, Barcelona, Spain.}
\address{$^{2}$ CIMNE – Centre Internacional de M\`etodes Num\`erics en 
Enginyeria, Parc Mediterrani de la Tecnologia, UPC, Esteve Terradas 5, 08860 
Castelldefels, Spain.}
\address{$^{3}$ Department of Fluid Mechanics. Universitat Polit\`ecnica de Catalunya, Eduard Maristany, 10-14, 08019, Barcelona, Spain.}
 
\thanks{The authors want to thank Jes\'us Bonilla, Oriol Colom\'es, Eric Neiva, Hieu Nguyen, Marc Olm, V\'ictor Sande, and Francesc Verdugo (in alphabetical order) for their strong commitment to the \FEMPAR{} project, the implementation of some of the software described in this work, and their thorough review of preliminary versions of this document. The resources needed to develop a scientific library library like \FEMPAR{} would have not been feasible without excellent research funding. In this sense, SB sincerely thanks the support of the European Research Council through the Starting Grant No. 258443 - \emph{COMFUS: Computational Methods for Fusion Technology} under the  the FP7 Program and the two related Proof of Concept Grant No. 640957 - \emph{FEXFEM: On a free open source extreme scale finite element software} and Proof of Concept Grant No. 737439 - \emph{NuWaSim: On a Nuclear Waste Deep Repository Simulator} under the H2020 Program. SB gratefully acknowledges the support received from the Catalan Government through the ICREA Acad\`emia Research Program.  E-mails: {\tt sbadia@cimne.upc.edu} (SB), {\tt amartin@cimne.upc.edu} (AM), {\tt principe@cimne.upc.edu} (JP)}

\date{\today}

\begin{abstract}
  \FEMPAR{} is an open source object oriented Fortran200X scientific software library for the high-performance scalable simulation of complex multiphysics problems governed by partial differential equations at large scales, by exploiting state-of-the-art supercomputing resources. It is a highly modularized, flexible, and extensible library, that provides a set of modules that can be combined to carry out the different steps of the simulation pipeline. \FEMPAR{} includes a rich set of algorithms for the discretization step, namely (arbitrary-order) grad, div, and curl-conforming finite element methods, discontinuous Galerkin methods, B-splines, and unfitted finite element techniques on cut cells, combined with $h$-adaptivity. The linear solver module relies on state-of-the-art bulk-asynchronous implementations of multilevel domain decomposition solvers for the different discretization alternatives and block-preconditioning techniques for multiphysics problems. \FEMPAR{} is a framework that provides users with out-of-the-box state-of-the-art discretization techniques and highly scalable solvers for the simulation of complex applications, hiding the dramatic complexity of the underlying algorithms. But it is also a framework for researchers that want to experience with new algorithms and solvers, by providing a highly extensible framework. In this work, the first one in a series of articles about \FEMPAR{}, we provide a detailed introduction to the software abstractions used in the discretization module  and the related geometrical module. We also provide some ingredients about the assembly of linear systems arising from finite element discretizations, but the software design of complex scalable multilevel solvers is postponed to a subsequent work.
\end{abstract}

\maketitle



\tableofcontents

\section{Introduction} \label{sec:introduction}


Even though the origins of the \ac{FE} method trace back to the 50s,
the field has drastically evolved during the last six decades, leading
to increasingly complex algorithms to improve accuracy, stability, and
performance. The use of the $p$-version of the \ac{FE} method and its
exponential convergence makes high-order approximations an excellent
option in many applications \cite{guo_h-p_1986}. Adaptive mesh
refinement driven by \emph{a posteriori} error estimates, i.e.,
$h$-adaptivity, is an essential ingredient to reduce computational
cost in an automatic way \cite{ainsworth_posteriori_2011}. For smooth
solutions, $p$-adaptivity or hybrid $hp$-adaptivity can further reduce
computational cost for a target level of accuracy
\cite{melenk_residual-based_2001}. Originally, \ac{FE} methods were
restricted to nodal Lagrangian bases for structural problems. The
extension of \ac{FE} methods to other applications, like porous media
flow or electromagnetism, motivated the design of more complex bases
and require different mappings from the reference to the physical
space, complicating the implementation of these techniques in standard
\ac{FE} codes. Saddle-point problems also require particular mixed
\ac{FE} discretizations for stability purposes
\cite{nedelec_mixed_nodate,brezzi_mixed_1991}. More recently, novel
\ac{FE} formulations have been proposed within the frame of exterior
calculus, e.g., for mixed linear elasticity problems
\cite{arnold_finite_2006}. Physics-compatible discretization are also
gaining attention, e.g., in the field of incompressible fluid
mechanics. Divergence-free mixed \acp{FE} satisfy mass conservation up
to machine precision, but their implementation is certainly
challenging \cite{neilan_stokes_2015}. During the last decade, a huge
part of the computational mechanics community has embraced
isogeometric analysis techniques \cite{hughes_isogeometric_2005}, in
which the discretization spaces are defined in terms of NURBS (or
simply splines), leading to smoother global spaces. In the opposite
direction, \ac{DG} methods have also been actively
developed, and novel approaches, like hybridizable \ac{DG} and
Petrov-Galerkin \ac{DG} methods, have been proposed
\cite{cockburn_unified_2009,demkowicz_class_2010}. As the
discretization methods become more and more complex, the efficient
implementation of these techniques is more complicated. It also poses
a challenge in the design of scientific software libraries, which
should be extensible and provide a framework for the (easy)
implementation of novel techniques, to be resilient to new algorithmic
trends.


The hardware in which scientific codes run evolves even faster.
During 40 years, core performance has been steadily increasing, as
predicted by Moore's law. In some years, supercomputers will reach 1
exaflop/s, a dramatic improvement in computational power that
will not only affect the extreme scale machines but radically
transform the whole range of platforms, from desktops to \ac{HPC}
clouds. The ability to efficiently exploit the forthcoming 100x boost
of computational performance will have a tremendous impact on
scientific discoveries/economic benefits based on computational
science, reaching almost every field of research. However, all the
foreseen exascale growth in computational power will be delivered by
increasing hardware parallelism (in distinct forms), and the efficient
exploitation of these resources will not be a simple task. \ac{HPC}
architectures will combine general-purpose fat cores, fine-grain
many-cores accelerators (GPUs, DSPs, FPGAs, Intel MIC, etc.), and
multiple-level disruptive-technology memories, with high
non-uniformity as common denominator~\cite{ang_abstract_2014}. This
(inevitable) trend challenges algorithm/software design. Traditional
bulk-synchronous \ac{MPI} approaches are likely to face significant
performance obstacles. Significant progress is already being made by
\ac{MPI}+X~\cite{wropp_2015_mpi_x_keynote} (with X=OpenMP, CUDA, OpenCL,
OmpSs, Kokkos, etc.) hybrid execution models. Going a step further,
asynchronous many-task execution models (e.g.,
Charm++\cite{kale_charm++:_1993},
Legion~\cite{Bauer:2012:LEL:2388996.2389086}, or HPX~\cite{Kaiser:2014:HTB:2676870.2676883}) and their supporting
run-time systems hold great promise~\cite{pi_asc_2015}. 


Traditionally, researchers in the field of scientific computing used
to develop codes with a very reduced number of developers, e.g., a
university department, and a limited life span. The software
engineering behind scientific codes was poor. Codes were rigid and
non-extensible, and developed for a target application and a specific
numerical method. However, the increasing levels of complexity both in
terms of algorithms and hardware make the development of scientific
software that can efficiently run state-of-the-art numerical
algorithms on \ac{HPC} resources a real challenge. Considering to start
from scratch a project of this kind has an ever increasing level of
complexity. Furthermore, due to the huge resources required to carry
out such a project, it is natural to develop a framework that will be
resilient to new algorithmic and hardware trends, in order to
maximize life time, and to be applicable to a broad range of
applications. In this sense, \ac{OO} programming, which
provides modularity of codes and data-hiding, is the key for the
software design of flexible and scalable (in terms of developers)
projects.


There is a number of open source \ac{OO} \ac{FE} libraries available through the
Internet, e.g., deal.II
\cite{bangerth_deal.ii-general-purpose_2007,bangerth_textttdeal.ii_2016},
FEniCS \cite{alnaes_fenics_2015}, GRINS \cite{bauman_grins:_2016},
Nektar++ \cite{cantwell_nektar++:_2015}, MOOSE \cite{_moose_????},
MFEM \cite{_mfem_????}, FreeFem++ \cite{hecht_new_2012}, and DUNE
\cite{dedner_construction_2012}. In general, these libraries aim to
provide all the machinery required to simulate complex problems
governed by \acp{PDE} using \ac{FE} techniques. In any case, every
library has its main goal and distinctive features. Some libraries,
like FreeFem++ or FEniCS, have extremely simple user
interfaces. FEniCS has its own domain specific language for weak forms
to automatically generate the corresponding \ac{FE} code
(preventing $p$-adaptivity) and includes a collection of Python wrappers
to provide user-friendly access to the services of the library. Other sophisticated libraries like
deal.II or DUNE have a slightly more demanding learning curve. In general, parallel
adaptivity is at most partially supported; as far as we know, none of
the libraries above have support for parallel $hp$-adaptivity, unless
\ac{DG} methods are being used. Some libraries are restricted to a
particular cell topology, e.g., deal.II is limited to
hexahedral/quadrilateral (n-cubes) meshes, while FEniCS only supports
simulations on triangular/tetrahedral (n-simplices) meshes.


In general, these libraries provide modules for some of the different
steps in the simulation pipeline, which involves the set-up of the
mesh, the construction of the \ac{FE} space, the integration and
assembly of the weak form, the solution of the resulting linear
system, and the visualization of the computed solution. The solution
of the linear system is clearly segregated from the discretization
step in all the scientific software libraries described above (for
parallel computations); the linear system is transferred to a
general-purpose sparse linear algebra library, mainly PETSc
\cite{balay_petsc_2016,balay_petsc_2016-1,balay_efficient_1997}, Hypre
\cite{falgout_hypre:_2002}, and Trilinos
\cite{heroux_overview_2005,_trilinos_????}.  As a result, the coupling
between the discretization step and the linear solver step is somehow
weak, since they rely on general purpose solvers, which usually
involve simple interfaces. The strong point of these general purpose
numerical linear algebra libraries is to be problem-independent, but
it also limits their performance for specific applications, since they
cannot fully exploit the underlying properties of the \ac{PDE}
operator and the numerical discretization.\footnote{A paradigmatic
  example is the design of scalable solvers for the discretization of
  the Maxwell equations using edge elements, which involve the
  discretization of additional operators (discrete gradients) and
  changes of basis at the reference \ac{FE} level
  \cite{toselli_dual-primal_2006}.} This segregation has a clear
impact on the type of methods to be used. This black-box approach to
general-purpose linear solvers has favoured the use of algebraic
multigrid methods, the \emph{de facto} linear solver
\cite{falgout_hypre:_2002}. On the other hand, geometric multigrid
methods and \ac{DD} methods, which are very specific to mesh-based
\ac{PDE} solvers, are not common, even though they can be superior to
algebraic methods in many cases. A geometric multigrid method that
exploits the $hp$-adaptive structure of the \ac{FE} space is included
in deal.II, but it can only be used in the serial case. In parallel
scenarios, \ac{DD} methods have certainly evolved during the last
decade. Modern \ac{DD} methods do not (necessarily) rely on a static
condensation of the internal variables, which requires sparse direct
methods for the local subdomain problems. Instead, \emph{inexact}
solvers can be used, e.g., multigrid methods, and linear complexity
\ac{DD} preconditioners can be defined (see
\cite{dohrmann_approximate_2007,badia_scalability_2015}). The
definition of two-level \ac{DD} methods resembles the one of \ac{FE}
methods, by exchanging the \ac{FE} and subdomain concepts, and their
definition is strongly related to the one of multiscale \acp{FE}
\cite{efendiev_multiscale_2009}. Furthermore, multilevel extensions
can be naturally defined. In short, state-of-the-art multilevel
\ac{DD} methods can be understood (in their \emph{inexact} version) as
a non-conforming multigrid method. Even though the mathematical theory
of the \ac{DD} methods is very sound, high performance implementations
are quite recent (see
\cite{klawonn_toward_2015,badia_multilevel_2016,zampini_pcbddc:_2016}). On
the other hand, we are not aware of any general purpose \ac{FE} code
that integrates a \ac{DD} algorithm in the solution workflow. \ac{DD} methods require sub-assembled matrices to be used, and
are not supported by the majority of the existing advanced \ac{OO} \ac{FE}
libraries. Analogously, the use of block-preconditioning is in general
poorly supported, because it involves the discretization of additional
operators to define the approximated Schur complement, and the
corresponding block-based assembly of matrices. 

On the other hand, based on the supercomputing trends, the segregation
between time discretization, linearization, space discretization, and
linear system solve, will progressively blur. As an example, nonlinear
preconditioning and parallel-in-time solvers are two natural ways to
attain the higher levels of concurrency of the forthcoming exascale
supercomputers \cite{badia_space-time_2017,klawonn_toward_2015}. These
facts will complicate even more the rigid workflow of current advanced
\ac{FE} libraries. In this sense, current efforts in PETSc to
provide nonlinear preconditioning interfaces can be found in
\cite{brune_composing_2015}, relying on call-back functions, and the
XBraid solver \cite{falgout_parallel_2014} aims to provide time-parallelism in a
non-intrusive way.



\section{The \FEMPAR{} project} \label{sec:fempar_project}

In this work, we present \FEMPAR{}, an \ac{OO} \ac{FE} framework for the solution of \acp{PDE}, designed from inception to be
highly scalable on supercomputers and to easily handle complex
multiphysics problems.  The first public release of \FEMPAR{}
has almost 300K lines of code written in (mostly) \ac{OO} Fortran and makes
intensive use of the features defined in the 2003 and 2008
standards of the language. The source code that is complementary to this work
corresponds to the first public release of \FEMPAR{}, i.e., version 1.0.0. 
It is available at a git repository~\cite{FEMPAR}. In particular, the first public release 
was assigned the git tag \mytexttt{FEMPAR-1.0.0}, in accordance with the ``Semantic Versioning'' system.\footnote{Available at \href{http://semver.org/}{http://semver.org/}.}

\FEMPAR{} is very rich in terms of \ac{FE}
technology. In particular, it includes not only Lagrangian \acp{FE}, but also curl- and
div-conforming ones, e.g., N\'ed\'elec (edge) and Raviart-Thomas \acp{FE}. The
library supports n-cube and n-simplex  meshes, and arbitrary
high-order bases for all the \acp{FE} included. Continuous and
discontinuous spaces can be used, providing all the machinery for the
integration of \ac{DG} facet (i.e., edges in 2D and faces in 3D) terms. Recently, in a
beta version of the code, B-splines have also been added, together
with the support for cut cell methods (using XFEM-type techniques) and
$hp$-adaptivity, but we will not discuss these developments for the
sake of brevity.

Moreover, \FEMPAR{} has been developed with the aim to provide a
framework that will allow developers to implement complex techniques
that are not well-suited in the traditional segregated workflow
commented above. \FEMPAR{} also provides a highly scalable built-in
numerical linear algebra module based on state-of-the-art domain
decomposition solvers. \FEMPAR{} can provide partially assembled
matrices, required for \ac{DD} solvers; the multilevel BDDC solver in
\FEMPAR{} has scaled up to almost half a million cores and 1.75
million \ac{MPI} tasks (subdomains) in the JUQUEEN Supercomputer
\cite{badia_scalability_2015,badia_multilevel_2016}. It includes an
abstract framework to construct applications and preconditioners based
on multilevel nonoverlapping partitions. Even though every block
within the library preserves modularity, the interface between
discretization and numerical linear algebra modules within \FEMPAR{}
is very rich and focused on \ac{PDE}-based linear systems. In the path to
the exascale, \FEMPAR{} has been designed to permit an asynchronous
implementation of multilevel methods, both in terms of multiphysics
\acp{FE} and multilevel solvers, which have been exploited,
e.g., in \cite{badia_multilevel_2016}. It is a unique feature that is
not available in other similar libraries. The library also allows the
user to define blocks in multiphysics applications, that can be used
to easily implement complex block preconditioners
\cite{elman_finite_2005,badia_block_2014,cyr_teko:_2016}. All these
blocks are very customizable, which has already been used to develop 
scalable \ac{DD} solvers for electromagnetics problems and block
preconditioners for multiphysics problems, e.g.,
magnetohydrodynamics \cite{badia_block_2014}. These distinctive features of \FEMPAR{},
however, are not discussed in this article but in a forthcoming
one. A general discussion of the main ingredients of our
implementation of the discretization step using FE-like approximations
is first necessary, which is the purpose of this work.
 
\FEMPAR{} has already been successfully used in a wide set of applications by the authors of the library: simulation of turbulent flows and stabilized \ac{FE} methods \cite{colomes_assessment_2015,colomes_segregated_2016,colomes_mixed_2016,colomes_segregated_2017}, magnetohydrodynamics \cite{badia_unconditionally_2013,badia_unconditionally_2013-1,planas_approximation_2011,smolentsev_approach_2015,badia_analysis_2015}, monotonic \acp{FE} \cite{badia_discrete_2015,badia_monotonicity-preserving_2014,hierro_shock_2016,badia_monotonicity-preserving_2017,badia_differentiable_2017}, unfitted \acp{FE} and embedded boundary methods \cite{badia_robust_2017}, and additive manufacturing simulations \cite{badia_am}. It has also been used for the highly efficient implementation of \ac{DD} solvers \cite{badia_implementation_2013,badia_scalability_2015,badia_enhanced_2013,badia_multilevel_2016,badia_balancing_2016,badia_physics-based_2016,badia_highly_2014,badia_space-time_2017} and block preconditioning techniques \cite{badia_block_2014}.


This work is more than an overview article with the main features of
the library. It is a detailed description of the software abstractions
being used within \FEMPAR{} to develop an efficient, modular, and
extensible implementation of \ac{FE} methods and supporting modules in
a broad sense. To this end, we enrich the discussion with code
snippets that describe data structures, bindings, and examples of use.\footnote{The code snippets are written in advanced \ac{OO} Fortran
  200X~\cite{adams_fortran_2009}. There is a close relationship between these language
  features and those available in the C++ language
  \cite{rouson_scientific_2011} and we established some code style
  rules to emphasize it. In particular, Fortran modules in \FEMPAR{}
  are always named with the suffix \mytexttt{\_names}, to indicate the
  analogy with namespaces in C++. Derived types, analog to C structs
  or C++ classes, are always named with \mytexttt{\_t} to distinguish
  them from instances. However it should be kept in mind that, whereas
  structs in C++ are passive data containers and classes are used to
  carry also methods, Fortran derived data types are used in both
  cases since the introduction in the 2003 standard of the so called
  \acp{TBP}.}  This document is intended to be used as a guide for new
\FEMPAR{} \emph{developers} that want to get familiarized with its
software abstractions. But it can also be a useful tool for developers
of FE codes that want to learn how to implement \ac{FE} methods in an
advanced \ac{OO} framework. In any case, due to the size of the
library itself, many details cannot be exposed, to keep a reasonable
article length. The article can be read in different ways, since it is
not necessary to fully understand all the preceding sections to grasp the
main ideas of a section. For instance, the section about the abstract
implementation of polytopes in arbitrary dimensions and its related
algorithms is quite technical and a reader that is not particularly
interested in the internal design of this type and its bindings
implementations can skip it. Experienced \ac{FE} researchers can skip
the short section with the basics of \ac{FE} methods, and only look at
this one (if needed) when referred in subsequent sections.

The article is organized as follows. In Sect. \ref{sec:fe_meth} we
present a concise mathematical description of the \ac{FE} framework. The
main mathematical abstractions are expressed in software by means of a set of derived data types 
and their associated \acp{TBP}, which are described in subsequent sections.  In particular, 
the main software abstractions in \FEMPAR{} and their roles in the solution of the problem are:
\begin{itemize}
\item The polytope, which describes a set of admissible geometries and permits the
  automatic, dimension-independent generation of reference cells and
  structured domains. The mathematics underlying the polytope are
  presented in Sect.~\ref{sec:polytope_topology}, while its software implementation in 
  Sect.~\ref{sec:polytope_implementation}.
\item The polynomial abstraction and related data types, which are presented in
  Sect.~\ref{subsec:polynomial_spaces_construction} and Sect.~\ref{sec:polynomial}, respectively. 
  These sections describe how shape functions
  bases can be generated for arbitrary orders and for n-cube and n-simplex
  topologies.
\item The reference \ac{FE} in Sect.~\ref{sec:reference_fe}, which
  describes the reference cell and defines a set of basis functions
  and degrees of freedom on each cell.
\item The triangulation in Sect.~\ref{sec:triangulation}, which
  represents a discrete approximation of the physical domain
  $\Omega$.
\item A set of tools required to perform numerical integration
  (e.g., quadratures and geometrical maps) produced by the reference \ac{FE}
  and described in Sect.~\ref{sec:cell_integration} for cell
  integrals and in Sect.~\ref{sec:face_integration} for facet
  integrals.
\item The \ac{FE} space described in Sect.~\ref{sec:fe_space}, built from
  a triangulation and a set of reference \acp{FE}, which represents a global
  space of \ac{FE} functions.
\item The discrete integration, an abstract class to be extended by
  the user to define an affine \ac{FE} operator, which describes the
  numerical integration of the weak form of the problem to be solved,
  described in Sect.~\ref{sec:disc_int}.
\item The linear (affine) operator in Sect.~\ref{sec:building_fe_systems},
  whose root is the solution of the problem at hand, constructed using
  the \ac{FE} space and a discrete integration.
\item An example of a user driver in Sect.~\ref{sec:driver}, in which
  the different ingredients previously described are used to simulate a
  problem governed by \acp{PDE}, the Stokes system.
\end{itemize} 

\begin{figure}[tbp]
\tikzstyle{abstract}=[rectangle, draw=black, rounded corners, fill=blue!30,
        text centered, anchor=north, text=black, minimum height=1cm]
\tikzstyle{myarrow}=[-stealth]
        
\begin{center}
\begin{tikzpicture}[node distance=0.75cm]
    \node (Triangulation) [abstract, text width=3cm] {Triangulation $\mathcal{T}_h$};
    \node (AuxNode01)     [text width=4cm, below=of Triangulation] {};
    \node (ReferenceFE)   [abstract, text width=3cm, below=of AuxNode01] {Reference \ac{FE} $(\georef,\reffesp,\momentsref)$};
    \node (FEspace)       [abstract, right=of AuxNode01] {\ac{FE} space $\mathcal{X}_h$};
    \node (DiscInt)       [abstract, right=of FEspace]  {Discrete integration
      $ \, a_h(\cdot,v_h), \, \ell_h(v_h)$};
    \node (AuxNode02)     [text width=6cm, right=of ReferenceFE] {};
    \node (FEope)         [abstract, below=of AuxNode02] {\ac{FE} affine operator $\fematrix \fesol - \ferhs$};
    \node (AuxNode03)     [text width=1cm, above=of FEspace] {};
    \node (FEfun)         [abstract, right=of AuxNode03] {\ac{FE} function $u_h \in \mathcal{X}_h$};

    \draw[myarrow] (Triangulation.east) -- ++(0.5,0) |- (FEspace.west);
    \draw[myarrow] (ReferenceFE.east) -- ++(0.5,0) |- (FEspace.west);
    \draw[myarrow, dashed] (Triangulation.south)  -- (ReferenceFE.north);

    \draw[myarrow] (FEspace.south) -- ++(0,-0.5) -| (FEope.north);
    \draw[myarrow] (DiscInt.south) -- ++(0,-0.5) -| (FEope.north);

    \draw[myarrow] (FEspace.north)  --  ++(0,0.5) |- (FEfun.west);

\end{tikzpicture}
\end{center}

\caption{Main software abstractions in \FEMPAR{} and some of their relationships.}
\label{fig:main-structures}
\end{figure}
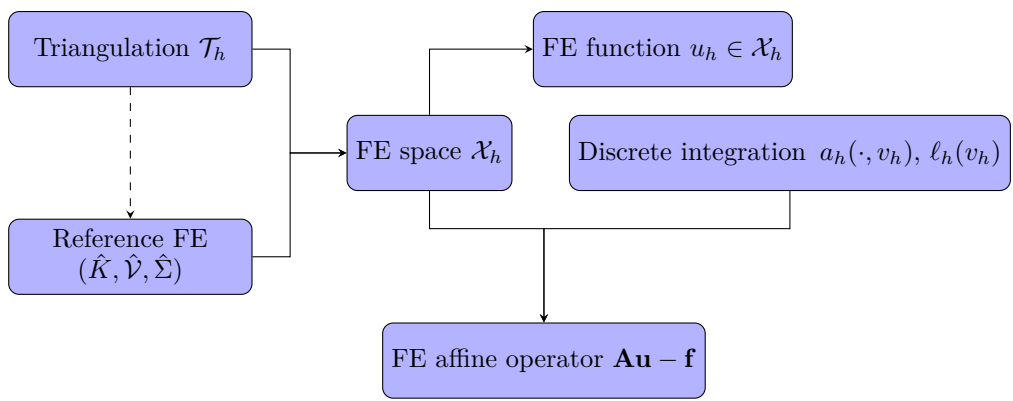

A (very simplified) graphical overview of the main software abstractions in \FEMPAR{} and some of 
their relationships is shown in Fig.~\ref{fig:main-structures}.

\def\naturals{\mathbb{N}}

\def\nodes{\mathcal{N}}

\def\R{\mathbb{R}}
\def\PDEop{{\mathit{A}}}
\def\bcop{{\mathit{B}}}
\def\force{g}
\def\trialsp{\mathcal{X}}
\def\testsp{\mathcal{Y}}
\def\blform#1#2{{a}(#1,#2)}
\def\blformh#1#2{{a_h}(#1,#2)}
\def\blformlocal#1#2#3{{a}_{#1}(#2,#3)}
\def\rhsform#1{\ell( #1) }
\def\rhsformh#1{\ell_h( #1) }
\def\domain{\Omega}
\def\boundary{{\Gamma}}
\def\Dirichlet{{\Gamma_{\rm D}}}
\def\Neumann{{\Gamma_{\rm N}}}
\def\triang{\mathcal{T}_h}
\def\x{\boldsymbol{x}}
\def\xh{\hat{\x}}
\def\xphy{\x}
\def\xref{\hat \x}

\def\cinf{\mathcal{D}}
\def\cinfdom{\mathcal{D}(\domain)}
\def\cinfbou{\mathcal{D}(\Neumann)}
\def\grad{{\boldsymbol{\nabla}}}
\def\strain{{\boldsymbol{\epsilon}}}
\def\dist{\varphi}

\def\normal{\boldsymbol{n}}
\def\half{\frac{1}{2}}

\def\dx{{\rm d}\x}
\def\ddom{{\rm d}\domain}
\def\dbou{{\rm d}\Gamma}

\def\integrandfe{\boldsymbol{\mathcal{F}}}
\def\quadrature{\mathrm{Q}}
\def\dxphy{{\rm d} \geophy}
\def\dxref{{\rm d} \georef}
\def\bsJ{\boldsymbol{J}}
\def\jacobian{\boldsymbol{J}_K}
\def\invjacob{\jacobian^{-1}}

\def\jacobf{\boldsymbol{J}_F}

\def\xref{\hat{\x}}
\def\gp{\rm gp}

\def\nodeset{\mathcal{S}}
\def\node{{\boldsymbol{s}}}
\def\lagpol{\ell}
\def\lagbasis{\mathcal{L}}

\def\order{{\boldsymbol{k}}}

\def\polsp{{\mathcal{Q}}}
\def\polspr{{\mathcal{P}}}

\def\ones{\boldsymbol{1}}

\def\ubs{\boldsymbol{u}}
\def\wbs{\boldsymbol{w}}
\def\vbs{\boldsymbol{v}}
\def\sbs{\boldsymbol{s}}
\def\ebs{\boldsymbol{e}}
\def\qbs{\boldsymbol{q}}
\def\fbs{\boldsymbol{f}}

\def\ddom{{\rm d}\domain}
\def\dbou{{\rm d}\Gamma}
\def\bs#1{\boldsymbol{#1}}

\def\jumpl{\lbrack\!\lbrack}
\def\jumpr{\rbrack\!\rbrack}
\def\jump#1{\jumpl #1 \jumpr}
\def\mean#1{\{\! \!\{ #1\}\! \!\}}
\def\faceref{{\hat F}}
\def\facephy{F}
\def\dxfacephy{{\rm d} \facephy}
\def\dxfaceref{{\rm d} \faceref}
\def\facejacobian{\boldsymbol{J}_F}
\def\jacobianplus{\boldsymbol{J}_{K^{+}}}
\def\jacobianminus{\boldsymbol{J}_{K^{-}}}

\section{The \ac{FE} framework}
\label{sec:fe_meth}

In this section, we briefly introduce all the mathematical
abstractions behind the \ac{FE} method for the discretization of
problems governed by \acp{PDE}. For a more detailed exposition of the
topics, we refer to
\cite{ern_theory_2004,brenner_mathematical_2010,quarteroni_numerical_2014,monk_finite_2003}.
The \acp{FE} described below (and many other not covered herein) can
be formulated and analyzed using the \emph{finite element exterior
  calculus} framework \cite{arnold_finite_2006}, which makes use of
exterior algebra and exterior calculus concepts. In this framework,
one can define \acp{FE}, e.g., div and curl-conforming ones, in
arbitrary space dimensions, using the concept of differential
$k$-forms. However, we have decided not to use such presentation of
\ac{FE} methods to simplify the exposition for readers not familiar
with these abstractions.

\subsection{The boundary value problem in weak form} \label{subsec:weak_form}

We are interested in problems governed by \acp{PDE} posed in a physical domain
$\domain \subset \mathbb{R}^d$ with boundary
$\boundary \doteq \partial \Omega$.  In practice $d=2,3$ but we are
also interested in $d>3$ for some particular applications (see
Sect.~\ref{sec:polytope_topology}).   Let us
consider a differential operator $\PDEop$, e.g., the Laplace operator
$-\Delta$, and a force term $f: \domain \rightarrow \R$. Let us also
consider a partition of $ \boundary$ into a Dirichlet boundary
$\Dirichlet$ and a Neumann boundary $\Neumann$, and the corresponding
boundary data $u_{\rm D}: \Dirichlet \rightarrow \R$ and
$g_{\rm N}: \Neumann \rightarrow \R$. The boundary value problem reads
as follows: find $u(\x)$ such that
\begin{align}\label{strongform} \PDEop u(\x) = f(\x) \quad \hbox{ in } \, \Omega, \qquad  \bcop_{\rm D} u(\x) = u_{\rm D}(\x) \quad \hbox{
  on } \, \Dirichlet, \qquad  \bcop_{\rm N} u(\x) = g_{\rm N}(\x) \quad \hbox{
  on } \, \Neumann.
\end{align}
The operator $\bcop_{\rm D}$ is a trace operator and $\bcop_{\rm N}$ is
the flux operator. Other boundary conditions, e.g., Robin (mixed)
conditions can also be considered. We assume that the unknown
$u(\x)$ in \eqref{strongform} can be a scalar, vector, or tensor
field. (The case of multi-field problems is considered in
Sect.~\ref{subsec:cartesian_product_FE_space}.)

For \ac{FE} analysis, we must consider the weak form of
\eqref{strongform}. The weak formulation can be stated in an abstract
setting as follows. Let us consider an abstract problem determined by
a Banach space $\trialsp$ (\emph{trial space}), a reflexive Banach
space $\testsp$ (\emph{test space}), a continuous bilinear form ${a} :
\trialsp \times \testsp \rightarrow \mathbb{R}$, and a continuous
linear form $\ell: \testsp \rightarrow \mathbb{R}$. The abstract problem is
stated as: find $u \in \trialsp$ such that
\begin{align}\label{problem1}
\blform{u}{v} = \rhsform{v}, \qquad \hbox{ for any } \, v \in \testsp.
\end{align}
The link between the two formulations is the following. Let
$\mathcal{D}(\domain)$ be the space of $\mathcal{C}^\infty$ functions
with compact support in $\domain$; the dual space
$\mathcal{D}(\domain)'$ is the space of distributions. We have that:
$$ \blform{u}{\dist} \doteq \langle \PDEop u, \dist \rangle_\domain
, \qquad   \rhsform{\dist} \doteq \langle g_{{\rm N}},  \dist \rangle_{\Neumann}
+ \langle f, \dist \rangle_\domain,
\quad \hbox{ for any } \dist \in \cinfdom,
$$ where the derivatives are understood in distributional sense. E.g.,
for the Laplace operator, the bilinear form reads $\blform{u}{v}
\doteq \int_\domain \grad u \cdot \grad v {\rm d}
\domain$. Furthermore, homogeneous Dirichlet boundary conditions,
i.e., $u = 0$ on $\Dirichlet$, are usually enforced in a
strong way; the functions in $\testsp$ satisfy these boundary
conditions. The extension to non-homogeneous boundary conditions is
straightforward. One can define an arbitrary extension $Eu_{\rm D}$ of the
Dirichlet data, i.e., $Eu_{\rm D} = u_{{\rm D}}$ on $\Dirichlet$. Next, we define the function
$u_0 \doteq u - Eu_{\rm D}$ with zero trace on $\Dirichlet$ and solve
\eqref{problem1} for ${u}_0$ with the right-hand side
\begin{align}\label{eq:dirichlet_bcs_rhs} \rhsform{v} -
\blform{Eu_{\rm D}}{v}.\end{align}
Let us consider two classical examples.

\begin{example}[Heat equation]
\label{poisson_problem}
Let us consider the Poisson problem $-\grad \cdot \boldsymbol{\kappa} \grad u = f$
with $u = u_{{\rm D}}$ on $\Dirichlet$ and $\partial_{\normal} u =
g_{\rm N}$; $\normal$ is the outward normal. Let us assume that
$\boldsymbol{\kappa} \in L^\infty(\Omega)^{d \times d}$, $f \in H^{-1}(\Omega)$,  $g_{{\rm N}} \in
H^{-\half}(\Neumann)$, and $u_{{\rm D}} \in H^\half({\Dirichlet})$. Let us
also consider an extension $Eu_{\rm D} \in H^{1}(\Omega)$ such that
$Eu_{\rm D} = u_{{\rm D}}$ on $\Dirichlet$. The weak form of the
problem reads as: find $u_0 \in H_0^1(\Omega)$ such that
$$
\int_\Omega \boldsymbol{\kappa} \grad u_0 \cdot \grad v \ddom  = \int_\Omega f v \ddom  + \int_\Neumann g v \dbou - \int_\Omega \boldsymbol{\kappa} \grad Eu_{\rm D} \cdot \grad v \ddom , \qquad \hbox{for any } \, v \in H_0^1(\Omega).
$$
The solution is $u \doteq u_0 + Eu_{\rm D}$. 
\end{example}

\begin{example}[Stokes problem]
  \label{stokes_problem}
The Stokes problem consists on finding a velocity field $\ubs$ and a pressure field $p$ such that 
\begin{equation}
-\grad \cdot (\mu \strain( \ubs ) ) + \grad p = \fbs,  \qquad 
 \qquad \grad \cdot \ubs = 0,
\end{equation}
and (for example) $\ubs=\ubs_{\rm D}$ on $\Gamma$, where $\strain(\ubs) = \frac{1}{2}(\grad \ubs + \grad \ubs^T)$ is the strain tensor. The weak form of the problem consists of finding $(\ubs_0,p) \in \trialsp \doteq \left[H^1_0(\Omega)\right]^d \times L_0^2(\Omega)$ such that
\begin{equation}
\mu \int_{\Omega} \strain( \ubs_0) :  \strain( \vbs )    -  \int_{\Omega} \grad \cdot \vbs p + \int_{\Omega} q \grad \cdot \ubs_0 =\int_{\Omega} \vbs \cdot \fbs -  \mu \int_{\Omega}  \strain( \boldsymbol{E}\ubs_{\rm D}) : \strain( \vbs)  - \int_{\Omega} q \grad \cdot \boldsymbol{E} \ubs_{\rm D}, 
\end{equation}for any $(\vbs,q) \in \trialsp $, where  $\boldsymbol{E}\ubs_{\rm D} \in \left[H^1_0(\Omega)\right]^d $ is an extension of the Dirichlet data, i.e., $ \boldsymbol{E}\ubs_{\rm D}=\ubs_{\rm D}$ on $\Gamma$. The solution is  $\ubs \doteq \ubs_0+ \boldsymbol{E}\ubs_{\rm D}$. 
\end{example}

\subsection{Space discretization with \acp{FE}} \label{subsec:space_discretization}

Problem \eqref{problem1} is an infinite-dimensional problem. In order to end up
with a computable one, we must introduce finite-dimensional
subspaces with some approximability properties. We restrict ourselves
to \ac{FE} schemes in a broad sense that involve conforming and
non-conforming spaces. Thus, our aim is to explicitly build spaces
$\trialsph$ (and $\testsph$) with some approximability properties. If
the discrete spaces are subspaces of the original ones (conforming), i.e.,
$\trialsph \subset \trialsp$ and $\testsph \subset \testsp$, the
discrete problem reads as: find $u_h \in \trialsph$ such that
\begin{align}\label{problem1h}
\blform{u_h}{v_h} = \rhsform{v_h}, \qquad \hbox{ for any } \, v_h \in \testsph.
\end{align}
This is the \emph{Petrov-Galerkin} problem. In the particular case
when $\trialsph = \testsph$, we have a \emph{Galerkin} problem. The
previous problem can be ill-posed for some choices of the \ac{FE}
spaces, e.g., using discrete spaces that do not satisfy the inf-sup
condition for indefinite problems \cite{brezzi_mixed_1991}. In some
cases, judiciously chosen perturbations of $\blform{\cdot}{\cdot}$ and $\rhsform{\cdot}$, represented with $\blformh{\cdot}{\cdot}$ and $\rhsformh{\cdot}$ respectively, 
can \emph{stabilize} the problem and make it stable and optimally
convergent, circumventing the inf-sup condition restriction. In the
most general case, we can describe any \ac{FE} space as: find $u_h \in \trialsph$ such that
\begin{align}\label{problem1h-d}
a_h(u_h,v_h) = \ell_h(v_h), \qquad \hbox{ for any } \, v_h \in \testsph,
\end{align}
replacing the continuous bilinear form by a general \emph{discrete}
bilinear form. One can also define the affine
operator \begin{align}\label{problem1h-op} \mathcal{F}_h(u_h) =
  a_h(u_h,\cdot) - \ell_h(\cdot) \in
  \testsph',\end{align} and state \eqref{problem1h-d} as: find
  $u_h \in \trialsph$ such that $\mathcal{F}_h(u_h) =
  0$. This statement is the one being used for the practical
  implementation of \ac{FE} operators in \FEMPAR{} (see
  Sect.~\ref{sec:building_fe_systems}).

In order to define \ac{FE} spaces, we require a triangulation $\triang$ of
the domain $\domain$ into a set $\{ \geophy \}$ of \emph{cells}. This
triangulation is assumed to be conforming, i.e., for two neighbour
cells $\geophy^+, \, \geophy^- \in \triang$, its intersection
$\geophy^+ \cap \geophy^-$ is a \emph{whole} $k$-face ($k<d$) of both
cells (note that $k$-face refers to a geometrical
entity, e.g. cells, faces, edges and vertices for $d=3$, see
Sect.~\ref{sec:polytope_topology}).  In practice, the cells must
be expressed as a particular type of mapping over a set of
admissible geometries (polytopes, see
Sect.~\ref{sec:polytope_topology}). Thus, for every element
$K \in \mathcal{T}_h$, we assume that there is a reference cell
$\georef_\geophy$ and a diffeomorphism
$\geomap_K: \georef \rightarrow \geophy$. In what follows, we usually use the
notation $\hat{\x} \doteq \geomap_K^{-1}(\x)$.

The definition of the functional space also relies on a reference
functional space as follows: 1) we define a functional space in the
reference cell $\georef$; 2) we define a set of functions in the
physical cell $\geophy$ via a function mapping; 3) we define the
global space as the assemble of cell-based spaces plus continuity
constraints between cells. In order to present this process, we
introduce the concept of {reference FE}, {FE}, and {FE
  space}, respectively.

\subsection{The \ac{FE} concept in the reference and physical spaces}
\label{subsec:fe_concept}

Using the abstract definition of Ciarlet, a \ac{FE} is represented by the
triplet $\{ \geophy, \lfesp, \moments \}$, where $\geophy$ is a
compact, connected, Lipschitz subset of $\mathbb{R}^d$, $\lfesp$ is a
vector space of functions, and $\moments$ is a set of linear
functionals that form a basis for the dual space $\lfesp'$. The
elements of $\moments$ are the so-called \acp{DOF} of
the FE. We denote the number of moments as $n_\Sigma$. The moments can
be written as $\sigma_a$ for
$a \in \mathcal{N}_\Sigma \doteq \{ 1, \ldots,n_\Sigma\}$. We can also
define the basis $\{\shapetest{a}\}_{a \in \mathcal{N}_\moments}$ for
$\lfesp$ such that $\sigma_a(\shapetest{b}) = \delta_{a b}$ for
$a,\, b \in \mathcal{N}_\moments$. These functions are the so-called
\emph{shape functions} of the FE, and there is a one-to-one mapping
between shape functions and \acp{DOF}. 
Given a function $v$, we define the \emph{local interpolator} for the \ac{FE} at hand as
\begin{align}\label{eq:local_interpolator}
\pi_K(v) \doteq \sum_{a \in \mathcal{N}_\moments } \sigma_a (v) \shapetest{a}.
\end{align}
It is easy to check that the interpolation operator is in fact a projection.

In the reference space, we build \emph{reference} \acp{FE}
$(\georef,\reffesp,\momentsref)$ as follows. First, we consider a
bounded set of possible cell geometries, denoted by $\georef$; see the
definition of polytopes in Sect.~\ref{sec:polytope_topology}. On $\georef$, we build a
functional space $\reffesp$ and a set of \acp{DOF} $\momentsref$. We
consider some examples of reference \acp{FE} in Sect.~\ref{subsec:h1_conforming_fes}, \ref{subsec:hdiv_conforming_fes}, and \ref{subsec:hcurl_conforming_fes}.

In the physical space, the \ac{FE} triplet $(\geophy,\lfesp,\moments)$ on a
mesh cell $\geophy \in \triang$ relies on: 1) a reference FE
$(\georef,\reffesp,\momentsref)$, 2) a geometrical mapping
$\geomap_\geophy$ such that $\geophy \doteq \geomap_\geophy(\georef)$,
and 3) a linear bijective function mapping
$\hat{\funmap}_\geophy: \reffesp \rightarrow \reffesp$. The functional
space in the physical space is defined as
$\lfesp \doteq \{ \hat{\funmap}_\geophy(\hat{v}) \circ
\geomap^{-1}_\geophy : \, \hat{v} \in \reffesp \}$; we will also use 
${\funmap}_\geophy: \reffesp \rightarrow \lfesp$ defined as
${\funmap}_\geophy(\hat{v}) \doteq \hat{\funmap}_\geophy (\hat{v})
\circ \geomap^{-1}_\geophy$. The set of \acp{DOF} in the physical space is defined as
$\moments \doteq \{ \hat{\sigma} \circ {\funmap}_\geophy^{-1} \, : \,
\hat{\sigma} \in \momentsref \}$. Given the set of shape functions
$\{ \shapetestref{a} : a \in \mathcal{N}_\momentsref \}$ in the
reference FE, it is easy to check that
$\{ \phi_\geophy^{a} \doteq \funmap_\geophy(
\shapetestref{a} ) : a \in \mathcal{N}_\momentsref \}$ are the set of
shape functions of the \ac{FE} in the physical space.


The reference \ac{FE} space $\reffesp$ is usually a polynomial
space. Thus, the first ingredient is to define bases of
polynomials; see Sect. \ref{subsec:polynomial_spaces_construction}. The analytical expression of the basis of shape functions is not straightforward for complicated definitions of moments; this topic is covered in Sect. \ref{subsec:shape_functions_construction}. After that, we will
consider how to build global (and conforming) \ac{FE} spaces in Sect. \ref{subsec:global_fe_space}, and how to integrate the bilinear forms in the corresponding weak formulation in Sect. \ref{subsec:numerical_integration}. We finally provide three examples of \acp{FE} in Sect.~\ref{subsec:h1_conforming_fes},~\ref{subsec:hdiv_conforming_fes}, and~\ref{subsec:hcurl_conforming_fes} .

\subsection{Construction of polynomial spaces} \label{subsec:polynomial_spaces_construction}

Local \ac{FE} spaces are usually polynomial spaces. Given an order $k \in \naturals$ and a set $\nodes_k$ of distinct points (nodes) in $\R$ (we will indistinctly represent nodes by their index $i$ or position $x_i$), we define the
corresponding set of Lagrangian polynomials $\{ \lagpol^k_0, \ldots,
\lagpol^k_k \}$ as:
\begin{equation}\label{eq:lagbas}
  \lagpol^k_m(x) \doteq \frac{ \Pi_{n \in \nodes_k \setminus \{m\} } (x - x_s) }{ \Pi_{n \in \nodes_k \setminus \{m\} } (x_m - x_s) }.
\end{equation}
We can also define the Lagrangian basis $\lagbasis^k = \{ \lagpol^k_i \, : \, 0 \leq i \leq k \}$.
This set of polynomials are a basis for $k$-th order polynomials. We note that $\lagpol^k_m(x_l) = \delta_{ml}$, for $0 \leq m, \, l \leq k$.

For multi-dimensional spaces, we can define the set of nodes as the
Cartesian product of 1D nodes. Given a $d$-tuple order $\order$, we
define the corresponding set of nodes for n-cubes as:
$\nodes^\order \doteq \nodes^{k_1} \times \cdots \times
\nodes^{k_d}$. Analogously, we define the multi-dimensional Lagrange
basis

\begin{equation}\label{eq:lagbasmd}
  \lagbasis^\order = \{ \lagpol^\order_{\boldsymbol{m}} \, : {\boldsymbol{m}} \in \nodes^\order \}, \qquad \hbox{ where } \quad \lagpol^\order_{\boldsymbol{m}}(\x) \doteq \Pi_{i=1}^d \lagpol^{k_i}_{m_i}(x_i).
  \end{equation}
Clearly, $\lagpol_{\boldsymbol{t}}^\order(\x_\node) = \delta_{\node \boldsymbol{t}}$, for $\node, \, \boldsymbol{t} \in \nodes^\order$.

This Cartesian product construction leads to a basis for the local \ac{FE} spaces usually used on n-cubes, i.e., the space of polynomials that are of degree less or equal to $k$ with respect to each variable $x_1, \ldots, x_d$. We can define monomials by a $d$-tuple $\mid$ as $p_{\mid}(\x) \doteq \Pi_{i=1}^d x_i^{\alpha_i}$, and the polynomial space of order $\order$ as $\polsp_\order = {\rm span} \{ p_{\mid}(\x) \, :  0 \leq \alpha_i \leq k_i, \, i = 1, \ldots, d  \}$. We have $\polsp_\order = {\rm span} \{ \ell \, : \, \ell \in \lagbasis^\order \}$.

The definition of polynomial spaces on n-simplices is slightly
different. It requires the definition of the space of polynomials of
degree equal or less than $k$ in the variables $x_1,\ldots,x_d$. It
does not involve a full Cartesian product of 1D Lagrange polynomials
(or monomials) but a truncated space, i.e., the corresponding
polynomial space of order $k$ is
$\polspr_k = {\rm span} \{ p_{\mid}(\x) \, : | \mid | \leq k \}$, with
$| \mid | \doteq \sum_{i=1}^d \alpha_i$. Analogously as for n-cubes, a
basis for the dual space of $\polspr_k$ are the values at the set of
nodes
$\tilde{\nodes}^{k} \doteq \{ \node \in \nodes^{k\boldsymbol{1}} \, :
\, |\node| \leq k \}$. It generates the  typical  grad-conforming \acp{FE} on n-simplices.


\subsection{Construction of the shape functions basis} \label{subsec:shape_functions_construction}

The analytical expression of shape functions can become very complicated for high order \acp{FE} and non-trivial definitions of \acp{DOF}, e.g., for electromagnetic applications. Furthermore, to have a code that provides a basis for an arbitrary high order, an automatic generator of shape functions must be implemented. When the explicit construction of the shape functions is not obvious, we proceed as follows.

Let us consider a \ac{FE} defined by $\{ \geophy, \lfesp, \moments \}$.\footnote{In this section, we do not make difference between reference and physical spaces, e.g., using the $\hat{\cdot}$ symbol. In any case, all the following developments are usually performed at the reference \ac{FE} level.} First, we generate a \emph{pre-basis} $\{ {\psi}^b \}_{b \in \Sigma}$ that spans the local \ac{FE} space $\lfesp$, e.g., a Lagrangian polynomial basis (see Sect.~\ref{subsec:polynomial_spaces_construction}). On the other hand, given the set of local \acp{DOF}, we proceed as follows. The shape functions can be written as ${\phi}^a = \sum_{b\in\mathcal{N}_\Sigma}  \bs{\Phi}_{ab} {\psi}^b$, where ${\psi}^b$ are the elements of the pre-basis. By definition, the shape functions must satisfy $\moment{a}({\phi}^b) = \delta_{ab}$ for $a, \, b \in\mathcal{N}_\Sigma$. As a result, let us define $\mathbf{C}_{ab} \doteq \moment{a}({\psi}^b)$. We have (using Einstein's notation):
$$
\moment{a}({\phi}^b) = \moment{a}( \bs{\Phi}_{bc} {\psi}^c ) = \moment{a} ({\psi}^c) \bs{\Phi}_{bc} = \delta_{ab},
$$
or in compact form, $\mathbf{C} \bs{\Phi}^T = I$, and thus $\bs{\Phi}^T = \mathbf{C}^{-1}$. As a result, $\bs{\Phi}_{ab} = \mathbf{C}^{-1}_{ba}$. The shape functions are computed as a linear combination of the pre-basis functions.


\subsection{Global \ac{FE} space and conformity} \label{subsec:global_fe_space}

Finally, we must define the \emph{global} \ac{FE} space. Conforming FE
spaces are defined as:
$\fesp \doteq \{ v \in \trialsp \, : \, v|_K \in \lfesp \}.$ The main
complication in this definition is to enforce the conformity of the FE
space, i.e., $\fesp \subset \trialsp$. In fact, the conformity
constraint is the one that motivates the choice of $\momentsref$ and
$\funmap$, and as a consequence, $\moments$. In practice, the
conformity constraint must be re-stated as a continuity constraint
over \ac{FE} \acp{DOF}. In general, these constraints are implicitly enforced
via a global \ac{DOF} numbering, even though it is not possible in general
for adaptive schemes with non-conforming meshes and/or variable order
cells, which require more involved constraints.

Let us define by $\mathcal{M}_h \doteq \{ (b,K) : b \in
\mathcal{N}_{\moments_\geophy}, \, \geophy \in \triang \}$ the Cartesian product of
local \acp{DOF} for all cells. We define the global \acp{DOF} as the quotient
space of $\mathcal{M}_h$ by an equivalence relation $\sim$. Using
standard notation, given $\sim$, the equivalence class of
$a \in \mathcal{M}_h$ with respect to $\sim$ is represented with
$[a] \doteq \{ b \in \mathcal{M}_h \, : \, a \sim b \}$, and the
corresponding quotient set is $\mathcal{N}_h \doteq \{ [a] \, : \,
a \in \mathcal{M}_h \}$. The set $\mathcal{N}_h$ is the set of global
\ac{DOF} and $[\cdot]$ represents the local-to-global \ac{DOF} map. We assume
that the equivalence relation is such that if two elements
$(b,K), \, (b',K') \in \mathcal{M}_h$ are such that
$(b,K) \sim (b',K')$, then $K \neq K'$.\footnote{This assumption in fact applies for \acp{FE} of any kind, since the local functional spaces are already conforming and do not require an equivalence class at the cell level.}
Using the one-to-one mapping between moments and shape functions, the
same operator allows one to define global shape functions
$\shapetest{a} = \sum_{(b,K) \sim a} \shapetest{b}_K$. We
assume that the choices above are such that they satisfy the
conformity constraint, i.e.,
$\fesp = {\rm span}\{ \shapetest{a} \}_{a \in \mathcal{N}_h} \subset
\trialsp$.

Let us consider an infinite-dimensional space $\tilde{\trialsp}$  such that 1) $\fesp \subset \tilde{\trialsp} \subset \trialsp$ and 2) for every function $v \in \tilde{\trialsp}$ and global \ac{DOF} $a \in  \mathcal{N}_h$, all the local \acp{DOF} $b, \, b' \in [a]$ are such that 
$\sigma_b(v) = \sigma_{b'}(v)$, i.e., local \acp{DOF} related to the same global \ac{DOF} are continuous among cells. The \emph{global interpolator} is defined as:
\begin{align}\label{eq:global_interpolator}
\pi_{\fesp}(v) \doteq  \sum_{\geophy \in  \triang} \pi_K(v)  = \sum_{\geophy \in  \triang} \sum_{b \in \mathcal{N}_{\Sigma_K} } \sigma_b(v) \shapetest{b}_K, \qquad \hbox{for } v \in \tilde{\trialsp}.
\end{align}
It is easy to check that it is in fact a projector. In any case, we use \emph{projection operator} to refer to other projectors that involve the solution of a global \ac{FE} system, e.g., based on the minimization of the $L^2$ or $H^1$ norm. 

Below, we provide details about how to choose the local \acp{DOF}
$\momentsref$, the function map $\funmap$, and the equivalence
relation $\sim$ such that the conformity property is satisfied for
grad, div, and curl-conforming \ac{FE} spaces. The case of non-conforming
methods, e.g., \ac{DG} methods, can readily be considered. In this case,
the conformity constraint is not required, which leads to much more
flexibility in the definition of \acp{DOF}. On the other side, these
schemes require numerical perturbations of the continuous bilinear and linear
forms in \eqref{problem1h-d} that involve integrals over the facets of
\acp{FE} to \emph{weakly} enforce the conformity. (Facets are
$(d-1)$-faces, e.g., faces in 3D and edges in 2D).

Once we have defined a basis for the \ac{FE} spaces $\trialsph$ and
$\testsph$ using the \ac{FE} machinery presented above, every \ac{FE}
function $u_h$ can be uniquely represented by a vector 
$\fesol \in \mathbb{R}^{|\mathcal{N}_h|}$ as
$u_h = \sum_{b \in \mathcal{N}_h} \shapetest{b} \fesol_b$.  In fact, problem
\eqref{problem1h-d} can be re-stated as: find $\fesol \in \mathbb{R}^{|\mathcal{N}_h|}$ such
that
$$ \blformh{\shapetest{b}}{\shapetrial{a}} \fesol_b = \rhsformh{\shapetrial{a}},
  \qquad \hbox{for any } \, a \in \mathcal{N}_h.
$$ We have ended up with a
finite-dimensional linear problem, i.e., a linear system. We note that in general, the trial space moments can be different
  from the ones of the test space, as soon as the cardinality is the
  same. In matrix form, the problem can be stated as:
\begin{equation}\label{eq:linear_system}
\mathrm{Solve} \ \ \ \fematrix \fesol = \ferhs, \qquad \hbox{with } \quad
  \fematrix_{ab} \doteq \blformh{\shapetest{b}}{\shapetrial{a}}, \quad
  \ferhs_{a} \doteq \rhsformh{\shapetrial{a}}.
\end{equation}
Assuming that the bilinear form can be split into cell contributions as $\blformh{\cdot}{\cdot} = \sum_{\geophy \in  \triang} \blformlocal{\geophy}{\cdot}{\cdot}$, e.g., by replacing $\int_\Omega$ by $\sum_{\geophy \in  \triang} \int_K$, the construction of the matrix is implemented through a cell-wise assembly process, as follows:
\begin{align} 
\label{eq:assembly}
  \fematrix_{[a][b]} = \sum_{\geophy \in  \triang} \sum_{a,b \in \mathcal{N}_{\Sigma_K} }\fematrix_{ab}^{\geophy} \doteq \sum_{\geophy \in  \triang} \sum_{a,b \in \mathcal{N}_{\Sigma_K}  } \blformlocal{\geophy}{\shapetest{b}_K}{\shapetrial{a}_K}.
\end{align}
The \ac{FE} affine operator \eqref{problem1h-op} can be represented as
$\mathcal{F}_h(u_h) \doteq \fematrix \fesol - \ferhs$, i.e., it can be represented with
a matrix and a vector of size ${|\mathcal{N}_h|}$.

\subsection{Numerical integration} \label{subsec:numerical_integration}
In general, the local
  bilinear form can be stated as:
  $$\blformlocal{\geophy}{\shapetest{b}_K}{\shapetrial{a}_K} = \int_\geophy \integrandfe(\x) \ddom,$$ 
where the evaluation of $\integrandfe(\x)$ involves the evaluation of shape function derivatives. Let us represent the Jacobian of the geometrical mapping with $ \jacobian \doteq \frac{\partial \geomap_K}{\partial \x}$. We can rewrite the cell integration in the reference cell, and next consider a quadrature rule ${\rm Q}$ defined by a set of points/weights $(\xh_{\rm gp}, {\rm w}_{\rm gp})$, as follows:
\begin{align}\label{fematrix}
  \int_\geophy \integrandfe(\x) {\rm d} \Omega
  = \int_\georef \integrandfe \circ \geomap(\x) |\jacobian| {\rm d}\Omega
  = \sum_{{\xh_{\rm gp} \in \quadrature}} \integrandfe \circ \geomap(\xh_{\rm gp}) {\rm w}(\xh_{\gp}) |\jacobian(\xh_{\gp})|.
  \end{align}
  Here, the main complication is the evaluation of
  $\integrandfe \circ \geomap(\xh_{\rm gp})$. By construction, the
  evaluation of this functional only requires the evaluation of
  $\partial_{\mid}\shapetest{b}_K \circ \geomap(\xh_{\rm gp})$ for
  some values of the multi-index $\mid$ (idem for the test functions). Usually,
  $|\mid| \leq 2$ in $\mathcal{C}^0$ \acp{FE}, since
  higher-order derivatives would require higher inter-cell
  continuity. The second derivatives, which only have sense for
  \emph{broken} cell-wise integrals, are in fact only needed for some
  method based on \emph{stabilization} techniques based on the
  pointwise evaluation of residuals in the interior of cells
  \cite{colomes_assessment_2015}.

Let us consider the case of zero and first derivatives, i.e., the
evaluation of $\shapetest{b}_K \circ \geomap_\geophy(\xh_{\rm gp})$ and
$\grad \shapetest{b}_K \circ \geomap_\geophy(\xh_{\rm gp})$.
The values of the shape functions (times the geometrical mapping) on
the quadrature points is determined as follows:
\begin{align}\label{eq:shfunx}
  \shapetest{b}_\geophy \circ \geomap_\geophy(\xh_{\rm gp})= \hat{\funmap}(\shapetestref{b})(\xh_{\rm gp}),
\end{align}
whereas shape function gradients are computed as:
\begin{align}\label{shape_grad}
\grad \shapetest{b}_\geophy \circ \geomap_\geophy(\xh_{\rm gp})  = \grad ( \hat{\funmap}(\shapetestref{b}) \circ \geomap_\geophy^{-1}) \circ \geomap_\geophy(\xh_{\rm gp})  = \grad_{\hat{\x}} \hat{\funmap}(\shapetestref{b})(\xh_{\rm gp}) \invjacob(\xh_{\rm gp}),
\end{align}
where we have used some elementary differentiation rules and the inverse function theorem 
in the last equality; $\grad_{\hat{\x}}$ represents the
gradient in the reference space. Thus, one only needs to provide the
values of the Jacobian, its inverse, and its determinant, from one
side, and the value of the shape functions $\funmap(\shapetestref{b})$
and their gradients $\grad_{\hat{\x}} \funmap(\shapetestref{b})$ in
the reference space, on the other side, at all quadrature points, to
compute all the entries of the \ac{FE} matrices; second order derivatives
can be treated analogously.

Quadrature rules for $\georef$ being an n-cube can readily be obtained
as a tensor product of a 1D quadrature rule, e.g., the Gauss-Legendre
quadrature. Symmetric quadrature rules on triangles and tetrahedra for
different orders can be found, e.g., in \cite{ern_theory_2004}. In any
case, to create arbitrarily large quadrature rules for n-simplices,
one can consider the so-called Duffy transformation
\cite{duffy_quadrature_1982,dunavant_high_1985}.

As it is well known, considering n-cube topologies for $\georef$,
Gauss quadratures with $n$ points per direction can integrate
\emph{exactly} $2n-1$ order polynomials. E.g., for a Lagrangian
reference \ac{FE} of order $p$ and an affine geometrical map, we
choose $n=p+ {\rm ceiling}( 1/2 ) = p+1$ per direction to integrate
exactly a mass matrix. For n-simplex meshes, we use either symmetric
quadratures (if available) or tensor product rules plus the Duffy
transformation \cite{duffy_quadrature_1982,dunavant_high_1985}.  The
latter case is based on introducing a change of variables that
transform our n-simplex integration domain into an n-cube, and
integrate on the n-cube using tensor product quadratures. It is worth
noting that this change of variables introduces a non-constant
Jacobian. The determinant of the Jacobian is of order at most $d-1$
with respect to each variable. To integrate a mass matrix exactly, we
must be able to integrate exactly polynomials of order $2p+d-1$.
Therefore, we need to take $n=p+ {\rm ceiling}( d/2 )$ to exactly integrate
mass matrices.

\subsection{Grad-conforming \acp{FE}: Lagrangian (nodal) elements} \label{subsec:h1_conforming_fes}

In this section, we consider one characterization of the abstract FE
technology above. First, we are interested in the so-called nodal \acp{FE},
based on Lagrange polynomials and \acp{DOF} based on nodal values.

Let us consider the same order for all components, i.e.,
$k \ones \doteq (k ,\ldots, k)$. When the reference geometry $\georef$
is an n-cube, we define the reference \ac{FE} space as
$\lfesp_k \doteq \polsp_{k \ones}$. The set of nodes $\nodes^{k\ones}$
can be generated, e.g., from the equidistant Lagrangian nodes. Let us
define the bijective mapping $\mathtt{i}(\cdot )$ from the set of
nodes $\nodes^{k\ones}$ to
$\{1, \ldots, |\nodes^{k\ones}| \} \equiv \nodes_\moments$,
i.e., the local node numbering. The set of local \acp{DOF}
$\mathcal{N}_{\moments_\geophy}$ are the nodal values, i.e.,
$\sigma_{\mathtt{i}(\boldsymbol{s})} \doteq v(\x_{\boldsymbol{s}})$,
for $\boldsymbol{s} \in \nodes^\order$. Clearly, the reference FE
shape functions related to these \acp{DOF} are
$\shapetest{\mathtt{i}(\boldsymbol{s})} \doteq
\lagpol_{\boldsymbol{s}}^{k\ones}$. On the other hand, we simply take
$\hat{\funmap}(v) \doteq v$.

For n-simplices, we consider the reference \ac{FE} space $\polspr_k$
spanned by the pre-basis
$\{ p_{\mid}(\x) \, : 0 \leq \alpha_i \leq k, \, i = 1, \ldots, d \}$
and the set of nodes $\tilde{\nodes}^{k}$ (see
Sect.~\ref{subsec:polynomial_spaces_construction}). The set of local
\acp{DOF} $\mathcal{N}_{\moments_\geophy}$ are the nodal values. Since
the pre-basis elements are not shape functions, we proceed as in
Sect.~\ref{subsec:shape_functions_construction} to generate the
expression of the shape functions basis for arbitrary order reference
\acp{FE} on n-simplices.

The global \ac{FE} space is determined by the following equivalence
relation. The set of local \acp{DOF} for n-cubes is
$\mathcal{M}_h \doteq \{ (\sbs,K) \, : \, \sbs \in \nodes^{k \ones}, K \in
\triang \}$  due to the one-to-one mapping between \acp{DOF}
and nodes; we replace the set of nodes by 
$\tilde{\nodes}^{k}$ for n-simplices. Furthermore, we say that $(\sbs,K) \sim (\sbs',K')$ iff
$\x_{\sbs} = \x_{\sbs'}$. The implementation of this equivalence relation, and
thus, the global numbering, relies on the ownership relation between
n-faces and \acp{DOF} (e.g., in 3D we can say whether a \ac{DOF}
belongs to a vertex, edge, or face) and a permutation between the
local node numbering in $K^+$ to the one in $K ^-$ for nodes on
$F$. See Sect.~\ref{sec:polytope_topology} for more details. With such
global \ac{DOF} definition, it is easy to check that the global
\ac{FE} space functions are $\mathcal{C}^0$ and thus grad-conforming.

Since Lagrangian moments involve point-wise evaluations of functions and 
$H^1_0(\Omega) \not\subset \mathcal{C}^0(\Omega)$ for $d>1$, 
 the interpolator \eqref{eq:global_interpolator} is not defined in such space. Instead, we consider that 
 functions to be interpolated belong, e.g., to the space $\tilde{\trialsp} \doteq  \mathcal{C}^0(\Omega)$.

When one has to deal with vector or tensor fields, we can generate
them as a Cartesian product of scalar spaces as follows. We define the
local \ac{FE} space $\boldsymbol{\lfesp}_k \doteq [\polsp_{k \ones}]^d$
and the function map $\hat{\funmap}(\vbs) \doteq \vbs$. In the vector
case, the local \acp{DOF} set is represented with
$\mathcal{M}_h \doteq \{ (i,\sbs,K) \, : \, 1 \leq i \leq d, \, \sbs
\in \nodes^{k \ones}, K \in \triang \}$, and
$(i,\sbs,K) \sim (i',\sbs',K')$ iff $i = i'$ and $\x_{\sbs} =
\x_{\sbs'}$.  Analogously, shape functions are computed as
${\phi}^a \doteq \sum_{(i,\sbs,K) \sim a} \ell_{\sbs}^{k \ones}
\vec{\ebs}_i$; $\vec{\ebs}_i$ represents the $i$-th canonical basis vector of $\R^d$. We proceed analogously for n-simplices.

The verification that two nodes are in the same position is not
straightforward. First, for every node $\sbs$ in $K$, we can assign an
n-face owner $F$ (e.g., a vertex, edge, face, or cell); cell \acp{DOF} are
not replicated. Given a node $\sbs \in \nodes^{k \ones}$ of cell $K$
that belongs to the n-face $F$, it can be determined by an index
$\sbs_F$ with respect to $F$ and $K$. Analogously, another node that
belongs to the same n-face but cell $K'$, is represented by
$\sbs_F'$. On the other hand, one can define a permutation mapping
\begin{equation}\label{eq:permmap}\mathtt{p}_F(F,K,K';\cdot),\end{equation} that, given the local index of a node within 
the n-face $F$ with respect to $K$, it provides the index in the n-face $F$
with respect to $K'$ (see
Sect.~\ref{sec:facet_int} and \ref{sec:polytope_rotations_and_permutations} for more details). Thus,
$\x_{\sbs} = \x_{\sbs'}$ iff $\mathtt{p}_F(F,K,K';\sbs_F) = \sbs_F'$.
\subsection{Div-conforming \acp{FE}} \label{subsec:hdiv_conforming_fes}

\def\lfespbs{{\boldsymbol{\lfesp}}} 
\def\facetref{{\hat{F}}}
\def\facetrefb{\facetref_0}
\def\georef{{\hat K}}

We present the so-called Raviart-Thomas \acp{FE} for vector fields \cite{brezzi_mixed_1991};
the implementation of Brezzi-Douglas-Marini \acp{FE} is analogous. In
this case, the order being used is different at every space
dimension. Let us start with Raviart-Thomas \acp{FE} on n-cubes. In
2D, the space reads as
$\lfespbs_{k} \doteq \polsp_{(k+1,k)} \times \polsp_{(k,k+1)}$,
whereas in 3D it reads as
$\lfespbs_{k} \doteq \polsp_{(k+1,k,k)} \times \polsp_{(k,k+1,k)}
\times \polsp_{(k,k,k+1)}$; the Raviart-Thomas element can in fact be
considered for any dimension. The basis for $\moments$ in 3D is
composed of two types of \acp{DOF}, boundary and interior \acp{DOF},
defined as
\begin{equation}\label{eq:rt-moments}
 \frac{1}{\| {\hat{F_0}} \|} \int_{\facetref_0} \vbs \cdot \normal\circ \geomap_\facetref  \, q   \dbou, \quad q \in  \polspr_{k}, \quad  \frac{1}{\| {\hat{K}} \|}\int_{\hat{K}} \vbs \cdot \ \qbs \ddom, \quad \qbs \in \polsp_{(k-1,k,k)} \times \polsp_{(k,k-1,k)} \times \polsp_{(k,k,k-1)},
\end{equation}
respectively\footnote{The test function spaces in the definition of the moments are always considered with respect to the corresponding domain of integration.}; the 2D case is straightforward, replacing the space of shape functions for the interior moments by $\polsp_{(k-1,k)} \times \polsp_{(k,k-1)}$. The definition of the boundary facets involves mappings
from a reference facet $\facetrefb$ to all facets $\facetref$ of the
\ac{FE} $K$, i.e., $\geomap_\facetref: \facetrefb \rightarrow
\facetref$. Every boundary moment can be associated to a function in a
Lagrangian space, and thus, a node index. As a result, the boundary
\acp{DOF} can be indexed with a node in $\nodes^{k\ones}$ (for $d=2$) on
the corresponding facet $F$, i.e.,
$\mathcal{M}_h^\partial \doteq \{ (F,\sbs,K) \, : \, F \hbox{ are
  facets of } K, \, \sbs \in \nodes^{k \ones}, K \in \triang \}$. We
say that $(F,\sbs,K) \sim (F',\sbs',K')$ iff $F = F'$ and
$\x_{\sbs} = \x_{\sbs'}$.  To check whether $\x_{\sbs} = \x_{\sbs'}$
holds, we can proceed similarly as for Lagrangian elements. The shape
functions are built as in
Sect.~\ref{subsec:shape_functions_construction}. We consider a
Lagrangian pre-basis for $\lfesp$, and compute the shape functions via
a change-of-basis. The function mapping reads as follows:
\begin{align}\label{eq:contrapiola}
\hat{\funmap}_K(\vbs) \doteq \frac{1}{|\jacobian|} \jacobian \vbs;
\end{align}
the mapping $\hat{\funmap}_K \circ \geomap^{-1}_K$ is the so-called
contravariant Piola transformation. One can check that the definition
of this mapping together with the assembly defined above leads to a
global \ac{FE} space that is div-conforming; i.e., its functions have
continuous normal component across inter-cell facets. Thus, $ \fsp_h\subset H({\rm div}, \Omega)$ \cite{brezzi_mixed_1991}.

On n-simplices, the reference \ac{FE} space is
$\lfespbs_{k} \doteq [\polspr_{k}]^d \times \x \polspr_{k}$, for
$k = 0, 1, 2, \ldots$, and the basis for $\moments$ is composed of the following boundary and interior \acp{DOF}:
\begin{equation}\label{eq:rt-moments-simplex}
 \frac{1}{\| {\hat{F_0}} \|} \int_{\facetref_0} \vbs \cdot \normal\circ \geomap_\facetref  \, q   \dbou, \quad q \in \polspr_{k}, \quad  \frac{1}{\| {\hat{K}} \|}\int_{\hat{K}} \vbs \cdot \ \qbs \ddom, \quad \qbs \in [\polspr_{k-1}]^d.
\end{equation}
In this case, the generation of the pre-basis is not a Lagrangian FE space of functions, but it can easily be expressed as the span of vector functions with components in a selected subset of $\polspr_{k+1}$.

\subsection{Curl-conforming \acp{FE}} \label{subsec:hcurl_conforming_fes}

The weak formulation of electromagnetic problems involve the
functional space $\Hcurl$. Conforming \ac{FE} spaces for $\Hcurl$ must
preserve the continuity of the tangential component of the field. The
so-called edge elements (or N\'ed\'elec elements) are curl-conforming \acp{FE} \cite{monk_finite_2003}. As Raviart-Thomas
elements, the edge elements pre-basis on n-cubes involves different orders per
dimension and per component. In 2D, the space reads as
$\lfespbs_{k} \doteq \polsp_{(k-1,k)} \times \polsp_{(k,k-1)}$,
whereas in 3D it reads as
$\lfespbs_{k} \doteq \polsp_{(k-1,k,k)} \times \polsp_{(k,k-1,k)}
\times \polsp_{(k,k,k-1)}$. The basis for $\moments$ is composed of
three types of \acp{DOF} (in 3D), namely edge, face, and interior \acp{DOF}, defined as:
\begin{align}
  &  \frac{1}{\| {\hat{E_0}} \|} \int_{{\hat{E}_0}} ( \vbs \cdot \boldsymbol{\tau} ) \circ \geomap_{\hat{E}} q \, {\rm d} \Lambda, \quad \forall q \in \mathcal{P}_{k-1}, \\
  &  \frac{1}{\| {\hat{F}_0} \|} \int_{{\hat{F}_0}}  ( \bsJ_{\hat{F}}^T (\vbs \times \normal) ) \circ \geomap_\facetref \cdot  \boldsymbol{q}  \,{\rm d}\Gamma, \quad \forall \boldsymbol{q} \in \mathcal{Q}_{(k-2,k-1)}\times \mathcal{Q}_{(k-1,k-2)}, \\
 &  \frac{1}{\| \hat{K} \|} \int_{\hat{K}}  \vbs  \cdot \boldsymbol{q} \, {\rm d} \Omega, \quad  \forall \boldsymbol{q} \in \mathcal{Q}_{(k-1,k-2,k-2)}\times \mathcal{Q}_{(k-2,k-1,k-2)} \times \mathcal{Q}_{(k-2,k-2,k-1)},
\end{align}
respectively, where the edge map $\geomap_{\hat{E}}$ is defined as the one for the face.  The boundary \acp{DOF} can be indexed by a triplet
$(F,\sbs,K)$, where $F$ can be an edge or a face in 3D, following the
same ideas as for Raviart-Thomas elements. In this case, the function mapping reads as follows:
\begin{align}\label{eq:copiola}
\hat{\funmap}_K(\vbs) \doteq \jacobian^{-T} \vbs;
\end{align}
the mapping $\hat{\funmap}_K \circ \geomap^{-1}_K$ is the so-called
covariant Piola transformation, which leads to a global \ac{FE} space that
is curl-conforming \cite{monk_finite_2003}, i.e., its functions have continuous tangential
component across inter-cell facets.

On n-simplices, the space reads as:
\begin{equation}\label{eq:edge-prebasis}
\lfespbs_{k} \doteq [ \polspr_{k} ]^d + \boldsymbol{\mathcal{S}}_{k}, \, \hbox{ where } \, \boldsymbol{\mathcal{S}}_{k} \doteq \{ \boldsymbol{v} \in [\polspr_{k+1}]^d \, : \,  \boldsymbol{v} (\x) \cdot \x = 0 \, \forall \, \x \in \georef \}.\end{equation}
The basis for $\moments$ in 3D is composed of the following boundary and interior \acp{DOF}:\footnote{We note that we can take $\bsJ_{\hat{F}}^T \vbs$ instead of $\bsJ_{\hat{F}}^T (\vbs \times \normal )$ in the definition of the face moments, since the rows of the Jacobian matrix are the transformation of the axes in the reference face $\hat{F_0}$ to the actual face $\hat{F}$ of the reference cell and the space of test functions is invariant to rotations.}
\begin{align}
  &  \frac{1}{\| {\hat{E_0}} \|} \int_{{\hat{E}_0}} ( \vbs \cdot \boldsymbol{\tau} ) \circ \geomap_{\hat{E}} q \, {\rm d} \Lambda, \quad \forall q \in \mathcal{P}_{k-1}, \\
  &  \frac{1}{\| {\hat{F}_0} \|} \int_{{\hat{F}_0}} ( { \bsJ}^T_{\hat{F}} (\vbs \times \normal) ) \circ \geomap_\facetref \cdot \boldsymbol{q}  \,{\rm d}\Gamma, \quad \forall \boldsymbol{q} \in [ \mathcal{P}_{k-2} ]^2 \\
 &  \frac{1}{\| \hat{K} \|} \int_{\hat{K}}  \vbs  \cdot \boldsymbol{q} \, {\rm d} \Omega, \quad  \forall \boldsymbol{q} \in  [ \mathcal{P}_{k-3} ]^3.
\end{align}
In 2D, only the first two types of DOFs are required, where the first one is now related to facets (edges in 2D) and the second one are interior \acp{DOF} owned by the cell. As for Raviart-Thomas elements, the pre-basis functions are not Lagrangian shape functions, but they can again be expressed as the span of vector functions with components in a selected subset of $\polspr_{k+1}$. We refer to \cite{olm-hts} for a discussion about the actual generation of a pre-basis for the space \eqref{eq:edge-prebasis} in \FEMPAR{}.

\subsection{Cartesian product of \acp{FE} for multi-field problems} \label{subsec:cartesian_product_FE_space}

Many problems governed by \acp{PDE} involve more than one field, e.g., the
Navier-Stokes equations or any multi-physics problem. Let us consider
a \ac{PDE} that involves a set of unknown fields
$(\ubs_1 , \ldots , \ubs_n) \in \trialsp^1 \times \ldots \times
\trialsp^n$, defined as the Cartesian product of functional spaces. We
can proceed as above, and define a \ac{FE} space for every field
space separately, leading to a global \ac{FE} space
$ \trialsp^1_h \times \ldots \times \trialsp^n_h$ defined by
composition of \ac{FE} spaces. To define the global numbering of \acp{DOF} in
the multi-field case, we consider that two \acp{DOF} are equivalent if they
are related to the same field and satisfy the equivalence relation of
the \ac{FE} space of this field.

The Cartesian product of \ac{FE} spaces is enough to define
volume-coupling multi-physics problems governed on the same physical
domain, i.e., the different physics are defined on the whole domain
and coupled through volume terms in the formulation. However, many
multi-physics problems are interface-based, i.e., the coupling between
different physics that are defined on different subdomains is through
transmission conditions on the interface. This is the case, e.g., of
fluid-structure problems (see, e.g., 
\cite{badia_modular_2008,badia_fluidstructure_2008,badia_splitting_2008,badia_robinrobin_2009}). In
these cases, different \ac{FE} spaces could be
defined on different parts of the global mesh, i.e., one must describe
the set of subdomains $( \Omega_1, \ldots, \Omega_n )$ of the whole
domain $\Omega$ in which the corresponding \ac{FE} spaces are defined.
  
\subsection{Non-conforming methods}\label{sec:nonconfm}

Up to now, we have considered a global \ac{FE} space that is conforming,
i.e., $\trialsp_h \subset \trialsp$. Alternatively, one can consider
\ac{FE} schemes that are not conforming. Since the original bilinear form
has no  sense in general for a non-conforming \ac{FE} space
$\trialsp_h$, one shall consider a \emph{stabilized} bilinear form
$a_h$ that is well-posed (stable and continuous) in the discrete
setting. In general, these schemes replace the required inter-cell
continuity for conformity by a weak imposition of such
continuity. Thus, the inter-cell continuity is imposed weakly
through penalty-like terms. \ac{DG} methods are schemes of this type~\cite{quarteroni_numerical_2014}.

In one sense, non-conforming \ac{FE} spaces are simpler than conforming
ones, since the conformity is not required; one has more
flexibility in the definition of local \acp{DOF} and the equivalence class
concept is not needed, since a \ac{DOF} never belongs to more than one
element. However, the bilinear form usually requires the integration
of facet terms, i.e., terms of the type:
$$
\sum_{F \in \mathcal{F}_h} \int_F \integrandfe(\x) \ddom.
$$
The integration of facet terms is far more complicated than cell terms.

Let us first briefly illustrate a simple application of non-conforming
methods, namely the \ac{FE} discretization of the Poisson problem using the
so-called interior penalty (IP) family of \ac{DG}
formulations~\cite{quarteroni_numerical_2014}. Dirichlet boundary
conditions constraints, say $u(x)=u_{{\rm D}}(x)$ on the whole
boundary $\boundary$ of the domain $\domain$, are to be weakly
imposed, as it is natural in such kind of formulations.  The
global discrete trial space $\trialsph$ is composed of functions that
are continuous within each cell, but discontinuous across cells, i.e.,
$\trialsph = \{ u_h \in L_2(\Omega): u_h|_K \in \trialsph|_K \subset
H^1(K),\ K \in \triang\}$, and the discrete test space
$\testsph=\trialsph$.  If we denote $\mathcal{F}^{\Omega}_{h}$ and
$\mathcal{F}^{\boundary}_{h}$ as the set of interior and boundary
facets of $\triang$, respectively, the discrete weak form underlying
this family of methods reads as: find $u_h \in \trialsph$ such that
\begin{equation} \label{eq:ip_dg_formulations}  
\begin{split}
& \sum_{K\in \triang} \int_K \grad u_h \cdot \grad v_h  
 - \sum_{\facephy \in \mathcal{F}^{\Omega}_{h}} \int_\facephy \jump{v_h} \cdot \mean{\grad u_h}  
 - \tau \sum_{\facephy \in \mathcal{F}^{\Omega}_{h}} \int_\facephy \jump{u_h} \cdot \mean{\grad v_h}  
 + \sum_{\facephy \in \mathcal{F}^{\Omega}_{h}} \gamma |\facephy|^{-1} \int_\facephy  \jump{u_h} \cdot \jump{v_h} \\
& -  \sum_{\facephy \in \mathcal{F}^{\boundary}_{h}} \int_\facephy v_h \grad{u_h} \cdot \normal
 -  \tau  \sum_{\facephy \in \mathcal{F}^{\boundary}_{h}} \int_\facephy u_h \grad{v_h} \cdot \normal
 +  \sum_{\facephy \in \mathcal{F}^{\boundary}_{h}} \gamma |\facephy|^{-1} \int_\facephy  u_h v_h  \\
 & = \sum_{K\in \triang} \int_K f v_h 
 - \tau \sum_{\facephy \in \mathcal{F}^{\boundary}_{h}} \int_\facephy u_{{\rm D}} \grad{v_h} \cdot \normal
 + \sum_{\facephy \in \mathcal{F}^{\boundary}_{h}} \gamma |\facephy|^{-1} \int_\facephy u_{{\rm D}} v_h \quad \quad \quad  
 \forall v_h \in \testsph,
\end{split}
\end{equation}   
where $\tau$ is a fixed constant that characterizes the particular method at hand, $\gamma$
is a facet-wise positive constant referred to as penalty parameter, and $|\facephy|$ denotes the surface
of the facet; $\tau$ and $\gamma$ should be suitably chosen such that the bilinear
form $a_h(u_h,v_h)$ on the left-hand side of~\eqref{eq:ip_dg_formulations}  is well-posed (stable and continuous) in the discrete setting, and the resulting \ac{FE} formulation enjoys optimal rates of 
convergence~\cite{quarteroni_numerical_2014}. Finally, if we denote as $K^+$ and $K^-$ the
two cells that share a given facet, then $\mean{w_h}$ and $\jump{w_h}$ denote mean values and
jumps of $w_h$ across cells facets:
\begin{equation} \label{eq:mean_and_jump_ops}
\mean{w_h} = \frac{w_h^++w_h^-}{2}, \quad  \quad \jump{w_h} = w_h^+ \normal^+ + w_h^- \normal^-,
\end{equation}
with $\normal^+$, $\normal^-$ being the facet outward unit normals, and 
$w_h^+$, $w_h^-$ the restrictions of $w_h$ to the facet, both from either the 
perspective of $K^+$ and $K^-$, respectively.

The computation and assembly of \acp{DOF} related to interior nodes is straightforward. With regard to the facet terms, assuming that we are sitting on an interior facet $\facephy \in \mathcal{F}^{\Omega}_{h}$, four facet-wise matrices, namely $\fematrix^{\facephy}_{K^+ K^+}$, $\fematrix^{\facephy}_{K^+ K^-}$, $\fematrix^{\facephy}_{K^- K^+}$, and  $\fematrix^{\facephy}_{K^- K^-}$, are computed. (The case of boundary facets $\facephy \in \mathcal{F}^{\boundary}_{h}$ is just a degenerated case of the one corresponding to interior facets where only a single facet-wise matrix $\fematrix^{\facephy}_{K^+ K^+}$ has to be computed; we omit this sort of facets from the discussion in order to keep the presentation short.) These hold all partial contributions of the facet to the corresponding global entries of the coefficient matrix. The entries of, e.g., $\fematrix^{\facephy}_{K^+ K^-}$, are defined  (for our particular problem at hand) as:
\begin{equation} \label{eq:local_to_face_matrix}
\left(\fematrix^{\facephy}_{K^+ K^-}\right)_{ab} = -\int_\facephy \jump{\shapetest{b}_{K^-}} \cdot \mean{\grad \shapetest{a}_{K^+}}  - 
                                        \tau \int_\facephy \jump{\shapetest{a}_{K^+}} \cdot \mean{\grad \shapetest{b}_{K^-}} +
                                         \gamma |\facephy|^{-1} \int_\facephy  \jump{\shapetest{a}_{K^+}} \cdot \jump{\shapetest{b}_{K^-}},
\end{equation}
with indices $a$ and $b$ ranging from 1 to the number of shape functions $\nodes_\moments$ of $K^+$ and $K^-$, respectively.

\subsection{Facet integration}\label{sec:facet_int}

As mentioned in Sect.~\ref{subsec:numerical_integration} for the case of
cell integrals, facet integrals involved in the computation of the
facet-wise  matrix \eqref{eq:local_to_face_matrix}
cannot be in general computed analytically. These are instead computed
 using quadrature rules. In general, the bilinear form that contains the facet terms can be stated as
$$\blformlocal{F}{{\shapetest{b}_{K^{+}}}}{{\shapetrial{a}_{K^{-}}}} = \int_F \integrandfe(\x) {\rm d}F.$$
We can consider a reference facet $\hat{F}$, and a mapping
$\geomap_F : \hat{F} \rightarrow F$ from the reference to the physical
space. Let us represent the Jacobian of the geometrical mapping with
$ \jacobf \doteq \frac{\partial {\geomap}_F}{\partial \x}$, which has values  in $\mathbb{R}^{(d-1) \times d}$. We can
rewrite the facet integral in the reference facet, and next consider
a quadrature rule ${\rm Q}$ on $\hat{F}$ defined by a set of points/weights
$(\xh_{\rm gp}, {\rm w}_{\rm gp})$, as follows:
\begin{align}\label{facematrix}
 \left(\fematrix^{\facephy}_{K^+ K^-}\right)_{ab}  =
  \int_F \integrandfe(\x) {\rm d} \Omega
  = \int_{\hat{F}} \integrandfe \circ \geomap_F(\x) |\jacobf| {\rm d}F
  = \sum_{{\xh_{\rm gp} \in \quadrature}} \integrandfe \circ \geomap_F(\xh_{\rm gp}) {\rm w}(\xh_{\gp}) |\jacobf(\xh_{\gp})|.
\end{align} $|\facejacobian|$ is defined as:
\begin{equation} \label{eq:face_jacobian_measure}
|\facejacobian|=\left \| \frac{{\rm d}\geomap_\facephy}{{\rm d} x} \right \|_2 \
\mbox{ and } \ |\facejacobian|=\left \| \frac{\partial\geomap^1_\facephy}{\partial \xh} \times \frac{\partial\geomap_\facephy^2}{\partial \xh} \right \|_2,
\end{equation}
for $d=2,3$, respectively. 

The expression of the shape functions and their gradients in the
physical space in terms of the ones in the reference space are
computed by using the cell-wise maps. Thus, two mappings
$\geomap_{K^+}$ and $\geomap_{K^-}$ among the reference cell $\georef$
and the cells $K^+$ and $K^-$ in physical space, respectively, are
involved in the numerical evaluation of interior facet integrals. We
can also consider the reference facet $\hat F$ and a map $\geomap_F$
from this reference facet to $F$ (analogously as $\geomap_K$ and $K$
but in one dimension less in the reference space). We can define a quadrature rule
$(\xh_{\rm gp},{\rm w}_{\rm gp})$ in $\hat F$. We can also define the
reference facet $\hat{F}^\pm$ of $\hat K$ such that
$\geomap_{K^\pm}( \hat{F}^\pm ) = F$, and the map
$\geomap_{\hat F^\pm}$ from $\hat F$ to $\hat{F}^\pm$. With this map,
we can define the quadrature
$(\xh_{\rm gp}^\pm \doteq \geomap_{\hat F^\pm}(\xh_{\rm gp}), {\rm
  w}_{\rm gp})$ with respect to the reference cell $\hat{K}$. 

\begin{figure}
\centering
\def\svgwidth{9cm}
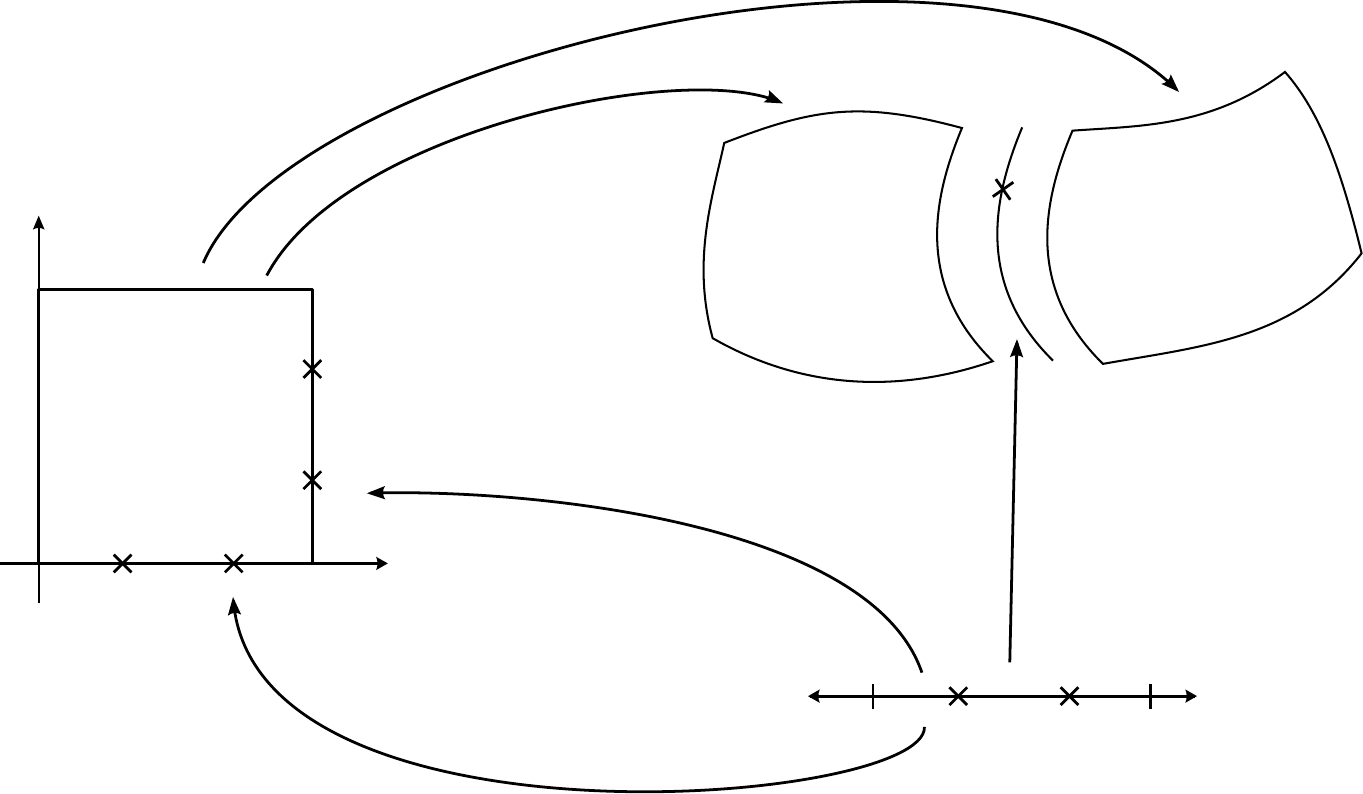
\caption{Mappings required for facet integration. The (only) quadrature point shown in the physical space is located at $\x=\geomap_F(\xh_{\rm 1})=\geomap_{K ^+}(\xh_{\rm 1}^+) = \geomap_{K ^-}(\xh_{\rm 2}^-) $, that is, $\Pi(1)=2$ in this case.}
\label{fig:facet-maps}
\end{figure}

However, the same facet $F$ has (in general) a different orientation
depending on the cell used as reference, and so, a different index
might be assigned to the same facet quadrature points from the
perspective of either cell, i.e.,
$\geomap_{K^+}(\xh_{\rm gp}^+) \neq \geomap_{K^-}(\xh_{\rm
  gp}^-)$ in general. We adopt the convention that facet quadrature points
identifiers are in the local numbering space of $K^+$, and these local
identifiers are translated into the local numbering space of
$K^-$. This is represented by the
permutation $\Pi({\rm gp})$ such that
$$\geomap_{K ^-}(\xh_{\Pi({\rm gp})}^-) = \geomap_{K ^+}(\xh_{\rm gp}^+) = \geomap_F(\xh_{\rm gp}).$$
The logic underlying this translation is equivalent to the one
discussed in
Sect.~\ref{sec:polytope_rotations_and_permutations}; see Fig.~\ref{fig:facet-maps} for an explanatory illustration.
 As a result, we have
$$\geomap_{\geophy^{+}}^{-1} \circ \geomap_F (\xh_{{\rm gp}}) = \xh_{{\rm gp}}^+, \qquad \hbox{and} \qquad \geomap_{\geophy^{-}}^{-1} \circ \geomap_F (\xh_{{\rm gp}}) = \xh_{\Pi({\rm gp})}^-.$$

Let us consider the evaluation of zero and first order derivatives on
facets, i.e., the evaluation of
$\shapetest{b}_{K^\alpha} \circ \geomap_{\geophy^{\alpha}}(\xh_{\rm
  gp})$ and
$\grad \shapetest{b}_{K^\alpha} \circ
\geomap_{\geophy^{\alpha}}(\xh_{\rm gp})$ for $\alpha \in \{+,-\}$,
where the quadrature points belong to a quadrature in the reference
facet $\hat{F}$. We note that the
introduction of $\hat{\funmap}$ is not needed for non-conforming methods, since there is no
continuity to be enforced, and we will consider it to be the identity
operator for simplicity. The values of the shape functions (times the
geometrical mapping) on the facet quadrature points is evaluated as
follows:
\begin{align}
  \shapetest{b}_{\geophy^{\alpha}} \circ \geomap_F (\xh_{\rm gp})=
  \shapetestref{b} \circ \geomap_{\geophy^{\alpha}}^{-1} \circ \geomap_F (\xh_{\rm gp}),
\end{align}
whereas shape function gradients are computed as:
\begin{align}
\grad \shapetest{b}_{\geophy^{\alpha}} \circ \geomap_F(\xh_{\rm gp})   = \grad_{\hat{\x}} \shapetestref{b}(\xh_{\rm gp}) \invjacob \circ \geomap_{\geophy^{\alpha}}^{-1} \circ \geomap_F (\xh_{\rm gp}).
\end{align}

Without loss of generality, let us focus on the first integral in \eqref{eq:local_to_face_matrix}. Replacing the mean value and jump operators by their definition in \eqref{eq:mean_and_jump_ops}, and taking into account
that $\shapetest{b}_{K^-}$ and $\grad \shapetest{a}_{K^+}$ vanish on $K^+$ and $K^-$ (by construction of $\trialsph$ and $\testsph$), respectively, we end up with the following integral to be computed numerically:
\begin{equation}
-\frac{1}{2} \int_F \shapetest{b}_{K^-} \normal^- \cdot \grad \shapetest{a}_{K^+} \dxfacephy \ .
\end{equation}
This integral is first mapped back to the reference facet
$\faceref \subset \mathbb{R}^{\mbox{\scriptsize d-1}}$, and then it is
approximated by the following sum over quadrature points:
\begin{equation} \label{eq:face_integral_mapped_back_and_approximated}
\begin{split}
   & -\frac{1}{2} \int_\facephy (\shapetest{b}_{K^-} \normal^-) \cdot \grad \shapetest{a}_{K^+} \dxfacephy  \\
   & =   -\frac{1}{2} \int_{\faceref} \shapetestref{b}_{K^-} \circ \geomap_{\geophy^{-}}^{-1} \circ \geomap_F (\xh_{\rm gp})  \normal^- \circ \geomap_F (\xh_{\rm gp}) \cdot  \grad_{\hat{\x}} \shapetestref{b}(\xh_{\rm gp}) \invjacob \circ \geomap_{\geophy^{+}}^{-1} \circ \geomap_F (\xh_{\rm gp})  |\facejacobian| \dxfaceref \\
   & \approx -\frac{1}{2} \sum_{gp \in {\rm Q} }
   \shapetestref{b}_{K^-}(\xref^{-}_{\Pi({\rm gp})}) \normal^-(\x_{\rm gp})
     \cdot \grad_{\hat{\x}} \shapetestref{b}(\xh_{\rm gp}^+) \invjacob(\xref^{+}_{q}) |\facejacobian(\xh_{\rm gp})| {\rm w}_{\rm gp}.
\end{split}
\end{equation}
Using these ideas, we can compute all the terms related to facet integrals. Furthermore, outward normals on facets can be computed as:
\begin{equation}  \label{eq:normal_out}
  \normal^\alpha = (-1)^{o_\alpha}
  \frac{ \frac{{\rm d}\geomap_\facephy}{{\rm d} x}
  }{\left \| \frac{{\rm d}\geomap_\facephy}{{\rm d} x} \right \|_2
  } \
  \mbox{ and } \ \normal^\alpha = (-1)^{o_\alpha}
    \frac{
\frac{\partial\geomap^1_\facephy}{\partial \xh} \times \frac{\partial\geomap_\facephy^2}{\partial \xh}
 }{
\left \| \frac{\partial\geomap^1_\facephy}{\partial \xh} \times \frac{\partial\geomap_\facephy^2}{\partial \xh} \right \|_2    
  },
\end{equation}
for $d=2,3$, respectively,  and $\alpha \in \{+,-\}$; $o$ is 0 or 1 and is used to enforce the normal to be outwards. Tangent vector(s) for a given facet can be easily computed out of the normal vector.

\def\TBD#1{{\color{red} [\textbullet]}}
\def\Eq#1{(\ref{eq:#1})}
\def\x{\boldsymbol{x}}
\def\s{\boldsymbol{s}}
\def\b{\boldsymbol{b}}
\def\top{{\rm top}}
\def\T{\boldsymbol{T}}
\def\zeros{\boldsymbol{0}}
\def\ones{\boldsymbol{1}}
\def\C{\boldsymbol{C}}
\def\t{\boldsymbol{t}}
\def\top{{\mathtt{t}}}
\def\ext{{\mathtt{e}}}
\def\zerob{{\mathtt{0}}}
\def\oneb{{\mathtt{1}}}
\def\c{\boldsymbol{c}}
\def\v{\boldsymbol{v}}
\def\ev{\boldsymbol{e}}
\def\e{{\bf{e}}}
\def\anc{{\mathtt{v}}}
\def\exan{{\mathtt{ev}}}
\def\ej{{\vec{\ev}_j}}
\def\es{{\vec{\ev}_s}}
\def\squarebf{\boldsymbol{\square}}
\def\nface{\square}
\def\extr#1#2#3#4{{#4}_{(#1;#2,#3)}}

\subsection{Polytopes} \label{sec:polytope_topology}

One of the motivations of \FEMPAR{} is to develop a framework that
can deal with arbitrary space dimensions. It permits to readily
implement space-time formulations, which are posed in 4D. Other
higher-dimensional applications include systems of \acp{PDE} posed in the
phase space, e.g., the 7D (including time) Vlasov-Maxwell equations
for the simulation of plasmas.

In this section, we provide the mathematical abstraction of cell topologies based on the concept
of \emph{polytope}. This abstract concept is of practical importance,
because it allows us to develop algorithms and codes that can be
applied to any topology that fits into the framework. The framework
developed herein is very general and includes triangles and
quadrilaterals in 2D, and tetrahedra, hexahedra, prysms, and pyramids
in 3D. Furthermore, it can also be extended to arbitrary dimensions,
to define not only n-cubes and n-simplices but many other topologies. A polytope is
mathematically defined as the convex hull of a finite set of
points. As a consequence, a polytope is a polyhedron. In the frame
of \FEMPAR{}, we consider polytopes that can be expressed as the
image of the composition of two operators. The definition of
topologies for reference \acp{FE} based on this idea can be found
in \cite{dedner_construction_2012}.

The main topological information consumed by \ac{FE} codes is the
description of the $d$-dim polytope boundary as the assemble of
$(d-1)$-dim polytopes, proceeding recursively till 0-dim objects are
obtained (vertices); we use the contraction $k$-\emph{dim} object to
say object of dimension $k$. These lower-dimensional entities
describing the polytope boundary are denoted herein as
\emph{n-faces}. Usually, the nomenclature used to describe n-faces in
\acp{FE} is restricted to 3D problems. In \FEMPAR{} and in the
following exposition, we use a dimension-independent nomenclature in
order to accommodate higher-dimensional problems. We consider the space
dimension $d \in \N^+$ and a $d$-dim polytope. We define the $d$-face
as the polytope itself. The set of $(d-1)$-dim polygons that compose
the boundary of the polytope are its $(d-1)$-faces; $(d-1)$-faces are
usually denoted as \emph{facets}. We can proceed recursively, i.e.,
defining the $(k-1)$-faces of the polytope as the set of facets of its
$k$-faces till reaching 0-faces. In 3D, 3-faces are called
\emph{cells}, 2-dim faces are \emph{faces}, 1-dim faces are
\emph{edges}, and 0-dim faces are \emph{vertices}. Herein, we use the
term n-faces to denote all these objects. In this work, we denote by
\emph{vefs} the set of n-faces of dimension lower than the space
dimension, e.g., it only includes vertices, edges, and faces in 3D.

Let us introduce some notation. We represent the set of bitmaps of
size $m$ with $\Bit^m$. The bitmaps $(1,1,\ldots,1)$ and
$(0,0,\ldots,0)$ are represented with $\oneb$ and $\zerob$,
respectively. Given a domain $\square \subset \R^d$ we use the
notation $\alpha \square + \b$, $\alpha \in \R$, $\b \in \R^d$ to
denote the domain $\{ \alpha \x + \b \, : \, \x \in \square
\}$. $\vec{\ev}_j$ represents the $j$-th canonical basis vector of
$\R^d$.

Let us define first the \emph{directional extrusion}
$\extr{j}{\alpha}{\beta}{\square}$ of $\square$ with respect to the
direction $\ej$ of type $(\alpha,\beta)$. $\alpha$ determines the
topology of the extrusion, namely a prysm-type extrusion (1) or a
pyramid-type extrusion (0) (see
also \cite{dedner_construction_2012}). $\beta$ determines whether we
want to perform the $\alpha$-extrusion (1) or do-nothing (0). Based on
this, we have the following definition.

\begin{definition}[Directional extrusion]
Given a domain $\square \subset \R^{d}$, we define
$\extr{j}{\alpha}{\beta}{\square} \subset \R^{d}$, with
$\beta, \, \alpha \in \{0,1\}$ and $j = 1, \ldots, d$, as
$$
\extr{j}{\alpha}{0}{\square} \doteq \square, \qquad 
\extr{j}{0}{1}{\square} \doteq \{ (1-z)\square + z \vec{\ev}_j \, : \, z \in [0,1] \}, \qquad
\extr{j}{1}{1}{\square} \doteq \{ \square + z \vec{\ev}_j : \, z \in [0,1] \}.
$$
\end{definition}

The directional extrusion can be used recursively to define polytopes
and their n-faces. An n-face is determined by a topology
$\top\in \Bit^d$, an extrusion $\ext\in \Bit^d$, and an anchor vertex
$\v \in \R^d$, using a recursive procedure as follows. The use of directional extrusions to get different polytopes and n-faces is illustrated in Figs.~\ref{fig:hex-nfacecode}-\ref{fig:tet-nfacecode}. One can observe how all the lower  dimensional n-faces after directional extrusion lead to one dimension larger n-faces for different values of $\alpha$.

\begin{definition}[n-face]
Given $\top, \, \ext \in \Bit^d$ and $\v \in \R^d$, we can define the
n-face $\square$ in a recursive way as follows. Let
$\nface^0 \doteq \{ \v \}$; we define $\nface \doteq \nface^d$ based
on the following recursion:
\begin{align}\label{eq:chain_n_faces}
\nface^0  \rightarrow 
\nface^1 \doteq \extr{1}{\top(0)}{\ext(0)}{\nface^0} \rightarrow \ldots
\rightarrow \nface^{i+1} \doteq \extr{i+1}{\top(i)}{\ext(i)}{\nface^{i}} \rightarrow \ldots
\rightarrow \nface^d \doteq \extr{d}{\top(d-1)}{\ext(d-1)}{\nface^{d-1}}.
\end{align}
\end{definition}

For our purposes, the anchor vertex $\v$ has only 0/1 entries, and
thus, it can be represented as an element $\anc$ of $ \Bit^d$.  As a
result, an n-face can be uniquely represented with
$(\top;\ext,\anc)$. Based on this definition, we can define a set of
$d$-dim polytopes by recursion. $d$-dim polytopes are given by $\top$,
and represented as n-faces with $(\top, \oneb, \zerob)$, i.e., using
the origin $\zeros$ as anchor vertex and performing extrusions in all
directions. On the other hand, a vertex $\v$ (with only 0/1
coordinates) is an n-face with $(\top,\zerob,\anc)$. Some examples of
n-face constructions using this procedure can be found in
Figs.~\ref{fig:hex-nfacecode}-\ref{fig:tet-nfacecode}. Furthermore, in
these figures we show all  n-faces of the 3-cube and 3-simplex, with
all the $\ext$  and $\anc$  values. In our implementation of polytopes,
we use Hasse diagrams based on the composition of extrusion and anchor
vertex bitmaps to label the different n-faces of a polytope.

\begin{figure}[htbp]
\begin{tabular}{ccc}
    \scalebox{0.65}{
\begin{tikzpicture}
      \input{points}



	\fill[gray!70] (A2) -- (A3) -- (A6) -- (A7) -- cycle; 
    \draw[fill=black] (barycentric cs:A2=1,A3=1,A6=1,A7=1) circle (0.15em);
	\node at (barycentric cs:A2=1,A3=1,A6=1,A7=1)[below] { $[011|000]$};
	
	\fill[gray!50] (A3) -- (A4) -- (A5) -- (A6) -- cycle; 
    \draw[fill=black] (barycentric cs:A3=1,A4=1,A5=1,A6=1) circle (0.15em);
	\node at (barycentric cs:A3=1,A4=1,A5=1,A6=1)[above] { $[101|000]$};
	
	\fill[gray!30] (A5) -- (A6) -- (A7) -- (A8) -- cycle; 
    \draw[fill=black] (barycentric cs:A5=1,A6=1,A7=1,A8=1) circle (0.15em);
	\node at (barycentric cs:A5=1,A6=1,A7=1,A8=1)[above] { $[110|000]$};

	\draw[thick] (A5) -- (A6);
	\draw[thick] (A3) -- (A6);
	\draw[thick] (A7) -- (A6);




	\draw[thick] (A5) -- (A6);
    \draw[fill=black] (barycentric cs:A5=1,A6=1) circle (0.15em);
	\node at (barycentric cs:A5=1,A6=1)[above ] { $[100|000]$};
	
	\draw[thick] (A3) -- (A6);
    \draw[fill=black] (barycentric cs:A3=1,A6=1) circle (0.15em);
	\node at (barycentric cs:A3=1,A6=1)[above left] { $[001|000]$};
	
	\draw[thick] (A7) -- (A6);
    \draw[fill=black] (barycentric cs:A6=1,A7=1) circle (0.15em);
	\node at (barycentric cs:A6=1,A7=1) [above right]{ $[010|000]$};
	
	
	\draw[thick] (A3) -- (A4);
    \draw[fill=black] (barycentric cs:A3=1,A4=1) circle (0.15em);
	\node at (barycentric cs:A3=1,A4=1)[above left] { $[100|001]$};

	\draw[thick] (A7) -- (A8);
    \draw[fill=black] (barycentric cs:A7=1,A8=1) circle (0.15em);
	\node at (barycentric cs:A7=1,A8=1)[above right] { $[100|010]$};


	\draw[thick] (A2) -- (A3);
    \draw[fill=black] (barycentric cs:A2=1,A3=1) circle (0.15em);
	\node at (barycentric cs:A2=1,A3=1)[below left] { $[010|001]$};

	\draw[thick] (A2) -- (A7);
	\node at (barycentric cs:A2=1,A7=1)[below right]  { $[001|010]$};
    \draw[fill=black] (barycentric cs:A2=1,A7=1) circle (0.15em);

	\draw[thick] (A4) -- (A5);
    \draw[fill=black] (barycentric cs:A4=1,A5=1) circle (0.15em);
	\node at (barycentric cs:A4=1,A5=1) [above left]{ $[001|100]$};

	\draw[thick] (A8) -- (A5);
    \draw[fill=black] (barycentric cs:A5=1,A8=1) circle (0.15em);
	\node at (barycentric cs:A8=1,A5=1)[above right] { $[010|100]$};
	

\draw[fill=black] (A2) circle (0.15em)
	    node[below] { $[000|011]$};
\draw[fill=black] (A3) circle (0.15em)
	    node[below left] { $[000|001]$ };
\draw[fill=black] (A4) circle (0.15em)
	    node[above left] { $[000|101]$};
\draw[fill=black] (A5) circle (0.15em)
	    node[above ] { $[000|100]$};
\draw[fill=black] (A6) circle (0.15em)
	    node[below ] { $[000|000]$};	
\draw[fill=black] (A7) circle (0.15em)
	    node[below right] { $[000|010]$};
\draw[fill=black] (A8) circle (0.15em)
	    node[above right] {$[000|110]$ };

\end{tikzpicture}} & \hfill &  
    \scalebox{0.65}{
\begin{tikzpicture}
      \input{points}


	 \fill[gray!70] (A1) -- (A8) -- (A7) -- (A2) -- cycle; 
	\draw[fill=black] (barycentric cs:A1=1,A8=1,A7=1,A2=1) circle (0.15em);
	 \node at (barycentric cs:A1=1,A8=1,A7=1,A2=1) [below right] { $[101|010]$};
	\fill[gray!50,opacity=0.2] (A1) -- (A2) -- (A3) -- (A4) -- cycle; 
	\draw[fill=black] (barycentric cs:A1=1,A2=1,A3=1,A4=1) circle (0.15em);
	\node at (barycentric cs:A1=1,A2=1,A3=1,A4=1) [below left] {$ [110|001]$};
	\fill[gray!90,opacity=0.2] (A1) -- (A4) -- (A5) -- (A8) -- cycle; 
	\draw[fill=black] (barycentric cs:A1=1,A4=1,A5=1,A8=1) circle (0.15em);
	\node at (barycentric cs:A1=1,A4=1,A5=1,A8=1) [above] { $[011|100]$};


%
%
%
	\draw[thick] (A1) -- (A2);
	\draw[fill=black] (barycentric cs:A1=1,A2=1) circle (0.15em);
	\node at (barycentric cs:A1=1,A2=1)[below left]  { $[100|011]$};
	
	\draw[thick] (A1) -- (A4);
	\draw[fill=black] (barycentric cs:A1=1,A4=1) circle (0.15em);
	\node at (barycentric cs:A1=1,A4=1)[below left] { $[010|101]$};

	\draw[thick] (A1) -- (A8);
	\draw[fill=black] (barycentric cs:A1=1,A8=1) circle (0.15em);
	\node at (barycentric cs:A1=1,A8=1)[below right] { $[001|110]$};

%
%
%
	

\draw[fill=black] (A1) circle (0.15em)
	    node[below] { $[000 | 111]$};

          \end{tikzpicture}}
\end{tabular}
\caption{$\ext$ and $\anc$ values for all the n-faces  (with the exception of the volume) of the 3-cube, with topology $\top = (111)$. }\label{fig:hex-nfacecode}
\end{figure}

\begin{figure}[htbp]
\begin{tabular}{ccc}
    \scalebox{0.65}{
\begin{tikzpicture}
   \input{points}


	\fill[gray!70] (A3) -- (A6) -- (A7) -- cycle; 
    \draw[fill=black] (barycentric cs:A3=1,A6=1,A7=1) circle (0.15em);
	\node at (barycentric cs:A3=1,A6=1,A7=1)[above] { $[011|000]$};
	
	\fill[gray!50] (A3) -- (A5) -- (A6) -- cycle; 
    \draw[fill=black] (barycentric cs:A3=1,A6=1,A5=1) circle (0.15em);
	\node at (barycentric cs:A3=1,A5=1,A6=1) [below]{ $[101|000]$};
	
	\fill[gray!30] (A5) -- (A6) -- (A7) -- cycle; 
    \draw[fill=black] (barycentric cs:A7=1,A6=1,A5=1) circle (0.15em);
	\node at (barycentric cs:A5=1,A6=1,A7=1)[below] { $[110|000]$};
	
	\draw[thick] (A5) -- (A6);
    \draw[fill=black] (barycentric cs:A5=1,A6=1) circle (0.15em);
    \node at (barycentric cs:A6=1,A5=1) [above] { $[100|000]$};
    
	\draw[thick] (A3) -- (A6);
    \draw[fill=black] (barycentric cs:A3=1,A6=1) circle (0.15em);
    \node at (barycentric cs:A6=1,A3=1) [above left]{ $[001|000]$};
    
	\draw[thick] (A7) -- (A6);
    \draw[fill=black] (barycentric cs:A7=1,A6=1) circle (0.15em);
    \node at (barycentric cs:A7=1,A6=1)[above right] { $[010|000]$};
        


	\draw[thick] (A3) -- (A5);
    \draw[fill=black] (barycentric cs:A5=1,A3=1) circle (0.15em);
    \node at (barycentric cs:A3=1,A5=1) [above left]{ $[100|001]$};
    
	\draw[thick] (A3) -- (A7);
    \draw[fill=black] (barycentric cs:A3=1,A7=1) circle (0.15em);
    \node at (barycentric cs:A3=1,A7=1) [below] { $[010|001]$};
    
	\draw[thick] (A5) -- (A7);
    \draw[fill=black] (barycentric cs:A5=1,A7=1) circle (0.15em);
    \node at (barycentric cs:A7=1,A5=1) [above right]{ $[100|010]$};
        

\draw[fill=black] (A3) circle (0.15em)
	    node[below left] { $[000 | 001]$};
\draw[fill=black] (A5) circle (0.15em)
	    node[above ] { $[000 | 100]$};
\draw[fill=black] (A6) circle (0.15em)
	    node[below ] { $[000 | 000]$};	
\draw[fill=black] (A7) circle (0.15em)
	    node[below right] { $[000 | 010]$};	
	    
\end{tikzpicture}}& \hfill &  
    \scalebox{0.65}{
\begin{tikzpicture}
   \input{points}


	\fill[gray!70] (A3) -- (A5) -- (A7) -- cycle; 
    \draw[fill=black] (barycentric cs:A3=1,A5=1,A7=1) circle (0.15em);
	\node at (barycentric cs:A3=1,A5=1,A7=1)[above] { $[110|001]$};

\end{tikzpicture}}
\end{tabular}
\caption{$\ext$ and $\anc$ values for all the n-faces (with the exception of the volume) of the 3-simplex, with topology $\top = (000)$. }\label{fig:tet-nfacecode}
\end{figure}

In codes, like in \FEMPAR{} , the topology can be coded with the bitmap
$\top$ (e.g., one 32-bit integer). \FEMPAR{} can use any geometry that can be
defined this way, for an arbitrary space dimension. This polytope
definition leads to the following geometries. The 1-dim line segment
topology is $\top = (0)$ or $(1)$; this ambiguity in 1D is inherited
to higher dimensions. In 2D, the triangle topology is $\top = (00)$
(or (01)) and the quadrilateral topology $\top = (10)$ (or (11)). In
3D, cubes are represented by $\top = (1,1,0)$ (or $(1,1,1)$),
tetrahedra $\top= (0,0,0)$ (or (0,0,1)), prysms by $\top=(1,0,0)$ (or
(1,1,1)), and pyramids by $\top=(0,1,0)$ (or (0,1,1)). Cosserats in 4D
are represented by $\top=(1,1,1,0)$ (or (1,1,1,1)). In general,
$2^{k-1}$ types of $k$-dim topologies are possible. n-cubes are
expressed by $\top = \oneb$ and n-simplices by $\top = \zerob$.

Given a bitmap $\top$ and a bit $\alpha$, we define the bit operation
that modifies the $j$ bit of $\top$ to $\alpha$ with
$\top.o_j(\alpha)$.  Given the chain on n-faces \Eq{chain_n_faces},
let us assume that $\square_{i-1}$ is represented by
$(\top,\ext',\anc)$. The extrusion
$\square_{i} \doteq \extr{i}{*}{\alpha}{\square_{i-1}}$ is
defined by $(\top,\ext'.o_{i-1}(\alpha),\anc)$. Thus, the
chain \Eq{chain_n_faces} can be represented as follows. Given a
topology $\top$, an extrusion $\ext$, and an anchor vertex $\anc$, we
start with $(\top,\ext',\anc) \doteq (\top,\zerob,\anc)$ and proceed
recursively:
\begin{align}\label{eq:chain_n_faces_bits}
(\top,\ext',\anc) \rightarrow (\top,\ext'.o_0(\ext(0)),\anc) \rightarrow \ldots \rightarrow
(\top,\ext'.o_{i}(\ext(i)),\anc) \rightarrow \ldots \rightarrow (\top,\ext'.o_{d-1}(\ext(d-1)),\anc) \equiv (\top,\ext,\anc).
\end{align}
E.g., in 3D, the polytope itself (or 3-face) is determined by $\top =
(1,1,1)$ and $(\ext,\anc) = ((1,1,1),(0,0,0))$. The chain \Eq{chain_n_faces_bits} in this
case reads as follows: (We omit $\top$ in the chain since it is the
same for all elements in the recursion.) 
$$
((0,0,0),(0,0,0)) \rightarrow
((0,0,1),(0,0,0)) \rightarrow
((0,1,1),(0,0,0)) \rightarrow
((1,1,1),(0,0,0)).
$$
Using the definition of the n-face, every element of the chain has a
geometrical representation. We start with the vertex $\zeros$, next
obtain the line segment $\{(x,0,0) \, : \, x \in [0,1]\}$, next the
square $\{(x,y,0) \, : \, x, \, y \in [0,1]\}$, and finally the unit
cube. The previous definition is not only useful to represent $d$-dim objects \emph{but all its n-faces}. See Figs.~\ref{fig:hex-nfacecode}-\ref{fig:tet-nfacecode}.

For a given n-face $\nface \equiv
(\top,\ext,\anc)$, we want to define the set $\mathcal{S}_{\nface}$ of
all n-faces of $\nface$. In order to do so, we introduce the following
concepts.
\begin{definition}[Oriented set extrusion]
Given a set $\mathcal{S} = \{ \nface : \nface \in \R^d \}$, we
define $\extr{j}{\alpha}{\beta}{\mathcal{S}}$, with
$\beta, \, \alpha \in \{0,1\}$ and $j = 1, \ldots, d$ as:
$$
\extr{j}{\alpha}{0}{\mathcal{S}} \doteq \mathcal{S}, \qquad
\extr{j}{0}{1}{\mathcal{S}} \doteq \{ \square, \zeros+\ej,
\extr{j}{0}{1}{\square} \, : \, \square \in \mathcal{S} \}, \qquad
\extr{j}{1}{1}{\mathcal{S}} \doteq \{ \square, \square + \ej, \extr{j}{1}{1}{\square} \, : \, \square \in \mathcal{S} \}. 
$$
\end{definition}

\begin{definition}[Set of n-faces]
Given an n-face $(\top,\ext,\anc)$, we can obtain all its n-faces
recursively as follows. Let $\mathcal{S}^0 \doteq \{ \v \}$; we define
$ \mathcal{S} \doteq \mathcal{S}^{d}$ based on the following
recursion:
\begin{align}\label{eq:n_faces_set}
\mathcal{S}^0  \rightarrow
\mathcal{S}^1 \doteq \extr{1}{\top(0)}{\ext(0)}{\mathcal{S}^0} \rightarrow \ldots \rightarrow 
\mathcal{S}^{i+1} \doteq \extr{i+1}{\top(i)}{\ext(i)}{\mathcal{S}^i} \rightarrow \ldots
\rightarrow
\mathcal{S}^{d} \doteq \extr{d}{\top(d-1)}{\ext(d-1)}{\mathcal{S}^{d-1}}.
\end{align}
\end{definition}
All the resulting n-faces can also be written with the
$(\top,\ext,\anc)$ notation commented above. In order to define this
chain as in \Eq{chain_n_faces_bits} (i.e., only based on the bitmap notation),
we note the following. Given the n-face $\square \equiv
(\top,\ext,\anc)$, the n-face $\square + \ej \equiv
(\top,\ext,\anc.o_j(1))$. With this ingredient, we can implement the generator of all n-faces of an n-face using the bitmap notation.

We also want to know the facets of an n-face. We use the following
statement. Given an n-face $\square \equiv (\top,\ext,\anc)$ and its
corresponding chain \Eq{chain_n_faces_bits}, the $i$-th element boundary
$\partial \nface^{i} \doteq \partial \extr{i}{\top(i-1)}{\ext(i-1)}{\nface^{i-1}}$ is
the following:
\begin{align}
\label{eq:facetsnface}
\partial \nface^i &= \partial \nface^{i-1}, \hbox{ if } \ext(i-1) = 0, \\
\partial \nface^i &= \{ \nface^{i-1}, \partial \extr{i}{0}{1}{\nface^{i-1}}
\}, \hbox{ if } \top(i-1) = 0, \, \ext(i-1) = 1, \\
\partial \nface^i &= \{ \nface^{i-1}, \nface^{i-1} + \hat\e_i, \partial \extr{i}{1}{1}{\nface^{i-1}} \} \hbox{ if } \top(i-1) = 1, \, \ext(i-1) = 1, 
\end{align}
with $\partial \nface^1 = \{ \nface^0, \nface^0 + \hat\e_1 \}$.


Using this definition of facets for the 3D cube, we get the following
faces: $((1,1,0);(0,0,0))$ and $((1,1,0);(0,0,1))$ faces ($x=0$ and
$x=1$ faces), $((1,0,1);(0,0,0))$ and $((1,0,1);(0,1,0))$ faces
($y=0$ and $y=1$ faces), $((0,1,1);(0,0,0))$ and $((0,1,1);(1,0,0))$
faces ($z=0$ and $z=1$ faces), having 6 faces in total. For every one
of these faces, we can use the same definition above, to obtain the
$(d-2)$-faces that are in the boundary of every $(d-1)$-face. All
these ideas can be used for any polytope, not only n-cubes. The only
difference is the type of extrusion being used in every case.

\subsection{Node generation and indexing} \label{sec:polytope_nodes}

\ac{FE} spaces are polynomial spaces, e.g., Lagrangian
polynomials. (Let us note that div- and curl-conforming \acp{FE} also
rely on Lagrangian polynomials for the definition of the pre-bases and
the definition of the equivalence classes.) In order to express these
polynomials, one must define a set of points (nodes). In the
following, we define a node generator for a given order on an arbitrary
polytope, using lexicographical notation.\footnote{We note that in
  fact the order $k$ is not a scalar but a vector
  $\boldsymbol{k} \in \R^d$. In principle, the use of a vector-valued
  order only has sense for n-cubes. The implementation in \FEMPAR{}
  makes use of a vector-valued order, even though all entries should
  be the same for polytopes that are not n-cubes. We note that the use
  of different orders in different directions is basic to define high
  order Raviart-Thomas and N\'ed\'elec elements on n-cubes. In the
  following presentation, we consider the scalar order case for
  simplicity.}
\begin{definition}[Set of nodes]
Let us consider a polytope $\square \in \R^d$ represented by
$(\top,\oneb,\zerob)$. Its set $\nodes^k$ of equidistant Lagrangian
nodes of order $k$, in lexicographical notation, are generated
recursively as follows: $\nodes^k \doteq \nodes_{d}^k$, where
\begin{align}\label{eq:nodes_set}
\nodes_{m+1}^p = \{ (\compijk,\beta) : \compijk \in \nodes_{m}^{p- \beta (1-\top(m))}, \beta \in \nodes_1^p \}, \quad \hbox{with } \,
\nodes_1^q = \{ \alpha \in \N^+ : \alpha \leq q \}.
\end{align}
\end{definition}
Given a node $\compijk \in \N^d$ in lexicographical notation and
assuming an equidistant distribution of nodes, its space coordinates
$x_\compijk \in \R^d$ can readily be obtained,
$x_\compijk \doteq \compijk /k$. We note that for n-cubes we recover
the typical tensor product definition of nodes and the corresponding
truncated subset of nodes for n-simplices. Other node generators can also
be considered, especially for very high-order elements (e.g., Fekete
points).

It is basic in \ac{FE} analysis to have an \emph{ownership} relation between
n-faces and nodes. In particular, it is basic to enforce continuity
between \acp{FE} by enforcing continuity of nodal values. In order to
generate the nodes of the polytope that belong to an n-face, we use
the following construction.

First, we generate the local set of nodes, using the definition above, 
for the n-face. Given a $k$-face $(\top,\ext,\anc)$ in $\R^d$, we
consider the \emph{reference} $k$-dim polytope $(\top',\oneb,\zerob)$, where $\top'$ is the restriction of $\top$ to the components that are
extruded, i.e., $\top' \doteq \top \circ m_{\ell g}$ with the mapping
$m_{\ell g} : \{1, \ldots, k\} \rightarrow \{ { j \in \{1, \ldots,
d\}: \ext(j) = 1} \}$. Next, we define the local nodes of the n-face
as the nodes of the reference polytope. It defines the n-faces nodes
and their local coordinates. Finally, we define the linear mapping
from the reference $k$-dim polytope to the $k$-face. The map can be
defined with $k+1$ independent conditions. It can be defined by
enforcing that the mapping maps the anchor vertex of the reference
polytope to the one of the $k$-face and the same for the extrusion of
the anchor vertex to all directions:
$$
m(\zeros) = \v, \qquad m(\es) = \vec{\ev}_{m_{\ell g}(s)}, \, \hbox{ if } \top'(s) = 0, \, \qquad m(\es) = \v + \vec{\ev}_{m_{\ell g}(s)}, \,
\hbox{ if }  \top'(s) = 1.
$$
Since the mapping is linear, it can be written as:
\def\alpbf{\boldsymbol{\alpha}}
$$
m(\x) = \alpbf_0 + x_1 \alpbf_1  + \ldots + x_k \alpbf_k.  
$$
Form the first constraints we get that $\alpbf_0 = \v$. For the other constraints, we get:
$$
m(\es) = \v \top'(s) + \ev_{m_{\ell g}(s)} = \v + \alpbf_s \longrightarrow \alpbf_s = \v(\top'(s)-1) + \ev_{m_{\ell g}(s)}.
$$
Thus, we get:
\begin{align}\label{eq:mapping_poly_n_face}
m(\x) = \v + \sum_{s=1}^k x_s \v(\top'(s)-1) + x_s \ev_{m_{\ell g}(s)} + x_s,
\end{align}
and thus:
$$
m(\x)_i = v_i (1 - \sum_{\substack{\{ s = 1, \ldots, k : \\  \top'(s) = 0 \}} } x_s) + x_{m_{\ell g}^{-1}(i)}.
$$
We could also obtain the expression for the inverse of the mapping $m$ analogously. 
We can readily use the mapping for lexicographical coordinates. As a
result, given a $k$-face, we can define its nodes with a local
numbering based on the lexicographical label of the reference
$k$-face. The local-to-global lexicographical label (where global is
the label of the $d$-dim base polytope) is obtained by applying the
mapping \Eq{mapping_poly_n_face}.

\subsection{Global DOF numbering and conformity}\label{sec:polytope_rotations_and_permutations} 

A basic ingredient in \ac{FE} analysis is the imposition of continuity
among \acp{FE} in order to build conforming global \ac{FE}
spaces. This process is mathematically defined with equivalence
classes on \acp{DOF} (see Sect.~\ref{subsec:global_fe_space}). For example,
functions in the Lagrangian \ac{FE} space are related to geometrical
nodes, and to impose continuity of a function among \acp{FE} is
equivalent to impose continuity of nodal values in the same spatial
position (see Sect.~\ref{subsec:h1_conforming_fes}). In the following,
we provide a mechanism to identify nodes in two different cells that
share the same position to implement the required equivalence
class. The situation is slightly more involved for div-conforming and
curl-conforming \ac{FE} spaces. In these cases, one can still
determine a \ac{DOF} with a node plus n-face ownership (see
Sect.~\ref{subsec:hdiv_conforming_fes} and
\ref{subsec:hcurl_conforming_fes}, respectively). Thus, the
equivalence class in these situations can be formulated as in
Lagrangian \acp{FE} (determine nodes with the same position) at every
n-face separately.

Following Sect.~\ref{subsec:global_fe_space}, a 
node within a cell of our triangulation can be represented as $(b,K)$,
where $b$ is the local cell-wise index of the node and $K$ is the cell
global index. Given an n-face $F$ of the cell, the same node can be
represented with $(b',F,K)$, where $b'$ is an n-face-based local
index. For example, node 8 (cell-wise local index) in the cell of 
Fig. \ref{fig:reference_fe_lagrangian_dofs}
can also be determined as the node 1 (facet-wise local index) of the
n-face 8 (see Fig.~\ref{fig:reference_fe_topology}). This facet-wise local index is determined
by the coordinate system being used at the n-face. For example, the nodes of
n-face 8 in Fig. \ref{fig:reference_fe_lagrangian_dofs} are ordered as 
$(8,12)$ (i.e., first 8 and then 12).  On the other hand,
node indices are represented with the coordinates in a lexicographical
coordinate system, as presented in \eqref{eq:nodes_set}. For example, node with $b=8$
($b'=1$ in n-face 8) is represented with the coordinates $\sbs=(4,1)$
($\sbs'=(1)$ in the n-face).

Let us consider an n-face $F$ in our triangulation, two cells $K^{+}$
(source cell) and $K^{-}$ (target cell) sharing the n-face, and nodes
$(\sbs_+',F,K^{+})$ and $(\sbs'^{-},F,K^{-})$ (with n-face-wise local
indices). The question that must be answered is: are nodes
$(\sbs_+',F,K^{+})$ and $(\sbs'^{-},F,K^{-})$ in the same spatial position?  This
question can be answered with the map $\mathtt{p}_F$ in
\eqref{eq:permmap} that, given the position of the node in the
coordinate system of $F$ in $K^{+}$, provides the one in $K^{-}$.

We note that this mapping is trivial when using structured (possibly
locally adapted) n-cube meshes, since the
local ordering of nodes in an n-face based on increasing local index
leads to the same ordering for all cells containing that n-face; we
say that the mesh is \emph{properly oriented} in this case. However,
2D or 3D unstructured mesh generators might not return properly
oriented meshes, and thus the \ac{FE} code has to deal with the
explicit construction and application of permutations. We also note
that one can always end up with oriented meshes for n-simplices by
simple cell-wise permutations (see, e.g.,
\cite[Sect. 5.5]{monk_finite_2003} and
\cite{rognes_efficient_2009}). After reading n-simplex meshes, these
meshes are always properly oriented in \FEMPAR{} before proceeding to any
computation. While this is also true for 2D n-cube meshes, 3D n-cube
meshes cannot be properly oriented in
general~\cite{agelek_orientation_2017}.

\def\a{\boldsymbol{a}}
\def\B{\boldsymbol{B}}

Let us consider the reference polytope $\hat{K}$ associated to $K^{+}$
and $K^{-}$. In general, the n-face $F$ has a different n-face local
index with respect to the two cells; its corresponding reference
n-face is represented with $\hat{F}^+$ and $\hat{F}^{-}$ for $K^{+}$
and $K^{-}$, respectively. In general, the map between nodes of these
two n-faces can be defined by using \Eq{mapping_poly_n_face}, which is
invertible (since it is linear and full rank). Using this approach,
the map can be generated for arbitrary dimension and polytope topology. 
However, for the particular case of 2D/3D n-cube meshes, we have
implemented this procedure in a more computationally efficient manner. In particular, the
required permutations (mappings) are expressed in terms of a set of
tables, which are stored and set up (filled) by the so-called
{\mytexttt{reference\_fe\_t}} abstract data type in \FEMPAR{}. We refer to
Sect. \ref{subsec:dof_set} for detailed implementation details. 
(Recall that n-simplices meshes do not
actually require this procedure as they can always be properly (re)oriented.)

\def\rotmap{\boldsymbol{\theta}}
\def\orimap{\boldsymbol{\varphi}}

Let us consider the case of 3D n-cube meshes. Vertices are trivial
because there is only one node and no permutation is needed. For edges
and faces, we rely on the three following concepts:
\begin{itemize}
\item Rotation index: Provides the local index of the anchor vertex of
  $F^{-}$ with respect to the coordinate system of $F_+$. When \acp{FE}
  are sharing two edges, we have the following situations. The edge
  can have the same anchor vertex seen from both elements, or not. For
  faces, the anchor vertex can be in 4 positions. It is called
  \emph{rotation} because it represents a map that keeps invariant the
  reference face $\hat{F}^{-}$ and makes the anchor vertices of the source
  and target cells coincide.
\item Orientation index: Given two cells sharing an n-face with the
  same anchor vertex, the orientation index codes the map from the
  coordinate system of the n-face with respect to the first cell to
  the one with respect to the second one.\footnote{In the following,
    one can consider two unit cubes sharing a face. Since all the
    concepts are logical one does not have to take into account the
    real shape of the cells in the physical space. On the other hand,
    we note that the orientation index is invariant to which of the
    two cells sharing the face we select as first and second cell.}.
  For edges, this map is always the identity, because two cells
  sharing an edge with the same anchor node provide the same edge-wise
  node coordinates to its nodes. For faces, the situation is more
  complex, because it involves 2 different possible situations. The
  orientation index is equal to 0 for the identity permutation and 1
  when we have to swap indices. We denote the base face as the face
  with the lowest local index (face $[011|000]$ in
  Fig.~\ref{fig:hex-nfacecode}). Next, we consider two cubes that
  share a face, restricted to the following scenario: 1) the face is
  the base face in at least one of the cubes; 2) the face has the same
  anchor vertex in the two cubes. It is trivial to compute the
  orientation index in these cases. The orientation index in the more
  general case of two cubes sharing a face only restricted to 2),
  i.e., two arbitrary faces with the same anchor vertex, can be
  obtained by composition as follows. If two faces have the same
  orientation index with the base face, they have an orientation index
  equal to 1, and 0 otherwise.
\item Permutation index: An index obtained by composition of the
  rotation and orientation indices (i.e., it ranges from 1 and 2, and
  1 and 8 for edges and faces, respectively), that codifies the final
  mapping between coordinates of two cells as the composition of a
  rotation and a orientation map. We note that the composition of all
  possible rotations and orientations cover all the possible relative
  positions of cells for a conforming mesh.
\end{itemize} 
 
\section{Implementation of \texttt{polytope\_t} and related data types}\label{sec:polytope_implementation}

In \FEMPAR{}, the reference \ac{FE} cell geometry is defined by the
\mytexttt{polytope\_t} data type; see Listing
\ref{lst:polytope_t}. The input needed to define the polytope is the
space dimension \mytexttt{num\_dimensions} and the topology $\top$ in
the 32-bit integer \mytexttt{topology}.

\begin{lstlisting}[language={[03]Fortran},float=h!,caption=The \mytexttt{polytope\_t} data type. ,label=lst:polytope_t, escapeinside={(*@}{@*)}]
type polytope_t
 private
 integer(ip)              :: num_dims
 integer(ip)              :: topology
 integer(ip)              :: num_n_faces 
 integer(ip), allocatable :: n_face_array(:)     
 integer(ip), allocatable :: ijk_to_index(:)
contains
 procedure          :: create                => polytope_create 
 procedure          :: create_facet_iterator => polytope_create_facet_iterator
 procedure          :: free                  => polytope_free 
 procedure, private :: fill_polytope_chain 
 ... ! Rest of getter TBPs that return the values of the member variables or
     ! processed data out of them
end type polytope_t
\end{lstlisting}

Using the ideas in \eqref{eq:chain_n_faces}, \eqref{eq:chain_n_faces_bits}, and \eqref{eq:n_faces_set}, we create the set of all n-faces of
the polytope $(\top,\ext,\anc)$ in the
(private) \mytexttt{fill\_polytope\_chain} \ac{TBP}, which is in turn invoked by
the (public) \mytexttt{create} \ac{TBP}. All n-faces of the polytope have the same topology,
and can be uniquely determined by a 32-bit integer that represents the composition
of $(\ext,\anc)$. We note that the
ordering of n-faces based on $(\ext,\anc)$ mixes n-faces of different
dimensions and it is non-consecutive in general. Thus, we consider an
ordering based first on the n-face dimension, and next by
$(\ext,\anc)$. The set of all n-faces generated by the recursion \Eq{n_faces_set}
are stored in \mytexttt{n\_face\_array}, an array of size \mytexttt{number\_n\_faces}.
This array in particular provides the $(\ext,\anc)$ associated to each n-face. The
inverse mapping (from $(\ext,\anc)$ to the actual numbering) is stored
in the \mytexttt{ijk\_to\_index} array.

It is also possible to iterate over facets
of an n-face, based on \eqref{eq:facetsnface}. The  \mytexttt{create\_facet\_iterator} \ac{TBP}
of \mytexttt{polytope\_t} creates a \mytexttt{facet\_iterator\_t} instance for
a given n-face. \mytexttt{facet\_iterator\_t} is defined in Listing~\ref{lst:facet_iterator}. The n-face $(\ext,\anc)$ is stored in \mytexttt{root}, the
topology can be extracted from its \mytexttt{polytope} pointer member variable. The
iterator over facets is described by two integers, \mytexttt{component}
and \mytexttt{coordinate}, using the ideas in \eqref{eq:facetsnface}. The complexity of the
traversal over facets is coded in \mytexttt{facet\_iterator\_next} and \mytexttt{facet\_iterator\_has\_finished}.

\begin{lstlisting}[language={[03]Fortran},float=h!,caption=The \texttt{facet\_iterator\_t} data type. ,label=lst:facet_iterator, escapeinside={(*@}{@*)}]
type facet_iterator_t
 private 
 type(polytope_t), pointer :: polytope
 integer(ip)               :: root
 integer(ip)               :: component
 integer(ip)               :: coordinate
contains
 ...
 procedure, non_overridable :: first         => facet_iterator_first
 procedure, non_overridable :: next          => facet_iterator_next
 procedure, non_overridable :: has_finished  => facet_iterator_has_finished
 ...
end type facet_iterator_t
\end{lstlisting}

\def\nodes{\mathcal{N}}

With regard to the implementation of nodes within \FEMPAR{}, we provide
the \mytexttt{node\_array\_t} data type to represent the set of nodes
defined in \Eq{nodes_set}; see Listing~\ref{lst:node_array}. It is constructed from a \mytexttt{polytope} and the
order. It provides a \mytexttt{create} \ac{TBP}, where we perform
\Eq{nodes_set} and fill all the resulting nodes in the
\mytexttt{node\_array} array member variable. We number the nodes using a consecutive numbering with increasing lexicographical index. The node array provides the
lexicographical label in one integer. The inverse is stored in
\mytexttt{ijk\_to\_index}. The total number of nodes is stored in
\mytexttt{num\_nodes}. Finally, the space coordinates of nodes are
stored in \mytexttt{coordinates}.

\begin{lstlisting}[language={[03]Fortran},float=h!,caption=The \texttt{node\_array\_t} data type.,label=lst:node_array, escapeinside={(*@}{@*)}]
type node_array_t
 private
 type(polytope_t), pointer      :: polytope
 integer(ip)                    :: order(SPACE_DIM)
 integer(ip)                    :: num_nodes
 integer(ip), allocatable       :: node_array(:)
 integer(ip), allocatable       :: ijk_to_index(:)
 integer(ip), allocatable       :: coordinates(:,:)
contains
  procedure                  :: create                   => node_array_create
  procedure                  :: free                     => node_array_free
  procedure                  :: create_node_iterator     => node_array_create_node_iterator
  procedure                  :: get_num_nodes         => node_array_get_num_nodes
  procedure        , private :: fill                     => node_array_fill
  procedure, nopass, private :: fill_permutations        => node_array_fill_permutations
  procedure, nopass, private :: compute_num_rot_and_perm => node_array_compute_num_rot_and_perm
end type node_array_t
\end{lstlisting}

We also provide the \mytexttt{node\_iterator\_t} object (see Listing~\ref{lst:node_iterator}), which iterates over the nodes of an n-face (stored in \mytexttt{n\_face}) using \Eq{nodes_set} and \Eq{mapping_poly_n_face}. 
 It has a pointer to
the \mytexttt{node\_array} of the base polytope.  Internally, it goes through the nodes of \mytexttt{n\_face} (using \Eq{nodes_set}) (the current node being stored in \mytexttt{displacement}), which can be translated to the base polytope node numbering using \Eq{mapping_poly_n_face} (stored in \mytexttt{coordinate}); the \mytexttt{coordinate} is computed on demand by calling the \ac{TBP} \mytexttt{node\_iterator\_current\_ijk}. The \mytexttt{own\_boundary} logical allows one to iterate over the nodes considering the n-face as an open or closed set. We note that the \mytexttt{create} TBP of \mytexttt{node\_array\_t} relies on \mytexttt{node\_iterator\_t}.

\begin{lstlisting}[language={[03]Fortran},float=h!,caption=The \texttt{node\_iterator\_t} data type.,label=lst:node_iterator, escapeinside={(*@}{@*)}]
type node_iterator_t
 private 
 type(node_array_t), pointer :: node_array
 logical                     :: own_boundary
 integer(ip)                 :: n_face
 integer(ip)                 :: topology
 integer(ip)                 :: displacement(0:SPACE_DIM-1)
 integer(ip)                 :: coordinate(0:SPACE_DIM-1)
 ...
contains
 ...
end type node_iterator_t
\end{lstlisting}

\section{The \texttt{polynomial\_t} abstraction}  \label{sec:polynomial}

\def\mtt#1{{\mytexttt{#1}}}

In \FEMPAR{}, the definition of shape functions is not hard-coded, as usually done in most \ac{FE} codes. Such approach has severe limitations: 1) it is not practical for high order discretizations, and the code cannot be written for an arbitrary order; 2) it involves a huge number of code lines with the analytical expression of shape functions for a given set of available orders (see the discussion in \cite{bangerth_data_2009}); and 3) it does not allow for dimension-independent code. Instead, we consider a framework based on the concepts in Sect.~\ref{subsec:shape_functions_construction}, in which one considers a pre-basis, defines the moments, and performs a change of basis. The pre-basis is defined using the product of 1D functions (e.g., the Cartesian product), and the 1D function generator is written in terms of the (arbitrary) order. Our machinery for the generation of 1D functions has been restricted for the moment to polynomial functions in one variable, namely Lagrangian polynomials, monomials, and B-splines, but the implementation can be extended to other choices. The product of 1D functions can be a Cartesian product of 1D Lagrange polynomials (or monomials), to define $\polsp_\order$ spaces on n-cubes, or a reduced combination of monomials to define $\polspr_k$ spaces on n-simplices.

The definition of the reference \ac{FE} functional space relies on the \mytexttt{polynomial\_t} data type in Listing~\ref{lst:polynomial}, which represents a polynomial in one variable, i.e., $p(x) = \sum_{i=0}^k a_ix^k$. Thus, a 1D polynomial is defined in terms of its order $k$ and a set of $k+1$ coefficients $\{a_i\}_{i=0}^k$, stored in \mytexttt{order} and the \mtt{coefficients} array, respectively. Different type extensions of \mytexttt{polynomial\_t} have been considered so far, namely \mtt{lagrange\_polynomial\_t} and \mtt{monomial\_t}. The first one generates a Lagrangian polynomial as in Sect.~\ref{subsec:polynomial_spaces_construction}, in which the \mtt{coefficients} array has in its first \mtt{order} entries the coordinates of the nodes and in the last entry the coefficient $\frac{1}{ \Pi_{n \in \nodes_k \setminus \{m\} } (x_m - x_s) }$ in \eqref{eq:lagbas}. The \mtt{monomial\_t} extension represents $x^k$ where $k$ is its order. It is just a trivial case of \mytexttt{polynomial\_t} for optimization purposes that is uniquely defined by the order (the coefficients array is not needed). We also consider the \mtt{polynomial\_basis\_t} data type, which is just a set (array) of (polymorphic) polynomials. 

\begin{lstlisting}[float=htbp,language={[03]Fortran},escapechar=@,caption=The
\mytexttt{polynomial\_t} data type and related data types.,
label={lst:polynomial}]
type, abstract :: polynomial_t
   private
   integer(ip)           :: order
   real(rp), allocatable :: coefficients(:)
 contains
   ...
   procedure(...)        , deferred :: get_values
   procedure(...), nopass, deferred :: generate_basis
end type polynomial_t
  
type, extends(polynomial_t) :: lagrange_polynomial_t
 contains
   procedure          :: get_values      => lagrange_polynomial_get_values
   procedure, nopass  :: generate_basis  => lagrange_polynomial_generate_basis
end type lagrange_polynomial_t
  
type, extends(polynomial_t) :: monomial_t
 contains
   procedure          :: get_values      => monomial_get_values
   procedure, nopass  :: generate_basis  => monomial_generate_basis
end type monomial_t
  
type :: polynomial_basis_t
   private
   class(polynomial_t), allocatable :: polynomials(:)
 contains
   ... 
end type polynomial_basis_t
  
type :: tensor_product_polynomial_space_t
   private
   integer(ip)                    :: num_dims
   integer(ip)                    :: num_pols_dim(SPACE_DIM)
   type(polynomial_basis_t)       :: polynomial_1D_basis(SPACE_DIM)
   type(allocatable_array_rp3_t)  :: work_shape_data(SPACE_DIM)
 contains
   ...
   procedure :: evaluate => tensor_product_polynomial_space_evaluate
end type tensor_product_polynomial_space_t
  
type, extends(tensor_product_polynomial_space_t) :: truncated_tensor_product_polynomial_space_t
 contains
   ... 
end type truncated_tensor_product_polynomial_space_t
\end{lstlisting}

Up to this point, we have defined Lagrange polynomials and monomials in one variable. \mtt{lagrange\_polynomial\_t} and \mtt{monomial\_t} also provide the binding \mtt{generate\_basis} that generates a Lagrangian and monomial basis of polynomials, for a given order $k$. The result of this subroutine is a \mtt{polynomial\_basis\_t} that includes as many polynomials as the polynomial space dimension. In the case of the Lagrangian basis, it implements the basis $\mathcal{L}^k$ in Sect.~\ref{subsec:polynomial_spaces_construction}, whereas the binding for monomials simply implements $\{ x^i \}_{i=0}^k$.

The next step is to generate higher dimensional spaces. We consider two types of spaces. The first one is a space that can be generated as the Cartesian product of 1D spaces, implemented in the data type \mtt{tensor\_product\_polynomial\_space\_t}. This data type is defined through the number of space dimensions and as many \mtt{polynomial\_basis\_t} as space dimensions. This data type can be applied to any combination of 1D spaces. E.g., in the case of 1D Lagrange bases (possibly with different order and nodes per dimension), it leads to the multi-dimensional basis in \eqref{eq:lagbasmd}. Thus, with this data type and Lagrangian 1D bases we generate the Lagrangian \ac{FE} spaces on top of n-cube cells, i.e., the $\polsp_{\order}$ space of polynomials.\footnote{Analogously, one could generate serendipity elements only by changing the generation of the multi-dimensional space in terms of 1D ones.}

Furthermore, we also consider the
\mtt{truncated\_tensor\_product\_polynomial\_space\_t} extension that
generates Lagrangian \ac{FE} spaces on n-simplices, i.e., the
$\polspr_k$ space of polynomials. In this case, the
\mtt{generate\_basis} \acp{TBP} of \mtt{monomial\_t} should be used to create the monomial
1D bases per direction and the order should also be the same for all
directions. Otherwise, the resulting multi-variable function would
have no sense. Next, the combination of 1D monomials only
involves terms such that $ | \mid | \leq k $ (see
Sect.~\ref{subsec:polynomial_spaces_construction}), to generate a
pre-basis for \ac{FE} spaces on tetrahedra, i.e., the $\polspr_k$
space of polynomials. 

We note that with these abstract representations of polynomial spaces one can define the reference \ac{FE} local space. However, unless one considers 1D Lagrangian basis and tensor product polynomials on n-cubes, the resulting basis is not the shape functions basis. Even in the case of Lagrangian n-simplices, a change-of-basis is needed, using the procedure in Sect.~\ref{subsec:shape_functions_construction} taking  nodal values as moments. In Sect.~\ref{sec:reference_fe_implementors}, we show how we can define the shape function basis for the case of div-conforming \acp{FE} of arbitrary order. The same ideas apply for grad-conforming Lagrangian \acp{FE} on n-simplices and curl-conforming \acp{FE} in general, but are not included for the sake of brevity. 

\section{The \texttt{reference\_fe\_t} abstraction}  \label{sec:reference_fe}

In this section, we introduce the \mytexttt{reference\_fe\_t} data type.
This data
type is the \ac{OO} representation of the standard mathematical definition
of a reference \ac{FE} presented in Sect.~\ref{subsec:fe_concept},
namely, a reference cell geometry $\georef$, a functional space
$\reffesp$, and a set of \acp{DOF} $\momentsref$ defined on top of it. The
\mytexttt{reference\_fe\_t} is a central abstraction in a \ac{FE} library and
must be judiciously designed to be extensible and reusable. In particular,
it must not only  accommodate Lagrangian \acp{FE}, but also other (more
involved/general) spaces like Raviart-Thomas or edge \acp{FE}, \ac{DG} methods, and B-spline
patches. An extensible and reusable design of \mytexttt{reference\_fe\_t} should allow one to,
e.g., easily incorporate new local functional spaces that were not
originally considered, and to do so without having to rewrite (and thus
recompile) any code that is grounded on the set of methods provided by \mytexttt{reference\_fe\_t}. 
To this end, in \FEMPAR{}, \mytexttt{reference\_fe\_t} 
is an {\em abstract} data type that serves as a template equipped with a set of member
variables and deferred bindings that subclasses have to set up and
implement (i.e., override), respectively, in order to complete the
description of the concrete \ac{FE} space at hand.
The definition of the \mytexttt{reference\_fe\_t} data type, a 
classification of its member variables into three different categories 
(corresponding to the three ingredients in Ciarlet's definition), 
and an enumeration of its most relevant regular and deferred bindings, 
are shown in Listing~\ref{lst:reference_fe}.

\begin{lstlisting}[float=htbp,language={[03]Fortran},escapechar=@,caption={The \mytexttt{reference\_fe\_t} abstract type, a classification of its member variables, and an enumeration of its most relevant regular and deferred bindings.},label={lst:reference_fe}]
type, abstract :: reference_fe_t
  private
  ! Member variables of reference_fe_t can be grouped into three different categories:
  ! 1. Description of topology of the reference cell @{\color{blue} $\hat{K}$}@ @{\color{blue}(see Section~\ref{subsec:reference_fe_topology})}@
  integer(ip)                                :: num_dims                  @\label{loc:top_start}@
  character(:)                 , allocatable :: topology
  integer(ip)                                :: num_n_faces
  integer(ip)                                :: ptr_n_faces_x_dim(SPACE_DIM+2)
  type(list_t)                               :: vertices_n_face
  type(list_t)                               :: facets_n_face                   @\label{loc:top_stop}@

  ! 2. Description of finite element space on top of @{\color{blue} $\reffesp$}@ @{\color{blue}(see Section~\ref{subsec:reference_fe_space})}@
  character(:)                 , allocatable :: fe_type                         @\label{loc:fe_start}@
  integer(ip)                                :: order                           
  character(:)                 , allocatable :: field_type
  integer(ip)                                :: num_shape_functions             @\label{loc:fe_stop}@

  ! 3. Description of the set degrees of freedom @{\color{blue} $\momentsref$}@ @{\color{blue} (see Section~\ref{subsec:dof_set})}@
  logical                                    :: conformity                      @\label{loc:dofs_start}@
  logical                                    :: continuity
  type(list_t)                               :: dofs_n_face  
  type(list_t)                               :: own_dofs_n_face
  type(allocatable_array_ip2_t), allocatable :: own_dofs_permutations(:)        @\label{loc:dofs_stop}@
contains 
  public
  ! ****Regular bindings****
  procedure, non_overridable :: get_num_dims
  procedure, non_overridable :: get_topology
  ... ! Rest of getter TBPs that return either the values of the member variables 
      ! themselves, a reference (pointer) to them (to avoid the overhead associated 
      ! to copies), or other processed data out of them
  procedure                  :: permute_dof_lid_n_face                  ! @{\color{blue} See Section \ref{subsec:dof_set}}@ @\label{loc:permute_dof_lid_n_face}@
  
  ! ****Deferred bindings****
  procedure(...), deferred   :: create                                  ! @{\color{blue} See Section \ref{subsec:reference_fe_creation}}@
  
  ! Cell-integration related bindings
  procedure(...), deferred   :: create_quadrature                       ! @{\color{blue} See Section \ref{subsec:numerical_quadrature}}@
  procedure(...), deferred   :: create_interpolation                    ! @{\color{blue} See Section \ref{subsec:interpolation}}@
  procedure(...), deferred   :: apply_cell_map                          ! @{\color{blue} See Section \ref{subsec:cell_integrator}}@
  procedure(...), deferred   :: get_default_quadrature_degree           ! @{\color{blue} See Section \ref{sec:fe_space}}@

  ! Facet-integration related bindings
  procedure(...), deferred   :: create_facet_quadrature                  ! @{\color{blue} See Section \ref{subsec:face_quadrature}}@
  procedure(...), deferred   :: create_facet_interpolation               ! @{\color{blue} See Section \ref{subsec:face_mapping}@
  procedure(...), deferred   :: create_interpolation_restricted_to_facet ! @{\color{blue} See Section \ref{subsec:face_integrator}}@
  procedure(...), deferred   :: fill_qpoints_perm                        ! @{\color{blue} See Section \ref{subsec:face_integrator}}@

  ! Extract data out of interpolation_t into user-friendly data structures 
  ! @{\color{blue}(See Section~\ref{subsec:cell_integrator} and~\ref{sec:face_integration})}@
  procedure(...), deferred :: get_values_scalar => cell_integrator_get_values_scalar
  procedure(...), deferred :: get_values_vector => cell_integrator_get_values_vector
  procedure(...), deferred :: get_values_tensor => cell_integrator_get_values_tensor
  generic                  :: get_values => get_values_scalar, get_values_vector, get_values_tensor
  
  ... ! Extract shape functions gradients out of interpolation_t
  ... ! Extract shape functions hessians out of interpolation_t
  ... ! Extract shape functions curls out of interpolation_t
  ... ! Extract shape functions divergences out of interpolation_t

  ! Restriction of global FE functions to cells and facets
  ! @{\color{blue}(See Section~\ref{subsec:fe_function})}@
  procedure(...), deferred :: evaluate_fe_function_scalar   @\label{loc:evaluate_fe_function_start}@
  procedure(...), deferred :: evaluate_fe_function_vector
  procedure(...), deferred :: evaluate_fe_function_tensor
  generic                  :: evaluate_fe_function_ => evaluate_fe_function_scalar, ...vector, ...tensor

  ... ! Variants of the above three deferred bindings for gradients, hessians, etc. @\label{loc:evaluate_fe_function_stop}@

  ! Generation of a global DOF numbering
  procedure(...), deferred   :: fill_own_dofs_permutations               ! @{\color{blue} See Section \ref{subsec:dof_set}}@
end type reference_fe_t
\end{lstlisting}

This section is structured as follows. The member variables in each of the three
aforementioned categories are covered in detail in 
Sect.~\ref{subsec:reference_fe_topology}-\ref{subsec:dof_set}, respectively. 
In Sect.~\ref{subsec:reference_fe_creation}, we discuss the \ac{OO} design
pattern chosen in \FEMPAR{} for the creation of
\mytexttt{reference\_fe\_t} polymorphic instances, and describe the
arguments that uniquely define a subclass of this data type; these  are in line with
its mathematical definition. In Sect.~\ref{subsec:reference_fe_subclasses}, we enumerate and briefly describe 
the subclasses of \mytexttt{reference\_fe\_t} currently available in \FEMPAR{}. We note that the section is not self-contained as most of the deferred
bindings of \mytexttt{reference\_fe\_t} are not covered here. These involve interactions 
with other data types in our \ac{OO} design, and will be described in the sections in which these
interactions are exposed. Code comments in Listing~\ref{lst:reference_fe} serve as a 
table of contents with the article sections in which these deferred bindings are covered.

\subsection{The reference cell topology} \label{subsec:reference_fe_topology}

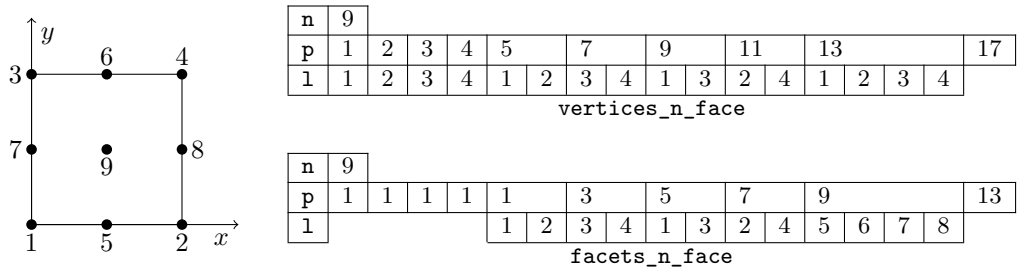
\begin{figure}[htbp]
\begin{tabular}{ccc}
\begin{minipage}{0.2\textwidth}
\begin{tikzpicture}
\coordinate (1) at (0,0);
\coordinate (2) at (2,0);
\coordinate (3) at (0,2);
\coordinate (4) at (2,2);
\coordinate (5) at ($(1)!.5!(2)$);
\coordinate (6) at ($(3)!.5!(4)$);
\coordinate (7) at ($(1)!.5!(3)$);
\coordinate (8) at ($(2)!.5!(4)$);
\coordinate (9) at ($(7)!.5!(8)$);
\foreach \i in {1,2,5,9}
    \fill (\i) circle (2pt) node [below] {\i};
\foreach \i in {4,6}
    \fill (\i) circle (2pt) node [above] {\i};
\fill (3) circle (2pt) node [left] {3};
\fill (7) circle (2pt) node [left] {7};
\fill (8) circle (2pt) node [right] {8};
\draw (1) --(2) --(4) --(3) --(1);
\draw[->] (0,0) -- (2.75,0) node[anchor=north east] {$x$};
\draw[->] (0,0) -- (0,2.75) node[anchor=north west] {$y$};
\end{tikzpicture}
\end{minipage} & &
\begin{minipage}{0.65\textwidth}
\small
\begin{tabular}{|c|c|c|c|c|c|c|c|c|c|c|c|c|c|c|c|c|c|}
\cline{1-2}
\mytexttt{n} & 9 & \multicolumn{16}{c}{} \\
\hline
\mytexttt{p} & 1 & 2 & 3 & 4 & \multicolumn{2}{l|}{5} &\multicolumn{2}{l|}{7} & \multicolumn{2}{l|}{9} & \multicolumn{2}{l|}{11} & \multicolumn{4}{l|}{13} & 17    \\
\hline
\mytexttt{l} & 1 & 2 & 3 & 4 & 1 & 2 & 3 & 4 & 1 & 3 & 2 & 4 & 1 & 2 & 3 & 4 & \multicolumn{1}{c}{}  \\
\cline{1-17}
\multicolumn{18}{c}{\mytexttt{vertices\_n\_face}} \\
\multicolumn{18}{c}{}\\
\cline{1-2}
\mytexttt{n} & 9 & \multicolumn{16}{c}{} \\
\hline
\mytexttt{p} & 1 & 1 & 1 & 1 & \multicolumn{2}{l|}{1} &\multicolumn{2}{l|}{3} & \multicolumn{2}{l|}{5} & \multicolumn{2}{l|}{7} & \multicolumn{4}{l|}{9} & 13    \\
\hline
\mytexttt{l} & \multicolumn{4}{c|}{} & 1 & 2 & 3 & 4 & 1 & 3 & 2 & 4 & 5 & 6 & 7 & 8 & \multicolumn{1}{c}{}  \\
\cline{1-1} \cline{6-17}
\multicolumn{18}{c}{\mytexttt{facets\_n\_face}} \\
\end{tabular}
\end{minipage}
\end{tabular}
\caption{Numbering convention for n-faces with $\georef$ a quadrilateral (left) and the status of 
\mytexttt{vertices\_n\_face} and \mytexttt{facets\_n\_face} corresponding
to that numbering (right). \mytexttt{n}, \mytexttt{p(n+1)}, and \mytexttt{l(p(n+1)-1)} are private member
variables of \mytexttt{type(list\_t)} storing the number of entities,
the start position in \mytexttt{l(:)} of the list associated to each entity, and 
the identifiers associated of all lists gathered in a single array, respectively. \label{fig:reference_fe_topology}}
\end{figure}

The reference cell $\georef$ is a polytope. Therefore, following Sect.~\ref{sec:polytope_topology}, 
it can be described with the topology,
coded as a set of $d$ bits, where $d$ is the dimension of the
polytope.  The reference cell topology is generated using
\mytexttt{polytope\_t} described in Sect.~\ref{sec:polytope_implementation},
which offers methods like composition and local numbering of
n-faces. Polytope topologies include triangles and quadrilaterals in
2D, and tetrahedra, hexahedra, prysms, and pyramids in 3D. The member
variables in charge of the description of the reference cell topology $\georef$
are shown in Lines~\ref{loc:top_start}-\ref{loc:top_stop} of Listing~\ref{lst:reference_fe}.
The user must provide the topology and dimension of the polytope
to define $\georef$, stored in the member variables
\mytexttt{topology} and \mytexttt{num\_dimensions}, respectively. A set of
\emph{getters} return this basic information, and other related data that
can be generated out of them, e.g., the number of n-faces in the boundary
of the cell is stored in the
\mytexttt{num\_n\_faces} member variable. The list of vertex identifiers per each n-face
and the list of facets (of dimension $n-1$) per each n-face are stored in
\mytexttt{vertices\_n\_face} and \mytexttt{facets\_n\_face}, respectively;
see Fig.~\ref{fig:reference_fe_topology} for an illustration of
these member variables and the data type \mytexttt{list\_t} used in
\FEMPAR{} to store and traverse lists.

The \FEMPAR{} data type \mytexttt{list\_t} stores a set of
(variable-sized) lists of integer identifiers, one per each entity; in
this particular scenario, entities are n-faces. As shown in Fig.~\ref{fig:reference_fe_topology}, 
the current implementation of this data type uses a compressed
storage layout as, e.g., in compressed storage formats for sparse
graphs. In order to preserve encapsulation and data hiding,
\mytexttt{list\_t} offers a rich set of \acp{TBP} that lets users to set up
(step by a step) a new \mytexttt{list\_t} instance; this type also
provides a \mytexttt{list\_iterator\_t} type that lets them to
sequentially read/write each of the integer identifiers of the list
associated to an entity. The  code snippet in Listing~\ref{lst:vertices_n_face_example} 
illustrates how to iterate and print the identifiers of those vertices belonging to the
n-face with identifier \mytexttt{n\_face\_lid}.

\begin{lstlisting}[float=htbp,language={[03]Fortran},escapechar=@,caption={User-level code that illustrates how to print to screen those (local within cell) vertex identifiers belonging to n-face with (local within cell) 
identifier \mytexttt{n\_face\_lid}.},label={lst:vertices_n_face_example}]
subroutine print_vertex_lids_in_n_face(reference_fe, n_face_lid)
  class(reference_fe_t), intent(in) :: reference_fe
  integer(ip)          , intent(in) :: n_face_lid
  type(list_iterator_t) :: iterator
  iterator = reference%create_vertices_n_face_iterator(n_face_lid)
  do while (.not. iterator%has_finished())
     write(*,*) 'vertex LID (within cell) ', iterator%current(), & 
                'in n_face LID (within cell) ', n_face_lid
     call iterator%next()
  end do
end subroutine print_vertex_lids_in_n_face
\end{lstlisting}

The number of n-faces of any dimension can be easily computed from
\mytexttt{ptr\_n\_faces\_x\_dim}.  We note that
\mytexttt{ptr\_n\_faces\_x\_dim} is not a \mytexttt{list\_t} instance, since 
we adopt the convention that n-faces are numbered from the lowest to highest 
dimension, and thus only the $\mytexttt{p}$ array of the list is actually needed (see Fig.~\ref{fig:reference_fe_topology}). In the example in
Fig.~\ref{fig:reference_fe_topology}, the value of this array is $\{1,5,9,10\}$,
since we have 4 vertices (dimension 0), 4 facets or edges (dimension
1), and 1 cell (dimension 2).

\subsection{The reference \ac{FE} space} \label{subsec:reference_fe_space}

For a given cell topology, different definitions of functional spaces
and sets of \acp{DOF} are possible, e.g., the ones of the nodal Lagrangian
grad-conforming reference \ac{FE} in Sect. \ref{subsec:h1_conforming_fes}, the Raviart-Thomas
div-conforming reference \ac{FE} in Sect. \ref{subsec:hdiv_conforming_fes}, or the
curl-conforming N\'ed\'elec reference \ac{FE} in Sect. \ref{subsec:hcurl_conforming_fes}. The
member variables of \mytexttt{reference\_fe\_t}
required to describe the functional space $\reffesp$ with support
on $\georef$ are encompassed within Lines~\ref{loc:fe_start}-\ref{loc:fe_stop} of 
Listing~\ref{lst:reference_fe}.

The local \ac{FE} space $\reffesp$ is determined by the member variables
\mytexttt{fe\_type}, (in some cases) \mytexttt{field\_type}, and
\mytexttt{order}.  \mytexttt{fe\_type} uniquely identifies the concrete FE
space at hand. Possible values are provided by means of the public
parameter constants \mytexttt{fe\_type\_lagrangian},
\mytexttt{fe\_type\_raviart\_thomas}, and \mytexttt{fe\_type\_nedelec} corresponding to the
\mytexttt{reference\_fe\_t} implementors currently supported in
\FEMPAR{}; see Sect.~\ref{subsec:reference_fe_subclasses} for
additional details on those. \mytexttt{field\_type} identifies the
``type'' of physical field being discretized, i.e., whether it is
scalar, vector-valued, etc. There are \ac{FE} spaces that are inherently
vector-valued such as, e.g., Raviart-Thomas and edge \acp{FE}. However,
Lagrangian \acp{FE} can be either used to discretize scalar, vector, or
tensor-valued fields, and \mytexttt{field\_type} must be provided.  We
assume that $\reffesp$ can be parameterized with respect to an order,
which is stored in \mytexttt{order}. Out of these values, we can
generate additional data, e.g., the number of shape functions is
stored in \mytexttt{num\_shape\_functions}. For example, for
(scalar-valued) bi-quadratic (2D) and tri-quadratic (3D) Lagrangian
\acp{FE}, the \mytexttt{field\_type} is scalar, \mytexttt{num\_components} is
equal to 1, \mytexttt{order} is equal to 2, and
\mytexttt{num\_shape\_functions} is equal to 9 and 27, respectively. 

\subsection{The set of local \acp{DOF}}\label{subsec:dof_set}

Additional data is required to describe the set of \acp{DOF}
$\momentsref$ for $\reffesp$. In particular, the member variables 
encompassed within Lines~\ref{loc:dofs_start}-\ref{loc:dofs_stop} of 
Listing~\ref{lst:reference_fe} serve this purpose. 

The \mytexttt{conformity} member variable
determines whether the global \ac{FE} space $\trialsph$ is conforming with
respect to the infinite-dimensional space $\trialsp$, i.e.,
whether $\trialsph \subset \trialsp$ or not. It is used to describe the n-face
that owns every \ac{DOF}, which is required to enforce conformity of the
global \ac{FE} space through equivalence classes (see Sect.
\ref{sec:fe_meth}). E.g., for Lagrangian \acp{FE}, setting it to \mytexttt{.true.}
results in a grad-conforming global \ac{FE} space, whereas setting it to
\mytexttt{.false.} it results in a discontinuous space for \ac{DG} methods.
It is conceptually possible to set it to \mytexttt{.true.} on some cells and false on others,
leading to the CDG method in \cite{badia_adaptive_2013}. On the other hand, the
\mytexttt{continuity} member variable is only determined by $\trialsp$,
and tells us whether $\trialsp$ admits a trace operator. Roughly
speaking, it tells us whether we must enforce some type of continuity
at the discrete level to preserve conformity, e.g., full, tangential,
or normal traces for $H^1(\Omega)$, $\Hcurl$, and $\Hdiv$, respectively. The
value of \mytexttt{continuity} is \mytexttt{.false.} when $\trialsp = L^2(\Omega)$,
since no continuity is required. When \mytexttt{continuity} is \mytexttt{.false.},
\mytexttt{conformity} must be \mytexttt{.true.}. \mytexttt{continuity} is barely used 
(see discussion in next paragraph).

\begin{figure}[htbp]
\begin{tabular}{ccc}
\begin{minipage}{0.2\textwidth}
\begin{tikzpicture}
\coordinate (1) at (0,0);
\coordinate (2) at (2,0);
\coordinate (3) at (0,2);
\coordinate (4) at (2,2);
\coordinate (5) at ($(1)!0.33!(2)$);
\coordinate (6) at ($(1)!.66!(2)$);
\coordinate (7) at ($(3)!0.33!(4)$);
\coordinate (8) at ($(3)!.66!(4)$);
\coordinate (9) at ($(1)!0.33!(3)$);
\coordinate (10) at ($(1)!.66!(3)$);
\coordinate (11) at ($(2)!0.33!(4)$);
\coordinate (12) at ($(2)!.66!(4)$);
\coordinate (13) at ($(10)!0.33!(12)$);
\coordinate (14) at ($(10)!.66!(12)$);
\coordinate (15) at ($(9)!0.33!(11)$);
\coordinate (16) at ($(9)!.66!(11)$);

\fill (1) circle (2pt) node [below] {1};
\fill (5) circle (2pt) node [below] {2};
\fill (6) circle (2pt) node [below] {3};
\fill (2) circle (2pt) node [below] {4};

\fill (3) circle (2pt) node [left] {13};
\fill (7) circle (2pt) node [above] {14};
\fill (8) circle (2pt) node [above] {15};
\fill (4) circle (2pt) node [above] {16};

\fill (9) circle (2pt) node [left] {5};
\fill (10) circle (2pt) node [left] {9};
\fill (11) circle (2pt) node [right] {8};
\fill (12) circle (2pt) node [right] {12};

\fill (13) circle (2pt) node [right] {10};
\fill (14) circle (2pt) node [right] {11};
\fill (15) circle (2pt) node [right] {6};
\fill (16) circle (2pt) node [right] {7};
\draw (1) --(2) --(4) --(3) --(1);
\draw[->] (0,0) -- (2.75,0) node[anchor=north east] {$x$};
\draw[->] (0,0) -- (0,2.75) node[anchor=north west] {$y$};
\end{tikzpicture}
\end{minipage} & &
\begin{minipage}{0.65\textwidth}
\small
\begin{tabular}{|c|c|c|c|c|c|c|c|c|c|c|c|c|c|c|c|c|c|}
\cline{1-2}
\mytexttt{n} & 9 & \multicolumn{16}{c}{} \\
\hline
\mytexttt{p} & 1 & 2 & 3 & 4 & \multicolumn{2}{l|}{5} &\multicolumn{2}{l|}{7} & \multicolumn{2}{l|}{9} & \multicolumn{2}{l|}{11} & \multicolumn{4}{l|}{13} & 17    \\
\hline
\mytexttt{l} & 1 & 4 & 13 & 16 & 2 & 3 & 14 & 15 & 5 & 9 & 8 & 12 & 6 & 7 & 10 & 11 & \multicolumn{1}{c}{}  \\
\cline{1-17}
\multicolumn{18}{c}{\mytexttt{own\_dofs\_n\_face (conformity==.true.)}} \\
\multicolumn{18}{c}{}\\
\cline{1-2}
\mytexttt{n} & 9 & \multicolumn{16}{c}{} \\
\hline
\mytexttt{p} & 1 & 1 & 1 & 1 & 1 & 1 & 1 & 1 & \multicolumn{8}{l|}{1} & 17 \\
\hline
\mytexttt{l} & \multicolumn{8}{c|}{} & 1 & 2 & 3 & 4 & 5 & 6 & ... & 16 & \multicolumn{1}{c}{}  \\
\cline{1-1} \cline{10-17}
\multicolumn{18}{c}{\mytexttt{own\_dofs\_n\_face (conformity==.false.)}} \\
\end{tabular}
\end{minipage}
\end{tabular}
\caption{Numbering convention for the \acp{DOF} of an (scalar-valued) bi-cubic Lagrangian \ac{FE} on top of a quadrilateral (left) and the status of 
\mytexttt{own\_dofs\_n\_face} for this \mytexttt{reference\_fe\_t} in its CG (right,top) and \ac{DG} forms (right,bottom). \label{fig:reference_fe_lagrangian_dofs}}
\end{figure}
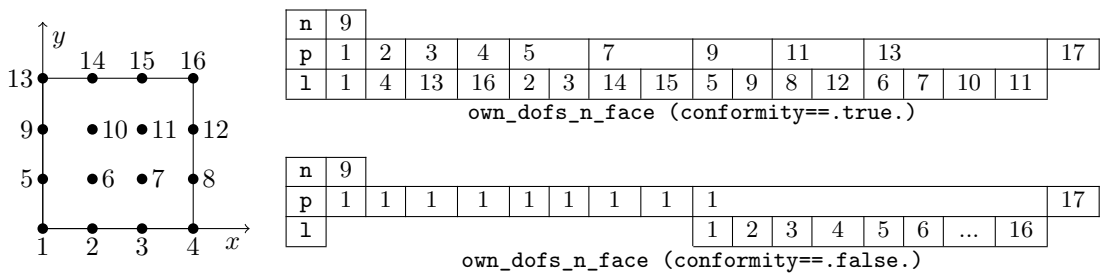

The value of \mytexttt{conformity} is used to generate the 
\mytexttt{own\_dofs\_n\_face} member variable of type \mytexttt{list\_t}.
This member variable stores, for every n-face, the \acp{DOF} it
owns; see Fig.~\ref{fig:reference_fe_lagrangian_dofs}. For CG
methods, the notion of ownership is related to the geometrical
location. For \ac{DG} \acp{FE}, although node functionals are still
geometrically located on the boundary of the cell, they are
nevertheless owned by the cell, and considered as interior \acp{DOF}, since
there is no global conformity to be enforced. This array is heavily
used to generate the global \ac{DOF} numbering.\footnote{We can consider three levels of \ac{DOF} numbering: the cell-wise \ac{DOF} numbering (referred to as local \acp{DOF}), the subdomain-wise \ac{DOF} numbering (referred to as global \acp{DOF}), and a full domain global \acp{DOF}. The latter numbering is never created/required in \FEMPAR{}. In serial environments, the latter two match.} On the other hand, the
\mytexttt{dofs\_n\_face} member variable, determines, for a given n-face, the set of
\acp{DOF} such that their respective shape functions are non-zero on the
n-face. The \mytexttt{continuity} member variable is (currently) only used for \ac{DG} methods in parallel
distributed-memory environments. In particular, in order to decide whether to associate or not 
a global \ac{DOF} identifier to nodes on the interface facets of ghost cells 
(and thus to be able to define non-singular 
sub-assembled matrices for the \ac{DD} methods in \cite{dryja_bddc_2007} for \ac{DG} discretizations). The \mytexttt{dofs\_n\_face}
member variable is used when \mytexttt{continuity} is \mytexttt{.true.} and a 
global \ac{DOF} numbering is to be generated, 
and also might be used by triangulation subclasses (see Sect.~\ref{sec:triangulation}) 
in order to extract the coordinates of those nodes on top of a vertex, edge, or face  
(using the \mytexttt{dofs\_n\_face} member variable of the  \mytexttt{reference\_fe\_t} 
instance that describes the geometry of the cell). For example, in
Fig.~\ref{fig:reference_fe_lagrangian_dofs}, the list corresponding
to n-face with identifier 8 in \mytexttt{dofs\_n\_face} is
\{4,8,12,16\}.

The \mytexttt{reference\_fe\_t} data type plays a crucial role in the algorithm in charge 
of assigning global \ac{DOF} identifiers to node functionals distributed over the interior of the 
triangulation cells and their boundary n-faces. (This algorithm, which is is covered in detail in Sect.~\ref{sec:fe_space}, 
is grounded on the notion of equivalence classes introduced in Sect.~\ref{sec:fe_meth}.) In particular, the function-like 
(regular) binding 
referred to as \mytexttt{permute\_dof\_lid\_n\_face} (see Line~\ref{loc:permute_dof_lid_n_face} of 
Listing~\ref{lst:reference_fe}) implements the mapping 
$\mathtt{p}_F$ in \eqref{eq:permmap}. This function takes as input the so-called permutation index 
in Sect.~\ref{sec:polytope_rotations_and_permutations}, the local index of a node within an n-face of given dimension
(e.g., in 3D, either 0 for vertices, 1 for edges, and 2 for faces) from the perspective of a source cell, and returns the 
local index of a node within that n-face from the perspective of the target cell.\footnote{We note that the responsibility of determining
the permutation index does not lay on \mytexttt{reference\_fe\_t}, but on the abstraction of 
\FEMPAR{} that represents the mesh of the computational domain; see Sect.~\ref{sec:triangulation}.} 
This is in particular the transformation
that we have to apply when global \ac{DOF} identifiers have been already assigned to n-face nodes in the source cell, and we want
to transfer them to n-face nodes in the target cell; see Sect.~\ref{subsec:global_dof_numbering}.
This binding, implemented in \mytexttt{reference\_fe\_t}, ultimately relies on  
its \mytexttt{own\_dof\_permutations(:)} member variable; see Line~\ref{loc:dofs_stop} in 
Listing~\ref{lst:reference_fe}. This allocatable array is indexed with the n-face dimension
(i.e., 1 for edges, and 2 for faces). For each n-face dimension larger than 0, it contains a rank-2
allocatable array (i.e., \mytexttt{type(allocatable\_array\_ip2\_t)} is the base type of the array),
which serves as a lookup table for the implementation of the aforementioned transformation. In particular,
the rows are indexed with the local index of the node identifier on top of the n-face from the perspective
of the source cell, and the columns with the permutation index; see Sect.~\ref{sec:polytope_rotations_and_permutations}.   
The entry in the corresponding row and column of the table provides the local index of the node 
within the n-face from the perspective of the target cell. These lookup tables are filled within the \mytexttt{fill\_own\_dofs\_permutations} deferred binding of \mytexttt{reference\_fe\_t}. We note that this latter binding, 
and \mytexttt{permute\_dof\_lid\_n\_face},
are declared as overridable bindings in Listing~\ref{lst:reference_fe} on purpose. This lets, e.g., subclasses of \mytexttt{reference\_fe\_t} to be used in conjunction with ({\em properly oriented}; see Sect.~\ref{sec:polytope_rotations_and_permutations}) 
n-simplex meshes to implement the former such that the \mytexttt{own\_dof\_permutations(:)}
member variable is not allocated nor filled, and the latter such that always returns the identity transformation.

\subsection{Creating \mytexttt{reference\_fe\_t} polymorphic
  instances} \label{subsec:reference_fe_creation}

Central to any \ac{OO} software system relying on abstract data types is
the approach chosen to create polymorphic instances at runtime.  For
simplicity, \FEMPAR{} follows the so-called simple factory design
pattern~\cite{freeman_head_2004}. It takes the form of a single
stand-alone function, called \mytexttt{make\_reference\_fe}, which
selects the dynamic type of the polymorphic instance to be returned at
runtime based on the values of its dummy arguments \mytexttt{topology}
and \mytexttt{fe\_type}. (For example, assuming the topology of an
hexahedron and \mytexttt{fe\_type\_lagrangian}, then it will select
its dynamic type to be \mytexttt{hex\_lagrangian\_reference\_fe\_t},
i.e., the concrete data type implementing Lagrangian-type \ac{FE} spaces on
top of n-cubes.)  Before returning, it calls a deferred binding of
\mytexttt{reference\_fe\_t}, called \mytexttt{create},
which is responsible to leave the \mytexttt{reference\_fe\_t} in a fully
functional state.  The interface of this deferred binding is shown
in Listing~\ref{lst:reference_fe_create}. 
\begin{lstlisting}[float=htbp,language={[03]Fortran},escapechar=@,caption={The signature of the \mytexttt{create} binding of \mytexttt{reference\_fe\_t}.},label={lst:reference_fe_create}]
abstract interface
 subroutine reference_fe_create(this, topology, num_dims, order, &
      &                         field_type, conformity, continuity)
  class(reference_fe_t), intent(inout) :: this 
  integer(ip)          , intent(in)    :: num_dims
  character(*)         , intent(in)    :: topology
  integer(ip)          , intent(in)    :: order
  character(*)         , intent(in)    :: field_type
  logical              , intent(in)    :: conformity
  logical              , intent(in)    :: continuity
 end subroutine reference_fe_create
...
end interface
\end{lstlisting}

We remark that \mytexttt{field\_type} is only a free parameter for
Lagrangian \acp{FE} (i.e., for a particular \mytexttt{reference\_fe\_t} subclass). 
In other words, it must be \mytexttt{field\_type\_vector} for Raviart-Thomas and edge elements. We note that 
despite its fix set of dummy arguments interface, it has been proven to be sufficient to fully
describe all subclasses currently available in \FEMPAR{}; see Sect.~\ref{subsec:reference_fe_subclasses}.
However, in the event that it is needed, and with extensibility in mind, 
a single parameter dictionary of <{\em
  key},{\em value}> pairs might have been used instead; \FEMPAR{}
indeed relies on an implementation of this data type where {\em
  key} is a string (typically denoting the name of the parameter),
and {\em value} a scalar or arbitrary rank array of intrinsic or even
user-defined types.\footnote{This data type is implemented within the \mytexttt{FPL} software package~\cite{FPL}.}

\subsection{Enumeration of \mytexttt{reference\_fe\_t} subclasses} \label{subsec:reference_fe_subclasses}

There is a rather complex data type hierarchy rooted at \mytexttt{reference\_fe\_t} in \FEMPAR{}, which has been judiciously designed with code re-use as the main driver. (For example, Lagrangian FE spaces on top of n-cubes and n-simplices share member variables and code
that can be gathered into a common base data type.) For the sake of brevity, in this work we do not
cover in full detail the implementation of the data types in this hierarchy (except those details given
in Sect. \ref{sec:polynomial} and \ref{sec:reference_fe_implementors}). 
However, for completeness, it is convenient to enumerate those \mytexttt{reference\_fe\_t} subclasses
that, at present, are available in this hierarchy. These subclasses, which lay at the leaves
of the hierarchy, are the following ones:

\begin{itemize}
\item \mytexttt{hex} and  \mytexttt{tet\_lagrangian\_reference\_fe\_t}. Space
      of polynomials of arbitrary degree $k$ on top of n-cubes (i.e., tensor-product like spaces
      $\polsp_\order$) and n-simplices (i.e., $\polspr_\order$),
      respectively, for the discretization of either scalar-valued, vector-valued or 
      tensor-valued fields; see Sect.~\ref{subsec:h1_conforming_fes}.
      By selecting the ownership relationship among node functionals and n-faces  appropriately (see Sect.~\ref{subsec:dof_set}), 
      this FE space can be either globally continuous, or entirely discontinuous across cell boundaries.

    \item \mytexttt{hex} and
      \mytexttt{tet\_raviart\_thomas\_reference\_fe\_t}. The
      vector-valued Raviart-Thomas \ac{FE} of arbitrary degree $k$ on
      top of n-cubes, and n-simplices, resp., suitable for the
      mixed Laplacian problem and some fluid flow problems. Global FE
      functions of this space (in its conformal variant) have
      continuous normal components across cell faces; see
      Sect.~\ref{subsec:hdiv_conforming_fes} for details.

\item \mytexttt{hex} and  \mytexttt{tet\_nedelec\_reference\_fe\_t}. The vector-valued curl-conforming N\'ed\'elec \ac{FE}
      of arbitrary degree $k$ on top of  n-cubes, and n-simplices, resp., suitable for electromagnetic problems.
      Global FE functions of this space (in its conformal variant) 
      have continuous tangential components across cell faces; see Sect.~\ref{subsec:hdiv_conforming_fes} for details.

\item \mytexttt{void\_reference\_fe\_t}. A software artifact that represents a FE space with no \acp{DOF} at all, neither at 
      the cell interiors, nor at their boundary n-faces. This sort of software resource has been proven extremely efficient
      for: 1) the numerical solution of a \ac{PDE} on a subdomain of our original discretized domain (which thus has to be aligned
      with the cells boundaries); 2) the numerical solution of a \ac{PDE} using XFEM-like discretization techniques (which are grounded
      on FE spaces that do not assign \acp{DOF} to cells exterior to the embedded domain); 3) to simplify the implementation of 
      discretization methods for \ac{PDE} problems that involve coupling at the interface level, e.g., fluid-structure interaction.
\end{itemize}

Apart from these \mytexttt{reference\_fe\_t} subclasses, there are
already concluded developments within this hierarchy in a beta version
of the code, such as B-splines \cite{hughes_isogeometric_2005}, and
other scheduled developments, such as div-conforming \acp{FE}
\cite{neilan_stokes_2015}.

\section{The description of the physical domain: the 
  \texttt{triangulation\_t} abstraction} \label{sec:triangulation}

A central abstraction in all \ac{FE} numerical simulation codes is the one that describes
the triangulation/mesh $\mathcal{T}_h$ of
the physical domain $\Omega \subset \mathbb{R}^d$ in which our problem
is posed. (In practice, the mesh generation for $\Omega$ introduces a
geometrical error, and the mesh is in fact over an approximated domain
$\Omega_h$.) In \FEMPAR{}, this abstraction is called \mytexttt{triangulation\_t}.
With flexibility, and code reuse in mind, this is an abstract data type.
In Sect.~\ref{subsec:abstract_triangulation}, we introduce  \mytexttt{triangulation\_t}, and the mechanism that
it provides to its subclasses in order to preserve encapsulation and data hiding,
while still letting subclasses to store and access to data efficiently.  For completeness,
in Sect.~\ref{subsec:static_triangulation}, we introduce  details underlying the implementation of a particular
concrete subclass of \mytexttt{triangulation\_t}.

\subsection{An abstract triangulation representation and its software implementation} \label{subsec:abstract_triangulation}

In this section, we present an abstract (conceptual) representation of a triangulation
that \FEMPAR{} exposes to user-level applications and other library software abstractions that are
grounded on it (see, e.g., Sect.~\ref{sec:fe_space}). This conceptual representation is 
provided by a set of abstract derived data types (and the methods bounded to them) to which we have converged as
a result of our experience in accommodating a wide range of state-of-the-art \ac{FE} discretizations and solver techniques within a single framework, 
from desktops/laptops, to high-end distributed-memory supercomputers (see Sect.~\ref{sec:fempar_project}).

For the sake of brevity, in this work we restrict ourselves to a
subset of this representation that only provides support to the
implementation of high-order conforming and non-conforming \ac{FE}
discretizations grounded on {\em conforming meshes} in a serial
computing environment.  We stress, however, that the actual (complete)
representation also incorporates concepts to express the mesh in a
distributed-memory environment (e.g., the set of cells of a subdomain
is divided into local cells and a layer of cells owned by remote
subdomains, which we denote as {\em ghost cells}). It also provides
support to the implementation of high-order $hp$-adaptive (i.e., on
locally refined, non-conforming meshes) conforming and non-conforming
\acp{FE} (using hanging node constraints \cite{bangerth_data_2009} and
subface integration over a facet between cells of different refinement
level, respectively) and to the implementation of XFEM-type techniques
(see \cite{badia_robust_2017} and references therein); provided an
implicit representation of the geometry of the domain, a background
mesh is able to know whether a cell is interior, exterior or cut by
the domain, and in the latter case, to provide the coordinates of the
intersection points. This extra expressivity comes in the form of
additional data types and an extended set of methods for those data
types that are covered in this section. We stress, however, that
neither the former nor the latter ones will be covered in this
section.

Although our abstract representation of a triangulation has been
proven to have high expressivity, we do not claim, however, that our
triangulation representation is universally applicable to the
implementation of arbitrary numerical discretization and solver
techniques. It indeed has been designed such that extra extensions are
foreseen to satisfy further requirements.

The triangulation representation encompasses both topological and
geometric data.  A triangulation is conceived as a partition of
$\Omega$ into a set of cells ($d$-faces). Each cell is uniquely
identified by a global identifier in the range
$\mytexttt{cell\_gid}=1,\ldots,\mytexttt{num\_cells}$.\footnote{
We note that the actual conceptual representation of the triangulation
in \FEMPAR{} differences among local (to subdomain) cell identifiers and global cell identifiers 
(among the whole triangulation of the domain) in a distributed-memory
  context. The second sort of identifiers are coded as long precision integers, i.e.,
\mytexttt{integer(igp)}, in order to accommodate simulations with more than 
$2^{31}-1$ global cells.}  Apart from
the cells, a triangulation is also composed by a set of lower
dimensional objects, i.e., a set of $k$-faces, for
$k=0,\ldots,d-1$. We will also refer to elements in this set as
``vefs'', provided that in the $d=3$ case, it is composed of vertices,
edges, and faces. Each of the objects in this set is uniquely
identified by a global identifier in the range
$\mytexttt{vef\_gid}=1,\ldots,\mytexttt{num\_vefs}$.\footnote{As mentioned in the case of cells, the actual conceptual representation of the triangulation
in \FEMPAR{} differences among local (to a subdomain) vef identifiers and global vef identifiers 
(among the whole triangulation of the domain) in a distributed-memory
  context. Again the latter ones are long precision integers.}.

Apart from the cells and vefs, a triangulation also encompasses adjacency data.
This sort of data describes how n-faces in a mesh are related to each other. 
We denote by $F$ the set of all n-faces in the mesh, by $F^k$ the set
of all $k$-faces, and by $F_i$ and $F^k_i$ the $i$-th n-face (of arbitrary dimension) 
and the $i$-th $k$-face (of fixed dimension $k$), respectively. In conforming meshes, there are mainly two relevant types 
of adjacency relationships, namely \emph{composition}
($m$-faces that are part of a $k$-face for $m<k$) and
\emph{neighbourhood} ($m$-faces around a given $k$-face for $m>k$).
Following \cite{beall_general_1997}, the
set of $m$-faces adjacent to $F^k_i$, is denoted by
$F^k_i \langle F^m \rangle$ (i.e., the operator $\langle \cdot \rangle$
selects from the set the $m$-faces adjacent to the one in the left).
A triangulation conforming with \FEMPAR{} abstract representation
should be able to provide the composition data
$F^3_i \langle F\rangle$, and the neighbourship data
$F_i \langle F^3 \rangle$, that is, n-faces that compose each
cell and cells around n-faces. 

A triangulation also includes geometry data. Cell geometries are
represented by a map $\geomap_K$ of a polytope $\georef$ in the
reference space to the physical space (see Sect.~\ref{sec:fe_meth}).
This map is represented as a function of a \emph{scalar} \ac{FE} space
(e.g., grounded on high-order Lagrangian \acp{FE} or B-splines), with
its \ac{DOF} values being the vectors of node coordinates (i.e., \mytexttt{point\_t} instances)  in the physical space.

At the core of the software design in charge of providing the triangulation-related data covered so far is an abstract data type 
named \mytexttt{triangulation\_t}. (The rationale behind this data type being abstract will be made clear in the course
of this section.) This data type is defined as shown in Listing~\ref{lst:triangulation}. \mytexttt{triangulation\_t}
is conceived as a template to which all subclasses have to conform. On the one hand, it is composed by a ({\em minimal}) set of member variables encompassing data common to any triangulation. 
In particular, any triangulation is embedded in a \mytexttt{num\_dimensions}-dimensional space, and is composed of a total number of \mytexttt{num\_cells} 
(\mytexttt{num\_dimensions}-dimensional) cells and \mytexttt{num\_vefs} vefs, respectively; see Lines~\ref{loc:tria_num_dims}-\ref{loc:tria_num_vefs} of 
Listing~\ref{lst:triangulation}, respectively. On the other hand, \mytexttt{triangulation\_t} is equipped
with a set of {\em deferred} methods that the subclasses of \mytexttt{triangulation\_t} must implement; see Lines~\ref{loc:tria_create_cell_iterator}-\ref{loc:tria_free_vef_iterator}.  
The rationale underlying these methods requires further elaboration, to be discussed in the sequel. 

\begin{lstlisting}[float=htbp,language={[03]Fortran},escapechar=@,label={lst:triangulation}, caption={The \mytexttt{triangulation\_t} abstract data type.}]
type, abstract :: triangulation_t
 private
 integer(ip) :: num_dims @\label{loc:tria_num_dims}@
 integer(ip) :: num_cells
 integer(ip) :: num_vefs       @\label{loc:tria_num_vefs}@
 ... ! Rest of member variables of the full conceptual triangulation representation
contains  
 procedure :: get_num_dims  => triangulation_get_num_dims
 procedure :: get_num_cells => triangulation_get_num_cells
 procedure :: get_num_vefs  => triangulation_get_num_vefs
 !subroutine (this(in)::class(triangulation_t),cell(inout)::class(cell_iterator_t),allocatable) @\label{loc:tria_create_cell_iterator}@
 procedure(...), deferred :: create_cell_iterator
 !subroutine (this(in)::class(triangulation_t),cell(inout)::class(cell_iterator_t),allocatable)
 procedure(...)  , deferred :: free_cell_iterator
 !subroutine (this(in)::class(triangulation_t),vef(inout)::class(vef_iterator_t),allocatable)
 procedure(...) , deferred :: create_vef_iterator 
 !subroutine (this(in)::class(triangulation_t),vef(inout)::class(vef_iterator_t),allocatable)   
 procedure(...) , deferred :: free_vef_iterator @\label{loc:tria_free_vef_iterator}@
 ... ! Rest of regular and deferred bindings of the full conceptual triangulation representation
end type triangulation_t
\end{lstlisting}

In order to construct a conceptual view of \mytexttt{triangulation\_t} suitable for the user (and library) code needs, \FEMPAR{} relies on the so-called {\em iterator} \ac{OO} design pattern~\cite{gamma_e._design_1995}. Iterators are data types that provide sequential traversals over the {\em full sets of objects} that all together (conceptually) comprise \mytexttt{triangulation\_t} as a mesh-like container. There are several different iterators available, each one related to a different set of objects to be traversed. For example, \mytexttt{cell\_iterator\_t} provides traversals over the set composed of all cells, while \mytexttt{vef\_iterator\_t} over the one composed of all vefs.\footnote{For completeness, let us mention that \mytexttt{triangulation\_t} also offers traversals over {\em subsets of objects} conveniently selected for acceleration purposes. For example, \mytexttt{triangulation\_t} provides an iterator over vertices, edges, and faces that lay on the interface among subdomains, called \mytexttt{itfc\_vef\_iterator\_t} (i.e., a subset of the set of objects traversed by \mytexttt{vef\_iterator\_t}) for those subclasses suitable for parallel distributed-memory environments.} In our software design, iterators are created and freed by a set of public \acp{TBP} provided by \mytexttt{triangulation\_t}; see Lines~\ref{loc:tria_create_cell_iterator}-\ref{loc:tria_free_vef_iterator} of Listing~\ref{lst:triangulation}. Thus, for example, the expression \mytexttt{call triangulation\%create\_cell\_iterator(cell)} creates an iterator on the \mytexttt{cell} client-space instance, while \mytexttt{call triangulation\%free\_cell\_iterator(cell)}
frees it. Iterators sequentially traverse objects in increasing order by their global identifiers. However, we note that \mytexttt{triangulation\_t} subclasses are completely free to decide how to internally label these objects.\footnote{Thus, e.g., a \mytexttt{triangulation\_t} subclass that internally labels the global identifiers of vefs by their dimension in increasing order would result in a traversal with such an order. This is however a potentially changing over time low-level implementation detail that user programs relying on \mytexttt{triangulation\_t} and its associated iterators should not assume nor rely on.}

As the reader might have already noted from the minimal set of member variables in Listing~\ref{lst:triangulation} (among others), our software design is such that we want to provide {\em complete flexibility} to concrete subclasses of \mytexttt{triangulation\_t} with respect to how do they internally layout the (topology and geometry) data to be provided. To this end,  \mytexttt{triangulation\_t} is an abstract class that defers this decision to its subclasses. There is a clear separation among how the data is handled (i.e., stored and accessed) by the {\em private data structures} (member variables) underlying \mytexttt{triangulation\_t} subclasses, and the conceptual/abstract view of \mytexttt{triangulation\_t} exposed to \FEMPAR{} users. This view renders \mytexttt{triangulation\_t} easily accessible and understandable. Whereas the public interface of \mytexttt{triangulation\_t} being used by client codes is designed to be stable over time, the internals of \mytexttt{triangulation\_t} subclasses, however, are allowed to (and are subject to) change over time (e.g., in order to accommodate further optimizations, additional requirements, etc.). At the price of dynamic run-time polymorphism, \mytexttt{triangulation\_t} subclasses might be designed such that they strongly strive to preserve encapsulation and data hiding while {\em still storing and accessing to data efficiently}. Thus, e.g., a \mytexttt{triangulation\_t} subclass in charge of handling structured/uniform meshes of simple domains may decide to not explicitly store the cell-wise global vef identifiers, nor the vertex coordinates of the mesh, but instead to provide them implicitly on demand as a function of the global cell identifier. 

Apart from encompassing the logic underlying the actual traversal over objects of the set at hand, iterators also have the following crucial responsibility. Following the software concept of ``accessors'' presented in~\cite{bangerth_deal.ii-general-purpose_2007}, they are able to tease out the data related to the current object on which they are seated from the global arrays and rest of private data structures that comprise the internals of the corresponding 
\mytexttt{triangulation\_t} subclass. They therefore do not explicitly store, e.g., the global vef identifiers of the current cell. Instead, they know how to fetch them from the corresponding \mytexttt{triangulation\_t} subclass into data structures suitable for the user needs. Provided that it is the responsibility of \mytexttt{triangulation\_t} subclasses to decide how to internally layout data, iterators are abstract data types as well, and most of its \acp{TBP} are deferred/virtual. This also justifies why the methods in the Lines~\ref{loc:tria_create_cell_iterator}-\ref{loc:tria_free_vef_iterator} of Listing~\ref{lst:triangulation} are deferred, and why the corresponding iterator dummy arguments, polymorphic allocatable. It is ultimately the responsibility of the concrete subclass of \mytexttt{triangulation\_t} to decide on execution time the dynamic type of the polymorphic variable being created. 

Let us next discuss the rationale underlying the design of iterators over cells and vefs. These data types are defined in Listing~\ref{lst:set_iterator}, where \mytexttt{set} must be actually replaced by the corresponding name uniquely identifying the set of objects to be traversed by the iterator at hand, i.e., either \mytexttt{cell} or \mytexttt{vef}. In Fig.~\ref{fig:set_iterator_implementation}, 
we illustrate the implementation of a partial (selected) subset of the bindings of these data types. 
\begin{lstlisting}[float=htbp,language={[03]Fortran},escapechar=@,caption=\mytexttt{triangulation\_t} ``\mytexttt{set}'' (either \mytexttt{cell} or \mytexttt{vef}) iterators., label={lst:set_iterator}]
type, abstract :: set_iterator_t
  private
  class(triangulation_t), pointer :: triangulation => NULL()
  integer(ip)                     :: gid
contains
  public
  procedure :: create            => set_iterator_create
  procedure :: free              => set_iterator_free
  procedure :: first             => set_iterator_first
  procedure :: next              => set_iterator_next
  procedure :: has_finished      => set_iterator_has_finished
  procedure :: get_gid           => set_iterator_get_gid
  procedure :: set_gid           => set_iterator_set_gid
  procedure :: get_triangulation => set_iterator_get_triangulation
  ... ! Set of deferred TBPs providing access to data items in the corresponding set
end type set_iterator_t
\end{lstlisting}

The \mytexttt{create} binding of \mytexttt{set\_iterator\_t} takes as input a polymorphic \mytexttt{triangulation\_t} instance to be traversed,  and leaves the iterator positioned in the first object of the \mytexttt{set},
i.e., in a state ready to start the sequential traversal over all of its objects; see Fig.~\ref{fig:set_iterator_implementation}. This method (like {\mytexttt{free}) is 
not intended to be directly called by the user. Instead, \mytexttt{triangulation\_t} clients should rely on the deferred bindings of \mytexttt{triangulation\_t}
presented in Listing~\ref{lst:triangulation}. The \mytexttt{init}, \mytexttt{next}, and \mytexttt{has\_finished} bindings let clients 
to position the iterator on the first object of the  \mytexttt{set}, move to its next object, and check whether all of its objects have been 
already traversed or not, respectively; see Fig.~\ref{fig:set_iterator_implementation}.

\begin{figure}[htbp]
\begin{center}
\begin{tabular}{ccc}
\begin{minipage}{0.45\textwidth}
\begin{lstlisting}[language={[03]Fortran}]
subroutine set_..._create(this,triang) 
 class(set_iterator_t), inout  :: this
 class(trian..._t), target, in :: triang
 call this%free()
 call this%triangulation=>triang
 call this%first()
end subroutine set_iterator_create

subroutine set_iterator_first(this)
  class(set_iterator_t), inout  :: this
  this%gid = 1
end subroutine set_iterator_first
\end{lstlisting}
\end{minipage} & &
\begin{minipage}{0.45\textwidth}
\begin{lstlisting}[language={[03]Fortran}]
function set_iterator_has_finished(this) 
 class(set_iterator_t), in :: this
 logical :: set_iterator_has_finished
 assert (associated(this%triangulation))
 set_iterator_has_finished = & 
  this%gid > this%triang...%get_num_sets()
end function set_iterator_has_finished

subroutine set_iterator_next(this)
  class(set_iterator_t), inout  :: this
  this%gid = this%gid+1
end subroutine set_iterator_nest
\end{lstlisting}
\end{minipage}
\end{tabular}
\end{center}
\caption{\label{fig:set_iterator_implementation} Implementation of a partial (selected) subset of the bindings of \mytexttt{set\_iterator} (see Listing~\ref{lst:set_iterator}).}
\end{figure}

The actual set of (deferred) \acp{TBP} of a \mytexttt{triangulation\_t} iterator highly depends on the type of object being pointed. 
We now briefly discuss those \acp{TBP} in the set corresponding to cell and vef iterators that provide support to the subset of the triangulation
conceptual representation we are focusing on. These are in particular enumerated in Listing~\ref{lst:cell_vef_iterator}.

\begin{lstlisting}[float=htbp,language={[03]Fortran},escapechar=@, caption=A subset of the deferred \acp{TBP} of the \mytexttt{cell\_iterator\_t} and \mytexttt{vef\_iterator\_t} data types (follow-up to Listing~\ref{lst:set_iterator})., label={lst:cell_vef_iterator}]
type, abstract :: cell_iterator_t
  ... ! Member variables (see Listing@~\ref{lst:set_iterator}@)
contains
  public
  ... ! Create/free/traversal TBPs (see Listing@~\ref{lst:set_iterator}@)

  ! Topology-data related TBPs (i.e., @{\color{blue} $F^3_i \langle F\rangle$}@)
  procedure(...), deferred :: get_num_vefs @\label{loc:cell_iterator_beg_topo}@
  procedure(...), deferred :: get_vef     
  procedure(...), deferred :: get_vef_lid  
  procedure(...), deferred :: get_vef_gid  
  procedure(...), deferred :: get_vefs_gid @\label{loc:cell_iterator_end_topo}@

  ! Transformation among n-face-wise node local numberings (see Section@~\ref{sec:polytope_rotations_and_permutations}@)
  procedure(...), deferred :: get_permutation_index @\label{loc:permutation_index}@

  ! Geometry-related data
  procedure(...), deferred :: get_reference_fe @\label{loc:cell_iterator_beg_geo}@
  procedure(...), deferred :: get_num_nodes
  procedure(...), deferred :: get_nodes_coordinates @\label{loc:cell_iterator_end_geo}@
  
  ! Set IDs-related
  procedure(...), deferred :: get_set_id @\label{loc:cell_iterator_beg_get_id}@
  procedure(...), deferred :: set_set_id @\label{loc:cell_iterator_beg_set_id}@

  ... ! Rest of TBPs of full conceptual triangulation presentation (e.g., is_local() vs is_ghost())
end type cell_iterator_t

type, abstract :: vef_iterator_t
   ... ! Member variables (see Listing@~\ref{lst:set_iterator}@)
contains
  public
  ... ! Create/free/traversal TBPs (see Listing@~\ref{lst:set_iterator}@)
  
  ! Topology-data related TBPs (i.e., @{\color{blue} $F_i \langle F^3\rangle$}@)
  procedure(...), deferred :: get_num_cells_around  @\label{loc:vef_iterator_beg_topo}@
  procedure(...), deferred :: get_cell_around       @\label{loc:vef_iterator_end_topo}@

  ! Geometry related-data TBPs
  procedure(...), deferred :: get_num_nodes @\label{loc:vef_iterator_beg_geo}@ 
  procedure(...), deferred :: get_nodes_coordinates @\label{loc:vef_iterator_end_geo}@

  ! Misc TBPs
  procedure(...), deferred :: is_at_interior @\label{loc:vef_iterator_beg_misc}@ 
  procedure(...), deferred :: is_at_boundary
  procedure(...), deferred :: get_dim
  procedure(...), deferred :: get_set_id
  procedure(...), deferred :: set_set_id @\label{loc:vef_iterator_end_misc}@  
  
  ... ! Rest of TBPs  of the full conceptual triangulation presentation (e.g., is_at_interface())
end type vef_iterator_t

  
\end{lstlisting}

The \acp{TBP} in Lines~\ref{loc:cell_iterator_beg_topo}-\ref{loc:cell_iterator_end_topo} of Listing~\ref{lst:cell_vef_iterator} are in charge of providing data related to the composition relationship $F^3_i \langle F\rangle$. In particular,
the \mytexttt{get\_num\_vefs} binding returns the number of vefs on the boundary of the mesh (i.e., the cardinality of
the composition relationship). Given the local index of a vef in a cell (within
the range $1,\ldots,\mytexttt{num\_vefs}$), \mytexttt{get\_vef} positions the \mytexttt{vef\_iterator\_t} instance on input such that it points to this vef, while \mytexttt{get\_vef\_gid}, returns its global identifier; \mytexttt{get\_vef\_lid} performs the inverse translation to the one of \mytexttt{get\_vef\_gid}. Finally, \mytexttt{get\_vefs\_gid} let the client obtain the global identifier of all vefs of the current cell in one shot provided a user-space pointer to integer array. The semantics of this last \ac{TBP} are such that subclasses of \mytexttt{cell\_iterator\_t} are not allowed to allocate the provided pointer, but to associate it with existing (internal) memory (for increased performance and memory leaks avoidance). 

The \ac{TBP} in Line~\ref{loc:permutation_index} of Listing~\ref{lst:cell_vef_iterator} provides support to the implementation of the transformation procedure described in Sect.~\ref{sec:polytope_rotations_and_permutations}. In particular, this binding has to be invoked on a \mytexttt{cell\_iterator\_t} instance positioned in the source cell, and given a \mytexttt{cell\_iterator\_t} positioned on the target cell, and the n-face local identifier within the former and latter cells, returns the permutation index;  see Sect.~\ref{sec:polytope_rotations_and_permutations}. We stress that both the rotation and orientation indices can be always computed using the TBPs in the previous paragraph. For example, in order to determine the rotation index, one can extract the global id of the anchor vertex of the n-face in the target cell (by calling \mytexttt{get\_vef\_gid}), and then searching for this global id in the set of vertices that comprise the n-face in the target cell (using an iterator over the corresponding sublist in \mytexttt{vertices\_n\_face}; see Sect.~\ref{subsec:reference_fe_topology}). However, we preferred to provide a specialized deferred binding for such purpose in order to leave room for optimizations in \mytexttt{triangulation\_t} subclasses. For example, in the case of a subclass that works with oriented meshes, then  \mytexttt{get\_permutation\_index} may be implemented such that it always returns the permutation index corresponding to the identity transformation. In the case of a subclass of \mytexttt{triangulation\_t} that is intended
to remain static (or to be adapted very infrequently) during the course of the simulation process (see, e.g., Sect.~\ref{subsec:static_triangulation}), then it might be beneficial for performance to precalculate all possible permutation indices during set up into lookup tables, and re-use them all the way through without having to perform the aforementioned searches over and over again.


The \acp{TBP} in Lines~\ref{loc:cell_iterator_beg_geo}-\ref{loc:cell_iterator_end_geo} are in charge of providing the cell geometry
related-data. In particular, \mytexttt{get\_reference\_fe} returns a polymorphic pointer to the \mytexttt{reference\_fe\_t}
instance that describes the space of functions  to which the mapping $\geomap_K$ belongs. \mytexttt{get\_num\_nodes} and \mytexttt{get\_nodes\_coordinates} return the number of nodes describing the geometry of the cell, and its associated coordinates in physical space, respectively. Instead of a pointer to an user-space array to be associated with internal storage (as \mytexttt{get\_vef\_gids}), 
\mytexttt{get\_nodes\_coordinates} takes a user-space 
(pre-allocated) array of type \mytexttt{point\_t} instances, and fills it 
(because of reasons made clear in Sect.~\ref{subsec:mapping}). Assuming that  \mytexttt{reference\_fe\_t} is a bi-linear
Lagrangian \ac{FE} on a quadrilateral, then \mytexttt{get\_num\_nodes} would return 4 (one node per cell-vertex),
while \mytexttt{get\_nodes\_coordinates} the coordinates in physical space of its vertices.

Any \mytexttt{triangulation\_t} subclass should let its clients to classify the cells into sets. Each set is globally identified
by an integer number, named \mytexttt{set\_id}. The methods \mytexttt{get\_set\_id} and \mytexttt{set\_set\_id} let the caller to associate a set to the current cell, or 
to retrieve the set to which the cell is currently associated. Cells set identifiers are primarily (although not only) used by \mytexttt{fe\_space\_t} during its set-up; see Sect.~\ref{sec:fe_space}.
In particular, they instruct the latter to determine which \mytexttt{reference\_fe\_t} instances to use on top of the cells belonging to the same set.
For example, assuming that we want to solve a scalar, single-field \ac{PDE} problem on a subdomain of our original domain (that we assume to be aligned with the cells boundaries), 
we would use two different sets. The first for the cells that are interior to the subdomain, and the second for those that are exterior.
Then we could associate e.g., a linear Lagrangian reference \ac{FE} to cells in the first set, and 
\mytexttt{void\_reference\_fe\_t} on those cells of the second set; see Sect.~\ref{subsec:reference_fe_subclasses}.
 
Sitting on a given vef, the \acp{TBP} in Lines~\ref{loc:vef_iterator_beg_topo}-\ref{loc:vef_iterator_end_topo}
are in charge of providing data related to the adjacency relationship $F_i \langle F^3\rangle$. 
In particular, \mytexttt{get\_num\_cells\_around} returns its cardinality, while \mytexttt{get\_cell\_around} returns
a cell in this set. To be more precise, the latter \ac{TBP} positions the instance of \mytexttt{cell\_iterator\_t} on input such that it points to a cell in this set identified with an index within the range
$1,\ldots,\mytexttt{get\_num\_cells\_around()}$. The order in which the cells around a vef are listed can be
arbitrary, so that codes relying on \mytexttt{triangulation\_t} should not assume, e.g., that they are 
ordered increasingly by their global cell identifiers. On the other hand, \mytexttt{get\_num\_nodes} and 
\mytexttt{get\_nodes\_coordinates} return the number of points on top of the vef (including those on top
of the lower-dimensional ones on its boundary), and its associated coordinates in physical space, respectively;
see Lines~\ref{loc:vef_iterator_beg_geo}-\ref{loc:vef_iterator_end_geo}.
We adopt the convention that these nodes are (locally) labeled (within the input/output array of point coordinates 
to be filled) according to the reference coordinate system of the {\em first cell} around the vef, i.e., the
cell obtained as \mytexttt{vef\%get\_cell\_around(1,cell)}.

The \acp{TBP} in Lines~\ref{loc:vef_iterator_beg_misc}-\ref{loc:vef_iterator_end_misc} let the client to determine whether
the vef is at the interior of the domain or on its boundary, the vef
dimension (e.g., in 3D, it would return 0, 1, and 2 for vertices, edges, and faces, respectively) and to  
retrieve the set to which the vef is currently associated, or associate a new set to it, respectively.
Sets in the case of vefs are primarily used to codify the boundary conditions of the \ac{PDE} problem at hand,
as discussed in Sect.~\ref{subsec:fe_space_strong_boundary_conditions}.

At this point we are already in position to show user-level code that exploits the software design covered so far.
In particular, Listing~\ref{lst:triangulation_iterators} splits the whole set of triangulation cells into two disjoint sets, those that are in contact to the boundary of the domain, and those that are in its interior.

\begin{lstlisting}[float=htbp,language={[03]Fortran},escapechar=@, caption=User-level code illustrating the usage of the data types and its associated \acp{TBP} supporting \FEMPAR{} conceptual triangulation representation., label={lst:triangulation_iterators}]
type(my_triangulation_t) :: triangulation
class(cell_iterator_t), allocatable :: cell
class(vef_iterator_t) , allocatable :: vef
integer(ip), parameter :: interior_cell_set_id = 1
integer(ip), parameter :: boundary_cell_set_id = 2

... ! Set up triangulation using the TBPs of my_triangulation_t
    ! (my_triangulation_t extends triangulation_t)
call triangulation%create_cell_iterator(cell)
call triangulation%create_vef_iterator(vef)
do while ( .not. cell%has_finished() )
  call cell%set_set_id(interior_cell_set_id)
  do vef_lid=1,cell%get_num_vefs()
    call cell%get_vef(vef_lid,vef)
    if (vef%is_at_boundary()) then
       call cell%set_set_id(boundary_cell_set_id)
       exit
    end if
  end do
  call cell%next()
end do
... ! Execute code which consumes set IDs for whatever purpose
call triangulation%free_cell_iterator(cell)
call triangulation%free_vef_iterator(vef)
... ! Free triangulation using the TBPs of my_triangulation_t
\end{lstlisting}


\subsection{An example \mytexttt{triangulation\_t} subclass and rationale} \label{subsec:static_triangulation}

In this section, we discuss how a particular subclass of
\mytexttt{triangulation\_t} is internally organized in order to
efficiently provide triangulation-related data by means of the
software abstractions presented in
Sect.~\ref{subsec:abstract_triangulation}.  This subclass is \mytexttt{static\_triangulation\_t}.  A
\mytexttt{static\_triangulation\_t} codifies a {\em conforming mesh},
which is set up from scratch at the beginning of the simulation, and
remains unaltered during the whole process. On the other hand,
\mytexttt{static\_cell\_iterator\_t} and
\mytexttt{static\_vef\_iterator\_t} are two non-abstract data type
extensions of \mytexttt{cell\_iterator\_t} and
\mytexttt{vef\_iterator\_t}, respectively. By overriding the
set of deferred methods of the former ones, the latter ones tease out the data related
to the current object on which they are seated from the global arrays
and rest of private data structures that comprise the internals of
\mytexttt{static\_triangulation\_t}.

There is no single approach to layout the data within a given triangulation subclass. The seek of an
 acceptable trade-off among memory consumption, computational time required 
to set up, update (if it applies), access to triangulation data, and the frequency on which 
these operations are performed should guide its internal organization. 
For example, in \cite{beall_general_1997}, two storage layouts are presented, and its memory and computational
cost for the computation of any possible adjacency relationship is evaluated in 3D. 
The first one, called \emph{one-level} representation, is defined by
$F^1_i \langle F^0 \rangle$, $F^2_i \langle F^1 \rangle$, and
$F^3_i \langle F^2 \rangle$, and by $F^0_i \langle F^1 \rangle$,
$F^1_i \langle F^2 \rangle$, and $F^2_i \langle F^3 \rangle$
(neighbourhood information). In other words, it stores vertices of each
edge, edges of each face, and faces of each cell, together with edges
around vertices, faces around edges, and cells around faces. The
second one, called \emph{circular} representation, is defined by the
composition information $F^1_i \langle F^0 \rangle$,
$F^2_i \langle F^1 \rangle$, $F^3_i \langle F^2 \rangle$ (as above),
together with the neighbourhood information $F^0_i \langle F^3 \rangle$ (cells
around vertices). An important property of these two storage layouts is
their \emph{completeness}, i.e., the possibility to determine any
adjacency without a loop over the entire mesh.
 The storage requirements
for a uniform mesh of a cube domain with $N_c$ cells are $48 N_c$ (for
hexahedra) and $24 N_c$ (for tetrahedra) in the former, and $32 N_c$
(for hexahedra) and $16 N_c$ (for tetrahedra) in the latter. However,
the operation count for determining some adjacencies, although
independent of $N_c$, is high. For example, in the case of the one-level
representation, to obtain the cells around a vertex requires 48 (for
hexahedra) and 140 (for tetrahedra) operations, whereas only one
operation is needed to obtain cells around facets. In the case of the
circular representation, these queries involve one and 148 (for
hexahedra) or 299 (for tetrahedra) operations, respectively
\cite{beall_general_1997}. (We recall that both kind of adjacencies
are required by \FEMPAR{} as presented in Sect.~\ref{subsec:abstract_triangulation}.)

Another quite different storage data layout is the one followed by the
triangulation in the deal.II library
\cite{bangerth_deal.ii-general-purpose_2007}, essentially defined by
the composition data $F^1_i \langle F^0 \rangle$,
$F^2_i \langle F^1 \rangle$, and $F^3_i \langle F^2 \rangle$ (referred
as hierarchical cell representation by the authors of the library),
and the neighbourhood data $F^3_i \langle F^2 \rangle$ {\em stored
  cell-wise} (i.e., a given cell stores the identifiers of its cell
neighbours across each face within the cell).  
 Besides, the
(potentially non-conforming) triangulation in this library is
conceived (and explicitly represented) as a collection of trees, where
the cells of a coarsest conforming mesh (generated by deal.II itself
for simple domains, or read from a file from several file formats)
form the roots, and the children branch off their parent cells, thus
forming binary-trees, quad-trees and oct-trees in $d=1,2,$ and $3$
spatial dimensions, respectively
\cite{bangerth_deal.ii-general-purpose_2007}.  While both the
ancestors (i.e., the so-called ``inactive'' cells) and leaf cells of
the tree (i.e., the so-called ``active'' cells) are stored, only the
latter ones actually form the partition of the domain.  Apart from a
hierarchy of cells, the deal.II triangulation also maintains a
hierarchy of $k$-faces for $k=1,\ldots,d-1$. Such quite complex data
structure is justified by the authors for two reasons. First, it
allows for an efficient implementation of adaptive mesh adaptation
(including coarsening and refinement). The hierarchy of n-faces
aids in the process of handling the so-called hanging node
constraints required to build conforming \ac{FE} spaces on top of
non-conforming meshes. The second reason is the implementation of
(geometric) multigrid preconditioners grounded on the adaptivity tree. In particular, such preconditioners require that \acp{DOF}
  are also associated to inactive cells. Thus, also inactive n-faces
  have to explicitly exist in the triangulation. In any case, such
structure is hard to generate and maintain, and does not fit well when
integrated with parallel octree libraries
\cite{bangerth_algorithms_2012}, like \mytexttt{p4est}
\cite{burstedde_p4est_2011}. The whole hierarchy must be generated
from scratch on each mesh adaptivity step. However,
based on our own experience, {\em such hierarchy is not really needed for
an efficient implementation of adaptive refinement}. The second reason,
i.e., the implementation of a serial hierarchical multigrid solver in
deal.II, would probably be more complicated without such a hierarchical
representation of the mesh.

While the hierarchical cell representation in deal.II has been proven
to be successful in the implementation of highly complex $hp$-adaptive
\ac{FE} discretization~\cite{bangerth_data_2009} and reduces memory
consumption over $F^3_i \langle F\rangle$, the restriction of the
global vef identifiers to a cell (a very frequent operation in \ac{FE}
codes), becomes significantly more expensive in this storage
layout as this operation requires permutations among the reference
coordinate system of the cell that owns the vef to the one to which
we are restricting to; the same applies to the restriction of global
\ac{DOF} identifiers to a cell when the \acp{DOF} are stored
n-face-wise. Furthermore, it is a {\em non-complete} storage
layout. In particular, neighbourship data $F_i \langle F^3 \rangle$ 
has to be computed by the user by means of a loop over all
cells. Besides, it prevents library support to loops over the facets
of the mesh, and access to the neighbouring cells, a natural operation
in the implementation of \ac{DG} methods. In our
experience, facet-loop based integration of \ac{DG} terms (versus cell-loop
based) leads to a software that is significantly easier to use, as it
might be designed such that most of the complexity underlying facet
integration can be hidden to the user (see
Sect.~\ref{sec:face_integration}). Finally, although it is very
efficient for hierarchical and local mesh adaptation (within each
subdomain), the most severe drawback is its costly set up (from
scratch) for a given initial conforming coarse mesh (this can be
mitigated by reducing the coarse mesh resolution, at the price of
potentially losing geometry modelling accuracy), and, in a
distributed-memory environment, the even more costly regeneration
 of an adapted non-conforming forest of trees after a re-distribution
step among processes for dynamic load-balancing \cite{burstedde_p4est_2011}. Indeed, in
~\cite{bangerth_algorithms_2012}, the latter is reported as the second more costly operation in
the simulation pipeline, only below the linear solver step.

The \mytexttt{static\_triangulation\_t} data type {\em explicitly}
stores the composition data $F^3_i \langle F\rangle$, and the
neighbourship data $F_i \langle F^3 \rangle$ within its internal
(private) member variables.\footnote{We note that
  $F^{3}_i \langle F^3 \rangle$ is simply $F^{3}_i$ and is not
  stored.}  The memory consumption of such {\em complete} storage
layout is $52 N_c$ (hexahedra) and $28 N_c$ (tetrahedra), which is
less than twice the one of the one-sided and circular
representations~\cite{beall_general_1997}.
At the price of this increased memory consumption,
\mytexttt{static\_triangulation\_t} is able to provide the required
adjacency data with $\mathcal{O}(1)$ arithmetic complexity.  Besides,
the cell-based storage of the composition relationship is perfectly
suited for its migration in parallel distributed-memory
environments. On the other hand, the amount of {\em permanent storage}
of this data layout can be reduced if one exploits the fact that
neighbourship data is only required in very specific parts of the code.
For example, unstructured mesh generators usually provide only the
composition data $F^3_i \langle F^0 \rangle$. In such a case,
\mytexttt{static\_triangulation\_t} requires the neighbourship data
$F^0_i \langle F^3 \rangle$ (plus the reference cell topology data
encompassed within the \mytexttt{reference\_fe\_t} instance mapped to
each cell; see Sect.~\ref{subsec:reference_fe_topology}) in order to
set up the composition data $F^3_i \langle F^1 \rangle$ and
$F^3_i \langle F^2 \rangle$.  It is also needed in
\mytexttt{triangulation\_t} subclasses suitable for distributed-memory
computers, among others, to set up the data structures required to
perform nearest neighbour exchanges of \acp{DOF} nodal values among
subdomains. (We stress that this process requires to globally identify
interface \acp{DOF} consistently among subdomains sharing such
\acp{DOF} .)  In this latter scenario, this adjacency data is only
required for n-faces that lay on the inter-subdomain interface (and
not for those on the interior). The evaluation of facet integrals (as
designed in \FEMPAR{}, see Sect.~\ref{sec:face_integration}) also
requires at least $F^2_i \langle F^3 \rangle$ and
$F^1_i \langle F^2 \rangle$, in 2D and 3D, respectively.  The use of
the full adjacency data can be needed for the implementation of
advanced numerical discretization schemes, e.g., for the
implementation of nodal-based \emph{shock detectors} for monotonic
\acp{FE}
\cite{badia_monotonicity-preserving_2017,badia_differentiable_2017}. Due
to the aforementioned reasons, we decided to design
\mytexttt{static\_triangulation\_t} such that it permanently stores
such data, but we stress that our software design is such that a
triangulation subclass is always free to offer methods that set up and
destroy these data on demand to reduce the amount of permanent data
storage.

The \mytexttt{static\_triangulation\_t} data type, together with a
selected set of its bindings, is defined as shown in
Listing~\ref{lst:static_triangulation}. Before going into more detail,
there are two main points to remark with respect to how this type
internally layouts its data. First, it relies all the way through on
intrinsic Fortran allocatable arrays. These sort of data structures
are perfectly suited for the particular case of
\mytexttt{static\_triangulation\_t}, due to its static nature. We
stress, however, that more efficient data structures (i.e., able to
mitigate the effect of frequent/costly allocatable array
re-allocations) would be convenient if it also had to support mesh
adaptation (e.g., a linked list, or even better for data locality, a
data structure with semantics close to \mytexttt{std:vector} of the
C++ standard template library, which in fact is already in \FEMPAR{}
but not included for brevity). Second, for increased data locality
during cell and vef sequential traversals (and thus a more efficient
on the memory hierarchy of modern computer architectures) the data is
not stored into cell-wise or vef-wise local arrays, but into global
arrays that are indexed either by the global cell or vef identifiers.

\begin{lstlisting}[float=htbp,language={[03]Fortran},escapechar=@,
label={lst:static_triangulation}, caption={The internals of \mytexttt{static\_triangulation\_t} and a 
selected set of its bindings.}]
type, extends(triangulation_t) :: static_triangulation_t
  private
  ! Container of polymorphic reference_fe_t instances (See Section@~\ref{sec:reference_fe}@)
  type(p_reference_fe_t)   , allocatable :: reference_fes(:) @\label{loc:st_reference_fes}@

  ! Mapping of cell GiDs to reference_fes(:)
  integer(ip)              , allocatable :: cell_to_ref_fes(:) @\label{loc:st_cell_to_ref_fes}@

  ! Composition data @{\color{blue} $F^3_i \langle F\rangle$}@
  integer(ip)              , allocatable :: ptr_vefs_x_cell(:)  @\label{loc:st_comp_beg}@
  integer(ip)              , allocatable :: lst_vefs_gids(:)    @\label{loc:st_comp_end}@

  ! Neighborship data @{\color{blue}$F_i \langle F^3 \rangle$}@
  integer(ip)              , allocatable :: ptr_cells_around(:) @\label{loc:st_neig_beg}@
  integer(ip)              , allocatable :: lst_cells_around(:) @\label{loc:st_neig_end}@
  
  ! Geometry interpolatory nodes global numbering + nodes coordinates in physical space
  integer(ip)              , allocatable :: ptr_nodes_x_cell(:)   @\label{loc:st_ptr_nodes_x_cell}@
  integer(ip)              , allocatable :: lst_nodes_gids(:)     @\label{loc:st_ptr_nodes_gids}@
  type(point_t)            , allocatable :: nodes_coordinates(:)  @\label{loc:st_nodes_coordinates}@

  ! Cell and vef set IDs, vefs at boundary? 
  integer(ip)              , allocatable :: cells_set_ids(:)      @\label{loc:st_cells_set_ids}@
  integer(ip)              , allocatable :: vefs_set_ids(:)       @\label{loc:st_vefs_set_ids}@
  logical                  , allocatable :: vefs_at_boundary(:)   @\label{loc:st_vefs_at_boundary}@

 contains  
  public
  procedure :: create  => static_triangulation_create @\label{loc:st_create}@
  procedure :: free    => static_triangulation_free
  ... ! TBPs overriding those which are deferred in triangulation_t (See Listing@~\ref{lst:triangulation}@)
  ... ! Comprehensive set of private TBPs providing support to the 
      ! implementation of the public TBPs above. These are hierarchically organized into 
      ! short, auxiliary re-usable subroutines/functions for code readability purposes
end type triangulation_t
\end{lstlisting}

A collection of \mytexttt{reference\_fe\_t} polymorphic instances is stored in the \mytexttt{reference\_fes(:)} array (see Line~\ref{loc:st_reference_fes} of Listing~\ref{lst:static_triangulation}). 
These instances are uniquely identified (within the local scope of \mytexttt{static\_triangulation\_t}) by their position in this array. For a given cell with global identifier \mytexttt{cell\_gid},
the  \ac{FE} space of functions to which the cell mapping $\geomap_K$ belongs, is described by the \mytexttt{reference\_fe\_t} instance with identifier \mytexttt{cell\_to\_ref\_fes(cell\_gid)} in the collection;
see Line~\ref{loc:st_cell_to_ref_fes}. The member variables used to store the composition data $F^3_i \langle F\rangle$ are encompassed within Lines~\ref{loc:st_comp_beg}-\ref{loc:st_comp_end} of 
Listing~\ref{lst:static_triangulation}.  As stated above, the global vef identifiers are stored cell-wise, in the \mytexttt{lst\_vefs\_gids(:)} array, which is in turn (indirectly) addressed by the \mytexttt{ptr\_vefs\_x\_cell(:)} array. In particular, the ones assigned to the vefs on cell \mytexttt{cell\_gid} start and end in position \mytexttt{ptr\_vefs\_x\_cell(cell\_id)} and \mytexttt{ptr\_vefs\_x\_cell(cell\_id+1)-1} of \mytexttt{lst\_vefs\_gids(:)}, respectively. Thus, e.g., the implementation of the (overridden) \mytexttt{get\_num\_vefs} \ac{TBP} in \mytexttt{static\_cell\_accessor} (see Listing~\ref{lst:triangulation_iterators}), just determines the number of vefs on the boundary of the current cell as \mytexttt{ptr\_vefs\_x\_cell(cell\_id+1)}-\mytexttt{ptr\_vefs\_x\_cell(cell\_id)}. On the other hand,  the member variables used to store the adjacency data $F_i \langle F^3\rangle$ are encompassed within Lines~\ref{loc:st_neig_beg}-\ref{loc:st_neig_end} of Listing~\ref{lst:static_triangulation}. The global identifiers of the cells around a vef \mytexttt{vef\_gid} start and end in position \mytexttt{ptr\_cells\_around(vef\_gid)} and \mytexttt{ptr\_cells\_around(vef\_gid+1)-1} of \mytexttt{lst\_cells\_around(:)}, respectively.

The geometry-related data is handled by the member variables in Lines~\ref{loc:st_ptr_nodes_x_cell}-\ref{loc:st_nodes_coordinates}. In particular, during the set up of \mytexttt{static\_triangulation\_t} a global numbering of the nodes of the global \ac{FE} space describing the geometry of the mesh is internally built. (The process that generates such numbering is identical to the one described in Sect.~\ref{subsec:global_dof_numbering}, so that we omit it here to keep the presentation short.) In particular, the global node identifiers restricted to cell \mytexttt{cell\_gid} start and end in position \mytexttt{ptr\_nodes\_gids(cell\_id)} and \mytexttt{ptr\_nodes\_gids(cell\_id+1)-1} of \mytexttt{lst\_nodes\_gids(:)}, respectively. These global node identifiers are used to (indirectly) address the global array of nodes coordinates in Line~\ref{loc:st_nodes_coordinates}. The \mytexttt{cells\_set\_ids(:)} and \mytexttt{vefs\_set\_ids(:)} arrays are used to store the user-provided cell and vef set identifiers (see Sect.~\ref{subsec:abstract_triangulation}), respectively, while \mytexttt{vefs\_at\_boundary(:)}, whether the corresponding vef lays on the boundary of the domain or not.
 
Finally, the \mytexttt{static\_triangulation\_create} binding sets up
a new \mytexttt{static\_triangulation\_t} instance. There are two
options for creating a \mytexttt{static\_triangulation\_t} in
\FEMPAR{}, depending on whether the mesh is structured or
unstructured. In the first case, \FEMPAR{} provides the machinery for
the automatic generation of a triangulation on simple domains (e.g., a unit cube), 
currently of brick (quadrilateral or hexahedral) cells. This function is
implemented exploiting a tensor product structure of the space,
numbering cells and vefs using lexicographical order.
The second way to create a
\mytexttt{static\_triangulation\_t} instance is from a mesh data file,
e.g., using the GiD mesh generator \cite{_gid_2016}.

\section{Evaluation of cell integrals} \label{sec:cell_integration}

In this section, we describe the data structures required to perform
the numerical integration of the local matrices. In order to compute
cell integrals \eqref{fematrix}, one needs (among others)
functionality to evaluate the shape functions and their derivatives at
the quadrature points in the physical cell and the determinant of the
Jacobian at the quadrature points in the reference cell. In turn, the
evaluation of the shape functions and derivatives in the physical cell
rely on their evaluation (and possibly the evaluation of the Jacobian) in the reference
cell (see, e.g., \eqref{eq:shfunx} and \eqref{shape_grad}). We note
that the evaluation of $\hat{\funmap}$ does not require any additional
information; it is the identity for Lagrangian elements and only
requires the Jacobian in the reference cell for vector-valued shape
functions (see \eqref{eq:contrapiola} and \eqref{eq:copiola}). In the
following, we present a set of data types that contain all this
information.

The evaluation of cell integrals involves the data type
\mytexttt{quadrature\_t} that represents the quadrature ${\rm Q}$,
\mytexttt{interpolation\_t}, that stores the values of the shape
functions and its first derivatives (either in the reference or
physical space) at the quadrature points of ${\rm Q}$, and a
\mytexttt{cell\_map\_t} that describes the mapping from a reference to a
physical cell $\geomap_\geophy$ (e.g., Jacobian-related
data). Additionally, the data type \mytexttt{cell\_integrator\_t}
provides the machinery to compute the \mytexttt{interpolation\_t}
corresponding to the physical space from the one at the reference
space and the \mytexttt{cell\_map\_t} at every cell of the
triangulation. In the following sections, we cover in detail these
software abstractions.

\subsection{Numerical quadrature} \label{subsec:numerical_quadrature}

The data type that in \FEMPAR{} represents an arbitrary 
quadrature rule is called \mytexttt{quadrature\_t} and is defined as
shown in Listing~\ref{lst:quadrature}.

\begin{lstlisting}[float=htbp,language={[03]Fortran},escapechar=@, label={lst:quadrature}, caption={The \mytexttt{quadrature\_t} data type.}]
type quadrature_t
 private
 integer(ip)           :: num_dims 
 integer(ip)           :: num_quadrature_points
 real(rp), allocatable :: coordinates(:,:)
 real(rp), allocatable :: weights(:)                         
contains
 public
 ! subroutine(this(inout)::class(quadrature_t),num_dims(in),num_quadrature_points(in)::integer(ip))
 procedure, non_overridable :: create => quadrature_create
 ! subroutine(this(inout)::class(quadrature_t))
 procedure, non_overridable :: free   => quadrature_free
 procedure, non_overridable :: fill_tet_gauss_legendre => quadrature_fill_tet_gauss_legendre @\label{loc:fill_tet_gp}@
 procedure, non_overridable :: fill_hex_gauss_legendre => quadrature_fill_hex_gauss_legendre @\label{loc:fill_hex_gp}@
 ... ! Rest of getter TBPs to preserve encapsulation
end type quadrature_t
\end{lstlisting}

In Listing ~\ref{lst:quadrature}, \mytexttt{coordinates(:,gp)} and \mytexttt{weights(gp)} store,
respectively,
$\xref_{gp} \in \mathbb{R}^{\mbox{\tt \scriptsize num\_dims}}$ and
$\mathrm{w}_{gp}$, for $gp=1,\ldots,\mbox{\tt num\_quadrature\_points}$.  It
might readily be observed from the interface of its {\tt create}
binding that \mytexttt{quadrature\_t} is designed to be simply a
placeholder for the quadrature points coordinates and its associated
weights. Indeed, this binding essentially allocates \mytexttt{coordinates(:,:)} and \mytexttt{weights(:)}. The code that ultimately decides how to distribute
the quadrature points over $\georef$ and set up its associated weights
is actually bounded to the \mytexttt{reference\_fe\_t} implementors
through the deferred binding with interface shown in Listing~\ref{lst:reference_fe_create_quadrature}.

\begin{lstlisting}[float=htbp,language={[03]Fortran},escapechar=@, label={lst:reference_fe_create_quadrature}, caption={The interface of the \mytexttt{create\_quadrature} deferred binding of \mytexttt{reference\_fe\_t}.}]
abstract interface
 ...
 subroutine reference_fe_create_quadrature ( this, quadrature, degree )
  class(reference_fe_t)        , intent(in)    :: this 
  type(quadrature_t)           , intent(inout) :: quadrature
  integer(ip)          optional, intent(in)    :: degree
 end subroutine reference_fe_create_quadrature
 ...
end interface
\end{lstlisting}

All \mytexttt{reference\_fe\_t} subclasses currently available in
\FEMPAR{} select by default a Gaussian quadrature that exactly
integrates mass matrix terms (within their implementation of the
binding in Listing~\ref{lst:reference_fe_create_quadrature}) by invoking \mytexttt{fill\_*\_gauss\_legendre} methods at lines Lines~\ref{loc:fill_tet_gp} and \ref{loc:fill_hex_gp} in Listing ~\ref{lst:quadrature}. This
quadrature can be solely determined from the attributes of the
\mytexttt{reference\_fe\_t} implementor at hand (its
topology and order).\footnote{As it is well known, considering n-cube
  topologies for $\georef$, for a Lagrangian reference \ac{FE} of order $p$
  and an affine geometrical map, we need a 1D Gaussian quadrature with $p+1$ points. For tetrahedral
  meshes with the Duffy transformation, we need to take
  $n=p+ {\rm ceiling}(d/2)$ to integrate exactly mass matrices (see Sect.~\ref{subsec:shape_functions_construction} for more details).}
However, in other more demanding situations, e.g., the integration of a trilinear
weak form, the user can provide
the desired quadrature degree through the \mytexttt{degree} optional
dummy argument. If more general scenarios to the ones currently
covered (e.g., a non-Gaussian quadrature) are to be addressed, then
the interface might be modified such that an optional parameter
dictionary is passed instead.

\subsection{Evaluation of reference cell shape functions} \label{subsec:interpolation}
As commented in the introduction of this section, to compute cell integrals \eqref{fematrix}, one needs to evaluate shape functions and their derivatives in the physical cell, which in turn rely on their evaluation in the reference cell (see, e.g., \eqref{eq:shfunx} and \eqref{shape_grad}). The values of the shape functions and their first derivatives at a set of quadrature points provided by a \mytexttt{quadrature\_t} instance are stored in the  \mytexttt{interpolation\_t} data type presented below. The same data type can be used to store this data in the reference or physical space.

Let us start with the evaluation of shape function in the reference space. The local \ac{FE} space on top of $\georef$ actually depends on the particular \mytexttt{reference\_fe\_t} implementor at hand. Consequently, this functionality has to be offered through a deferred binding of this abstract type. The interface of this binding is declared in Listing~\ref{lst:reference_fe_create_interpolation}. The subroutine overriding it in concrete subclasses is in charge of computing the shape functions values and derivatives at quadrature points in the reference space and stores them in a raw-data container of type \mytexttt{interpolation\_t} (to be discussed later in this section).

\begin{lstlisting}[float=htbp,language={[03]Fortran},escapechar=@, label={lst:reference_fe_create_interpolation}, caption={The interface of the \mytexttt{create\_interpolation} deferred binding of \mytexttt{reference\_fe\_t}.}]
abstract interface
 ...
 subroutine reference_fe_create_interpolation ( this, quadrature, reference_cell_interpolation )
   class(reference_fe_t), intent(in)    :: this 
   type(quadrature_t)   , intent(in)    :: quadrature
   type(interpolation_t), intent(inout) :: reference_cell_interpolation
 end subroutine reference_fe_create_interpolation
 ...
end interface
\end{lstlisting}

Let us remark several points related to this interface. First, this binding is typically called only once, and the data pre-computed and stored within the passed \mytexttt{interpolation\_t} dummy argument  is repeatedly re-used when transforming these values to an actual cell; see Sect.~\ref{subsec:cell_integrator}.  Second, this binding is designed such that all functions are evaluated at all quadrature points within a single call, instead of following a (much) finer granularity approach in which only one function is evaluated at a quadrature point per call.\footnote{Here (and in many other places) we try to maximize the granularity of each call to a deferred binding for efficiency reasons. The reader should be aware that calling to deferred bindings with the granularity of the latter approach would be very expensive, apart from preventing a number of potential compiler optimizations enabled by the former.} Third, we stress that the actual implementation of this deferred binding in \FEMPAR{} computes shape functions values and first derivatives in the reference space, whereas it lets the caller to selectively decide whether to compute or not the second derivatives of the shape functions, provided that they are expensive to compute and only required in very particular scenarios; see Sect.~\ref{subsec:numerical_integration}. Indeed, the code implementation of this feature is of cross-cutting nature, being reflected in several interfaces and data types in which the cell (and face) integration functionality is split. We will nevertheless omit here (and in the rest of sections) details regarding second derivatives (and its optional computation) in order to keep the presentation simple.

Let us now discuss on the rationale underlying \mytexttt{interpolation\_t}. This data type is not exposed at all to the user of \FEMPAR{}. It is instead used as an internal low-level container that lets the data types involved in the implementation of cell integrals exchange the sort of data subject to consideration. It is ultimately the responsibility of the concrete \mytexttt{reference\_fe\_t} subclass to decide how the data is actually laid out within the member variables of \mytexttt{interpolation\_t}. Thus, \mytexttt{reference\_fe\_t} is the only data type that can access or modify \mytexttt{interpolation\_t}. In its current flavour, \mytexttt{interpolation\_t} is a concrete (i.e., non-abstract) data type with a fixed set of multi-rank allocatable array member variables for storing shape function values and derivatives. For example, the one storing shape function values is a 3-rank array, where a \mytexttt{reference\_fe\_t} implementor may choose its indices, from left to right, to refer to the component of the shape function, the shape function, and the quadrature point, respectively. The \mytexttt{reference\_fe\_t} subclass is, however, completely free to lay out the data in these arrays, and it is in this flexibility where the extensibility of the software design to accommodate several \ac{FE} space realizations resides.  This, indeed has been proven to be sufficient to (efficiently) implement all \ac{FE} spaces currently available in \FEMPAR{}, including scalar, vector, and tensor-valued Lagrangian \acp{FE} (where higher-rank spaces are determined as the tensor product of the scalar spaces, and shape functions have only one non-zero component), and genuinely vector-valued \ac{FE} spaces (where more than one component of the shape function may be non-zero). 

\subsection{Geometrical mapping} \label{subsec:mapping}
A basic building block is the mapping $\geomap_K$ among the reference cell $\georef$ coordinate system and the one corresponding to an actual cell $K$ of the triangulation in the physical space; see Sect.~\ref{subsec:space_discretization} and~\ref{subsec:fe_concept}. For example, we are able to pull back the gradients of the shape functions from the reference to the physical space in \eqref{shape_grad} using the Jacobian of the transformation evaluated at quadrature points, or to evaluate the source term at quadrature points in real space. The Jacobian is also required to  the transform the integral from the physical to the reference space in \eqref{fematrix} and to compute the Piola transformations in div and curl-conforming \ac{FE} spaces (see \eqref{eq:contrapiola} and \eqref{eq:copiola}). The derived type \mytexttt{cell\_map\_t} in \FEMPAR{} is designed to be a placeholder for the data required to provide this sort of services. It is declared as shown in Listing~\ref{lst:fe_map}. The rationale underlying the inheritance relationship among \mytexttt{cell\_map\_t} and \mytexttt{base\_map\_t} will be made clear in Sect.~\ref{sec:face_integration}.

\begin{lstlisting}[float=htbp,language={[03]Fortran},escapechar=@, label={lst:fe_map}, caption={The \mytexttt{cell\_map\_t} data type.}]
type base_map_t
 private
 integer(ip)                 :: num_dims
 integer(ip)                 :: num_quadrature_points
 real(rp)     , allocatable  :: jacobian(:,:,:)  
 real(rp)     , allocatable  :: det_jacobian(:)  
 type(interpolation_t)       :: interpolation  
 type(point_t), allocatable  :: coordinates_quadrature_points(:)
 type(point_t), allocatable  :: coordinates_nodes(:)
contains
  public
  ! subroutine(this(inout)::class(base_map_t))
  procedure                  :: free => base_map_free
  ... ! Getter TBPs to preserve encapsulation;
end type base_map_t

type, extends(base_map_t) :: cell_map_t
 private       
 real(rp)     , allocatable  :: inv_jacobian(:,:,:)
 ... ! Member variables related to the hessian (2nd derivatives) of the mapping 
     ! (omitted to keep the presentation simple)  
contains
 public
 ! subroutine(this(inout)      :: class(cell_map_t),
 !            quadrature(in)   :: type(quadrature_t),
 !            reference_fe(in) :: class(reference_fe_t))
 procedure, non_overridable :: create => cell_map_create
 ! subroutine(this(inout)::class(cell_map_t),quadrature(in)::type(quadrature_t))
 procedure, non_overridable :: update => cell_map_update
 ! subroutine(this(inout)::class(cell_map_t))
 procedure, non_overridable :: free   => cell_map_free
 ... ! Rest of getter TBPs to preserve encapsulation;
end type cell_map_t
\end{lstlisting}

The \mytexttt{create} binding of \mytexttt{cell\_map\_t} takes as input a \mytexttt{quadrature\_t} instance with a set of integration points where $\jacobian(\xref_{gp})$, $\invjacob(\xref_{gp})$, and $|\jacobian(\xref_{gp})|$ are to be evaluated (see Listing~\ref{lst:fe_map}). These geometry-related data are stored in the \mytexttt{jacobian(:,:,gp)}, \mytexttt{inv\_jacobian(:,:,gp)}, and \mytexttt{det\_jacobian(gp)} allocatable array member variables of \mytexttt{cell\_map\_t}, respectively, and allocated during a call to this binding. Apart from a \mytexttt{quadrature\_t} instance, \mytexttt{cell\_map\_t} also requires a description of the (discrete) space of functions to which $\geomap_K$ belongs. \FEMPAR{} supports mappings $\geomap_K$ belonging to abstract \ac{FE} spaces (e.g., high-order polynomial \ac{FE} spaces or spline-based spaces). The \mytexttt{reference\_fe} dummy argument of polymorphic type \mytexttt{reference\_fe\_t} serves the purpose. (We note that dynamic run-time polymorphism in this particular context let us re-use \mytexttt{cell\_map\_t}, e.g., with an arbitrary cell topology.) It turns out that the only information that \mytexttt{reference\_fe\_t} has to provide to \mytexttt{cell\_map\_t} are its shape functions, first derivatives, and (on demand) second order derivatives at the quadrature points (in the reference space). The \mytexttt{interpolation} member variable (see Listing~\ref{lst:fe_map}) is used by \mytexttt{reference\_fe} to exchange this sort of data with \mytexttt{cell\_map\_t} via a call to the \mytexttt{create\_interpolation} binding of the former (see Listing~\ref{lst:reference_fe_create_interpolation}) during a call to the $\mytexttt{create}$ binding of the latter.

While the \mytexttt{create} \ac{TBP} of \mytexttt{cell\_map\_t} is designed to be called once, the \mytexttt{update} \ac{TBP} of \mytexttt{cell\_map\_t} is, however, designed to be called multiple times, once per every cell $K$ of the triangulation. A pre-condition of \mytexttt{update} is that the \mytexttt{nodes\_coordinates(:)}  scratch member variable (see Listing~\ref{lst:fe_map}) has been loaded with the coordinates in real space of the nodes describing the geometry of $K$ (stored into \mytexttt{point\_t} instances). Once this pre-condition is fulfilled, $\geomap_K$ can be expressed as a linear combination of the \mytexttt{reference\_fe\_t} shape functions with \mytexttt{nodes\_coordinates(:)} being the corresponding coefficients in the expansion. At this stage, \mytexttt{coordinates\_quadrature\_points(:)}, which stores the coordinates of quadrature points in real space, and \mytexttt{jacobian(:,:,:)}, can be easily computed. Finally, \mytexttt{inv\_jacobian(:,:,:)} and \mytexttt{det\_jacobian(:)} can be computed from \mytexttt{jacobian(:,:,:)} using straightforward numerical algorithms.

\subsection{Evaluation of shape functions in the physical space} \label{subsec:cell_integrator}

The user code that evaluates cell integrals in \eqref{fematrix}, may need the value, gradient, curl, and divergence of the shape functions at the integration points in the physical space, provided that  we want to unburden \FEMPAR{} users from the complexity of having to explicitly apply mapping transformations.  As commented in Sect. \ref{sec:fe_meth}, the mapping that transforms a shape function $\shapetestref{a}(\xref)$ in the reference \ac{FE} space into the one in the physical space $\shapetest{a}(\x) = \hat{\funmap}_\geophy(\shapetestref{a}) \circ \geomap^{-1}_\geophy$, depends on the particular \ac{FE} space at hand; see Sect.~\ref{subsec:h1_conforming_fes},~\ref{subsec:hdiv_conforming_fes}, and~\ref{subsec:hcurl_conforming_fes} for details.  For this reason, the actual code that performs these transformations is not actually bounded to \mytexttt{cell\_map\_t}, but to \mytexttt{reference\_fe\_t}, through the deferred binding with interface declared in Listing~\ref{lst:reference_fe_map}.

\begin{lstlisting}[float=htbp,language={[03]Fortran},escapechar=@, label={lst:reference_fe_map}, caption={The interface of the \mytexttt{apply\_cell\_map} deferred binding of \mytexttt{reference\_fe\_t}.}]
abstract interface
 ...
 subroutine reference_fe_apply_cell_map ( this, cell_map, &
                               interpolation_reference_cell, interpolation_real_cell )
  class(reference_fe_t), intent(in)    :: this 
  type(cell_map_t)     , intent(in)    :: cell_map
  type(interpolation_t), intent(in)    :: interpolation_reference_cell
  type(interpolation_t), intent(inout) :: interpolation_real_cell
 end subroutine reference_apply_cell_map
 ...
end interface
\end{lstlisting}

The \mytexttt{interpolation\_reference\_cell} input dummy argument of \mytexttt{apply\_cell\_map} (see Listing~\ref{lst:reference_fe_map}) must have been obtained from a call to the binding in Listing~\ref{lst:reference_fe_create_interpolation} invoked on the same \mytexttt{reference\_fe\_t} instance. The output dummy argument \mytexttt{interpolation\_real\_cell} holds the shape functions and their derivatives evaluated at quadrature points in physical space (see \eqref{eq:shfunx} and \eqref{shape_grad}). It is also assumed that, on input, \mytexttt{interpolation\_real\_cell} already contains the data that does not have to be re-computed on each mesh cell, e.g., the value of the shape functions on integration points for Lagrangian \acp{FE}; see the discussion related to the \mytexttt{update} binding below for the strategy that we follow in order to fulfill this requirement. This leaves room for optimization in the implementation of this deferred binding (on subclasses), since these quantities do not have to be re-computed on each cell. The \mytexttt{reference\_fe\_t} subclass uses the \mytexttt{cell\_map\_t} instance (passed to the \mytexttt{apply\_cell\_map} binding, see Listing~\ref{lst:reference_fe_map}) as a placeholder for the data required to provide the mapping transformations required. 

We stress, however, that {\tt interpolation\_t} is a low level structure that is not designed as a data type that \FEMPAR{} users have to interact with, for reasons made clear in Sect.~\ref{subsec:interpolation}. Therefore, we need to introduce an additional data type in our software design, called \mytexttt{cell\_integrator\_t}, that, among other services, is able to fetch raw data from {\tt interpolation\_t} into field data types (i.e., scalars, vectors, and tensors)  the user can be easily familiarized with. This data type is declared as shown in Listing~\ref{lst:cell_integrator}.

\begin{lstlisting}[float=htbp,language={[03]Fortran},escapechar=@, label={lst:cell_integrator}, caption={The \mytexttt{cell\_integrator\_t} data type.}]
type cell_integrator_t 
 private
 ... ! num_dims, num_quadrature_points, num_shape_functions
 class(reference_fe_t), pointer :: reference_fe
 type(interpolation_t)          :: interpolation_reference_cell
 type(interpolation_t)          :: interpolation_real_cell
contains
 public
 ! subroutine(this(inout)      :: class(cell_integrator_t),
 !            quadrature(in)   :: type(quadrature_t),
 !            reference_fe(in) :: class(reference_fe_t))
 procedure, non_overridable :: create => cell_integrator_create
 ! subroutine(this(inout) :: class(cell_integrator_t),
 !            cell_map(in)  :: type(cell_map_t))
 procedure, non_overridable :: update => cell_integrator_update
 ! subroutine(this(inout) :: class(cell_integrator_t))
 procedure, non_overridable :: free   => cell_integrator_free
 
 ! Evaluation of shape functions values
 procedure, non_overridable, private :: get_values_scalar => cell_integrator_get_values_scalar
 procedure, non_overridable, private :: get_values_vector => cell_integrator_get_values_vector
 procedure, non_overridable, private :: get_values_tensor => cell_integrator_get_values_tensor
 generic :: get_values => get_values_scalar, get_values_vector, get_values_tensor

 ... ! Evaluation of shape functions gradients
 ... ! Evaluation of shape functions hessians
 ... ! Evaluation of shape functions divergences
 ... ! Evaluation of shape functions curls
 ... ! Rest of TPBs to preserve encapsulation
end type cell_integrator_t
\end{lstlisting}

An instance of \mytexttt{cell\_integrator\_t} is created from a quadrature rule (where the shape functions and their derivatives are to be evaluated) and a polymorphic \mytexttt{reference\_fe\_t} instance describing the reference \ac{FE} space at hand; see interface of the \mytexttt{create} binding in Listing~\ref{lst:cell_integrator}. During this stage, \mytexttt{reference\_fe} creates the \mytexttt{interpolation\_reference\_cell} member variable of \mytexttt{cell\_integrator\_t} via \mytexttt{create\_interpolation}; see Listing~\ref{lst:reference_fe_create_interpolation}. It also clones \mytexttt{interpolation\_reference\_cell} into \mytexttt{interpolation\_real\_cell}, and copies the contents of the former into the latter. This lets \mytexttt{cell\_integrator\_t} to fulfill later on the pre-condition on the last dummy argument of \mytexttt{apply\_cell\_map}. The \mytexttt{create} binding also associates  its 
polymorphic pointer \mytexttt{reference\_fe} member variable to the \mytexttt{reference\_fe\_t} instance provided to it on input. This pointer is required later on by the \mytexttt{update} and \mytexttt{get\_*}
bindings (see discussion in the sequel).

The \mytexttt{update} binding of \mytexttt{cell\_integrator\_t} simply invokes \mytexttt{apply\_cell\_map} on its polymorphic \mytexttt{reference\_fe} member variable, using the instance of \mytexttt{cell\_map\_t} provided on input to \mytexttt{update}, and the two \mytexttt{interpolation\_t} member variables as actual arguments, respectively; see Listings~\ref{lst:reference_fe_map} and \ref{lst:cell_integrator}. It leaves the \mytexttt{cell\_integrator\_t} instance on which it is invoked in a state such that it is able to provide the services it was primarily designed for. These are offered through the \mytexttt{get\_values}, \mytexttt{get\_gradients}, \mytexttt{get\_divergences}, \mytexttt{get\_curls}, etc., {\em generic bindings}. We note that \mytexttt{cell\_integrator\_t} is designed such that it can handle either scalar, vector, or tensor-valued \mytexttt{reference\_fe\_t} instances (see Sect.~\ref{subsec:reference_fe_space}). With this purpose in mind, each of the aforementioned generic bindings are overloaded with subroutines that have appropriate interfaces for these three types of \acp{FE}. For example, the subroutine overloading \mytexttt{get\_gradients} in the case of scalar-valued \acp{FE} is declared and implemented as shown in Listing~\ref{lst:get_gradients_scalar}, with \mytexttt{vector\_field\_t} representing a $d$-dimensional rank-1 tensor; the interface of the one corresponding to vector-valued \acp{FE} only differs from the one above on the base type of the \mytexttt{gradients} allocatable array dummy argument, which is of base type \mytexttt{tensor\_field\_t} (i.e., data type representing a $d$-dimensional rank-2 tensor).

\begin{lstlisting}[float=htbp,language={[03]Fortran},escapechar=@, label={lst:get_gradients_scalar}, caption={The code implementing the \mytexttt{get\_gradients\_scalar} binding of  \mytexttt{cell\_integrator\_t} ultimately relies on a deferred binding of \mytexttt{reference\_fe\_t} with the same name.}]
subroutine cell_integrator_get_gradients_scalar (this, gradients)
 implicit none
 class(cell_integrator_t)              , intent(in)    :: this
 type(vector_field_t),      allocatable, intent(inout) :: gradients(:,:)
 call this%reference_fe%get_gradients(this%interpolation_real_cell,gradients) @\label{loc:get_gradients}@
end subroutine cell_integrator_get_gradients_scalar
\end{lstlisting}

Let us remark some important points with respect to the subroutines overloading the generic bindings of \mytexttt{cell\_integrator\_t}. First, we note that the actual argument passed in place of, e.g., the \mytexttt{gradients(:,:)} dummy argument in Listing~\ref{lst:get_gradients_scalar}, is intended to be actually declared in code written by the user of \FEMPAR{}. Provided that \FEMPAR{}  can support variable degree \acp{FE} on top of different triangulation cells (see Sect.~\ref{sec:fe_space}), the \mytexttt{allocatable} attribute of the \mytexttt{gradients(:,:)} dummy argument not only unburdens the user from the complexity of having to (pre)allocate this array, but even from the one associated to variable degree \acp{FE}. For example, if on input, the size of \mytexttt{gradients(:,:)} is not sufficient to hold the data to be provided by the \mytexttt{cell\_integrator\_t} instance corresponding to the \mytexttt{reference\_fe\_t} on top of the current triangulation cell, then it can be re-allocated to the appropriate size. Second, this binding is designed such that all functions are evaluated at all quadrature points within a single call, justifying why the dummy argument has to be a rank-2 allocatable array.\footnote{This represents another design decision in the seek of maximizing the granularity of the calls to deferred bindings for code efficiency reasons.} At this point, let us note that all subroutines subject to consideration ultimately rely on (deferred bindings of) \mytexttt{reference\_fe\_t}; see, e.g., line~\ref{loc:get_gradients} in Listing~\ref{lst:get_gradients_scalar}. We recall that \mytexttt{reference\_fe\_t} must mediate in any process that requires retrieving data from \mytexttt{interpolation\_t}; see Sect.~\ref{subsec:interpolation}.

\subsection{Cell integration user code example} \label{sec:cell_integration_example}
At this point of the discussion, we are already in position to show user code that evaluates the entries of 
the (current cell) local matrix for the Example~\ref{poisson_problem} presented in Sect.~\ref{subsec:weak_form}.
This code is sketched in
Listing~\ref{lst:element_matrix_entries_computation}. This code
would be bounded to a subclass of the \mytexttt{discrete\_integration\_t} abstract data type 
presented in Sect.~\ref{sec:disc_int} suitable for the Galerkin discretization of the 
Poisson problem.

\begin{lstlisting}[float=htbp,language={[03]Fortran},escapechar=@, label={lst:element_matrix_entries_computation}, caption={User-level code illustrating the usage of cell integration data structures in order to compute the element matrix for the Example~\ref{poisson_problem} presented in Sect.~\ref{subsec:weak_form}.}]
type(vector_field_t), allocatable :: shape_gradients(:,:)
real(rp)            , allocatable :: element_matrix(:,:)
... ! Declaration + variable initialization
! Loop over all cells K
... ! Update cell integration data structures to reflect the status corresponding to current cell
element_matrix = 0.0_rp
call cell_integrator%get_gradients(shape_gradients)
do q = 1, quadrature%get_num_quadrature_points()
   detJ_x_wq = cell_map%get_det_jacobian(q) * quadrature%get_weight(q)
   do b = 1, num_shape_functions
      do a = 1, num_shape_functions
         element_matrix(a,b) = element_matrix(a,b) + & 
                               shape_gradients(a,q) * shape_gradients(b,q) * detJ_x_wq
      end do
   end do
end do
... ! Assemble element_matrix into global linear system coefficient matrix
! End Loop over all cells K
\end{lstlisting}

The reader may note from Listing~\ref{lst:element_matrix_entries_computation} that \FEMPAR{} also offers an expression syntax that lets its users code weak forms in a way that resembles their mathematical expression. The user is in charge of explicitly writing the expression of the numerical integration in the reference cell, i.e., of explicitly implementing the quadrature point summation (loop) and handling the determinant of the Jacobian and the quadrature point weighting in \eqref{fematrix}. However, the evaluation of the shape function and their gradients, curls, etc., at the quadrature points in the physical space (e.g., expressions  \eqref{eq:shfunx} and \eqref{shape_grad}) are completely hidden to the user. This can be achieved using a feature of modern programming languages called {\em operator overloading}. (We refer to \cite{adams_fortran_2009} for a detailed exposition of this mechanism in Fortran2003.) Common (contraction) operations among tensors are provided by means of overloaded intrinsic and library-defined operators. For example, the \mytexttt{operator(*)} generic interface (corresponding to the \mytexttt{*} intrinsic operator) has to be overloaded with the single contraction of rank-1 tensors, and the multiplication of a rank-1 tensor by a scalar to let our code compile. A crucial design requirement in the seek of code efficiency is that no dynamic memory allocation/deallocation is involved as the partial evaluation of sub-expressions proceeds (in the order dictated by operator associativity and priority rules in Fortran). In order to fulfill this requirement, the data types representing vectors and tensors are declared such that their entries are stored in an array member variable {\em of size known at compilation time}. This size is stored in the library-level parameter constant \mytexttt{SPACE\_DIM}, defined as the maximum number of space dimensions of the physical space in which the physical problem is posed. By default, \FEMPAR{} is prepared to deal with 3D simulations, but the code is written such that a 2D simulation might also be performed if \mytexttt{SPACE\_DIM} is equal to 3, 
at the price of extra storage and computation.\footnote{In fact, 2D problems for \acp{PDE} that involve curl operators require \mytexttt{SPACE\_DIM} to be equal to 3.} Higher dimensional problems could be considered by compiling \FEMPAR{} with a larger value for \mytexttt{SPACE\_DIM}. Apart from avoiding dynamic memory allocation/deallocation during the evaluation of weak forms, this solution has the following advantages:  1) there is no need to explicitly have the number of dimensions as a member variable of the data types representing vectors and tensors; 2) the limits of the loops implementing tensor contraction operations are known at compilation time, enabling compiler optimizations. We finally stress that we preferred this solution over the usage of Fortran2003 parameterized data types~\cite{adams_fortran_2009} due to the lack of support of this feature in some of the most popular compilers widely available on high-end computing environments. 

\section{Evaluation of facet integrals} \label{sec:face_integration}


This section covers the data types (and their interactions) in which
the evaluation of integrals over the facets of the triangulation is
grounded on. The integration of facet-wise matrices and vectors (see,
e.g., \eqref{facematrix}) involves the evaluation of shape functions
and gradients of the neighbouring cells at the quadrature points
within the facet in the physical space and the Jacobian of the facet
map at the reference space. As described in
Sect.~\ref{sec:cell_integration}, the former quantities are computed
at every neighbouring cell from their values at the reference space
and the Jacobian of the cell mapping. The evaluation of interior facet
also requires the computation of the permutation $\Pi({\rm gp})$ (see 
\eqref{eq:face_integral_mapped_back_and_approximated}) provided that the coordinate systems of the
cells surrounding the facet might not be aligned in physical space.

In \FEMPAR{} the assembly process of the global linear system
underlying the discrete weak problem~\eqref{eq:ip_dg_formulations}
involves two loops, over all cells and facets, respectively.  In the
former loop, a cell-wise matrix $\fematrix^{K}$ and vector
$\ferhs^{K}$ are computed per each cell.  These hold the partial
contributions of the cell to the corresponding entries of the global
coefficient matrix and right-hand side vector, respectively.  The data
structures involved in their efficient computation have been already
covered in Sect.~\ref{sec:cell_integration}.  In the latter loop,
and assuming that we are sitting on an interior facet 
$\facephy \in \mathcal{F}^{\Omega}_{h}$, four facet-wise matrices,
namely $\fematrix^{\facephy}_{K^+ K^+}$,
$\fematrix^{\facephy}_{K^+ K^-}$, $\fematrix^{\facephy}_{K^- K^+}$,
and $\fematrix^{\facephy}_{K^- K^-}$ are computed (see Sect.~\ref{sec:nonconfm}).
 
We depict  in Fig.~\ref{fig:uml_face_integration_data_types} a complete UML class diagram of the data types involved in the
evaluation of facet integrals and their relationships. 
The data types the user has to
ultimately interact with are $\mytexttt{quadrature\_t}$, which holds
the facet quadrature points and weights, $\mytexttt{facet\_maps\_t}$,
which handles (i.e., stores, updates, provides) all the geometrical
related data of the facet and neighbouring cells $K^+$ and $K^-$,
and, finally, $\mytexttt{facet\_integrator\_t}$, which stores and
updates shape function values and first derivatives, and provides shape function values, gradients, curls, etc., of $K^+$ and $K^-$ evaluated at facet quadrature points in real space. The
rest of data types in Fig.~\ref{fig:uml_face_integration_data_types}
are auxiliary data types, not exposed to the user, which aid the
latter two in the implementation of their corresponding services. The
reader might readily observe in
Fig.~\ref{fig:uml_face_integration_data_types} that our software
design is such that the data types that
provide support to the evaluation of cell integrals, i.e.,
\mytexttt{quadrature\_t}, \mytexttt{cell\_map\_t}, and
\mytexttt{cell\_integrator\_t} (see
Sect.~\ref{sec:cell_integration}), can be re-used to a large extent
for the evaluation of facet integrals. As we will see in the rest of
the section, some of the methods to be invoked in order to control
their respective life cycles in the context of facet integrals are
nevertheless different from the ones to be invoked in the context of
cell integrals; see, e.g., the signature of the
\mytexttt{create\_restricted\_to\_facet} binding of \mytexttt{cell\_integrator\_t}
in Fig.~\ref{fig:uml_face_integration_data_types} compared to that
of its \mytexttt{create} binding in Listing~\ref{lst:cell_integrator}.

\tikzstyle{class}=[rectangle, draw=black, align=left, text=black, text justified,  minimum height=1cm] 
\tikzstyle{inherit}=[open triangle 60-]
\tikzstyle{assoc}=[-angle 60]
\tikzstyle{depen}=[dashed,-angle 60]
\tikzstyle{comp}=[diamond-angle 60]

\begin{figure}[htbp]
\begin{minipage}{\textwidth}
\begin{tikzpicture}[node distance=1.cm,auto]

    \node (map) [class, text width=2.8cm, rectangle split, rectangle split parts=3] { 
          \small \tt base\_map\_t 
          \nodepart{second} 
          \nodepart{third} ... };

    \node (aux1) [minimum width=0.0cm, minimum height=3.0cm, below=of map] {};

    \node (face_map) [class, text width=4.9cm, rectangle split, rectangle split parts=3, left=of aux1] { 
          \tt \small facet\_map\_t
          \nodepart{second} \footnotesize  $-$\mytexttt{outward\_unit\_normals(:,:)} \\ ...
          \nodepart{third}  \footnotesize  $+$\mytexttt{create(q,ref\_fe\_geo)} \\ 
                                           $+$\mytexttt{update(q,reorientation\_factor)} \\ ...};

    \node (fe_map) [class, text width=6.7cm, rectangle split, rectangle split parts=3, right=of aux1] { 
          \small \tt cell\_map\_t 
          \nodepart{second} ... See Listing~\ref{lst:fe_map}
          \nodepart{third} \footnotesize  $+$\mytexttt{create\_restricted\_to\_facet(q,ref\_fe\_geo, \&} \\
                            {\color{white}$+$\mytexttt{create\_restricted\_to\_facet(}}\mytexttt{facet\_lid)} \\
                                          $+$\mytexttt{update(q)} \\  ...};


    \node (face_maps) [class, text width=5.3cm, rectangle split, rectangle split parts=3, below=of face_map] { 
          \small \tt facet\_maps\_t 
          \nodepart{second} \footnotesize  $-$\mytexttt{is\_at\_boundary} \\ ...
          \nodepart{third}  \footnotesize  $+$\mytexttt{create(q,ref\_fe\_geo$^{+}$,ref\_fe\_geo$^{-}$)} \\ 
                                                  $+$\mytexttt{update(q,facet\_lids(2))} \\  ...};

    \node(fe_map_face_restriction) [class, text width=4.6cm, rectangle split, rectangle split parts=3, below=of fe_map] { 
          \small \tt cell\_map\_facet\_restriction\_t 
          \nodepart{second} \footnotesize  $-$\mytexttt{current\_facet\_lid} \\ ...
          \nodepart{third}  \footnotesize  $+$\mytexttt{create(q,ref\_fe\_geo)} \\
                                                  $+$\mytexttt{update(q,facet\_lid)} \\  ...};

    \node (reference_fe) [class, text width=2.8cm, rectangle split, rectangle split parts=3, below=of fe_map_face_restriction] { 
          \small \tt reference\_fe\_t 
          \nodepart{second} 
          \nodepart{third} ... };

    \node (aux2) [minimum width=0.0cm, minimum height=0.0cm, below=of face_maps] {};


    \node (face_integrator) [class, text width=5.8cm, rectangle split, rectangle split parts=3, below=of aux2] { 
          \small \tt facet\_integrator\_t 
          \nodepart{second} \footnotesize  $-$\mytexttt{is\_at\_boundary} \\  $-$\mytexttt{current\_permutation\_index} \\ 
                                           $-$\mytexttt{qpoints\_perm(:,:)}  \\ ...
          \nodepart{third}  \footnotesize  $+$\mytexttt{create(q,ref\_fe$^{+}$,ref\_fe$^{-}$)} \\
                                           $+$\mytexttt{update(facet\_maps,facet\_lids(2), \&} \\ 
                            {\color{white} $+$\mytexttt{update(}}\mytexttt{permutation\_index)} \\  ...};

    \node (cell_integrator_face_restriction) [class, text width=6.cm, rectangle split, rectangle split parts=3, right=of face_integrator] { 
          \small \tt cell\_integrator\_facet\_restriction\_t 
          \nodepart{second} \footnotesize  $-$\mytexttt{current\_facet\_lid} \\ ...
          \nodepart{third}  \footnotesize  $+$\mytexttt{create(q,ref\_fe)} \\
                                           $+$\mytexttt{update(cell\_map,facet\_lid)} \\  ... };


    \node (cell_integrator) [class, text width=6.9cm, rectangle split, rectangle split parts=3, below=of cell_integrator_face_restriction] { 
          \small \tt cell\_integrator\_t 
          \nodepart{second} 
          \nodepart{third} \footnotesize  $+$\mytexttt{create\_restricted\_to\_facet(q,ref\_fe,facet\_lid)} \\ 
                                                 $+$\mytexttt{update(cell\_map)} \\ ...   };


    \draw[inherit] (map.south)  -- ++(0,-0.5) -|  (face_map.north);
    \draw[inherit] (map.south)  -- ++(0,-0.5) -|  (fe_map.north);

    \draw[comp] (face_maps.north) -- node[near start,right]{\footnotesize \tt 1} (face_map.south);
    \draw[comp]  (fe_map_face_restriction.north)  -- node[midway,right]{\footnotesize \tt num\_facets} (fe_map.south);
    \draw[comp]  (face_maps.east) -- node[midway,above]{\footnotesize \tt 2} (fe_map_face_restriction.west);

    \draw[assoc] (face_maps.south)       |- node[near start,right]{\footnotesize \tt 2}  (reference_fe.west);
    \draw[assoc] (face_integrator.north) |- node[near start,right]{\footnotesize \tt 2}  (reference_fe.west);

    \draw[comp] (face_integrator.east) to node[near start,above]{\footnotesize \tt 2} (cell_integrator_face_restriction.west);
    \draw[comp] (cell_integrator_face_restriction.south) to node[midway]{\footnotesize \tt num\_facets} (cell_integrator.north);

\end{tikzpicture}

\caption{UML class diagram of the data types on which the numerical evaluation of facet integrals is grounded on.}
\label{fig:uml_face_integration_data_types}
\end{minipage}
\end{figure}
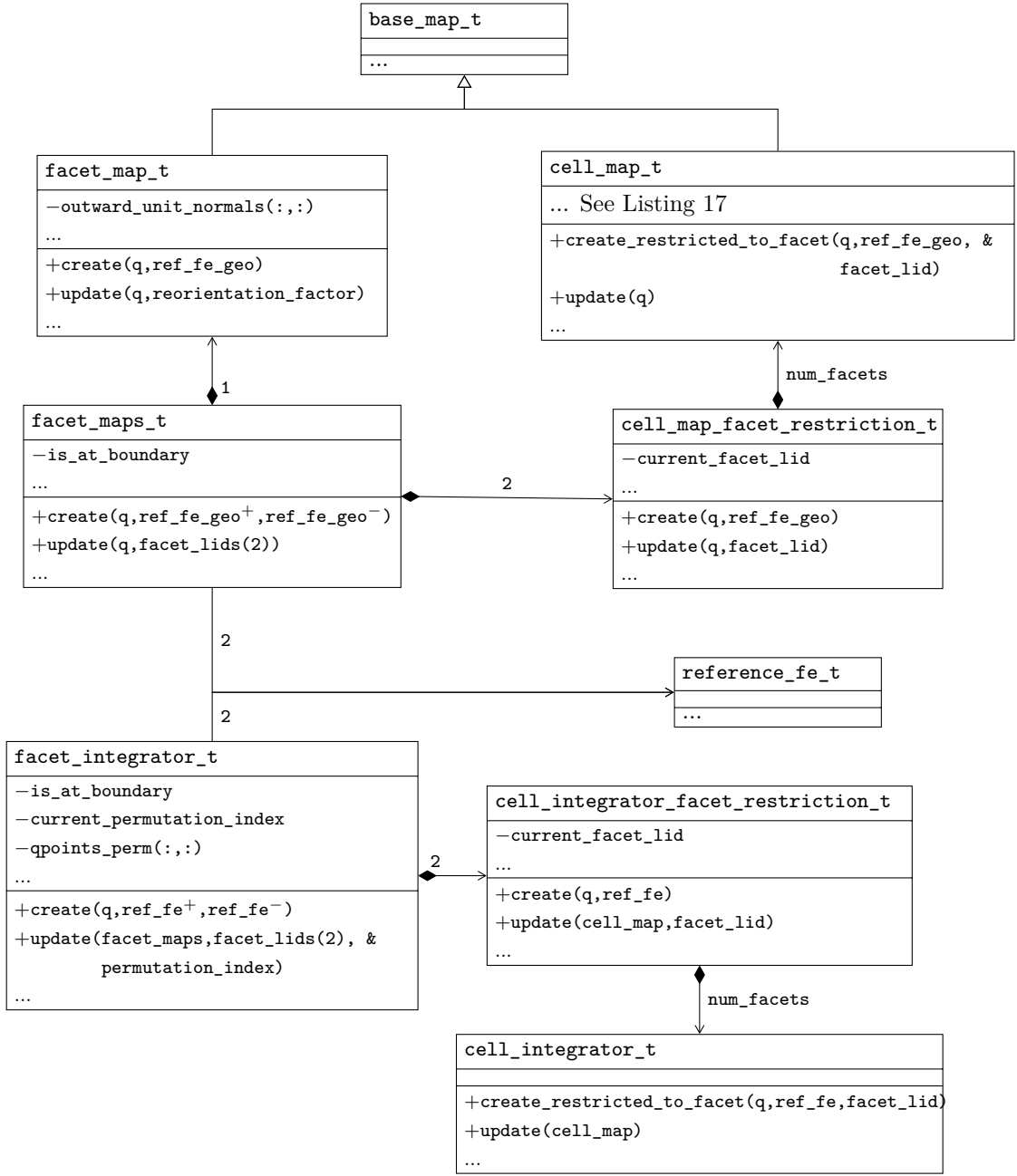

\subsection{Numerical quadrature} \label{subsec:face_quadrature}

The data type \mytexttt{quadrature\_t} is designed to be a placeholder for the {\em facet quadrature} points $\xref_q$ and its associated weights ${\rm  w}_{q}$. However, the code that ultimately decides how to distribute
$\xref_q$ over the reference facet $\faceref$ coordinate system, and
set up ${\rm w}_{q}$, is bounded to \mytexttt{reference\_fe\_t}, in particular through the deferred binding
with interface shown in Listing~\ref{lst:reference_fe_create_face_quadrature}.
We refer to Sect.~\ref{subsec:numerical_quadrature} for the
rationale underlying the \mytexttt{degree} optional dummy argument
of this deferred binding. 

\begin{lstlisting}[float=h!,language={[03]Fortran},escapechar=@, label={lst:reference_fe_create_face_quadrature}, caption={The interface of the \mytexttt{create\_facet\_quadrature} deferred binding of \mytexttt{reference\_fe\_t}.}]
abstract interface
 ...
 subroutine reference_fe_create_facet_quadrature ( this, quadrature, degree  )
  class(reference_fe_t)        , intent(in)    :: this 
  type(quadrature_t)           , intent(inout) :: quadrature
  integer(ip)          optional, intent(in)    :: degree
 end subroutine reference_fe_create_facet_quadrature
 ...
end interface
\end{lstlisting}

\subsection{Geometrical mappings} \label{subsec:face_mapping}
The \mytexttt{facet\_maps\_t} data type in Fig.~\ref{fig:uml_face_integration_data_types} handles the geometrical facet mapping and the two geometrical cell mappings. The facet mapping is represented by \mytexttt{facet\_map\_t}, whereas the cell mappings by \mytexttt{cell\_map\_t}; see Sect.~\ref{subsec:facet_mapping} and~\ref{subsec:neig_cell_mappings}, respectively.

\subsubsection{Facet mapping} \label{subsec:facet_mapping}
As illustrated in Fig.~\ref{fig:uml_face_integration_data_types},  \mytexttt{facet\_maps\_t}  is composed, among others, of a single instance of type \mytexttt{facet\_map\_t}. The member variables (and associated code) that are common to \mytexttt{facet\_map\_t} and \mytexttt{cell\_map\_t} are factored into a superclass \mytexttt{base\_map\_t} (see Listing~\ref{lst:fe_map}). \mytexttt{facet\_map\_t} handles all data related to the facet map $\geomap_\facephy$, including the facet outward unit normals (see Fig.~\ref{fig:uml_face_integration_data_types}). An extra 2-rank real allocatable array member variable, \mytexttt{outward\_unit\_normals(:,:)}, stores the facet outward unit normals (with respect to $K^+$ by convention) evaluated at facet quadrature points in real space, as required by \eqref{eq:face_integral_mapped_back_and_approximated}; $\normal^-(\x_{\rm gp})$ can be simply obtained as $\normal^-(\x_{{\rm gp}})=-\normal^+(\x_{\rm gp})$.

Let us now see how \mytexttt{facet\_maps\_t} controls the life
cycle of its \mytexttt{facet\_map\_t} instance. The \mytexttt{create} binding of \mytexttt{facet\_map\_t} takes a \mytexttt{quadrature\_t} instance with the facet quadrature points. $\facejacobian(\xref_{\rm gp})$ and  $|\facejacobian(\xref_{\rm gp})|$ are evaluated at these quadrature points and stored in the \mytexttt{jacobian} and \mytexttt{det\_jacobian} member variables, which are allocated during a call to this binding together with \mytexttt{outward\_unit\_normals(:,:)}. Apart from a \mytexttt{quadrature\_t} instance, \mytexttt{facet\_map\_t} also requires a description of the discrete, lower dimensional space of functions on top of the reference facet $\faceref$ to which  $\geomap_\facephy$ belongs. The \mytexttt{ref\_fe\_geo} dummy argument of \mytexttt{create}, of polymorphic type \mytexttt{reference\_fe\_t}, is provided for this purpose; in particular, \mytexttt{facet\_maps\_t} sends the \mytexttt{reference\_fe\_t} on top of $K^+$ as an actual argument to the \mytexttt{ref\_fe\_geo} dummy argument in order to comply with the above described convention for the normals. The \mytexttt{interpolation\_t} member variable of \mytexttt{facet\_map\_t} (see Listing~\ref{lst:fe_map}) is used by \mytexttt{ref\_fe\_geo} to exchange with \mytexttt{facet\_map\_t} the shape function values and their derivatives. To this end, \mytexttt{reference\_fe\_t} is equipped with the \mytexttt{create\_facet\_interpolation} deferred binding (see its signature in Listing~\ref{lst:reference_fe_create_face_interpolation}) that computes these quantities on top of the reference facet $\faceref$.

\begin{lstlisting}[float=htbp,language={[03]Fortran},escapechar=@, label={lst:reference_fe_create_face_interpolation}, caption={The signature of the \mytexttt{create\_facet\_interpolation} deferred binding of \mytexttt{reference\_fe\_t}.}]
abstract interface
 ...
 subroutine reference_fe_create_facet_interpolation ( this, quadrature, & 
                                                     reference_facet_interpolation )
   class(reference_fe_t) , intent(in)    :: this 
   type(quadrature_t)    , intent(in)    :: quadrature
   type(interpolation_t) , intent(inout) :: reference_facet_interpolation
 end subroutine reference_fe_create_facet_interpolation
 ...
end interface
\end{lstlisting}

The \mytexttt{update} binding of \mytexttt{facet\_map\_t} is intended
to be called once per facet loop iteration, i.e., once per each facet
of the triangulation. A pre-condition of this binding is that the
\mytexttt{nodes\_coordinates(:)} scratch member array of
\mytexttt{facet\_map\_t} (see Listing~\ref{lst:fe_map}) has been loaded
with the coordinates in real space of the nodes that lay on the
the facet.\footnote{This can be easily fulfilled by
  calling the \mytexttt{get\_nodes\_coordinates} binding of
  \mytexttt{vef\_iterator\_t} in Listing~\ref{lst:cell_vef_iterator}.}
The \mytexttt{update} binding takes as input dummy arguments a
\mytexttt{quadrature\_t} instance and the real parameter
\mytexttt{reorientation\_factor} in order to adjust the sign of the facet normals (see~\eqref{eq:normal_out}). 
Within \mytexttt{update},
\mytexttt{quadrature\_points\_coordinates(:)} and
\mytexttt{jacobian(:,:,:)} can be easily computed from the basis
shape functions and their first derivatives, respectively.  On the other hand,
\mytexttt{det\_jacobian(:)} and \mytexttt{outward\_unit\_normals(:,:)}
can be computed from \mytexttt{jacobian(:,:,:)}. The former as stated
in \eqref{eq:face_jacobian_measure}, while the latter as in
\eqref{eq:normal_out}.

\subsubsection{Neighbouring cells mappings}  \label{subsec:neig_cell_mappings}

The \mytexttt{facet\_maps\_t} data type is also composed by two instances of type \mytexttt{cell\_map\_facet\_restriction\_t}; see Fig.~\ref{fig:uml_face_integration_data_types}.
These instances handle all data related to $\geomap_{K^\alpha}$, with $\alpha$ being either $+$ or $-$. Let us thus refer to these instances as \mytexttt{cell\_map\_facet\_restriction}$^\alpha$, and to the 
polymorphic \mytexttt{reference\_fe\_t} instances on top of $K^\alpha$ 
as \mytexttt{ref\_fe\_geo}$^\alpha$.
In turn, \mytexttt{cell\_map\_facet\_restriction}$^\alpha$ are composed by as many \mytexttt{cell\_map\_t} instances as facets in $K^\alpha$. 
Provided that an actual facet $\facephy$ can {\em potentially} have local identifier $\facephy^\alpha$  in
$K^\alpha$ within the range $\facephy^\alpha=1,\ldots,\mbox{num\_facets}(K^\alpha)$, having as many 
\mytexttt{cell\_map\_t} instances as facets per surrounding cell let us hold and (pre)calculate within these instances the result of evaluating the $\georef^\alpha$ shape functions and their derivatives  
 at the facet quadrature points for all facets in the reference  system. 
To this end, the \mytexttt{create} binding of \mytexttt{cell\_map\_facet\_restriction}$^\alpha$ is invoked (from the one corresponding to \mytexttt{facet\_maps\_t}) 
with the facet quadrature $\mytexttt{q}$ and \mytexttt{ref\_fe\_geo}$^\alpha$ as input actual arguments.
It then walks over all possible local facet identifiers in the corresponding cell, and for each local facet identifier, invokes a specialized
version of the \mytexttt{create} binding of the corresponding \mytexttt{cell\_map\_t} instance, named \mytexttt{create\_restricted\_to\_facet} (that additionally requires the local facet identifier); see Fig.~\ref{fig:uml_face_integration_data_types}. The  \mytexttt{reference\_fe\_t} is  ultimately responsible to exchange this sort of data with \mytexttt{cell\_map\_t}. This service is in particular provided by the \mytexttt{create\_interpolation\_restricted\_to\_facet}  deferred binding of \mytexttt{reference\_fe\_t}, with signature defined in Listing~\ref{lst:create_interpolation_restricted_to_facet}.

\begin{lstlisting}[float=htbp,language={[03]Fortran},escapechar=@, caption={The signature of the \mytexttt{create\_interpolation\_restricted\_to\_facet} deferred binding of \mytexttt{reference\_fe\_t}}., label={lst:create_interpolation_restricted_to_facet}]
abstract interface
 ...
 subroutine ..._create_interpolation_restricted_to_facet ( this, quadrature, facet_lid, & 
                                                          ref_cell_interp_restricted_to_facet )
   class(reference_fe_t), intent(in)    :: this 
   type(quadrature_t)   , intent(in)    :: quadrature
   integer(ip)          , intent(in)    :: facet_lid
   type(interpolation_t), intent(inout) :: ref_cell_interp_restricted_to_facet
 end subroutine ..._create_interpolation_restricted_to_facet
 ...
end interface
\end{lstlisting}

As seen so far, the \mytexttt{create} binding of \mytexttt{facet\_maps\_t} is designed to be called right before the actual loop over all triangulation facets,
and it sets up all the scratch data. It does so by covering all possible scenarios corresponding to potential values of local facet identifiers within the two surrounding cells (even if some of
these scenarios are not actually exposed in the triangulation). 
The \mytexttt{update} binding of \mytexttt{facet\_maps\_t}, however, is intended to be called sitting on a particular facet $\facephy$ of the triangulation, and it has to
only update those two \mytexttt{cell\_map\_t} instances within \mytexttt{cell\_map\_facet\_restriction}$^\alpha$ 
corresponding to the particular scenario at hand, i.e., to the particular combination of local facet identifiers $\facephy^+$ and $\facephy^-$ of the facet on which it is
being updated. To this end, the
\mytexttt{update} binding of \mytexttt{facet\_maps\_t} receives these local identifiers in \mytexttt{facet\_lids} (see Fig.~\ref{fig:uml_face_integration_data_types}) and then calls the \mytexttt{update} binding of \mytexttt{cell\_map\_facet\_restriction}$^+$ and \mytexttt{cell\_map\_facet\_restriction}$^-$ with \mytexttt{facet\_lid=facet\_lid(1)} and
\mytexttt{facet\_lid=facet\_lid(2)}, respectively. The \mytexttt{update} binding of \mytexttt{cell\_map\_facet\_restriction\_t} 
picks up the \mytexttt{cell\_map\_t} corresponding to \mytexttt{facet\_lid} and invokes the \mytexttt{update} binding of the latter. 
We stress that no specialized version of this binding is required in the context of facet integration, i.e., the same version discussed in Sect.~\ref{subsec:cell_integrator} for
cell integration can be re-used here.\footnote{We note that, as in Sect.~\ref{subsec:mapping}, the \mytexttt{nodes\_coordinates(:)} member variable of these two \mytexttt{cell\_map\_t} instances has
to be loaded with the coordinates in physical space of the geometry nodes of the two cells surrounding the facet.} During the update process, 
\mytexttt{cell\_map\_facet\_restriction\_t} also registers in its \mytexttt{current\_facet\_lid} private
member variable, the value supplied to the \mytexttt{facet\_lid} dummy argument.  This lets \mytexttt{facet\_maps\_t} to
extract later on from \mytexttt{cell\_map\_facet\_restriction}$^\alpha$ the {\em updated} \mytexttt{cell\_map\_t} instances; see discussion
of \mytexttt{facet\_integrator\_t} in the sequel.

\subsection{Evaluation of shape functions in the physical space} \label{subsec:face_integrator}

The last data type that remains to be covered is \mytexttt{facet\_integrator\_t}; see Fig.~\ref{fig:uml_face_integration_data_types}. This data type is the counterpart 
of \mytexttt{cell\_integrator\_t} (see Sect.~\ref{subsec:cell_integrator}) for the case of facet integrals. In particular, 
it stores and updates shape function values and derivatives,  and provides the values, gradients, curls, and divergences of the respective fields 
for both $K^+$ and $K^-$ evaluated at facet quadrature points in real space. 
As can be observed from Fig.~\ref{fig:uml_face_integration_data_types}, its overall design is very close to the one of 
\mytexttt{facet\_maps\_t}, with \mytexttt{cell\_integrator\_facet\_restriction\_t} and the \mytexttt{cell\_integrator\_t} instances it is composed of, playing the role
of its counterparts in the scope of \mytexttt{facet\_maps\_t} (i.e., \mytexttt{cell\_map\_facet\_restriction\_t} and \mytexttt{cell\_map\_t}, respectively).
There are, however, two major differences among these two. First, \mytexttt{facet\_integrator\_t} deals with (e.g., it is created from) the two polymorphic \mytexttt{reference\_fe\_t} instances
(see \mytexttt{ref\_fe}$^\alpha$ dummy arguments of its \mytexttt{create} binding in  Fig.~\ref{fig:uml_face_integration_data_types})  on which the global \ac{FE} spaces of functions $\trialsp_h$, $\testsp_h$ are grounded on. 
For example, the \mytexttt{create} binding of \mytexttt{cell\_integration\_facet\_restriction}$^+$ invokes the \mytexttt{create\_restricted\_to\_facet} binding of the \mytexttt{cell\_integrator\_t} for all facets $\facephy^+$ within $K^+$. The latter computes at a given facet $\shapetestref{a}_{K^+}(\xref^{+}_{\rm gp})$, $\grad \shapetestref{a}_{K^+}(\xref^{+}_{\rm gp})$ through the deferred binding \mytexttt{create\_interpolation...to\_facet}  of \mytexttt{reference\_fe\_t} presented in Listing~\ref{lst:create_interpolation_restricted_to_facet}. Second, \mytexttt{facet\_integrator\_t} has to unburden the user from the complexity underlying the fact that
the coordinate systems of $K^+$ and $K^-$ might not be aligned in real space. To this end, it is equipped with a private lookup permutation table, called \mytexttt{qpoints\_perm(:,:)} in
Fig.~\ref{fig:uml_face_integration_data_types}, that lets it translate facet quadrature points identifiers from the local numbering space of $K^+$ into the one of $K^-$.    
This table is allocated and filled during the \mytexttt{create}
binding of \mytexttt{facet\_integrator\_t}, in particular by \mytexttt{reference\_fe\_t} through a deferred binding called \mytexttt{fill\_qpoints\_permutations}.
Given the facet quadrature identifier $\mytexttt{gp}$ and the facet permutation index $\mytexttt{pi}$ (see Sect.~\ref{sec:polytope_rotations_and_permutations}),  \mytexttt{qpoints\_perm(gp,pi)} stores the value of $\Pi({\rm gp})$ (see~\eqref{eq:face_integral_mapped_back_and_approximated}). The permutation index is stored within the \mytexttt{current\_permutation\_index} of \mytexttt{facet\_integrator\_t}, extracted from the \mytexttt{permutation\_index} dummy argument of the \mytexttt{update} binding. In turn, this parameter is extracted from the array \mytexttt{facet\_permutation\_indices(:)} of \mytexttt{fe\_space\_t}
 in Listing~\ref{lst:fe_space} (see Sect.~\ref{sec:fe_space}). We note that for n-simplices, we consider a renumbering such that all facets have the same orientation on both cells that share it, as commented in Sect.~\ref{sec:polytope_rotations_and_permutations}. In this case, \mytexttt{fill\_qpoints\_permutations}  fills the table with the identity permutation in all columns. We note that the re-orientation of the n-simplices can lead to mappings $\geomap_K$ such that $| \jacobian | < 0$, but this is not a problem as soon as one takes its absolute value, e.g.,  in\eqref{fematrix}.

\subsection{Facet integration user code example} \label{sec:face_integration_example}

In order to grasp how the data structures covered so far are actually used together in practice, the Fortran pseudocode snippet at Listing~\ref{lst:face_matrix_entries_computation} 
shows user's space code in charge of evaluating the first integral in \eqref{eq:local_to_face_matrix} for each interior facet in a loop over all facets. It would be bounded
to a subclass of the \mytexttt{discrete\_integration\_t} abstract data type presented
in Sect.~\ref{sec:disc_int} suitable for the non-conforming \ac{DG} discretization of the Poisson problem.

\begin{lstlisting}[float=htbp,language={[03]Fortran},escapechar=@, label={lst:face_matrix_entries_computation}, caption={User-level pseudocode illustrating the usage of facet integration data structures in order to compute the first integral in \eqref{eq:local_to_face_matrix} for each interior facet in a loop over all facets.}]
real(rp)            , allocatable :: shape_values_@$K^+$@(:,:),shape_values_@$K^-$@(:,:)  
type(vector_field_t), allocatable :: shape_gradients_@$K^+$@(:,:), shape_gradients_@$K^-$@(:,:)
real(rp)            , allocatable :: facet_matrix_@$K^+$@_@$K^-$@(:,:)
... ! Declaration of the rest of facet_matrices
type(vector_field_t)              :: outward_unit_normals(2)
integer(ip)         , parameter   :: @$K^+$@ = 1, @$K^-$@ = 2
... ! Declaration + variable initialization
! Loop over all facets F
   ... ! Update facet integration data structures to reflect the status corresponding to current facet @\label{loc:update_face_integration}@
   if (current_facet_interior) then
      facet_matrix_@$K^+$@_@$K^-$@ = 0.0_rp
      ... ! Initialize the rest of facet_matrices
      call facet_map%get_outward_unit_normals(outward_unit_normals)
      call facet_integrator%get_values(@$K^-$@,shape_values_@$K^-$@) @\label{loc:get_values_k_minus}@
      call facet_integrator%get_gradients(@$K^+$@,shape_gradients_@$K^+$@)
      ... ! Get values and gradients required to evaluate the rest of facet_matrices
      do gp = 1, facet_quadrature%get_num_quadrature_points()
         @$|\facejacobian|$@_x_wgp = facet_map%get_@$|\facejacobian|$@(gp) * facet_quadrature%get_weight(gp)
         ... ! Compute facet_matrix_@{\color{blue}$K^+$}@_@{\color{blue}$K^+$}@(:,:)
         ! Compute facet_matrix_@{\color{blue}$K^+$}@_@{\color{blue}$K^-$}@(:,:)
         do b = 1, num_shape_functions_@$K^-$@
            do a = 1, num_shape_functions_@$K^+$@
               facet_matrix_@$K^+$@_@$K^-$@(a,b) = facet_matrix_@$K^+$@_@$K^-$@(a,b) + & 
                    (0.5*(shape_values_@$K^-$@(b,gp)*outward_unit_normals(@$K^-$@)*shape_gradients_@$K^+$@(a,gp))+...)*@$|\facejacobian|$@_x_wgp
            end do
         end do
         ... ! Compute facet_matrix_@{\color{blue}$K^-$}@_@{\color{blue}$K^+$}@(:,:)
         ... ! Compute facet_matrix_@{\color{blue}$K^-$}@_@{\color{blue}$K^-$}@(:,:)
      end do
   else ! Current facet lays on the boundary
      ...
   end if
... ! Assemble facet_matrices into global linear system coefficient matrix
! End Loop over all facets F
\end{lstlisting}

There are a pair of worth noting remarks about Listing~\ref{lst:face_matrix_entries_computation}. 
First, the call to the \mytexttt{get\_values()} binding of \mytexttt{facet\_integrator\_t} in Line~\ref{loc:get_values_k_minus} already returns
the permuted $K^-$ shape function values, i.e., \mytexttt{shape\_values\_$K^-$(b,gp)} actually stores
$\shapetest{b}_{K^-}(\x^{-}_{\Pi({\rm gp})})$. Second,
it is the so-called \mytexttt{fe\_space\_t} abstraction (to be covered in Sect.~\ref{sec:fe_space}) the one in charge of creating the facet integration data structures on loop initialization and to update them at each facet
loop iteration (see Line~\ref{loc:update_face_integration}). Therefore, the user does not actually directly deals with all the data types bindings and their interactions
illustrated in Fig.~\ref{fig:uml_face_integration_data_types}. In this example, it becomes evident that facet-loop based integration is very convenient for the implementation of \ac{DG} methods, since it very much resembles the blackboard expressions (see, e.g., \eqref{eq:ip_dg_formulations}).

\subsection{Change-of-basis implementation in a \texttt{reference\_fe\_t} subclass} \label{sec:reference_fe_implementors}

In this section, we provide a detailed presentation of how the change-of-basis required to compute the shape functions basis is implemented in a \mytexttt{reference\_fe\_t} subclass. In particular, we show the implementation for the Raviart-Thomas div-conforming \ac{FE} on n-cubes in Sect.~\ref{subsec:shape_functions_construction} (see also Sect.~\ref{subsec:hdiv_conforming_fes} for details). The pre-basis, e.g., $\polsp_{(k+1,k,k)} \times \polsp_{(k,k+1,k)} \times \polsp_{(k,k,k+1)}$ in 3D, has to be generated before this subroutine is called; see, e.g., the evaluation of the pre-basis in Line~\ref{loc:rt:faceint} of Listing~\ref{lst:rt_shape_functions}.

\begin{lstlisting}[float=htbp,language={[03]Fortran},escapechar=@,caption={Implementation
of the change-of-basis required for Raviart-Thomas div-conforming
\acp{FE} on n-cubes, following the procedure presented in
Sect.~\ref{subsec:shape_functions_construction}.},label={lst:rt_shape_functions}]
subroutine hex_raviart_thomas_reference_fe_change_basis(this) 
  class(hex_raviart_thomas_reference_fe_t), intent(inout) ::  this
  type(hex_lagrangian_reference_fe_t) :: d_1_fe
  type(quadrature_t)                  :: d_1_quadrature
  type(interpolation_t)               :: d_1_interpolation, facet_interpolation
  real(rp)                            :: shape_test
  type(vector_field_t)                :: normal
  type(hex_lagrangian_reference_fe_t) :: d_fe_geo
  type(vector_field_t)                :: v_shape_trial
  integer(ip)                         :: ishape, jshape, gp
  real(rp)                            :: factor
  integer(ip)                         :: facet_id
  
  ... ! Allocate this%change_basis_matrix (this%num_shape_functions,this%num_shape_functions)
  ! Create a d-1-dim scalar lagrangian reference FE of order k
  call d_1_fe%create(topology          = this%get_topology(), & @\label{loc:rt:ref_fe}@
       num_dims = this%num_dims-1, order = this%order, &
       field_type        = field_type_scalar, conformity = .true. )
  ! Create a d-dim scalar lagrangian reference FE of order 1 (geometry)
  call d_fe_geo%create(topology          = this%get_topology(), & @\label{loc:rt:ref_fe_geo}@
       num_dims = this%num_dims, order = 1, &
       field_type        = field_type_scalar, conformity = .true. )
  ! Create a d-1 dimension quadrature from RT reference FE
  call this%create_facet_quadrature( d_1_quadrature ) @\label{loc:rt:face_quadrature}@
  call d_1_fe%create_interpolation( d_1_quadrature, d_1_interpolation )@\label{loc:rt:face_interpolation}@
  ! Initialize change of basis matrix
  this%change_basis_matrix = 0.0_rp
  d = 0
  do facet_id = this%get_first_facet_id(), this%get_first_facet_id() + this%get_num_facets()-1
     ! Map quadrature points in reference facet to cell facet and evaluate shape functions
     call this%create_facet_interpolation ( facet_id - d_fe_geo%get_first_facet_id()+1, & @\label{loc:rt:faceint}@
                                           d_1_quadrature, facet_interpolation )
     ! Integrate boundary moments int_Facet(u.n q), q \in Q_k
     do gp = 1, d_1_quadrature%num_quadrature_points
        factor = d_1_quadrature%get_weight(gp) / ( 2.0 ** d_1_fe%get_num_dims() )
        ... ! Compute normal such that (for oriented meshes) two elements sharing a facet have a
        ! moment determined with the same normal, in order to have conformity
        ! Integrate moments and assemble in this%change_basis_matrix
        do ishape=1, d_1_interpolation%num_shape_functions
           call d_1_fe%get_value(d_1_interpolation, ishape, gp, shape_test)
           do jshape=1, facet_interpolation%num_shape_functions
              call this%get_value(facet_interpolation, jshape, gp, v_shape_trial)
              this%change_basis_matrix(d+ishape,jshape) += & @\label{loc:rt:int}@
                   shape_test * v_shape_trial * normal * factor
           end do
        end do
     end do
     d = d + d_1_interpolation%num_shape_functions
  end do
  ... ! Compute number interior moments shape functions: Qk-1,k,k x Qk,k-1,k x Qk,k,k-1
  ! Compute interior moments (using same approach as above for facet moments)
  ! and assemble them in this%change_basis_matrix
  ! Invert change_basis_matrix
  call this%invert_change_basis_matrix() @\label{loc:rt:invm}@
  ... ! Transform type(list_t) member variables of this to reflect change of basis
  ... ! Deallocate dynamic memory
end subroutine hex_raviart_thomas_reference_fe_change_basis
\end{lstlisting}

We also present how to compute the boundary moments in
\eqref{eq:rt-moments} in Listing~\ref{lst:rt_shape_functions};
interior moments are simpler and omitted for the
sake of brevity. The implementation of the boundary moments requires:
1) to create the \mytexttt{reference\_fe\_t} that implements
$[ \polsp_{k \ones} ]^{d-1}$ in Line \ref{loc:rt:ref_fe}, 2) a facet
quadrature on the reference facet in Line
\ref{loc:rt:face_quadrature}, and 3) the evaluation of the reference
\ac{FE} in the quadrature points in the \mytexttt{interpolation\_t} in
Line \ref{loc:rt:face_interpolation}. We also require a Lagrangian
(first order) \ac{FE} that represents the geometry in Line
\ref{loc:rt:ref_fe_geo}. Next, we loop over all the facets of the cell
and compute the values of the shape functions of the cell in the facet
quadrature, stored in the \mytexttt{interpolation\_t} instance in Line
\ref{loc:rt:faceint}. With all these ingredients, we can compute the
boundary moments for the pre-basis functions (see line
\ref{loc:rt:int}) and assemble them in the change-of-basis
matrix. After doing the same for interior moments, we just need to
invert the change-of-basis matrix in Line \ref{loc:rt:invm}. At this
point, we have the shape functions basis as a linear combination of
pre-basis functions. Thus, when one calls the
\mytexttt{fill\_interpolation} binding of the corresponding reference
\ac{FE}, it creates the pre-basis \mytexttt{interpolation\_t} instance 
and next applies the change-of-basis matrix to compute the one for the
shape functions basis, i.e., the placeholder where the evaluation of
the shape functions and its derivatives (at the set of quadrature
points for which the interpolation has been created) are stored. We
note that the ownership of \acp{DOF} also changes in this process. The
boundary moments (integrals of functions on facets) belong to the
corresponding facet, whereas interior moments belong to the
cell. Vertices and edges do not have \acp{DOF} in this case. The
definition of the ownership is skipped for brevity.

\section{Integration and global \ac{DOF} handling: the \texttt{fe\_space\_t} abstraction} \label{sec:fe_space}

In this section, we introduce a software abstraction, referred to as \mytexttt{fe\_space\_t}, which represents (in the most general scenario) the mathematical concept of a global \ac{FE} space $\trialsp_h = \trialsp^1_h \times \ldots \times \trialsp^n_h$ obtained by means of the Cartesian product of global \ac{FE} spaces $\trialsp^{i}_h$ corresponding to each of the $i=1,\ldots,n_{\rm field}$ field unknowns involved in a system of \acp{PDE}; see Sect.~\ref{subsec:global_fe_space} and \ref{subsec:cartesian_product_FE_space}. Each $\trialsp^{i}_h$ is described as a combination of: 1)~an approximation $\Omega_h$ of the physical domain $\domain$ provided by \mytexttt{triangulation\_t}, i.e., a mesh-like container for the cells on which $\domain_h$ is partitioned, their boundary lower-dimensional objects, and their adjacency relationships; see Sect.~\ref{sec:triangulation}; 2)~a description of the $n_{\rm field}$  reference \acp{FE} associated to each triangulation cell grounded on \mytexttt{reference\_fe\_t}; see Sect.~\ref{sec:reference_fe}. 

These two basic building blocks equip \mytexttt{fe\_space\_t} with the tools required to provide the following two crucial services.\footnote{We stress, however, that the full set of services provided by \mytexttt{fe\_space\_t} is not actually restricted to only these two.} On the one hand, it is in charge of handling (i.e., generating, storing, fetching) a {\em global} enumeration of the \acp{DOF}  corresponding to each $\trialsp^{i}_h$ taking into account the notion of conformity; see e.g., Sect.~\ref{subsec:global_fe_space}  and~\ref{subsec:reference_fe_space}. On the other hand, it handles the data structures that are required to evaluate integrals over cells and facets (see Sect.~\ref{sec:cell_integration} and~\ref{sec:face_integration}, respectively). In particular, it judiciously sets up them, and orchestrates their respective life cycles and interactions, while unburdening the user (to a large extent) from the complexity (among others) inherent to high order \acp{FE}.

The \ac{OO} design of \mytexttt{fe\_space\_t} (as the one of many other data types in \FEMPAR{}, e.g., \mytexttt{triangulation\_t}) strongly strives to preserve encapsulation and data hiding while still storing and accessing data efficiently (i.e., in a way that leverages data locality for the efficient exploitation of modern computer memory architectures). The user-friendly view of \mytexttt{fe\_space\_t} is implicitly (re)constructed by the data types (associated interfaces and interactions) that will be covered in Sect.~\ref{subsec:fe_space_conceptual_view}. We now move on the approach that we follow for the internals of \mytexttt{fe\_space\_t}.

\subsection{The internal organization of \mytexttt{fe\_space\_t}} \label{subsec:fe_space_internals}
In this section, we {\em sketch} how the internals of \mytexttt{fe\_space\_t} are organized in order to efficiently deliver the two services outlined above. For simplicity, we restrict ourselves to a simplified version of \mytexttt{fe\_space\_t} that, to a large extent, captures the spirit of its actual counterpart in \FEMPAR{}. The declaration of this simplified data type is shown in Listing~\ref{lst:fe_space}.\footnote{We note that \mytexttt{fe\_space\_t} is not actually in \FEMPAR{}. It is a whole data type hierarchy rooted at \mytexttt{base\_fe\_space\_t}, not included here for simplicity. Within this hierarchy, we have, e.g., \ac{FE} space concretizations suitable for either serial or parallel distributed-memory environments. The one shown in the listing very much resembles \mytexttt{serial\_fe\_space\_t}.}
\begin{lstlisting}[float=htbp,language={[03]Fortran},escapechar=@, label={lst:fe_space}, caption={The internals of \mytexttt{fe\_space\_t} and a selected set of its bindings.}]
type :: fe_space_t
  private
  integer(ip)                            :: num_fields

  ! Container of polymorphic reference_fe_t instances (See Section@~\ref{sec:reference_fe}@)
  type(p_reference_fe_t)   , allocatable :: reference_fes(:)
  
  ! Mapping of (field_id,cell_id) pairs to reference_fes(:)
  integer(ip)              , allocatable :: field_cell_to_ref_fes(:,:)

  ! (Polymorphic) pointer to a triangulation it was created from (See Section@~\ref{sec:triangulation}@)
  class(triangulation_t)   , pointer     :: triangulation => NULL() @\label{loc:triangulation_pointer}@

  ! Descriptor of the block layout selected for the PDE system at hand (See Section@~\ref{subsec:global_dof_numbering}@)
  type(block_layout_t)     , pointer     :: block_layout => NULL() @\label{loc:fe_space_block_layout}@

  ! Global DoF enumeration-related data
  integer(ip)              , allocatable :: ptr_dofs_x_fe(:,:)  @\label{loc:global_dof_start}@
  integer(ip)              , allocatable :: lst_dofs_gids(:)
  integer(ip)              , allocatable :: num_dofs_x_field(:)         @\label{loc:global_dof_end}@

  ! Cell-related integration member variables (See Section@~\ref{sec:cell_integration}@)
  integer(ip)              , allocatable :: cell_quadratures_degree(:) @\label{loc:cell_integration_begin}@
  type(quadrature_t)       , allocatable :: cell_quadratures(:)        @\label{loc:cell_quadratures}@
  type(hash_table_t)                     :: cell_quadratures_position
  type(cell_map_t)         , allocatable :: cell_maps(:)               @\label{loc:cell_maps}@
  type(hash_table_t)                     :: cell_maps_position
  type(cell_integrator_t)  , allocatable :: cell_integrators(:)        @\label{loc:cell_integrators}@
  type(hash_table_t)                     :: cell_integrators_position  @\label{loc:cell_integration_end}@  
                                          
  ! Facet-related integration member variables (See Section@~\ref{sec:face_integration}@)
  integer(ip)              , allocatable :: facet_quadratures_degree(:) @\label{loc:face_integration_begin}@
  type(quadrature_t)       , allocatable :: facet_quadratures(:) 
  type(hash_table_t)                     :: facet_quadratures_position  
  type(facet_maps_t)       , allocatable :: facet_maps(:)
  type(hash_table_t)                     :: facet_maps_position         
  type(facet_integrator_t) , allocatable :: facet_integrators(:)
  type(hash_table_t)                     :: facet_integrators_position @\label{loc:face_integration_end}@
  
  ! Member variables to provide support to implementation of fe_facet_iterator_t (See Section@~\ref{subsec:fe_space_conceptual_view}@)
  integer(ip)              , allocatable :: facet_gids(:)              @\label{loc:face_gids}@
  integer(ip)              , allocatable :: facet_permutation_indices(:) @\label{loc:face_permutation_index}@ (See Section@~\ref{sec:polytope_rotations_and_permutations}@)


  ! Strong imposition of boundary conditions-related data (See Section@~\ref{subsec:fe_space_strong_boundary_conditions}@)
  integer(ip)                            :: num_fixed_dofs

  ... ! Remaining accelerator look-up tables, scratch data, other member variables 
      ! reflecting low-level implementations details, member variables providing support
      ! to other services of fe_space_t, etc.
contains
  public
  ! A specialized overload of the generic create binding where a single cell topology is assumed, 
  ! and each global space is ground on the same reference_fe_t on all cells. The latter might be 
  ! different among fields, as e.g., in Q2xQ1 inf-sup FEs for the Navier-Stokes problem
  procedure, private :: fe_space_create_same_reference_fes_on_all_cells
  ! The most general overload of the generic create binding, i.e., mixed cell topologies, and 
  ! (potentially) different per-field and per-cell reference_fe_t instances
  procedure, private :: fe_space_create_different_ref_fes_between_cells
  generic   :: create => fe_space_create_same_reference_fes_on_all_cells, &
                         fe_space_create_different_ref_fes_between_cells
 ...
  procedure :: generate_global_dof_numbering => fe_space_generate_global_dof_numbering
  procedure :: set_up_cell_integration       => fe_space_set_up_cell_integration
  procedure :: set_up_facet_integration      => fe_space_set_up_facet_integration
  procedure :: free                          => fe_space_free
  ... ! TBPs in charge of providing other fe_space_t services
  ... ! TBPs in charge of creating iterators on fe_space_t objects (See Section@~\ref{subsec:fe_space_conceptual_view}@)
  ... ! Full set of getter TBPs to preserve data hiding and encapsulation
  ... ! Comprehensive set of private TBPs providing support to the 
      ! implementation of the public TBPs above. These are hierarchically organized into 
      ! short, auxiliary re-usable subroutines/functions for code readability purposes
end type fe_space_t
\end{lstlisting}
A collection of \mytexttt{reference\_fe\_t} polymorphic instances is stored in the \mytexttt{reference\_fes(:)} array. These instances are uniquely identified (within the local scope of \mytexttt{fe\_space\_t}) by their position in this array. The global \ac{FE} space corresponding to a given field, with identifier \mytexttt{f\_id} in the range $1,\ldots,\mbox{\mytexttt{num\_fields}}$ (with \mytexttt{num\_fields} equal to $n_{\rm field}$ above), is described by: 1) the \mytexttt{triangulation} member variable (the rationale underlying it being polymorphic is made clear in Sect. \ref{subsec:fe_space_conceptual_view}; 2) its restriction to each cell provided by the reference \ac{FE} space defined by the \mytexttt{reference\_fe\_t} instance with identifier \mytexttt{field\_cell\_to\_ref\_fes(f\_id,c\_id)} in the collection; \mytexttt{c\_id} is assumed to be a positive integer in  $1,\ldots,\mytexttt{triangulation\%get\_num\_cells()}$ that uniquely identifies each cell.

The member variables used to handle the global \ac{DOF} numbering are encompassed within Lines~\ref{loc:global_dof_start}-\ref{loc:global_dof_end} of Listing~\ref{lst:fe_space}. 
The global \ac{DOF} identifiers are stored cell-wise, and field-wise within each cell, in the \mytexttt{lst\_dofs\_gids(:)} array, which is in turn (indirectly) addressed by the \mytexttt{ptr\_dofs\_x\_fe(:,:)} array. In particular, the ones assigned to the local \acp{DOF} related to field \mytexttt{f\_id} on cell \mytexttt{c\_id} start and end in position \mytexttt{ptr\_dofs\_x\_fe(f\_id,c\_id)} and \mytexttt{ptr\_dofs\_x\_fe(f\_id+1,c\_id)-1} of \mytexttt{lst\_dofs\_gids(:)}, respectively, if  $\mytexttt{f\_id} < \mytexttt{num\_fields}$, and in position  \mytexttt{ptr\_dofs\_x\_fe(f\_id,c\_id)} and  \mytexttt{ptr\_dofs\_x\_fe(1,c\_id+1)}, respectively, if $\mytexttt{f\_id} = \mytexttt{num\_fields}$. The number of \acp{DOF}  of the global \ac{FE} space corresponding to each field (excluding those that are subject to strong boundary conditions) is stored in the \mytexttt{num\_dofs\_x\_field(:)} array.

The member variable in Line~\ref{loc:fe_space_block_layout} stores a reference to a data type that describes the block layout {\em currently selected} (i.e., it can be changed on demand) for the global matrix and right-hand side vector of the linear system (or a sequence of them) required for the solution of the \ac{PDE} system at hand. The role of \mytexttt{block\_layout\_t} in the global \ac{DOF} numbering generation process will be illustrated in Sect.~\ref{subsec:global_dof_numbering}.

The data structures that let \mytexttt{fe\_space\_t} handle the evaluation of cell integrals are declared in Lines~\ref{loc:cell_integration_begin}-\ref{loc:cell_integration_end} of Listing~\ref{lst:fe_space}. The \mytexttt{set\_up\_cell\_integration} binding sets up them. The method is intended to be called by the user's program right before any cell integration loop. It ensures that any (scratch) data that can be computed on its final form in the reference cell is pre-computed {\em for any of the triangulation cells} while minimizing the number of integration data structures required for the particular scenario at hand. To this end, \mytexttt{fe\_space\_t} is equipped with three array containers of \mytexttt{quadrature\_t}, \mytexttt{cell\_map\_t} and  \mytexttt{cell\_integrator\_t} objects (see Lines~\ref{loc:cell_quadratures},~\ref{loc:cell_maps}, and~\ref{loc:cell_integrators}, respectively), which are indirectly addressed by the \mytexttt{hash\_table\_t} member variables with corresponding names.\footnote{The term hash table here reflects its usual meaning, i.e., an associative array that maps keys to values.} This is required because \mytexttt{fe\_space\_t} supports, e.g., non-conforming \ac{FE} spaces with variable order per cell. A {\em unique identifier} (dynamically generated within the scope of \mytexttt{fe\_space\_t}) is assigned to each of the integration objects that must be created. The \mytexttt{hash\_table\_t} instances let \mytexttt{fe\_space\_t} transform these unique identifiers into container array positions from which the integration objects can be fetched.  

The \mytexttt{set\_up\_cell\_integration} method loops over all cells. Sitting on a cell, it determines an appropriate quadrature  to be used on that cell and its associated unique identifier. (See discussion in the next paragraph for more details.) If this quadrature has not been generated yet (i.e., if the hash table lookup fails), then a new quadrature is created on the next free position of the \mytexttt{cells\_quadratures(:)} array container, and a new identifier-position pair is inserted into the hash table. Otherwise, the quadrature is fetched from this array. The same process is repeated for the \mytexttt{cell\_map\_t} and \mytexttt{cell\_integrator\_t} instances. The former ones are uniquely determined by the combination of the unique identifier \mytexttt{quadrature\_t} just created/fetched and that of the \mytexttt{reference\_fe\_t} instance on top of the current cell (see Sect.~\ref{sec:triangulation}). On the other hand, a \mytexttt{cell\_integrator\_t} instance has to be associated to each field within the current cell; the \mytexttt{cell\_integrator\_t} instance corresponding to a field is uniquely determined by the unique identifier of the \mytexttt{quadrature\_t} just created/fetched and the one of the \mytexttt{reference\_fe\_t} associated to that field (see Sect.~\ref{subsec:cell_integrator}). Therefore, the unique identifiers of the \mytexttt{cell\_map\_t} and \mytexttt{cell\_integrator\_t} instances required for the evaluation of cell integrals over the current cell can be easily determined combining the ones
corresponding to the instances from which they are created. We recall that the unique identifier of the \mytexttt{reference\_fe\_t} instance on top of the current cell, \mytexttt{c\_id}, 
for a given field, \mytexttt{f\_id}, can be retrieved from \mytexttt{reference\_fe\_id=field\_cell\_to\_ref\_fes(f\_id,c\_id)}, while the \mytexttt{reference\_fe\_t} instance 
itself from \mytexttt{reference\_fes(reference\_fe\_id)}.

The allocatable array member variable in
line~\ref{loc:cell_integration_begin} (with as many entries as
triangulation cells) can be used by the user in order to (optionally)
determine the degree of the quadrature to be used on each
triangulation cell. This member variable is allocated and initialized
(during \mytexttt{fe\_space\_t} creation) to a reserved flag that
instructs \mytexttt{set\_up\_cell\_integration} to use an automatic
(default) strategy to decide the degree of the quadrature to be used
on each cell. This default strategy relies on a deferred binding of
\mytexttt{reference\_fe\_t}, named
\mytexttt{get\_default\_quadrature\_degree}, which
typically returns the quadrature degree for which mass matrix terms
are integrated exactly (see
Sect.~\ref{subsec:numerical_quadrature}).\footnote{We stress,
  however, that each particular \mytexttt{reference\_fe\_t} subclass
  at hand has the freedom to implement a different strategy if
  required.} The strategy, in particular, walks over all
\mytexttt{reference\_fe\_t} instances on top of the cell, and the one
for which its (polynomial) reference cell functional space is of
maximum order becomes ultimately responsible of creating the
quadrature via an invocation to its \mytexttt{create\_quadrature}
deferred binding.  Alternatively, the user may explicitly select the
quadrature degree to be used on each cell. In such a case,
\mytexttt{create\_quadrature} is invoked to create a
quadrature with the degree given by the corresponding entry in the
\mytexttt{cell\_quadratures\_degree(:)}; see
Sect.~\ref{subsec:numerical_quadrature}. In any case (i.e., default
or explicit quadrature degree), both the unique identifier of the
\mytexttt{reference\_fe\_t} instance on top of the current
cell and the quadrature degree are used to
generate a unique identifier of the quadrature to be created/fetched.

On the other hand,
Lines~\ref{loc:face_integration_begin}-\ref{loc:face_integration_end}
of Listing~\ref{lst:fe_space} encompass those data structures required
for the evaluation of (both boundary and interior) facet integrals;
see Sect.~\ref{sec:face_integration}. A very close rationale to the
one underlying their cell counterparts is followed to set up these data structures.  The
\mytexttt{set\_up\_facet\_integration} binding loops over all
facets. Sitting on a facet, it determines an appropriate
facet \mytexttt{quadrature\_t} rule.  The quadrature
degree is either the default or a user-defined one (via the
allocatable array member variable in
Line~\ref{loc:face_integration_begin}). It also determines
the unique identifier of the quadrature and of the rest of the
facet-integration data structures, which are created as necessary,
while handling their interactions. Both the topology of the two cells
sharing the facet and the quadrature degree are used to generate a
unique identifier of facet quadratures. The member variables in Lines~\ref{loc:face_gids}-\ref{loc:face_permutation_index} provide support to the implementation on the so-called \mytexttt{fe\_facet\_iterator\_t} data type and will be covered in detail in Sect.~\ref{subsec:fe_space_conceptual_view}. Finally, the member variable \texttt{num\_fixed\_dofs} in Listing~\ref{lst:fe_space} is used by \mytexttt{fe\_space\_t} to count how many \acp{DOF} are subject to strong boundary conditions; see Sect.~\ref{subsec:fe_space_strong_boundary_conditions}.

\subsection{A conceptual view of \mytexttt{fe\_space\_t}} \label{subsec:fe_space_conceptual_view} 
Following the ideas presented in Sect.~\ref{subsec:abstract_triangulation}, \mytexttt{fe\_space\_t} offers a number of iterators to provide traversals over its objects,
and uniform data access to its internals. Apart from iterators over cells and vefs, \mytexttt{fe\_space\_t} also provides traversals over facets by 
means of the so-called \mytexttt{fe\_facet\_iterator\_t} data type. This iterator is essentially required to implement the evaluation of jump terms in, e.g., error estimators or \ac{DG} methods
in a user-friendly manner. For reasons made clear in the course of this section, a design goal to be fulfilled by \mytexttt{fe\_space\_t} iterators is that they are able to provide access to the same data as their counterpart  \mytexttt{triangulation\_t} iterators (see Sect.~\ref{subsec:abstract_triangulation}), and that
they are able to do so {\em efficiently} while avoiding duplication of code bounded to the latter ones. For example, \mytexttt{fe\_cell\_iterator\_t} should be designed such that it is also able to provide the coordinates (in physical space) of the nodes describing the geometry of the cell, apart from the global \ac{DOF} identifiers on top of it.

Let us first discuss the design of iterators over cells and vefs (as the one of both follows the same lines). These data types are defined in Listing~\ref{lst:fe_set_iterator}, where \mytexttt{set} must be actually replaced by either \mytexttt{cell} or \mytexttt{vef}.
\begin{lstlisting}[float=htbp,language={[03]Fortran},escapechar=@,caption=\mytexttt{fe\_space\_t} ``\mytexttt{set}'' (either \mytexttt{cell} or \mytexttt{vef}) iterators and the composition relationship with their counterpart \mytexttt{triangulation\_t} iterators (\mytexttt{set\_iterator\_t}). We note that additional scratch member variables (omitted from the code snippet) are required in order to avoid dynamic
memory allocation/deallocation for each object of the set traversal., label={lst:fe_set_iterator}]
type fe_set_iterator_t
  private
  class(set_iterator_t), allocatable :: set_iterator    @\label{loc:itdec}@
  class(fe_space_t)    , pointer     :: fe_space
contains
  public
  procedure :: create       => fe_set_iterator_create
  procedure :: free         => fe_set_iterator_free
  procedure :: first        => fe_set_iterator_first
  procedure :: next         => fe_set_iterator_next
  procedure :: has_finished => fe_set_iterator_has_finished
  ... ! TBPs providing access to data items in the corresponding set (either cell or vef)
end type fe_set_iterator_t
\end{lstlisting}
As shown in Listing~\ref{lst:fe_set_iterator}, \mytexttt{fe\_set\_iterator\_t} holds a {\em polymorphic} pointer to the \mytexttt{fe\_space\_t} instance to which it has to provide data access. Dynamic polymorphism is exploited here with extensibility and code reuse in mind. Any type extension of \mytexttt{fe\_space\_t} (e.g., the one suitable for distributed-memory environments), can also become the target of this polymorphic pointer, thus enabling reuse of data and code bounded to \mytexttt{fe\_set\_iterator\_t} with these extensions. 
Of special relevance in Listing~\ref{lst:fe_set_iterator} is the composition relationship among the data type being defined and \mytexttt{set\_iterator\_t}, i.e., its \mytexttt{triangulation\_t} iterator counterpart (see Sect.~\ref{subsec:abstract_triangulation}). This lets \mytexttt{fe\_set\_iterator\_t} to fulfill the aforementioned design goal, i.e., to provide a superset of data over the class it is composed of, while still being able to access to any data stored within the triangulation scope. \mytexttt{fe\_set\_iterator\_t} also reuses from  \mytexttt{set\_iterator\_t} the code underlying the sequential traversal over all objects of the \mytexttt{set}. Indeed, as many other \acp{TBP} of \mytexttt{fe\_set\_iterator\_t}, 
\mytexttt{init}, \mytexttt{next}, and \mytexttt{has\_finished} \acp{TBP} of \mytexttt{fe\_set\_iterator\_t} are simply implemented as wrappers of their counterparts in \mytexttt{set\_iterator\_t}. (We remark that this is possible provided that  \mytexttt{fe\_space\_t}  is deliberately set up such that it shares with \mytexttt{triangulation\_t} a consistent global numbering for cells and lower-dimensional objects.)  

At this point it is important to remark that the \mytexttt{set\_iterator\_t} instance that \mytexttt{fe\_set\_iterator\_t} aggregates is also {\em polymorphic} (see Line~\ref{loc:itdec} in Listing~\ref{lst:fe_set_iterator}).  As stated in Sect.~\ref{subsec:fe_space_internals} (in particular, see Line~\ref{loc:triangulation_pointer} of Listing~\ref{lst:fe_space}), a \mytexttt{fe\_space\_t} instance is created from a polymorphic \mytexttt{triangulation\_t} instance. The \mytexttt{create} binding of \mytexttt{fe\_set\_iterator\_t} extracts the latter from  \mytexttt{fe\_space\_t}, and then calls its \mytexttt{create\_cell\_iterator} binding
(see Sect.~\ref{subsec:abstract_triangulation}), which becomes ultimately in charge of determining the dynamic type of the \mytexttt{set\_iterator\_t} member variable of  \mytexttt{fe\_set\_iterator\_t} (apart from leaving the iterator positioned in the first object of the \mytexttt{set}). This lets \mytexttt{fe\_space\_t} (and its associated iterators) to 
be re-used with any type extension of \mytexttt{triangulation\_t} (e.g., the one suitable for distributed-memory computers and/or $h$-adaptivity). Likewise, the \mytexttt{free} binding of \mytexttt{fe\_set\_iterator\_t} relies
on the \mytexttt{free\_cell\_iterator} binding of \mytexttt{triangulation\_t} in order to safely deallocate any dynamic memory
allocation performed during creation. We stress that, as in the case of \mytexttt{triangulation\_t} iterators, both the \mytexttt{create} and  \mytexttt{free} \acp{TBP} are 
not intended to be directly called by the user. Instead, \mytexttt{triangulation\_t} provides a set of (public) \acp{TBP} (as many as different iterators) for this purpose.
For example, the expression \mytexttt{call fe\_space\%create\_fe\_cell\_iterator(fe\_cell\_iterator)} creates an iterator on the
{\em polymorphic} \mytexttt{fe\_cell\_iterator} client-space instance, 
while \mytexttt{call fe\_space\%free\_fe\_cell\_iterator(fe...)}
is in charge of safely deallocating this polymorphic instance.

The implementation of \mytexttt{fe\_facet\_iterator\_t} is based on a very close rationale to the one of cell and vefs iterators, with subtle differences though; see Listing~\ref{lst:fe_face_iterator}. 
Provided that \mytexttt{fe\_facet\_iterator\_t} is a kind of \mytexttt{fe\_vef\_iterator\_t}, it should provide the same set of data access methods of the latter (e.g., the cells sharing the facet). However, it should restrict the traversal to those vefs that are actually facets, and to be able to provide all data required for the implementation of jump terms over facets. As shown in Listing~\ref{lst:fe_face_iterator}, \mytexttt{fe\_facet\_iterator\_t} extends \mytexttt{fe\_vef\_iterator\_t}. This automatically equips the former with the data access methods of the latter. On the other hand, it overrides 
those methods controlling the sequential traversals over the items in the set such that it restricts to facets, i.e., \mytexttt{create/free/first/next/has\_finished} in Listing~\ref{lst:fe_face_iterator}. 
The implementation of these methods relies on its member variable \mytexttt{facet\_gid}, and the \mytexttt{facet\_gids(:)} member variable of \mytexttt{fe\_space\_t}; see Line~\ref{loc:face_gids} 
of Listing~\ref{lst:fe_space}. For a given facet with global identifier $\mytexttt{facet\_gid}$, \mytexttt{facet\_gids(facet\_gid)} holds the global vef identifier corresponding to the facet.

\begin{lstlisting}[float=htbp,language={[03]Fortran},escapechar=@,caption=The \mytexttt{fe\_facet\_iterator\_t} data type., label={lst:fe_face_iterator}]
type, extends(fe_vef_iterator_t) :: fe_facet_iterator_t
  private
  integer(ip) :: facet_gid
contains
  public
  ! create/free/first/next/has_finished TBPs of fe_facet_iterator_t override 
  ! the ones in fe_vef_iterator_t. This lets fe_facet_iterator_t to restrict the 
  ! traversal to the set of facets of the triangulation
  procedure :: create       => fe_facet_iterator_create
  procedure :: free         => fe_facet_iterator_free
  procedure :: first        => fe_facet_iterator_first
  procedure :: next         => fe_facet_iterator_next
  procedure :: has_finished => fe_facet_iterator_has_finished  
  ... ! TBPs providing access to facet-related data in fe_space_t (See Listing@~\ref{lst:fe_cell_face_iterator}@)
end type fe_facet_iterator_t
\end{lstlisting}

The actual set of \acp{TBP} of a \mytexttt{fe\_space\_t} iterator highly depends on the type of object being pointed to. 
For completeness, we now briefly discuss those \acp{TBP} in the set corresponding to cell and facet iterators, which provide support for the implementation of the two services of \mytexttt{fe\_space\_t} we are focusing on. These are in particular shown in Listing~\ref{lst:fe_cell_face_iterator}. This listing also includes the generic \acp{TBP} in Lines~\ref{loc:fe_cell_iterator_generic_assembly} and \ref{loc:fe_face_iterator_generic_assembly}, although they will be discussed in Sect.~\ref{sec:linalg}.

\begin{lstlisting}[float=htbp,language={[03]Fortran},escapechar=@, caption=The \mytexttt{fe\_cell\_iterator\_t} and its facet counterpart., label={lst:fe_cell_face_iterator}]
type :: fe_cell_iterator_t
  ... ! Member variables (see Listing@~\ref{lst:fe_set_iterator}@)
contains
  public
  ... ! Create/free/traversal TBPs (see Listing@~\ref{lst:fe_set_iterator}@)
  procedure :: update_integration    => fe_cell_iterator_update_integration @\label{loc:cell_update_integration}@
  
  ! Helpers for the generation of a global DoF numbering
  procedure, private :: count_own_dofs_cell                      @\label{loc:count_dofs_on_cell}@
  procedure, private :: count_own_dofs_vef
  procedure, private :: generate_own_dofs_cell                   @\label{loc:own_dofs_on_cell}@ 
  procedure, private :: generate_own_dofs_vef                    @\label{loc:own_dofs_on_vef}@
  procedure, private :: fetch_own_dofs_vef_from_source_fe        @\label{loc:dofs_on_vef_source}@

  ! Getter TBPs
  procedure :: get_fe_dofs           => fe_cell_iterator_get_fe_dofs     @\label{loc:cell2dof}@
  procedure :: get_vef               => fe_cell_iterator_get_vef
  procedure :: get_quadrature        => fe_cell_iterator_get_quadrature   @\label{loc:cell_quadrature}@
  procedure :: get_det_jacobian      => fe_cell_iterator_get_det_jacobian
  ! Evaluation of shape functions values
  procedure, private :: get_values_scalar => fe_cell_iterator_get_values_scalar
  procedure, private :: get_values_vector => fe_cell_iterator_get_values_vector
  procedure, private :: get_values_tensor => fe_cell_iterator_get_values_tensor
  generic :: get_values => get_values_scalar, get_values_vector, get_values_tensor
  ... ! Evaluation of shape functions gradients
  ... ! Evaluation of shape functions hessians
  ... ! Evaluation of shape functions divergences
  ... ! Evaluation of shape functions curls @\label{loc:cell_integrator}@
  
  ! Assembly TBPs (See Section @~\ref{sec:linalg}@)
  procedure, private :: fe_cell_iterator_assembly_array
  procedure, private :: fe_cell_iterator_assembly_matrix
  procedure, private :: fe_cell_iterator_assembly_matrix_array  
  procedure, private :: fe_cell_iterator_assembly_matrix_array_with_strong_bcs
  generic  :: assembly => fe_cell_iterator_assembly_array,                       & @\label{loc:fe_cell_iterator_generic_assembly}@ 
       &                  fe_cell_iterator_assembly_matrix,                      &
       &                  fe_cell_iterator_assembly_matrix_array,                &
       &                  fe_cell_iterator_assembly_matrix_array_with_strong_bcs
  ... ! Rest of getter TBPs  
end type fe_cell_iterator_t

type, extends(fe_vef_iterator_t) :: fe_facet_iterator_t
   ... ! Member variables (see Listing@~\ref{lst:fe_face_iterator}@)
contains
  public
  ... ! Create/free/traversal TBPs (see Listing@~\ref{lst:fe_face_iterator}@)
  procedure :: update_integration    => fe_facet_iterator_update_integration     @\label{loc:face_update_integration}@
  
  ! Getter TBPs
  procedure :: get_quadrature        => fe_facet_iterator_get_quadrature         @\label{loc:face_quadrature}@
  procedure :: get_det_jacobian      => fe_facet_iterator_get_det_jacobian
  ! Evaluation of shape functions values
  procedure, private :: get_values_scalar => fe_facet_iterator_get_values_scalar
  procedure, private :: get_values_vector => fe_facet_iterator_get_values_vector
  procedure, private :: get_values_tensor => fe_facet_iterator_get_values_tensor
  generic :: get_values => get_values_scalar, get_values_vector, get_values_tensor
  ... ! Evaluation of shape functions gradients
  ... ! Evaluation of shape functions hessians
  ... ! Evaluation of shape functions divergences
  ... ! Evaluation of shape functions curls
  procedure :: get_permutation_index => fe_facet_iterator_get_permutation_index    @\label{loc:face_integrator}@
  
  ! Assembly TBPs (See Section @~\ref{sec:linalg}@)
  procedure, private :: fe_facet_iterator_assembly_array
  procedure, private :: fe_facet_iterator_assembly_matrix
  procedure, private :: fe_facet_iterator_assembly_matrix_array
  procedure, private :: fe_facet_iterator_assembly_matrix_array_with_strong_bcs
  generic  :: assembly => fe_facet_iterator_assembly_array,                      & @\label{loc:fe_face_iterator_generic_assembly}@ 
       &                  fe_facet_iterator_assembly_matrix,                     &
       &                  fe_facet_iterator_assembly_matrix_array,               &
       &                  fe_facet_iterator_assembly_matrix_array_with_strong_bcs
  ... ! Rest of getter TBPs
end type fe_facet_iterator_t

  
\end{lstlisting}

The \acp{TBP} in  Lines~\ref{loc:cell_quadrature}-\ref{loc:cell_integrator}, and~\ref{loc:face_quadrature}-\ref{loc:face_integrator} of Listing~\ref{lst:fe_cell_face_iterator} let the user fetch from \mytexttt{fe\_space\_t} the integration data associated to 
the current cell and facet being pointed to, respectively. On the other  hand, the \mytexttt{update\_integration} bindings in 
Lines~\ref{loc:cell_update_integration} and~\ref{loc:face_update_integration} perform those computations required to update these data structures such that
they hold shape function values and derivatives evaluated at (current) cell and facet (quadrature points) in the physical space. The former binding is implemented as
shown in Listing~\ref{lst:fe_cell_iterator_update_integration}. Finally, the \mytexttt{get\_permutation\_index} TBP of \mytexttt{fe\_facet\_iterator\_t} lets the caller to 
obtain the permutation index (see Sect.~\ref{sec:polytope_rotations_and_permutations} and \ref{subsec:face_integrator} for further details). The implementation of this method relies on the \mytexttt{facet\_permutation\_indices(:)} member variable of \mytexttt{fe\_space\_t}; see Line~\ref{loc:face_permutation_index} 
of Listing~\ref{lst:fe_space}. For a given facet with global identifier $\mytexttt{facet\_gid}$, \mytexttt{facet\_permutation\_indices(facet\_gid)} holds the permutation index corresponding to the facet.
 We have decided to permanently store facet permutation indices for performance reasons. These can be reused over and over again (e.g., in a transient and/or nonlinear \ac{PDE} problem) without the overhead associated to its computation on each traversal over the facets of the triangulation.

\begin{lstlisting}[float=htbp,language={[03]Fortran},escapechar=@, label={lst:fe_cell_iterator_update_integration}, caption={Implementation of the \mytexttt{update\_integration} binding of \mytexttt{fe\_cell\_iterator\_t}.}]
subroutine fe_cell_iterator_update_integration( this )
  class(fe_cell_iterator_t), intent(inout) :: this 
  type(cell_map_t), pointer :: cell_map
  type(point_t) , pointer :: coordinates_nodes(:)
  ... ! Declaration of the rest of local variables

  ! Update the cell_map_t instance associated to current cell
  cell_map => this%get_cell_map()
  coordinates_nodes => cell_map%get_coordinates_nodes()
  call this%get_coordinates(coordinates_nodes)  @\label{loc:get_coordinates}@
  quadrature => this%get_quadrature()
  call cell_map%update(quadrature) @\label{loc:update_fe_map}@
  
  ! Update field-wise cell_integrator_t instances associated to current cell
  do field_id = 1, this%get_num_fields()
     cell_integrator => this%get_cell_integrator(field_id) 
     call cell_integrator%update(cell_map)  @\label{loc:update_cell_integrator}@
  end do
end subroutine fe_cell_iterator_update_integration

  
\end{lstlisting}

An update of the \mytexttt{cell\_map\_t} instance (associated to the cell pointed by the \mytexttt{fe\_cell\_iterator\_t} instance on which this subroutine
is invoked) is performed in Line~\ref{loc:update_fe_map} of Listing~\ref{lst:fe_cell_iterator_update_integration}. It is followed by a loop over the number of fields of the \ac{PDE} system at hand in order to update the \mytexttt{cell\_integrator\_t} for every field in Line~\ref{loc:update_cell_integrator}. 
The update of the former requires that its \mytexttt{nodes\_coordinates(:)} scratch member variable 
has been loaded with the coordinates in the physical space of the nodes describing the geometry
of the cell at hand (see Sect.~\ref{subsec:mapping}). This is in particular fulfilled in Line~\ref{loc:get_coordinates}. 
The coordinates fetched by this call are actually stored within the triangulation. However,
\mytexttt{fe\_cell\_iterator\_t} can satisfy this query provided that it is composed of a \mytexttt{cell\_iterator\_t} instance;
see Listing~\ref{lst:fe_set_iterator} and accompanying discussion. At this point, the reader should be already capable to grasp
how the \mytexttt{fe\_facet\_iterator\_t} counterpart of this subroutine is implemented, so that it is omitted here
in order to keep the presentation short.

Going back to Listing~\ref{lst:fe_cell_face_iterator}, the binding in Line~\ref{loc:cell2dof} lets the user fetch the field-wise global \ac{DOF} identifiers that \mytexttt{fe\_space\_t}
has associated to the node functionals on the current cell interior and its vefs.  
(The bindings in Lines~\ref{loc:count_dofs_on_cell}-\ref{loc:dofs_on_vef_source} of Listing~\ref{lst:fe_cell_face_iterator}, however, assist \mytexttt{fe\_space\_t} on the
generation of the global \ac{DOF} numbering and their usage will be illustrated in Sect.~\ref{subsec:global_dof_numbering}.) 
This binding is implemented in Listing~\ref{lst:fe_cell_iterator_get_cell2dof}. 

\begin{lstlisting}[float=htbp,language={[03]Fortran},escapechar=@, label={lst:fe_cell_iterator_get_cell2dof}, caption={Implementation of the \mytexttt{get\_fe\_dofs} binding of \mytexttt{fe\_cell\_iterator\_t}.}]
subroutine fe_cell_iterator_get_fe_dofs(this,fe_dofs)
  class(fe_cell_iterator_t) , intent(in)    :: this
  type(p_1D_ip_array_t)     , intent(inout) :: fe_dofs(:)
  ... ! Declaration of local variables
  assert (size(fe_dofs) == this%get_num_fields())
  do field_id = 1, this%get_num_fields()-1
    spos = this%fe_space%ptr_dofs_x_fe(field_id  ,this%get_id())
    epos = this%fe_space%ptr_dofs_x_fe(field_id+1,this%get_id())-1
    fe_dofs(field_id)%p => this%fe_space%lst_dofs_gids(spos:epos)
  end do
  ... ! Treat the last field_id special case
end subroutine fe_cell_iterator_get_fe_dofs

\end{lstlisting} 

In Listing~\ref{lst:fe_cell_iterator_get_cell2dof}, \mytexttt{p\_1D\_ip\_array\_t} is assumed to be a data type with a single member variable, called \mytexttt{p}, declared as a pointer to a rank-1 \mytexttt{integer(ip)} array. For each field, the subroutine locates the region within the \mytexttt{lst\_dofs\_gids(:)} member variable corresponding to that field within the current cell, and then it associates to it the corresponding pointer in \mytexttt{fe\_dofs(:)}. At the expense of sacrificing type safety (in Fortran there is no mechanism to declare a pointer to be read-only), we avoid
the costly re-allocation of user-level allocatable arrays that would be needed in the case of non-conforming \ac{FE} spaces with highly varying degree polynomial spaces among cells.

To end up,  the \mytexttt{get\_vef} binding in Listing~\ref{lst:fe_cell_face_iterator} sets up a \mytexttt{fe\_vef\_iterator\_t} instance to point to the corresponding vef within the cell. As a consequence, one may navigate over the cells, its vefs, cells around these vefs, etc., using \mytexttt{fe\_space\_t} iterators all the way round.

\subsection{Global \ac{DOF} numbering generation} \label{subsec:global_dof_numbering}
In this section, we discuss how \mytexttt{fe\_space\_t} coordinates the building blocks covered so far in order to generate a {\em global} enumeration of the \acp{DOF}  describing the global \ac{FE} space $\trialsp_h \doteq  \trialsp^1_h \times \ldots \times \trialsp^n_h$ for general multi-field systems of \acp{PDE}.  This process is encompassed within the \mytexttt{generate\_global\_dof\_numbering} binding of \mytexttt{fe\_space\_t} (see Listing~\ref{lst:fe_space}). The code of this method is shown in Listing~\ref{lst:global_dof_numbering}. The \mytexttt{block\_layout} dummy argument lets the caller to customize the global \ac{DOF} numbering to be generated.\footnote{We refer to Listing~\ref{lst:block_layout} and its accompanying text in Sect.~\ref{sec:building_fe_systems} for a full description of the member variables and \acp{TBP} of \mytexttt{block\_layout\_t}. In this section, we restrict ourselves to those that are relevant for the global \ac{DOF} numbering process.} On the one hand, this data type specifies in how many blocks the user wants to split the (discrete) \ac{PDE} system at hand. In particular, the user may select to generate a \ac{DOF} numbering suitable for monolithic or blocked storage linear algebra data structures, with \mytexttt{block\_layout\%get\_num\_blocks()} returning one and a number larger than one, respectively. On the other hand, \mytexttt{block\_layout\_t} specifies the mapping of fields into blocks, with \mytexttt{block\_layout\%get\_block\_id(field\_id)} returning the block identifier the field with identifier \mytexttt{field\_id} is mapped to.  Provided that blocked linear algebra data structures in \FEMPAR{} are addressed using row/column identifiers that are local to each block, \mytexttt{block\_layout} equips the subroutine with the input necessary to generate a block-aware global \ac{DOF} numbering, in which the \acp{DOF}  belonging to fields of the first block are numbered first, followed by the ones of the second, and so on. We note that \mytexttt{block\_layout\_t} also holds inside how many \acp{DOF}  are there per block (see Sect.~\ref{sec:fe_affine_operator}). These latter quantities are computed within \mytexttt{generate\_global\_dof\_numbering} (see discussion in the sequel).

\begin{lstlisting}[float=htbp,language={[03]Fortran},escapechar=@,caption=The \mytexttt{generate\_global\_dof\_numbering} binding of \mytexttt{fe\_space\_t}.,label={lst:global_dof_numbering}]
subroutine fe_space_generate_global_dof_numbering(this,block_layout)
 class(fe_space_t)            , intent(inout) :: this
 type(block_layout_t) , target, intent(inout) :: block_layout
 logical :: perform_numbering

 ! Is block_layout compatible with fe_space_t?
 assert ( this%num_fields == block_layout%get_num_fields() ) 
 
 perform_numbering = .not. associated(this%block_layout) @\label{loc:dof_numbering_predicate_1}@
 if (.not. perform_numbering) perform_numbering = .not. (this%block_layout == block_layout) @\label{loc:dof_numbering_predicate_2}@
 if (perform_numbering) then ! Generate block_layout-conforming global DoF numbering
    this%block_layout => block_layout
    ... ! Re-allocate this%num_dofs_x_field(:) to have size this%num_fields (see Listing@~\ref{lst:fe_space}@)
    call this%count_dofs() ! Count how many DoFs are there per field and block @\label{loc:dof_numbering_count_dofs}@
    call this%list_dofs()  ! Actually generate global DoF identifiers          @\label{loc:dof_numbering_list_dofs}@
 else
    call block_layout%copy_num_dofs_x_block(this%block_layout) 
 end if
end subroutine fe_space_generate_global_dof_numbering

\end{lstlisting}

The subroutine in Listing~\ref{lst:global_dof_numbering} starts checking whether it has to actually generate a global \ac{DOF} numbering. It has to do so if there is no global \ac{DOF} numbering available yet (see predicate in Line~\ref{loc:dof_numbering_predicate_1}), or if the one available is not suitable for the input \mytexttt{block\_layout} (see predicate in Line~\ref{loc:dof_numbering_predicate_2}). The bulk of \mytexttt{generate\_global\_dof\_numbering}
is concentrated in the private helper \acp{TBP} of \mytexttt{fe\_space\_t} called \mytexttt{fe\_space\_count\_dofs} and \mytexttt{fe\_space\_list\_dofs}; see Lines~\ref{loc:dof_numbering_count_dofs} and~\ref{loc:dof_numbering_list_dofs} of Listing~\ref{lst:global_dof_numbering}, respectively. The code of these bindings is shown in Listings~\ref{lst:dof_numbering_count_dofs} and~\ref{lst:dof_numbering_list_dofs}, respectively. While the former computes the number of \acp{DOF}  per field and block, the latter is in charge of the actual generation of the global \ac{DOF} identifiers.

\begin{lstlisting}[float=htbp,language={[03]Fortran},escapechar=@,caption=The \mytexttt{count\_dofs} binding of \mytexttt{fe\_space\_t}.,label={lst:dof_numbering_count_dofs}]
subroutine fe_space_count_dofs ( this ) 
  class(fe_space_t), intent(inout) :: this   
  ... ! Declaration of local variables
  
  ! Count #DoFs per field
  this%num_dofs_x_field = 0 @\label{loc:start_count_dofs_x_field}@
  
  ... ! Allocate owner_cell_x_vef_visited(:,:) to size num_fields x triangulation%get_num_vefs()
  owner_cell_x_vef_visited = .false.

  call this%create_fe_iterator(fe)
  do while ( .not. fe%has_finished())  @\label{loc:count_dofs_outer_loop}@
    do field_id=1, this%num_fields
      this%num_dofs_x_field(field_id) += fe%count_dofs_on_cell_interior(field_id) @\label{loc:count_own_dofs}@
      if ( this%conforming_fe_space(field_id) .and. this%continuous_fe_space(field_id) ) then @\label{loc:cg_conforming}@
         do vef_lid = 1, fe%get_num_vefs() @\label{loc:count_dofs_inner_loop}@
            vef_gid = fe%get_vef_gid(vef_lid)
            if ( .not. owner_cell_x_vef_visited(field_id,vef_gid) ) then
               num_own_dofs_vef = fe%count_own_dofs_vef(vef_lid,field_id)
               if (num_own_dofs_vef>0) then
                  owner_cell_x_vef_visited (field_id,vef_gid) = .true.
                  this%num_dofs_x_field(field_id) += num_own_dofs_vef  @\label{loc:count_own_dofs_on_vef}@
               end if
            end if
         end do
      end if
    end do
    call fe%next()
  end do 
  call this%free_fe_iterator(fe)
  ... ! Free owner_cell_x_vef_visited(:,:) @\label{loc:end_count_dofs_x_field}@

  ! Count #DoFs per block
  call this%block_layout%clear_num_dofs_x_block()  @\label{loc:start_count_dofs_x_block}@
  do field_id=1, this%get_num_fields()
    iblock = this%block_layout%get_block_id(field_id) 
    call this%block_layout%add_to_block_num_dofs(iblock,this%num_dofs_x_field(field_id))     
  end do                                           @\label{loc:end_count_dofs_x_block}@
  
end subroutine fe_space_count_dofs

\end{lstlisting}

Lines~\ref{loc:start_count_dofs_x_field}-\ref{loc:end_count_dofs_x_field} of Listing~\ref{lst:dof_numbering_count_dofs} are in charge of computing the number of \acp{DOF}  per field, while those in Lines-\ref{loc:start_count_dofs_x_block}-\ref{loc:end_count_dofs_x_block}, those per block. The latter lines just determine the number of \acp{DOF}  per block by accumulating those corresponding to fields mapped to the block (computed in the former lines). The former lines are grounded on the notion of {\em owner cell of a vef}; a cell is the owner of a vef if 1) the latter lays on the boundary of the former, 2) it is the {\em first cell} for which 1) holds in the order in which the iterator over all cells presents them, and 3) the vef owns at least one \ac{DOF} of the global \ac{FE} space subject to consideration.\footnote{The last requirement has been introduced to include the concept of void \acp{FE} for multi-field problems in which some fields are not defined on the whole domain (see Sect.~\ref{subsec:reference_fe_subclasses}).} The (logical) work array \mytexttt{owner\_cell\_per\_vef\_visited(:)} keeps track whether the owner cell of the vefs have been already visited (or not) as these are traversed in the nested loop over all cells (see outer loop in Line~\ref{loc:count_dofs_outer_loop}), and over all vefs within the current cell (see inner loop in Line~\ref{loc:count_dofs_inner_loop}). Sitting on a cell, the algorithm first counts those \acp{DOF}  associated to node functionals logically placed in the interior of the current cell (see line~\ref{loc:count_own_dofs}). It then loops over the vefs of the current cell. If the owner cell of the current vef has not been visited yet, and the current cell is its owner, then the current cell is registered as the owner of the cell, and the \acp{DOF}  associated to node functionals logically placed on this vef within the current cell are counted in Line~\ref{loc:count_own_dofs_on_vef}. Provided that non-conforming \ac{FE} spaces do not have \acp{DOF}  on vefs, we can skip the loop over the vefs of a cell and accelerate the process in this case (see the if clause in Line~\ref{loc:cg_conforming} of Listing~\ref{lst:dof_numbering_count_dofs}).

The algorithm shown in Listing~\ref{lst:dof_numbering_list_dofs} is in charge of the actual generation of the global \ac{DOF} identifiers. The work array \mytexttt{owner\_cell\_gid\_per\_vef(:,:)} is used to store the owner cell global identifier of the vefs. On the other hand, \mytexttt{vef\_lid\_in\_owner\_cell(:,:)} array is used as an accelerator lookup table that stores the vef local identifiers (i.e., \mytexttt{vef\_lid}) within their corresponding owner cells if they have been already visited, and \mytexttt{-1} otherwise. Both arrays are indexed using vef global identifiers (i.e., \mytexttt{vef\_gid}). Sitting on a cell, the algorithm first allocates global \ac{DOF} identifiers for all node functionals associated to the interior of the current cell starting from \mytexttt{fields\_current\_dof(field\_id)}, i.e.,
the next freely available global identifier; see Line~\ref{loc:dof_numbering_cell_interior}. It then loops over the vefs of the current cell. If the current vef has not been visited yet, then the current cell becomes its owner, and both the cell and the local identifier of this vef within the cell are registered in the corresponding work arrays. The global \ac{DOF} identifiers associated to node functionals on this vef within the owner cell are allocated in Line~\ref{loc:dof_numbering_vef} (as above starting from \mytexttt{fields\_current\_dof(field\_id)}). On the other hand, if the current vef has been visited, then the global \ac{DOF} identifiers associated to node functionals on this vef within the current cell are fetched from the corresponding ones within the owner cell in Line~\ref{loc:dof_numbering_vef_source_cell}. The binding called in this line encodes the permutations described in Sect.~\ref{sec:polytope_rotations_and_permutations}.

\begin{lstlisting}[float=htbp,language={[03]Fortran},escapechar=@,caption=The \mytexttt{list\_dofs} binding of \mytexttt{fe\_space\_t}.,label={lst:dof_numbering_list_dofs}]
subroutine fe_space_list_dofs ( this ) 
  class(fe_space_t), intent(inout) :: this   
  ... ! Declare all local variables

  ... ! Allocate owner_cell_gid_x_vef(:,:) to size num_fields x triangulation%num_vefs()
  owner_cell_gid_x_vef = -1
  
  ... ! Allocate vef_lid_in_owner_cell(:,:) to size num_fields x triangulation%num_vefs()
  vef_lid_in_owner_cell = -1
  
  ... ! Allocate blocks_current_num_dofs(:) work array to block_layout%get_num_blocks()
  blocks_current_num_dofs = 0
  
  ... ! Allocate fields_current_dof(:) work array to num_fields
  fields_current_dof = 0
  
  do field_id=1, this%num_fields
    block_id = this%block_layout%get_block_id(field_id)
    fields_current_dof(field_id) += blocks_current_num_dofs(block_id)
    blocks_current_num_dofs(block_id) += this%num_dofs_x_field(field_id)
  end do
  
  call this%create_fe_iterator(owner_fe)
  call this%create_fe_iterator(fe)
  do while ( .not. fe%has_finished())
   do field_id=1, this%get_num_fields()
     call fe%generate_dofs_on_cell_interior_numbering(field_id, fields_current_dof(field_id)) @\label{loc:dof_numbering_cell_interior}@
     do vef_lid = 1, fe%get_num_vefs()
       vef_gid = fe%get_vef_gid(vef_lid)
       if ( owner_cell_gid_x_vef(field_id,vef_gid) == -1) then
         previous_dof_block = fields_current_dof(field_id)
         call fe%generate_dofs_on_vef_numbering(vef_lid, field_id, fields_current_dof(field_id)) @\label{loc:dof_numbering_vef}@
         if (previous_dof_block < fields_current_dof(field_id)) then
           owner_cell_gid_x_vef(field_id, vef_gid) = fe%get_gid() 
           vef_lid_in_owner_cell (field_id, vef_gid) = vef_lid
         end if
       else 
         call owner_fe%set_gid(owner_cell_gid_x_vef(field_id, vef_gid))
         call fe%fetch_dofs_on_vef_numbering_from_source_cell(vef_lid, & @\label{loc:dof_numbering_vef_source_cell}@
                                                              owner_fe, &
                                                              vef_lid_in_owner_cell(field_id,vef_gid), &
                                                              field_id)
       end if
     end do
   end do
   call fe%next()
  end do
  call this%free_fe_iterator(owner_fe)
  call this%free_fe_iterator(fe)
  
  ... ! Deallocate all work arrays
subroutine fe_space_list_dofs ( this ) 

\end{lstlisting}

As the reader might observe, Listing~\ref{lst:dof_numbering_list_dofs} is grounded on several (private) helper bindings of \mytexttt{fe\_cell\_iterator\_t} that, at the cell level, aid in the generation of a global \ac{DOF} numbering; see Lines~\ref{loc:count_dofs_on_cell}-\ref{loc:dofs_on_vef_source} of Listing~\ref{lst:fe_cell_face_iterator}. These bindings ultimately rely on the \mytexttt{reference\_fe\_t} instances mapped to the cells of the triangulation; see Sect.~\ref{subsec:fe_space_internals}. In particular, sitting on a cell, \mytexttt{reference\_fe\_t} instructs \mytexttt{fe\_cell\_iterator\_t} with the association of its node functionals to the interior of the cell, and its lower-dimensional boundary objects according to the notion of conformity underlying the \ac{FE} space at hand; see Sect.~\ref{subsec:global_fe_space} and~\ref{subsec:reference_fe_space}. For example, the implementation of the \mytexttt{generate\_own\_dofs\_vef\_numbering} binding is implemented as shown in Listing~\ref{lst:generate_dofs_on_vef_numbering}. 

\begin{lstlisting}[float=htbp,language={[03]Fortran},escapechar=@,caption=Implementation of the \mytexttt{generate\_own\_dofs\_vef\_numbering} binding of \mytexttt{fe\_cell\_iterator\_t}.,label={lst:generate_dofs_on_vef_numbering}]
subroutine fe_cell_iterator_generate_own_dofs_vef (this, vef_lid, field_id, current_dof)
  class(fe_cell_iterator_t) , intent(inout) :: this
  integer(ip)               , intent(in)    :: vef_lid
  integer(ip)               , intent(in)    :: field_id
  integer(ip)               , intent(inout) :: current_dof
  
  class(reference_fe_t), pointer :: reference_fe
  type(list_iterator_t)          :: own_dofs_vef_iterator
  ... ! Declaration of the rest of local variables
  
  base_pos = this%fe_space%ptr_dofs_x_fe(field_id, this%get_gid())-1
  reference_fe => this%get_reference_fe(field_id)
  own_dofs_vef_iterator = reference_fe%create_own_dofs_n_face_iterator(vef_lid)
  do while (.not. own_dofs_vef_iterator%has_finished())
    dof_pos = base_pos+own_dofs_vef_iterator%current()
    if (this%fe_space%lst_dofs_gids(dof_pos)==0) then @\label{loc:generate_dofs_on_vef_check}@
      this%fe_space%lst_dofs_gids(dof_pos) = current_dof
      current_dof = current_dof + 1
    end if
    call own_dofs_vef_iterator%next()
  end do
end subroutine fe_cell_iterator_generate_own_dofs_vef
\end{lstlisting}

The code in Listing~\ref{lst:generate_dofs_on_vef_numbering} extracts a \mytexttt{list\_iterator\_t} from the \mytexttt{own\_dofs\_n\_face} member variable of the \mytexttt{reference\_fe\_t} instance used in the current cell for \mytexttt{field\_id}. 
This iterator lets it to traverse those node functionals owned by the vef with local identifier \mytexttt{vef\_lid} (see Sect.~\ref{subsec:reference_fe_space}), and thus determine the (relative) position in 
\mytexttt{lst\_dofs\_gids(:)} of the global \ac{DOF} identifiers to be allocated for such node functionals.
We note that the logical predicate in Line~\ref{loc:generate_dofs_on_vef_check} is evaluated to \mytexttt{.true.} if 
the \ac{DOF} at hand is actually free, i.e., not subject to boundary conditions imposed in strong form; 
see Sect.~\ref{subsec:fe_space_strong_boundary_conditions}.

Finally, we would like to stress that error checking statements and a major optimization that can be applied for the single-field single-block
case are not shown in the code listings of this section in order to keep the presentation as simple as possible. Both are present in \FEMPAR{}. In particular, for the aforementioned
case, the global \ac{DOF} numbering can be generated with a single loop over all cells (instead of two). The call in Line~\ref{loc:dof_numbering_count_dofs} of Listing~\ref{lst:global_dof_numbering} can be avoided, deferring the computation
of the number of \acp{DOF}  per field and block to the call in Line~\ref{loc:dof_numbering_list_dofs}. 

On the other hand, there is no need to generate a global \ac{DOF} numbering from scratch
when there is already one available, a permutation from the old to the new numbering could be computed and applied to \mytexttt{lst\_dofs\_gids(:)} by a single sweep over all cells. 
This optimization, however, is not present in \FEMPAR{}, as indeed we did not find frequent the case where an application requires to change on-the-fly the block-layout of the system of \acp{PDE}
at hand.


\subsection{Strong imposition of boundary conditions} \label{subsec:fe_space_strong_boundary_conditions}
In this section, we discuss the mechanisms that \mytexttt{fe\_space\_t} provides in order to support the strong imposition of boundary conditions.  In order to grasp why these mechanisms are needed and how \mytexttt{fe\_space\_t} is designed to provide them,  we must first briefly introduce the approach chosen by \FEMPAR{}
in order to handle this type of boundary conditions. We will use the term ``fixed \acp{DOF}'' to refer to those \acp{DOF}  sitting on the boundary whose values are constrained (i.e., subject to strong boundary conditions), and the term ``free \acp{DOF}'' to refer to the remaining ones. For simplicity, let us restrict ourselves to the Laplacian problem with inhomogeneous Dirichlet boundary conditions $u(x)=u_{{\rm D}}(x)$ on $\Dirichlet$ discretized with grad-conforming \acp{FE}.\footnote{We stress, however, that the approach discussed in the sequel to handle the strong imposition of boundary conditions is applicable to more complex problems and discretizations, e.g., the Maxwell equations discretized with curl-conforming \ac{FE} spaces.} The discrete solution $u_h \in \trialsp_h$ can be split into two parts as $u_h = \bar{u}_h + E_h u_{{\rm D}}$, where: 
\begin{equation}
\bar{u}_h   \ = \ \sum_{\mathclap{a \in {\rm \{free\ \acp{DOF} \}}}} \bar{u}_a \shapetest{a} \ \  + \ \  \sum_{\mathclap{a \in {\rm \{fixed\ \acp{DOF} \}}}} 0 \shapetest{a} \quad \quad {\rm and} \quad \quad 
E_h u_{{\rm D}} \ = \ \sum_{\mathclap{a \in {\rm \{free\ \acp{DOF} \}}}} 0 \shapetest{a} \ \  + \ \  \sum_{\mathclap{a \in {\rm \{fixed\ \acp{DOF} \}}}} u_{a}^{{\rm D}} \shapetest{a} .
\end{equation}
The nodal values $\bar{u}_a$ are the actual unknowns of the problem at hand. $E_h u_{{\rm D}}$ is a discrete Dirichlet data extension, which can be understood as the projection of a Dirichlet data extension $Eu_{\rm D}(x)$  introduced in Sect.~\ref{subsec:weak_form}. Its nodal values $u_{a}^{{\rm D}}$ are selected such that $E_h u_{{\rm D}}$ becomes a suitable boundary \ac{FE} approximation of $u_{{\rm D}}(x)$ (e.g., a boundary \ac{FE} interpolation).\footnote{It is assumed that the discrete Dirichlet data extension is zero on free \acp{DOF}, but other more general situations can also be accommodated.} The linear system to be solved  in order to compute the nodal values of $\bar{u}_h$  can be written as:
\begin{equation} \label{eq:dirichlet_bc_discrete_problem}
\sum_{\mathclap{b \in {\rm \{free\ \acp{DOF} \}}}} a(\shapetest{a},\shapetest{b}) \bar{u}_b = (\shapetest{a},f)-\sum_{\mathclap{c \in {\rm \{fixed\ \acp{DOF} \}}}} a(\shapetest{a},\shapetest{c}) u_{c}^{{\rm D}} \quad \forall a \in \{{\rm free \ \acp{DOF} }\},
\end{equation}
where {\em its coefficient matrix has as many rows as free \acp{DOF}}, and its right-hand side is the \ac{FE} discretization of the linear form in \eqref{eq:dirichlet_bcs_rhs}; see Sect.~\ref{subsec:weak_form}.

In order to assemble \eqref{eq:dirichlet_bc_discrete_problem}, the process described in Sect.~\ref{sec:cell_integration} has to be slightly modified. A sweep over all triangulation cells is still required. Sitting on a given cell $K$, the element matrix $\fematrix^{K}$ and vector $\ferhs^{K}$ are computed as usual. However, the rows/columns corresponding to fixed \acp{DOF}  in $\fematrix^{K}$ are not assembled into the global matrix. The same applies to the entries of  $\ferhs^{K}$. However, $\ferhs^{K}$  has to be updated before assembly in order to reflect the contributions of strong boundary conditions (see the right-hand side of \eqref{eq:dirichlet_bc_discrete_problem}). Fortunately, the users of \FEMPAR{} are unburdened from these subtleties. These are hidden within the \mytexttt{assembly} generic binding of \mytexttt{fe\_cell\_iterator\_t}; see Listing~\ref{lst:fe_cell_face_iterator} and~\ref{lst:fe_assembly}. Apart from adding the contributions of the current cell to the global coefficient linear system and right-hand side, this binding is in charge of computing the contribution to $\ferhs^{K}$ from strong Dirichlet boundary conditions. This poses two additional requirements on \mytexttt{fe\_space\_t}. In particular, 1)~it should handle a global enumeration of free and \emph{fixed} \acp{DOF}, while being able to distinguish among both kinds of \acp{DOF}; and 2)~it should offer a suitable set of bindings to project/interpolate  $u_{{\rm D}}(x)$ on the boundary to get $E_h u_{{\rm D}}$.

In order to satisfy 1), \mytexttt{fe\_space\_t} splits the whole set of \acp{DOF} into free and fixed \acp{DOF}, and the \acp{DOF} within each subset are labeled separately from each other as $\{1,2,\ldots,|\{{\rm free \ \acp{DOF} }\}|\}$, and $\{-1,-2,\ldots,-|\{{\rm fixed \ \acp{DOF} }\}|\}$, respectively. (This is nevertheless an implementation detail that is never exposed to \FEMPAR{} users.) In turn, free and fixed \ac{DOF} values are actually stored into different arrays, so that they can be addressed separately using the corresponding global identifiers in the former and latter set, respectively; see Sect.~\ref{subsec:fe_function}.


The process that associates global identifiers to free \acp{DOF}  has been already covered in Sect.~\ref{subsec:global_dof_numbering}. The one corresponding to fixed \acp{DOF}  very much resembles the one for free \acp{DOF}. It is, however, restricted to vertices, edges, and faces of the triangulation that lay at the boundary, and it has to be equipped with support from the user that lets the process become aware of which \acp{DOF}  sitting on the boundary are actually fixed. The fixed \acp{DOF}  global enumeration process occurs during the initial set-up of \mytexttt{fe\_space\_t}; see \mytexttt{create} generic binding in Listing~\ref{lst:fe_space}. This process is grounded on two different ingredients. On the one hand, the user can determine $\Gamma$ sub-regions through the sets associated to vefs sitting on the boundary (see Sect.~\ref{subsec:abstract_triangulation}). For example, the user may decide to use set identifier 1 and 2 to split the vefs in $\Gamma$ into those which belong to $\Dirichlet$ and $\Neumann$, respectively. On the other hand, an abstract data type, called \mytexttt{conditions\_t}, to be extended by \FEMPAR{} users, lets users to customize the strong imposition of boundary conditions. In particular, with regard to the fixed \acp{DOF}  global enumeration process, this data type offers a deferred binding that given a set identifier, provides a logical component mask. For each component of the \ac{PDE} system, this mask provides whether the \acp{DOF}  associated to vefs marked with this set identifier are fixed or free. For those \ac{FE} spaces for which there is no \ac{DOF}-to-component association (e.g., Raviart-Thomas or N\'ed\'elec \acp{FE}), only the first component in the mask is taken into account, and the rest neglected.

On the other hand, for 2),  \mytexttt{fe\_space\_t} provides a set of methods that let the user interpolate/project $u_{{\rm D}}(x)$ on the boundary to get $E_h u_{{\rm D}}$ in a number of suitable ways. $E_h u_{{\rm D}}$ is ultimately stored within an instance of the \texttt{fe\_function\_t} data type; see Section~\ref{subsec:fe_function}. Boundary projectors involve the solution of a boundary mass matrix problem where integrals over boundary facets have to be evaluated; see Sect.~\ref{sec:face_integration}. Again, all these bindings rely on the \mytexttt{conditions\_t} abstract data type. In particular, given  a boundary  vef set identifier, a deferred binding of this data type returns a user-defined 
(scalar-valued) function to be imposed for each component  of the \ac{PDE} system at hand. In the case of Raviart-Thomas or N\'ed\'elec \acp{FE}, the $d$ scalar-valued functions corresponding to its components are used to reconstruct the vector-valued function, whose tangential or normal component, respectively is to be imposed.

\subsection{Global \ac{FE} functions and their restriction to triangulation cells/facets} \label{subsec:fe_function}

In this section, we introduce a convenient software abstraction in our \ac{OO} design, referred to as \mytexttt{fe\_function\_t}, which represents a global \ac{FE} function $u_h \in \trialsp \doteq \trialsp^1_h \times \ldots \times \trialsp^n_h$. This data type and a subset of its \acp{TBP} (in particular, those that are relevant for the present section) are presented in Listing~\ref{lst:fe_function}.

\lstinputlisting[float=htbp,language={[03]Fortran},escapechar=@, label={lst:fe_function}, caption={The \mytexttt{fe\_function\_t} data type.}]{fe_function.f90}

In Listing~\ref{lst:fe_function}, the \mytexttt{free\_dof\_values} and \mytexttt{fixed\_dof\_values} are used to store $\bar{u}_h$ and $E_h u_{{\rm D}}$, respectively; see Sect.~\ref{subsec:fe_space_strong_boundary_conditions}. 
The former is a {\em polymorphic} member variable of type \mytexttt{array\_t}; see Section~\ref{sec:linalg}.
Relying on the set of deferred bindings offered by \mytexttt{array\_t}, the code bounded to \mytexttt{fe\_function\_t} can be written independently of how the entries within the concrete implementation of \mytexttt{array\_t} are laid out in memory, enabling code re-use to a large extent. For example, \mytexttt{scalar\_array\_t} is a concrete realization of \mytexttt{array\_t} that uses monolithic storage, while \mytexttt{block\_array\_t}  stores the entries organized 
into blocks (see Sect.~\ref{sec:linalg} for more details).
On the other hand, \mytexttt{fixed\_dof\_values} is a member variable of {\em static} type \mytexttt{scalar\_array\_t}; see Sect.~\ref{sec:linalg}.\footnote{In parallel environments, every processor only stores the fixed \ac{DOF} values that belong to its associated subdomain.} Fixed \acp{DOF}  belonging to different fields might be indeed assigned intermixed global identifiers, significantly simplifying the enumeration process. In particular, a single sweep over all boundary objects suffices, in contrast to Listing~\ref{lst:global_dof_numbering}, where two sweeps over all cells are required in order to generate a block-aware global numbering. From our experience, it turns out that 
neither blocked storage nor a data structure suitable for distributed-memory environments are strictly required to store $E_h u_{{\rm D}}$,
so that we can prevent the overhead associated to run-time polymorphism when dealing with \mytexttt{fixed\_dof\_values}.\footnote{Some of the algorithms in charge of 
computing $E_h u_{{\rm D}}$ may require a different storage layout from the one of \mytexttt{scalar\_array\_t} (e.g., blocked storage and/or suitable for distributed-memory computers), and/or 
restrict themselves to those fixed \acp{DOF}  of $E_h u_{{\rm D}}$ corresponding to a given field (or set of fields).
In such a case, $E_h u_{{\rm D}}$ is scattered in place back and forth into temporary work space with the appropriate layout for the algorithm at hand in charge of computing its entries (e.g., a serial or parallel distributed-memory boundary mass problem iterative solver). It turns out that it is not such a high performance penalty provided that such algorithms already require to perform a 
sweep over boundary facets (e.g., in order to assemble a boundary mass matrix). During this sweep, the fixed \acp{DOF}  in question can be already counted and identified.}


A \mytexttt{fe\_function\_t} instance is created from a \mytexttt{fe\_space\_t} instance (to which it belongs); see signature of the \mytexttt{create} binding in Listing~\ref{lst:fe_function}. 
This binding selects the dynamic type of \mytexttt{free\_dof\_values}, and therefore its storage layout, according to the one currently selected
for the \ac{PDE} system at hand; see \mytexttt{block\_layout} member variable in Listing~\ref{lst:fe_space}. 
The entries of \mytexttt{free\_dof\_values} can be determined in a number of ways. They might become the unknowns of a problem to be solved (e.g., by 
a preconditioned iterative linear solver or sparse direct solver), or computed from an expression involving other \mytexttt{fe\_function\_t} instances, 
e.g., $u_h=v_h$, or $u_h=v_h+w_h$, with $u_h,v_h,w_h \in \trialsp_h$. (Indeed, \FEMPAR{} offers an expression syntax for global \ac{FE} functions grounded on overloaded operators.) 
Apart from these, \mytexttt{fe\_space\_t} offers a pair of generic bindings, referred to as \mytexttt{interpolate} and \mytexttt{project}, to compute the \acp{DOF}  nodal values of ${u}_h$ by either interpolation  (using the expression in \eqref{eq:global_interpolator}) or projection (e.g., a global $L^2$ projection) into the \ac{FE} space of a user-defined function ${u}(x)$.\footnote{Analytical scalar, vector, and tensor-valued functions are also supported in \FEMPAR{} through the classes \mytexttt{scalar\_function\_t}, \mytexttt{vector\_function\_t}, and \mytexttt{tensor\_function\_t}, respectively. To implement an analytical scalar function $f(\x)$ in \FEMPAR{}, the user has to extend \mytexttt{scalar\_function\_t} methods \mytexttt{get\_value}, \mytexttt{get\_gradient} (if used), etc., with the analytical expression, for a given $\mytexttt{point\_t}$ that represents $\x$. We proceed analogously for vector and tensor fields. These data types are very simple and we omit their description here.} Each of these generic bindings is overloaded with three different regular bindings suitable for scalar, vector, and tensor-valued functions, respectively. The \mytexttt{interpolate} bindings in \mytexttt{fe\_space\_t} can be written independently of the reference \ac{FE} by using a TBP associated to \mytexttt{reference\_fe\_t} that computes the local interpolator in \eqref{eq:local_interpolator}.

Apart from the software representation of a global \ac{FE} function, \ac{FE} codes typically need a mechanism that, sitting on a cell or facet of the triangulation, provides the values, gradients, etc. of a global \ac{FE} function $u_h=u^1_h \times \ldots \times u^n_h$ evaluated at quadrature points in the physical space. To this end, \FEMPAR{} offers a set of data types, referred to as \mytexttt{cell\_fe\_function\_type\_t} and \mytexttt{facet\_fe\_function\_type\_t}, with \mytexttt{type=scalar,vector,tensor}, that represent the restriction of $u^i_h$ to a given triangulation cell and facet, respectively. The two code snippets in Fig.~\ref{fig:cell_and_face_fe_function} illustrate the usage of these data types, where we are assuming that $u^i_h$ belongs to a global \ac{FE} space of vector-valued functions.

\begin{figure}
\begin{center}
\begin{tabular}{ccc}
\begin{minipage}{0.45\textwidth}
\lstinputlisting[language={[03]Fortran}]{cell_fe_function.f90}
\end{minipage} & &
\begin{minipage}{0.45\textwidth}
\lstinputlisting[language={[03]Fortran}]{face_fe_function.f90}
\end{minipage}
\end{tabular}
\end{center}
\caption{\label{fig:cell_and_face_fe_function} User-level code snippets illustrating the usage of the \mytexttt{cell\_fe\_function\_type\_t} (left) and \mytexttt{facet\_fe\_function\_type\_t} (right) data types.}
\end{figure}

There are three worth noting remarks in these two code snippets. First, the \mytexttt{update} binding of both data types rely on the 
\mytexttt{gather\_nodal\_values} binding of \mytexttt{fe\_function\_t}; see Listing~\ref{lst:fe_function}. The latter 
equips cell/facet \ac{FE} functions with the ability to restrict (gather) the nodal values of $u^i_h$ from global to local
arrays (stored as private scratch data within cell/facet \ac{FE} function data types), {\em while taking care of strong boundary conditions}. Second, the \mytexttt{update} bindings require a procedure that, given the shape functions, first derivatives, etc.,
evaluated at quadrature points in physical space, and the nodal values $u^i_h$ restricted to the current cell, provides the shape
function values, gradients, curls, etc., of the \ac{FE} function at these quadrature points. This service is provided by 
\mytexttt{reference\_fe\_t} by the set of \mytexttt{evaluate\_fe\_function...} deferred bindings in Lines~\ref{loc:evaluate_fe_function_start}-\ref{loc:evaluate_fe_function_stop} of Listing~\ref{lst:reference_fe}. We note that \mytexttt{fe\_function\_t} can extract the first set of data from the \mytexttt{cell\_integrator\_t} and \mytexttt{facet\_integrator\_t} instances accessible through  \mytexttt{fe\_cell\_iterator\_t} and \mytexttt{fe\_facet\_iterator\_t} (provided on input to \mytexttt{update}), 
respectively. Third, facet \ac{FE} functions provide $u^i_h$ values, gradients, etc., at facet quadrature points from the perspective of its two surrounding cells. This make sense for functions $u^i_h$ belonging to non-conforming \ac{FE} spaces, which might be discontinuous across cell 
boundaries. Facet \ac{FE} functions should also cope with the fact that the coordinate systems of its surrounding cells might
not be aligned in physical space, so that a different local numbering might be assigned to facet quadrature points from the
perspective of either cell; see Sect.~\ref{subsec:face_integrator} for an exposition of the strategy followed to solve this issue.

\section{Building \ac{FE} affine operators} \label{sec:building_fe_systems}

In this section, we introduce the software abstractions on which the construction of the algebraic problem \Eq{linear_system} in Sect.~\ref{sec:fe_meth} relies. These software abstractions, and their relationship, are depicted in Fig.~\ref{fig:uml_fe_affine_operator_t}. The main design goal underlying
the proposed software architecture is as follows. In the seek of code reusability and extensibility, 
\FEMPAR{} users should have at their disposal {\em a unique entry point data type and associated bindings} 
in order to build their \ac{FE} linear system, no matter whether a scalar or a system of PDEs, no matter whether the linear algebra data structures holding
the linear system entries are either scalar (monolithic) or blocked, and no matter how they are laid out in memory (centralized, distributed-memory). 
In \FEMPAR{}, this unique entry point data type is referred to as \mytexttt{fe\_affine\_operator\_t}.
Mathematically, \mytexttt{fe\_affine\_operator\_t} represents the affine operator in \eqref{problem1h-op}, obtained from the discrete weak formulation of the linear(ized) problem~\eqref{problem1h-d}.
As introduced in Sect.~\ref{subsec:global_fe_space}, the operator can be represented (after defining bases for trial and test \ac{FE} spaces) with $\fematrix$ and $\ferhs$ defined in \eqref{eq:linear_system}. The solution of the \ac{FE} problem is the only root of this operator (as soon as the \ac{FE} problem is nonsingular). 

\begin{figure}[htbp]
\begin{minipage}{\textwidth}
\begin{center}
\scalebox{0.85}{
\begin{tikzpicture}[node distance=1.cm,auto]

    \node (disc_int) [class, text width=5.2cm, rectangle split, rectangle split parts=3] { 
          \it \small discrete\_integration\_t
          \nodepart{second} ...
          \nodepart{third} \footnotesize  $+${\it integrate\_galerkin}\mytexttt{(this,fe\_space,} \\ 
                           {\color{white} $+$\mytexttt{integrate\_galerkin(}}\mytexttt{assembler)} \\...};

    \node (fe_ope) [class, text width=3.3cm, rectangle split, rectangle split parts=3, right=of disc_int] { 
          \tt \small fe\_affine\_operator\_t
          \nodepart{second} ...
          \nodepart{third}  ...};

    \node (fe_space) [class, text width=2.cm, rectangle split, rectangle split parts=3, right=of fe_ope] { 
          \tt \small fe\_space\_t
          \nodepart{second} ...
          \nodepart{third}  ...};


    \node (assembler) [class, text width=1.9cm, rectangle split, rectangle split parts=3, below=of fe_ope] { 
          \small \it assembler\_t 
          \nodepart{second}  ...
          \nodepart{third} ...};

    \node(scalar_assembler) [class, text width=3.2cm, rectangle split, rectangle split parts=3, left=of assembler] { 
          \small \tt scalar\_assembler\_t 
          \nodepart{second} ...
          \nodepart{third} ...};

    \node(block_assembler) [class, text width=3.cm, rectangle split, rectangle split parts=3, right=of assembler] { 
          \small \tt block\_assembler\_t 
          \nodepart{second} ...
          \nodepart{third}  ...};

    \node (aux1) [minimum width=6.0cm, minimum height=1.5cm, text width=0.cm, below=of assembler] {};
    \node (aux2) [minimum width=0.0cm, minimum height=1.5cm, text width=0.cm, below=of assembler] {};


    \node (matrix) [class, text width=1.4cm, rectangle split, rectangle split parts=3, left=of aux1] { 
          \small \it matrix\_t 
          \nodepart{second} ...
          \nodepart{third}  ...};

    \node (array) [class, text width=1.2cm, rectangle split, rectangle split parts=3, right=of aux1] { 
          \small \it array\_t 
          \nodepart{second} ...
          \nodepart{third}  ...};



    \node (block_matrix) [class, text width=2.4cm, rectangle split, rectangle split parts=3, below left=of aux2] { 
          \small \tt block\_matrix\_t 
          \nodepart{second} ...
          \nodepart{third}  ...};

    \node (aux3) [minimum size=0.0cm, text width=0.cm, left=of block_matrix] {};

    \node (sparse_matrix) [class, text width=2.6cm, rectangle split, rectangle split parts=3, left=of aux3] { 
          \small \tt sparse\_matrix\_t 
          \nodepart{second} ...
          \nodepart{third}  ...};

    \node (scalar_array) [class, text width=2.4cm, rectangle split, rectangle split parts=3, below right=of aux2] { 
          \small \tt scalar\_array\_t 
          \nodepart{second} ...
          \nodepart{third}  ...};

    \node (aux4) [minimum size=0.0cm, text width=0.cm, right=of scalar_array] {};

    \node (block_array) [class, text width=2.4cm, rectangle split, rectangle split parts=3, right=of aux4] { 
          \small \tt block\_array\_t 
          \nodepart{second}  ...
          \nodepart{third}   ...};

    \draw[assoc] (fe_ope.west) -- node[midway,above]{\footnotesize \tt 1} (disc_int.east);
    \draw[assoc] (fe_ope.east) -- node[midway,above]{\footnotesize \tt 1/2} (fe_space.west);
    \draw[comp]  (fe_ope.south)   -- node[midway,right]{\footnotesize \tt 1} (assembler.north);
    \draw[inherit] (assembler.west)  -- (scalar_assembler.east);
    \draw[inherit] (assembler.east) -- (block_assembler.west);
    \draw[comp] (node cs:name=assembler, angle=-120) -- ++(0,-0.5) -| node[near start,below]{\footnotesize \tt 1} (matrix.north);
    \draw[comp] (node cs:name=assembler, angle=-60)  -- ++(0,-0.5) -| node[near start,below]{\footnotesize \tt 1} (array.north);

    \draw[inherit] (matrix.south) -- ++(0,-0.5) -| (sparse_matrix.north);
    \draw[inherit] (matrix.south) -- ++(0,-0.5) -| (block_matrix.north);
    \draw[inherit] (array.south) -- ++(0,-0.5) -| (scalar_array.north);
    \draw[inherit] (array.south) -- ++(0,-0.5) -| (block_array.north);

    \draw[comp] (block_matrix.west) -- node[midway,above]{\footnotesize \tt nblocks$^\texttt{2}$} (sparse_matrix.east);
    \draw[comp] (block_array.west) -- node[midway,above]{\footnotesize \tt nblocks} (scalar_array.east);
\end{tikzpicture}
}
\end{center}
\caption{UML class diagram of the \mytexttt{fe\_affine\_operator\_t} abstraction and its relationship with other \FEMPAR{} classes.}
\label{fig:uml_fe_affine_operator_t}
\end{minipage}
\end{figure}
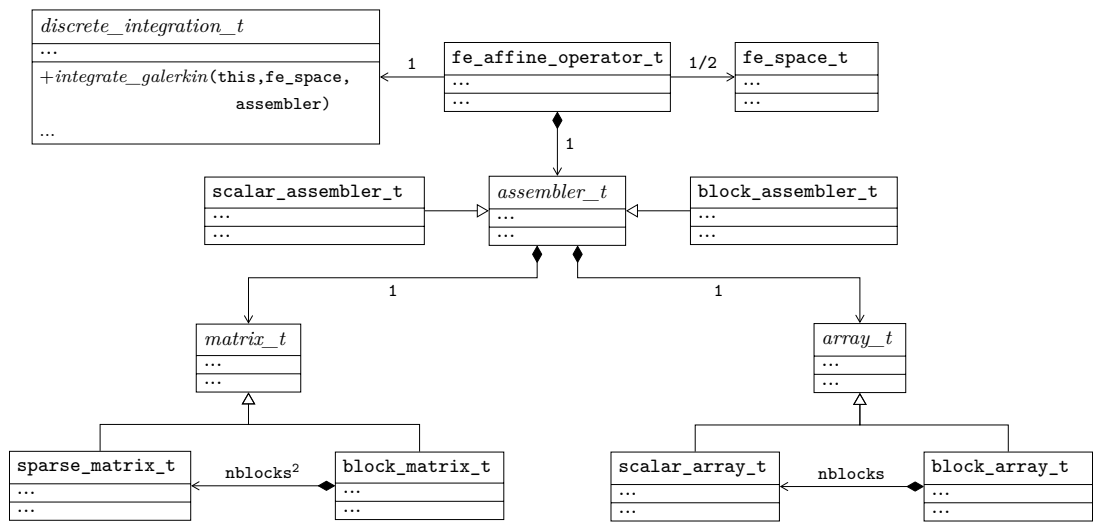

In order to seek the aforementioned goal, \mytexttt{fe\_affine\_operator\_t} relies on an abstract data type, 
referred to as \mytexttt{assembler\_t} (see Fig.~\ref{fig:uml_fe_affine_operator_t}).
In a nutshell, \mytexttt{assembler\_t} offers a set of {\em \ac{FE}-assembly tailored},
data structure neutral, deferred  \acp{TBP}, e.g., to assemble the contributions of a cell or facet integral into the linear system coefficient matrix $\fematrix$ and/or right-hand side $\ferhs$.
The subclasses of  \mytexttt{assembler\_t} are the ones ultimately responsible to deal with the details underlying the particular linear algebra data structures
at hand. The latter ones offer {\em \ac{FE}-assembly neutral} interfaces to inject new entries or add contributions to them, such that this software piece becomes reusable
and separable, e.g., to be used in third party software projects (not necessarily \ac{FE}-oriented) as a standalone software subsystem. \FEMPAR{} offers a rich set of linear algebra data structures, e.g., data structures organized by blocks, which enable the implementation of block preconditioners for multiphysics problems (see, e.g., \cite{elman_finite_2005,badia_block_2014,cyr_teko:_2016}). Apart from those required to handle 
the linear coefficient matrix and right-hand side of the system, \mytexttt{fe\_affine\_operator\_t}  also interacts with other data types required to deliver its life cycle (i.e., its auto-generation). In particular, $\fematrix$ and $\ferhs$ entries 
are computed according to the expressions in  \eqref{eq:linear_system}. These expressions involve a \ac{FE} space (\mytexttt{fe\_space\_t}) and the discrete (bi)linear forms of the problem at hand. To express in software 
this second ingredient, we introduce the \mytexttt{discrete\_integration\_t} abstraction; see Fig.~\ref{fig:uml_fe_affine_operator_t}.
 
We have structured this section as follows. In Sect.~\ref{sec:linalg}, we first present the \mytexttt{assembler\_t} abstract data type, and the rationale underlying the design of the linear algebra structures it is
grounded on. Next, in Sect.~\ref{sec:disc_int}, we introduce the \mytexttt{discrete\_integration\_t} abstract data type that ultimately is in charge of performing the integration of the (bi)linear forms and assembly of the discrete affine
operator. We show a particular implementation of this data type (i.e., a subclass) for the Galerkin approximation of the Stokes problem. Finally, the \mytexttt{fe\_affine\_operator\_t} data type is described in Sect.~\ref{sec:fe_affine_operator}.
  
\subsection{Linear algebra data structures and associated assemblers} \label{sec:linalg}
    
Linear algebra in \FEMPAR{} relies on a pair of data type hierarchies rooted at the mathematical
abstractions of a linear algebra operator and a vector, and represented in software by means of the \texttt{linear\_operator\_t}
and \texttt{vector\_t} abstract data types, respectively. 
These abstract data types let a number of linear algebra algorithms within \FEMPAR{} 
(e.g., iterative linear solvers and block preconditioners for PDE systems) to be expressed independently from the actual implementation of the concrete matrix and vector data structures 
being used, such as block layout (if any),  storage (e.g., dense or sparse storage format) or memory layout (e.g., local or distributed-memory), 
enabling code re-use and extensibility to a large extent. An abstract expression syntax that allows the construction of complex expressions involving
operations among operators and/or vectors is also provided. This enables the implementation of new algorithms in a compact manner. However, because 
these linear algebra algorithms are not discussed herein but postponed to a further work, the description of the data types and associated methods 
in these hierarchies will be restricted to what is necessary to describe the assembly of the \ac{FE} affine operator.

The \mytexttt{sparse\_matrix\_t} data type can be found at an intermediate level in the hierarchy rooted at \texttt{linear\_operator\_t}. This is a crucial data type
in \FEMPAR{}, which represents a scalar, non-distributed, sparse matrix. 
Its design follows the ideas presented in \cite{filippone_object-oriented_2012}. This design (re)uses the {\em state} \ac{OO} design pattern~\cite{gamma_e._design_1995} 
to hide the actual sparse matrix storage format to the user. Following this pattern, \mytexttt{sparse\_matrix\_t} is composed of a polymorphic member variable of (declared) type 
\mytexttt{base\_sparse\_matrix\_t}. Its dynamic type can be thus changed at runtime (via re-allocation). This dynamic type represents the storage at hand being used. 
Current subclasses of \mytexttt{base\_sparse\_matrix\_t} include \mytexttt{coo\_sparse\_matrix\_t}, \mytexttt{csr\_sparse\_matrix\_t}, \mytexttt{csc\_sparse\_matrix\_t}, 
corresponding to the coordinate list (COO), the compressed sparse row (CSR), and the compressed sparse column (CSC) sparse matrix storage formats~\cite{saad_iterative_2003}, respectively.

The life cycle of a \mytexttt{sparse\_matrix\_t} instance is as follows. The user first invokes its \texttt{create} \ac{TBP}, in which one solely specifies 
the size of the matrix, i.e., the number of rows and columns. This method, however, triggers a number of subsequent actions. In particular, {\em it allocates its dynamic type
to the one corresponding to the COO format}, and leaves it ready for the injection or addition of contributions to the entries of the matrix. 
Although not compressed, this format is ideally shaped for the injection or addition of contributions to the entries of the matrix. These are simply pushed back
into member arrays that can grow dynamically during the integration/assembly loop (via a judiciously reallocation strategy to trade off cost and memory). 
Other sparse storage formats, as the CSR storage implemented in 
the \mytexttt{csr\_sparse\_matrix\_t} data type (also a type extension of \mytexttt{base\_sparse\_matrix\_t}), although more memory efficient, 
require a predefined sparsity pattern, which has to be precomputed. They are not thus well suited for the dynamic build up process of the matrix. 
At this point the reader should note that, for such inflexible storage formats, {\em one typically needs an accurate estimation of the maximum number of nonzeros
per each row (or column) to be memory efficient}. This estimation, however, can only be a quite large upper bound in complex scenarios (e.g., $hp$-adaptive methods in 3D, among others).

Once the build up process finishes, the user can call a method specially designed to leave the \mytexttt{sparse\_matrix\_t} instance ready for being used (e.g., to perform
operations with it). This involves a compression process, in which duplicated entries are either summed up, or filtered (as selected by the user)  
and a transformation of the COO storage format into the
storage format that the user actually requires (e.g., CSR). For simplicity, we refer to this stage as the ``compression'' of the matrix. 
Once the \mytexttt{sparse\_matrix\_t} instance is in this final state, it is still possible to insert or add contributions
to its entries, as far as they belong to the sparsity pattern resulting from the first build up process.
Thus, e.g., if a transient and/or nonlinear problem is to be solved and the triangulation of the domain does not change, 
the assembly in COO format will only be performed at the first nonlinear iteration of the first time step. 

As shown so far, the software architecture of \mytexttt{sparse\_matrix\_t} is such that several (current and future) storage formats are possible within a single framework.
This flexibility is convenient for two main reasons. First, no given storage format is likely to be uniformly better in performance across all possible operations and
computer architectures. Second, \FEMPAR{} interoperability with external software dramatically increases. If a new library, that uses its own storage format, is to be
integrated, only a new extension of \mytexttt{base\_sparse\_matrix\_t} has to be added, while leveraging dozens of thousands of lines of code already written. Apart from \mytexttt{sparse\_matrix\_t}, there are other sparse matrix data types available, suitable to handle blocks and/or distributed-memory
computers. All these data types are essentially composed in some way or another of \mytexttt{sparse\_matrix\_t} instances. For example, \mytexttt{block\_sparse\_matrix\_t} is composed
of $\mytexttt{nblocks}^2$ \mytexttt{sparse\_matrix\_t} instances; see Fig.~\ref{fig:uml_fe_affine_operator_t}. 
It, however, provides a set of specialized \acp{TBP} that only apply in the blocked case, e.g., the \mytexttt{get\_block} \ac{TBP}
that lets a client to retrieve the \mytexttt{sparse\_matrix\_t} instance corresponding to a given block of the matrix.

The counterpart of \mytexttt{sparse\_matrix\_t} in the vector case is referred to as \mytexttt{scalar\_array\_t}. It represents a scalar, non-distributed, linear algebra vector, with its
entries stored explicitly in a simple (Fortran intrinsic) allocatable array. However, provided that it does not have to exploit sparsity, the code bounded to
this data type is significantly simpler to the one bounded to \mytexttt{sparse\_matrix\_t}. It is equipped with a pair of generic bindings, with signatures coming in different flavours,
in order to insert or add contributions to the vector.  Likewise, there are other vector-like data types available suitable to handle blocks and/or distributed-memory
computers. For example, \mytexttt{block\_array\_t} is composed of $\mytexttt{nblocks}$ \mytexttt{scalar\_array\_t} instances; see Fig.~\ref{fig:uml_fe_affine_operator_t}.

Apart from the linear algebra data structures so far, we need the additional data type \mytexttt{assembler\_t}, which offers \ac{FE}-assembly tailored signatures to \texttt{fe\_affine\_operator\_t}.
The interface of its deferred \acp{TBP}, which its extensions, e.g., \mytexttt{scalar\_assembler\_t} and \mytexttt{block...assembler\_t}, implement, are shown in Listing~\ref{lst:assembly}.
\mytexttt{assembler\_t} has to be ``general enough'' to handle many storage layouts and it is in charge to isolate \texttt{fe\_affine\_operator\_t} from implementation details.
With that purpose in mind, it is composed of a (polymorphic) \mytexttt{matrix\_t} and a (polymorphic) \mytexttt{array\_t} instance. These are in turn abstract data types rooted at all the
matrix and array data types seen so far, respectively. The set of deferred \acp{TBP} of these two abstract data types is designed (on purpose) to be not sufficiently rich to handle the whole
life cycle of the concrete matrix and array instances. The high heterogeneity of the concrete subclasses of  \mytexttt{matrix\_t} and \mytexttt{array\_t} precludes it. This set of \acp{TBP}
is, in particular, restricted to allocation of memory for its entries, initialization of its entries to a given value  (e.g., initialization to zero), and deallocation of any 
internal memory. These three operations are required by \texttt{fe\_affine\_operator\_t} during the deployment of its life cycle. 
The bulk of the life cycle of the concrete subclasses of \mytexttt{matrix\_t} and  \mytexttt{array\_t} is handled by the subclasses of \mytexttt{assembler\_t}.
This is how it should be, provided that \mytexttt{assembler\_t} subclasses are the ones aware of the concrete details of the corresponding \mytexttt{matrix\_t} and  \mytexttt{array\_t} subclasses.
Besides, by doing this, we can overcome the overhead associated to dynamic run-time polymorphism, provided that the binding of fine-grain calls to those \acp{TBP} injecting or adding
contributions to the matrix or the array {\em can be determined at compilation time}.

\lstinputlisting[float=htbp,language={[03]Fortran},escapechar=@,caption=The \texttt{assembler\_t} abstract data type and its deferred \acp{TBP}.,label={lst:assembly}]{assembler.f90}

Going back to Listing~\ref{lst:assembly}, observe that \mytexttt{assembly\_array} (resp., \mytexttt{assembly\_matrix}) takes an intrinsic Fortran array (resp., rank-2 array) as dummy argument for the element vector (resp., matrix). Besides, it also gets the global \acp{DOF} identifiers on top a single cell, or those corresponding to cells surrounding the facet (see Lines~\ref{loc:assembly_cell2dofs},~\ref{loc:assembly_cell2row_dofs} and \ref{loc:assembly_cell2col_dofs} in Listing~\ref{lst:assembly}). In the case of \mytexttt{scalar\_assembler\_t}, the implementation is made using the \acp{TBP} provided by \mytexttt{scalar\_array\_t} in order to add contributions to its entries and the corresponding \acp{TBP} of \mytexttt{sparse\_matrix\_t}. In the case of \mytexttt{block\_assembler\_t}, the implementation is made by looping through the blocks, obtaining a reference to the current block with the \mytexttt{get\_block} \ac{TBP}, and using the corresponding \acp{TBP} as in the previous case. The \mytexttt{assembly\_array} and \mytexttt{assembly\_matrix} \acp{TBP} are used by the \mytexttt{fe\_cell\_iterator\_t} and \mytexttt{fe\_facet\_iterator\_t} data types to implement their assembly \acp{TBP} (see Lines~\ref{loc:fe_cell_iterator_generic_assembly} and~\ref{loc:fe_face_iterator_generic_assembly} in Listing~\ref{lst:fe_cell_face_iterator} of Sect.~\ref{subsec:fe_space_conceptual_view}). For completeness, in Listing~\ref{lst:fe_assembly} we show the signature of the latter \acp{TBP}. These are the ones actually used by the user in the type extension of \mytexttt{discrete\_integration\_t}, as described in Sect.~\ref{sec:disc_int}. 

\lstinputlisting[float=htbp,language={[03]Fortran},escapechar=@,caption={The interfaces of the assembly \acp{TBP} of ``\mytexttt{set}'' (either \mytexttt{cell} or \mytexttt{facet}) iterators. We note that for \mytexttt{set}=\mytexttt{facet}, \mytexttt{elmat} and \mytexttt{elvec} dummy arguments are actually 4-rank and 2-rank assumed shape arrays to store all combinations of facet-wise arrays.},label={lst:fe_assembly}]{fe_assembly.f90}

Finally, the \mytexttt{compress\_storage} deferred \ac{TBP} of \mytexttt{assembler\_t} lets \texttt{fe\_affine\_operator\_t} to signal that the build up process of the linear algebra
data structures has already finished and that they can already be ``compressed'' into its final stage. 

We stress that the software architecture presented in this section provides uniform assembly interfaces to the client that are completely independent of the underlying implementation of linear algebra data structures. The subclasses of \mytexttt{assembler\_t} are in charge of the management of blocks (if any), whereas \mytexttt{sparse\_matrix\_t} is in charge of the management of the storage schemes.

\subsection{Discrete integration of \ac{FE} operators} \label{sec:disc_int}

In this section, we introduce the abstract data type \mytexttt{discrete\_integration\_t} (see Listing~\ref{lst:discrete_integration}). It defines the generic \mytexttt{integrate} binding, which is overloaded
by the \mytexttt{integrate\_galerkin} and \mytexttt{integrate\_petrov\_galerkin} deferred \acp{TBP}, depending on the number of \mytexttt{fe\_space\_t} instances being passed to them
(see, e.g.,  Line~\ref{loc:integrate_signature} of Listing~\ref{lst:discrete_integration} for the interface corresponding to the Galerkin case). A user that wants to implement a \ac{FE} problem must extend this data type and overwrite the \ac{TBP} to be used (Galerkin or Petrov-Galerkin) in the user-defined subclass. In the overridden method, the user must implement the evaluation of the entries of $\fematrix$ and  $\ferhs$ as the numerical integration of the discrete bilinear and linear forms as in \Eq{linear_system} (see Sect.~\ref{sec:fe_meth}).

Based on our experience, the integration part of a \ac{FE} code must exhibit a huge level of flexibility. Every time one wants to consider a new set of \acp{PDE} or a new expression of the discrete bilinear form, this component must be modified. It must also have the ability to integrate general time integration schemes that can require functions in an arbitrary number of steps, deal with nonlinear problems that involve the need to evaluating \ac{FE} functions in the integration of the discrete forms, or including variable physical coefficients of body force terms determined through analytical functions.
As a result, any rigidity at this level must be eliminated. Indeed, the \mytexttt{discrete\_integration\_t} abstract data type only forces its subclasses to adhere to the signatures of the deferred
\acp{TBP} overloading \mytexttt{integrate}, and has no member variables that subclasses are forced to handle. 
Using the design previously sketched, the user has absolute flexibility to design its own \mytexttt{discrete\_integration\_t}
subclass, adding the attributes and methods that can be required to integrate and assemble the discrete forms, e.g., by adding an arbitrary number of \mytexttt{fe\_function\_t} and \mytexttt{*\_function\_t}
instances (and corresponding setters to be used at the driver level) that can describe physical properties, previous time step values, the solution at the previous nonlinear iteration, etc.

  The integration of cell-wise terms of the (bi)linear forms is accomplished by traversing through all the cells using a \mytexttt{fe\_cell\_iterator} instance (see Sect.~\ref{subsec:fe_space_conceptual_view}), which has access to 1) all the cell integration data (see Sect.~\ref{sec:cell_integration}) required to compute the local cell contributions in \eqref{eq:assembly} and 2) the local-to-global \ac{DOF} numbering needed for the assembly in the global linear algebra data structures. Analogously, the integration of facet terms, e.g., the ones in \eqref{eq:ip_dg_formulations} for \ac{DG} formulations, requires the use of a \mytexttt{fe\_facet\_iterator\_t} instance to traverse through the facets and integrate the corresponding facet terms. The method \mytexttt{integrate} is called during the execution of the \mytexttt{numerical\_setup} \ac{TBP} of \mytexttt{fe\_affine\_operator\_t}. It is in fact the \mytexttt{fe\_affine\_operator\_t} the one that decides whether to invoke the Galerkin or Petrov-Galerkin integration, depending on whether one or two \ac{FE} spaces have been passed as actual arguments (the second one being optional) in its \mytexttt{create} binding (see Line~\ref{loc:fe_affine_operator_create} of Listing~\ref{lst:fe_affine_operator}). Analogously, the \ac{FE} space(s) are also passed as actual argument(s) to the \mytexttt{integrate\_*} bindings, since they will be needed at any integration step (see Line \ref{loc:integrate_signature} of Listing~\ref{lst:discrete_integration} for the Galerkin case).
\lstinputlisting[float=htbp,language={[03]Fortran},escapechar=@,caption=The abstract data type \mytexttt{discrete\_integration\_t} and its deferred \acp{TBP}.,label={lst:discrete_integration}]{discrete_integration.f90}

For illustration purposes, we present in Listing~\ref{lst:stokes_discrete_integration} an example extension of \mytexttt{discrete\_integration\_t}. It shows the implementation of the deferred procedure \mytexttt{integrate\_galerkin} for the approximation of the Stokes problem using a Galerkin method. This data types will be used in the example driver presented in Sect.~\ref{sec:driver} for the inf-sup stable Taylor-Hood mixed \ac{FE} method (see Listing~\ref{lst:stokes_discrete_integration}).\footnote{We note that the Stokes subclass of \mytexttt{discrete\_integration\_t} in Listing~\ref{lst:stokes_discrete_integration} implements the Galerkin approximation for this problem but it is independent of the \ac{FE} space being used. It can be re-used for any conforming inf-sup stable mixed \ac{FE} method, e.g., Taylor-Hood, conformal Crouzeix-Raviart, MINI element, etc. The choice of the mixed \ac{FE} space will be determined by the user in the driver, when building the Cartesian two-field \ac{FE} space.}
\lstinputlisting[float=tbph,language={[03]Fortran},escapechar=@,caption=The implementation of a binding that overrrides the \mytexttt{integration\_galerkin} \ac{TBP} of \texttt{discrete\_integration\_t} 
for the Galerkin approximation to the Stokes problem.,label={lst:stokes_discrete_integration}]{stokes_discrete_integration.f90}

As commented above, the integration of the (bi)linear forms requires the cell integration machinery, which is provided by \mytexttt{fe\_space\_t} through the creation of the \mytexttt{fe\_cell\_iterator\_t} in Line \ref{loc:fe_space_create_fe_iterator} of Listing~\ref{lst:stokes_discrete_integration}. Apart from controlling the loop over cells (Lines~\ref{loc:fe_iterator_to_loop_1} and \ref{loc:fe_iterator_to_loop_2}),  \mytexttt{fe\_cell\_iterator\_t} provides the numerical quadrature, which is in turn required to get the number of integration points (line \ref{loc:quad_num_points}), and its associated weights (line \ref{loc:quad_weight}). It also provides the determinant of the Jacobian of the cell map (line \ref{loc:quad_weight}), and the shape functions and gradients at Lines \ref{loc:integrator_get_shape_1} to \ref{loc:integrator_get_shape_2} (see \eqref{eq:shfunx} and  \eqref{shape_grad}). The implementation of the (bi)linear forms is very close to the blackboard expression, making it compact, simple, and intuitive. This is possible through the definition of the \mytexttt{vector\_field\_t}, and \mytexttt{tensor\_field\_t} data types, together with their corresponding expression syntax available in \FEMPAR{}. As it was carefully discussed in Sect.~\ref{sec:cell_integration_example}, it is achieved using operator overloading for different vector and tensor operations, e.g., the contraction and scaling operations. The \mytexttt{symmetric\_part} (used at Lines \ref{loc:symmetric_part_macro_1} and  \ref{loc:symmetric_part_macro_2}), \mytexttt{double\_contract} (used at line \ref{loc:double_contract_macro}) and \mytexttt{trace} helper stand-alone functions (used at Lines \ref{loc:trace_macro_1} and  \ref{loc:trace_macro_2}) are also offered to make tensor operations easy. We also note that this implementation is also \emph{efficient}, since all these operations are made without any dynamic memory allocation/deallocation.

Finally, the \mytexttt{fe\_cell\_iterator\_t} also offers a \ac{TBP} to assemble the element matrix and vector into the \mytexttt{assembler} and to impose strong Dirichlet conditions (line \ref{loc:assembly_elmat_elvec}) using the perturbation in \eqref{eq:dirichlet_bcs_rhs} (see \eqref{subsec:fe_space_strong_boundary_conditions}). The Dirichlet data is extracted from a \mytexttt{fe\_function\_t} that represents $E_h u_{{\rm D}}$, which must be an attribute of the concrete \mytexttt{discrete\_integration\_t}. For non-conforming \ac{FE} spaces, the formulation requires also a loop over the facets to integrate \ac{DG} terms. It can be written in a similar fashion using the tools described in Sect.~\ref{sec:face_integration}. In this example, the \mytexttt{stokes\_galerkin\_integration\_t} extension has the attribute $\mytexttt{force}$, which is used in Line \ref{loc:stokes_force} to integrate the right-hand side. It is a vector field described by an instance of the  \mytexttt{vector\_function\_t} data type.

\subsection{The \ac{FE} affine operator abstraction} \label{sec:fe_affine_operator}
A (simplified) declaration of the \mytexttt{fe\_affine\_operator\_t} data type is shown in Listing~\ref{lst:fe_affine_operator}.
 The \mytexttt{fe\_affine\_operator\_t} is created from a single \mytexttt{fe\_space\_t} instance, or even two for Petrov-Galerkin formulations; the second instance is optional and, when it is not passed, the Galerkin method is used, i.e., the same \ac{FE} space is used for trial and test spaces. The user can (optionally) configure a desired block layout. Given a Cartesian product \ac{FE} space $ \trialsp^1_h \times \ldots \times \trialsp^{n_{\rm field}}_h$ for a multi-field problem with $n_{\rm field}$ fields (see Sect.~\ref{subsec:cartesian_product_FE_space}), the block layout represents a partition of fields into subsets.\footnote{The actual ordering of the fields in the Cartesian \ac{FE} space is determined by the user in the creation of the multi-field \ac{FE} space, which must be consistent with the implementation of the discrete weak form. See, e.g., the creation of the mixed Taylor-Hood \ac{FE} space in Lines~\ref{loc:veloc_reference_fe}-\ref{loc:press_reference_fe} of Listing~\ref{lst:setup_reference_fes}, where the first field is the velocity field and the second one is the pressure field, and the integration of the weak form, e.g., in Lines~\ref{loc:loop_test1}, \ref{loc:loop_test2}, and \ref{loc:loop_test3} of Listing~\ref{lst:stokes_discrete_integration}, where this numbering is respected.} It is described through the argument array \mytexttt{field\_blocks} of size \mytexttt{num\_fields} equal to $n_{\rm field}$, which indicates the block to which each field is assigned; by default, the one-block case is used. E.g., for the Stokes problem in Example \ref{stokes_problem}, one can consider a monolithic block layout with only one block that includes both the velocity and pressure field (\mytexttt{field\_blocks}=[1,1]), or two one-field blocks (\mytexttt{field\_blocks}=[1,2] or [2,1]). Additionally, the user must provide additional information about the diagonal blocks, namely 1) whether the block is symmetric or not, 2) whether symmetric storage wants to be used for the block or not, and 3) whether the block is positive definite, semi-positive definite, or indefinite. The user can optionally provide the array of logicals \mytexttt{field\_coupling} (of size \mytexttt{num\_fields} $\times$ \mytexttt{num\_fields}); the position $(\mytexttt{i},\mytexttt{j})$  determines whether the matrix entries related to trial/test functions of the \ac{FE} space $\mytexttt{i}$ and \ac{FE} space $\mytexttt{j}$ are always zero (in this case, the coupling is false) or not. For the Stokes problem and the Galerkin method, the only entry that is false (no coupling) is the pressure-pressure entry. When this array is not provided, the case by default is that all fields are coupled. It only implies more memory consumption, e.g., to store the zero entries in the pressure-pressure block for the Stokes problem.

\lstinputlisting[float=htbp,language={[03]Fortran},escapechar=@,caption=The \mytexttt{fe\_affine\_operator\_t} data type.,label={lst:fe_affine_operator}]{fe_affine_operator.f90}

The block layout information is stored in the data type \mytexttt{block\_layout\_t}, sketched in Listing~\ref{lst:block_layout}, which stores the arrays \mytexttt{field\_blocks} and \mytexttt{field\_coupling}. It is created in the binding that creates the \mytexttt{fe\_affine\_operator\_t}. It also stores a block-wise \ac{DOF} numbering generated by the \mytexttt{fe\_space\_t} instance, which  is instructed to do so by passing the  \mytexttt{block\_layout\_t}\footnote{The block-wise numbering creates independently the \ac{DOF} numbering of every block. Thus, \acp{DOF} of different blocks can have the same block-wise \ac{DOF} label.} when calling its \ac{TBP} \mytexttt{generate\_global\_dof\_numbering}, described in Sect.~\ref{subsec:global_dof_numbering}. 
  \lstinputlisting[float=htbp,language={[03]Fortran},escapechar=@,caption=The \mytexttt{block\_layout\_t} data type.,label={lst:block_layout}]{block_layout.f90} 

  The \mytexttt{fe\_affine\_operator\_t} also holds a polymorphic pointer to an \mytexttt{assembler\_t} instance.  Its dynamic type is selected during the creation phase depending on the number of blocks, the storage layout required, and the (parallel or serial) environment. Finally, a polymorphic pointer to an instance of declared type \mytexttt{discrete\_integration\_t} is also stored (see line \ref{loc:discrete_integration_pointer} of Listing~\ref{lst:fe_affine_operator}). After the creation phase, the \mytexttt{fe\_affine\_operator\_t} is ready for its setup. Thanks to the design of the linear algebra data structures in \FEMPAR{}, it does not require a symbolic setup, i.e., to precompute a (potential) sparsity pattern. The \mytexttt{numerical\_setup} \ac{TBP} at line \ref{loc:fe_affine_operator_setup} of Listing~\ref{lst:fe_affine_operator} calls the  \mytexttt{integrate\_galerkin} \ac{TBP} of \mytexttt{discrete\_integration}  when the pointer to \mytexttt{trial\_fe\_space} is not associated or \mytexttt{integrate\_petrov\_galerkin} otherwise, as discussed in Sect.~\ref{sec:disc_int}.

\section{Driver example for the Stokes problem} \label{sec:driver}

In this section, we describe the software architecture of a driver program that approximates the solution of the Stokes problem. To this end, it implements a Galerkin \ac{FE} method grounded on a ``static'' (i.e., non-adaptable) conforming mesh and inf-sup stable \ac{FE} spaces. In particular, we consider a conforming \ac{FE} space $\boldsymbol{\mathcal{V}}_h \times \mathcal{Q}_h $, where $\boldsymbol{\mathcal{V}}_h$ is a grad-conforming Lagrangian space of order $k+1$, and $\mathcal{Q}_h$, a grad-conforming Lagrangian space of order $k$, i.e., the mixed Taylor-Hood \ac{FE} \cite{brezzi_mixed_1991}.\footnote{The pressure field belongs to $L^2(\Omega)$. Thus, a discontinuous pressure \ac{FE} space could have been also considered as well. 
It would still be $L^2(\Omega)$-conforming. This is the case of, e.g., the conformal Crouzeix-Raviart mixed \ac{FE}.}

It is up to \FEMPAR{} users to decide how to design the software architecture of their main driver program.  Any driver program has nevertheless to follow the typical stages needed in a simulation pipeline based on \acp{FE}. In the seek of uniformity, the architecture presented in Listing~\ref{lst:main_program} and~\ref{lst:stokes_driver} is recommended to \FEMPAR{} users. The main program unit relies on a number of driver-level module units, which are not part of the \FEMPAR{} library but developed by the user specifically for the problem at hand. Each of these modules defines a driver-level derived data type and its \acp{TBP}. A central derived data type, called \mytexttt{stokes\_driver\_t} in this example, is designed to drive all the necessary steps. In particular, it offers a public \ac{TBP}, called \mytexttt{run\_simulation}, on which the driver program relies to perform the actual simulation. The driver program is therefore as simple and concise as shown in Listing~\ref{lst:main_program}. 

\lstinputlisting[float=htbp,language={[03]Fortran},escapechar=@,caption=The main program for the solution of the Stokes problem.,label={lst:main_program}]{main_program.f90}

%

\lstinputlisting[float=htbp,language={[03]Fortran},escapechar=@,caption=The main data type of the Stokes driver.,label={lst:stokes_driver}]{stokes_driver.f90}

The main data type of the driver, \mytexttt{stokes\_driver\_t}, is shown in Listing~\ref{lst:stokes_driver}. It is equipped with a set of member variables of type already described in previous sections;
see comments on the right-hand side of each member variable. The data type \mytexttt{solver\_t} in Line \ref{loc:solver_t} does not exist in \FEMPAR{} as such.  There is actually a complete set of data types
that provide interfaces to high-end third party sparse direct solvers. Besides, we have developed our own abstract implementation of iterative linear solvers (including, e.g., the conjugate gradient or GMRES Krylov subspace solvers). The convergence of these solvers can be accelerated using advanced preconditioners grounded on the Multilevel Balancing Domain Decomposition by Constraints (MLBDDC) preconditioner~\cite{badia_scalability_2015,badia_multilevel_2016}. The description of the linear solvers software subsystem deserves considerable space and is postponed to a future work. In this example, it has to be understood as a data type that provides the necessary services required to implement the \mytexttt{solve\_system} \ac{TBP} at Line~\ref{loc:solve_system} of Listing~\ref{lst:stokes_driver}. The data type \mytexttt{stokes\_conditions\_t} at Line~\ref{loc:stokes_conditions} extends \mytexttt{conditions\_t} in Sect.~\ref{subsec:fe_space_strong_boundary_conditions}. It encodes the strong Dirichlet boundary conditions data for this particular operator. The member variable \mytexttt{parameter\_list} (see Line \ref{loc:parameter_list}) is a parameter dictionary of <{\em key},{\em value}> pairs. Its implementation is provided as a stand-alone external software library called \mytexttt{FPL} \cite{FPL}. The member variable \mytexttt{stokes\_parameters} (see Line \ref{loc:stokes_parameters}) is a user-defined data type that encapsulates the interaction with a command line parser provided by the \mytexttt{FLAP} software package \cite{FLAP}. Both of them are used to implement the \ac{TBP} in Line~\ref{loc:parse_command_line_parameters}, which parses the arguments given by the user in the command line, and transfers them into the aforementioned \mytexttt{parameter\_list} member variable.

The  \mytexttt{run\_simulation} \ac{TBP} (called from the main program in Line \ref{loc:driver_run_simulation} of Listing~\ref{lst:main_program}) is implemented with the help of the private 
\acp{TBP} in Lines \ref{loc:setup_triangulation}-\ref{loc:write_solution} of Listing~\ref{lst:stokes_driver}. The \mytexttt{setup\_triangulation} \ac{TBP} invokes the \mytexttt{create} \ac{TBP} 
of \mytexttt{static\_triangulation\_t}. Depending on the command-line parameter values, the user may select to automatically generate a structured/uniform 
triangulation for simple domains (e.g., a unit cube), 
currently of brick (quadrilateral or hexahedral) cells, or read it from a mesh data file,
e.g., using the GiD unstructured mesh generator \cite{_gid_2016}.
The \ac{FE} space is built in \mytexttt{setup\_fe\_space} \ac{TBP}, sketched in Listing~\ref{lst:setup_reference_fes}.

\lstinputlisting[float=htbp,language={[03]Fortran},escapechar=@,caption=The implementation of the \mytexttt{setup\_fe\_space} binding for the Stokes problem.,label={lst:setup_reference_fes}]{setup_fe_space.f90}

An array with base type \mytexttt{p\_reference\_fe\_t}, a data type that wraps a polymorphic pointer to a \mytexttt{reference\_fe\_t} instance, is allocated in Line \ref{loc:reference_fes_alloc} of Listing~\ref{lst:setup_reference_fes}.
The \mytexttt{reference\_fe\_t} instances for the velocity and pressure fields are created by calling \texttt{make\_reference\_fe} in Lines~\ref{loc:veloc_reference_fe} and~\ref{loc:press_reference_fe}, respectively; see Sect.~\ref{subsec:reference_fe_creation}. The interpolation order of the numerical scheme is read from command-line in  Line \ref{loc:interpolation_order}. We select \texttt{order} equal to $k+1$ and $k$ in Lines~\ref{loc:veloc_reference_fe} and~\ref{loc:press_reference_fe}, respectively. The dummy argument \mytexttt{continuity} determines whether $\trialsp$ admits a trace operator. In this particular
example, we could consider  \mytexttt{continuity=.false.} if we wanted to use a discontinuous pressure space.
The  \mytexttt{create} \ac{TBP} of \mytexttt{fe\_space\_t} (Line~\ref{loc:create_fe_space}) performs the composition of the reference \acp{FE} to build the Cartesian product space $\trialsp_h$.
Finally, we call the \mytexttt{set\_up\_cell\_integration} \ac{TBP} of \mytexttt{fe\_space\_t} in Line~\ref{loc:setup_fe_space_cell_integration} to set up all the data structures required to evaluate cell integrals in Listing~\ref{lst:discrete_integration}.

The implementation of the \mytexttt{setup\_fe\_affine\_operator} binding is shown in Listing~\ref{lst:setup_fe_affine_operator}. It first invokes the  \mytexttt{create} \ac{TBP} of \mytexttt{fe\_affine\_operator\_t} in Line~\ref{loc:setup_affine_create}. We state monolithic storage for the global coefficient matrix (Line~\ref{loc:setup_affine_field_blocks}), that it is symmetric (Line~\ref{loc:setup_affine_symmetric}), that we want symmetric storage, i.e., to only store its upper triangle (Line~\ref{loc:setup_affine_symmetric_storage}), and the fact that it is indefinite (Line~\ref{loc:setup_affine_sign}). The definition of \mytexttt{field\_coupling} in Line~\ref{loc:setup_affine_field_coupling} reflects that the pressure diagonal block is null.  We also pass an instance of \mytexttt{fe\_space\_t} in Line~\ref{loc:setup_affine_fe_space} and an instance of the subclass \mytexttt{stokes\_integration\_t} in Line~\ref{loc:setup_affine_discrete_integration}.

\lstinputlisting[float=htbp,language={[03]Fortran},escapechar=@,caption=The implementation of the \mytexttt{setup\_fe\_affine\_operator} binding for the Stokes problem.,
label={lst:setup_fe_affine_operator}]{setup_fe_affine_operator.f90}

Before we set up the operator in Line~\ref{loc:setup_affine_setup}, we create a \mytexttt{fe\_function\_t} instance in Line~\ref{loc:setup_affine_create_fe_function}. In Line~\ref{loc:setup_affine_interpolate}, by means of the services provided by \mytexttt{fe\_space\_t}, we interpolate the analytical function to be prescribed on the boundary for the velocity field (retrieved from \mytexttt{stokes\_conditions}), fixing the strong Dirichlet \acp{DOF} of the \mytexttt{fe\_function\_t} instance at hand. As a result, this FE function represents $E_h u_{\rm D}$, with the zero extension to free \acp{DOF}; see Sect. \ref{subsec:fe_space_strong_boundary_conditions}. This \ac{FE} function is passed to the \mytexttt{stokes\_integration\_t} instance in Line~\ref{loc:setup_affine_pass_fe_function}. Finally, we trigger the operator auto-construction in Line~\ref{loc:setup_affine_setup}.

The \mytexttt{solve\_system} \ac{TBP} (see Line~\ref{loc:solve_system} of Listing~\ref{lst:stokes_driver}) invokes either a direct or preconditioned iterative solver to obtain the free \acp{DOF} nodal values of our \ac{FE} function (see Sect.~\ref{subsec:fe_function}).
Provided that \texttt{this\%solution} on input to \mytexttt{solve\_system}  is such that it vanishes on free \acp{DOF} (see discussion in previous paragraph), a common
practice used in \FEMPAR{} drivers to save space is to re-use the space devoted for free \acp{DOF} in \texttt{this\%solution} to store the free \acp{DOF} nodal values of the solution of the problem at hand. We stress that all solvers in \FEMPAR{} are such that they only solve for free \acp{DOF}. In our experience, this decision dramatically simplifies the development of some preconditioners, provided that they can be developed without taking care of strong Dirichlet boundary conditions.

Finally, the \mytexttt{write\_solution} \ac{TBP} (see Line~\ref{loc:write_solution} of Listing~\ref{lst:stokes_driver}) is in charge of the generation of simulation results in data files 
for later visualization using, e.g., VisIt~\cite{HPV:VisIt} or Paraview~\cite{Ayachit:2015:PGP:2789330}. To this end,  \mytexttt{write\_solution} relies on a format independent, 
extensible abstraction, referred to as \texttt{output\_handler\_t}. It lets the user to register an arbitrary number of FE functions (together with the corresponding FE space these functions were generated from) and cell data arrays (e.g., material properties or error estimator indicators), to be output in the appropriate format for later visualization. Among its responsibilities, this (abstract) data type generates the data to be written to the (potentially parallel-distributed) file system in neutral, cell-oriented data structures, dealing with (potentially) non-conforming (discontinuous), and variable degree FE spaces among cells. The user may also select to apply a differential operator to the FE function, such as divergence, gradient or curl, which involve further calculations to be performed on each cell, or to customize those cells to be output (e.g., only those that belong to the interior of the geometry in unfitted \ac{FE} simulations) via their own implementation of cell iterators. 

The generation of the actual data files in the appropriate format is in charge of the implementations (extensions) of \texttt{output\_handler\_t}. \FEMPAR{} currently offers two implementations of \texttt{output\_handler\_t} (although many others could be implemented as well by the growing community of \FEMPAR{} developers given the extensible software architecture designed). \texttt{vtk\_output\_handler\_t} lets the user to generate their data in the standard-open model VTK~\cite{Schroeder:1998:VTO:272980}. It currently relies on \mytexttt{Lib\_VTK\_IO}~\cite{vtk_io}, which (by now) does not actually exploit parallel \ac{MPI} I/O but instead uses a naive single file per \ac{MPI} task scheme.  \texttt{vtk\_output\_handler\_t} is therefore the recommended option for serial computations or parallel computations on a moderate number of processors. The second one, \texttt{xh5\_output\_handler\_t}, lets the user generate their data in XDMF~\cite{xdmf}. XDMF separates the description of the raw data, referred to as ``light data'', from the data itself, referred to as ``heavy data''.  The light data is expressed using a set of XML-based constructs that are suited to represent the distributed-memory data structures in \FEMPAR{}. XDMF in turn supports the heavy data to be stored using HDF5~\cite{hdf5}. HDF5 is, among others, a data model and file format designed with the parallel I/O data challenge in mind. By means of a set of supporting open source libraries, referred to as parallel HDF5 libraries, \FEMPAR{} takes advantage of the underlying distributed file system without having to deal with the high complexity of other lower-level implementations, such as raw \ac{MPI} I/O. In particular, the latter service is provided by \mytexttt{XH5For}~\cite{xh5for}, a stand-alone software library, which we developed from scratch, and lets the user to read/write parallel partitioned FEM meshes taking advantage of the Collective/Independent \ac{MPI}-IO provided by the PHDF5 library for the efficient generation of the vast amount of data typically resulting from a large-scale scientific computing simulation. 

\section{Conclusions}

In this work, we have thoroughly described the approach that we have followed in \FEMPAR{} in order to abstract in software the numerical approximation of problems governed by \acp{PDE} using \ac{FE} methods.
The mathematical framework of \acp{FE} has been split into a number of (mathematically motivated) derived data types and their interaction, resulting into a well-separated, robust, and stable set of customizable software abstractions for the development of widely applicable \ac{FE} solvers. These tools equip \FEMPAR{} users with the machinery needed to perform all the steps in the simulation pipeline, including mesh import/generation, \acp{DOF} enumeration, evaluation/assembly of the algebraic system of linear equations via FE integration, solution of the linear system, and output of computational results in the appropriate format for later visualization. In order to achieve this goal, the software architecture of \FEMPAR{} has been thoroughly designed by means of advanced \ac{OO} software re-engineering techniques (including the recurrent application of \ac{OO} design patterns~\cite{gamma_e._design_1995,freeman_head_2004}) in order to increase its ease of use, extensibility, flexibility, and reusability. \FEMPAR{} software architecture has been implemented using the latest \ac{OO} features of the Fortran03/08 standard, namely, {\em information hiding and data encapsulation}, {\em inheritance via type extension}, and {\em dynamic run-time polymorphism}.  This version of the Fortran standard is already widely (and robustly) supported by most of the compilers typically available on high-end computing environments.  A judiciously set of programming techniques let us achieve a reasonable trade-off among extensibility and performance, while avoiding  in most cases the computational overheads frequently associated with abstract \ac{OO} software libraries. 

The software abstractions covered in this work include:
\begin{itemize}
  \item The definition of reference \ac{FE}s, which relies on the concept of polytopes to define the cell topology in arbitrary dimensions, a machinery to define multi-dimensional polynomial functions of arbitrary order in an easy and automatic way, and a general procedure for the generation of the shape function bases and local \ac{DOF}s.
  \item The global \ac{FE} space abstraction, which relies on reference \ac{FE}(s) and a triangulation of the physical domain. It is responsible to define the local-to-global \ac{DOF} numbering, which must respect conformity (if needed). The \ac{FE} space also provides tools for the numerical integration of (bi)linear forms, e.g., mappings from the reference to the physical space, etc., in cells and facets (for \ac{DG} methods).
  \item The \ac{FE} affine operator generated after the discretization of the original problem (probably after a linearization step). The \ac{FE} solution is the only root (as soon as the problem is well-posed) of this operator. This operator, once the trial and test functions and the discrete (bi)linear forms of the problem at hand are defined, is represented through a matrix and a vector whose entries can be computed by numerical integration using the \ac{FE} space.
\end{itemize}

\FEMPAR{} has been used for more than 4 years now by a team of about 10 researchers of different research institutions and universities. During the initial \ac{OO} re-design, derived data types (attributes and bindings) were gradually modified to accommodate new features that had not been considered, to fix expressivity limitations or even dependency knots of the original design. The software architecture to which we have converged, although certainly subject to future change, has been already proven to be capable to satisfy a number of users' software requirements, even when the application problems involved complex and advanced features (e.g., the development of growing geometries in 3D printing technology). 
 We consider that this steady regime, which has been attained after years of development, and a tremendous man-month power effort, is the proof that the software abstraction in \FEMPAR{} is of practical relevance not only for prospective users and developers, but also for researchers that want to learn about the \ac{OO} implementation of \ac{FE} methods. It has motivated the decision of the authors to promote the library as a community software project, to open it to external users and new collaborators, to publish the library in an public git repository \cite{FEMPAR}, and to write this article. In particular, the architecture described here corresponds to the first public release of \FEMPAR{}, to which we assigned the git tag \mytexttt{FEMPAR-1.0.0}.

The first public release of \FEMPAR{} has almost 300K lines of (mostly) Fortran code. Thus, a document like this one, with a quite detailed description of the services provided by the library and the motivation underlying our software design, can be a very valuable resource to complement the source code, which can become overwhelming in itself. In this paper, we have restricted ourselves to the construction of \ac{FE} operators for body-fitted \ac{FE} spaces. However, a major (and unique compared to other \ac{FE} scientific software packages available on the Internet) cornerstone of \FEMPAR{} is an abstract \ac{OO} framework for the implementation of widely applicable highly scalable multilevel \ac{DD} solvers.\footnote{Indeed, the multilevel \ac{DD} solvers within \FEMPAR{} are since 2014 in the High-Q club of the most scalable European codes, maintained by the J\"ulich supercomputing center~\cite{brommel_juqueen_2015}.} By letting this framework to be highly coupled with the numerical integration data structures of the application, on the one hand, and to be highly customizable, on the other, one can derive optimal preconditioners for the particular structure of the discrete operator at hand, and tackle new problems and challenges, while leveraging the distributed-memory implementation ideas~\cite{badia_multilevel_2016} on which the framework is grounded on. Customizable building blocks in the framework include the fine-grid to coarse-grid \acp{DOF} aggregation, the constraint matrix underlying the imposition of continuity of coarse \acp{DOF} functionals across coarse objects, the weighting operator underlying the injection among the continuous and discontinuous spaces, and the kind of solvers to be used for the Dirichlet, Neumann constrained local problems, and the coarsest-grid global problem~\cite{dohrmann_preconditioner_2003}. However, we postpone the discussion about solvers, preconditioners, data structures suitable for parallel distributed-memory computers, and other more exotic discretization techniques in \FEMPAR{}, like B-splines and XFEM methods, to subsequent works.

\section*{Compliance with Ethical Standards}
This study was funded by the European Research Council through the Starting Grant No. 258443 - \emph{COMFUS: Computational Methods for Fusion Technology} under the  the FP7 Program and the two related Proof of Concept Grant No. 640957 - \emph{FEXFEM: On a free open source extreme scale finite element software} and Proof of Concept Grant No. 737439 - \emph{NuWaSim: On a Nuclear Waste Deep Repository Simulator} under the H2020 Program. The authors declare that they have no conflict of interest.

\bibliographystyle{myabbrvnat}
\bibliography{art022}
\end{document}